\newcounter{myctr}

\documentclass{ws-acs}
\usepackage{url}

\usepackage{mathrsfs}
\usepackage[bookmarks=true,colorlinks,linkcolor=blue,anchorcolor=blue,citecolor=blue,unicode]{hyperref}

\begin{document}

\makeatletter
\def\@biblabel#1{[#1]}
\makeatother

\markboth{J.-A. Li, L. Wang, W.-J. Xie and W.-X. Zhou}{Evolving community structure in the international pesticide trade networks}

%
\catchline{}{}{}{}{}
%

\title{Evolving community structure in the international pesticide trade networks}

\author{\footnotesize Jian-An Li}

\address{School of Business, East China University of Science and Technology,\\ Shanghai 200237, China}

\author{Li Wang}

\address{School of Physics, East China University of Science and Technology,\\ 
Research Center for Econophysics, East China University of Science and Technology,\\
Shanghai 200237, China}

\author{Wen-Jie Xie}

\address{School of Business, East China University of Science and Technology,\\ 
Research Center for Econophysics, East China University of Science and Technology,\\
Shanghai 200237, China \\
wjxie@ecust.edu.cn}

\author{Wei-Xing Zhou}

\address{School of Business, East China University of Science and Technology,\\ 
Research Center for Econophysics, East China University of Science and Technology,\\
School of Mathematics, East China University of Science and Technology,\\
Shanghai 200237, China\\
wxzhou@ecust.edu.cn}

\maketitle

\begin{history}
\received{(10 July 2022)}
\revised{(revised date)}
\end{history}

\begin{abstract}
The statistical properties including community structure of the international trade networks of all commodities as a whole have been studied extensively. However, the international trade networks of individual commodities often behave differently. Due to the importance of pesticides in agricultural production and food security, we investigate the evolving community structure in the international pesticide trade networks (iPTNs) of five categories from 2007 to 2018. We unveil the community structures in the undirected and directed iPTNs exhibits regional patterns. However, the regional patterns are very different for undirected and directed networks and for different categories of pesticide. Moreover, the community structure is stabler in the directed iPTNs than in the undirected iPTNs. We also extract the intrinsic community blocks for the directed international trade networks of each pesticide category. It is found that the largest intrinsic community block is the stablest that appears in every pesticide category and contains important economies (Belgium, Germany, Spain, France, United Kingdom, Italy, Netherlands, and Portugal) in Europe. Other important and stable intrinsic community blocks are Canada and the United States in North America, Argentina and Brazil in South America, and Australia and New Zealand in Oceania. These findings imply the importance of geographic distance and the complementarity of important adjacent economies in the international trade of pesticides.
\end{abstract}


\section{Introduction}
\label{S1:Introduction}

Food crises have accompanied human history. According to the fifth edition of the Global Report on Food Crises released by the Food Security Information Network of the United Nations, there were around 155 million people in food crisis or worse in 55 countries/territories globally in 2020\footnote{https://www.wfp.org/publications/global-report-food-crises-2021, accessed 10 July 2022.}. Conflicts, climate disruption and economic shocks are main reasons causing food crises, which have been aggravated by the COVID-19 pandemic in recent years. Ensuring and increasing agricultural production are important means to relieve food shortages. In this sense, pesticides, as a main agricultural input, play an important role. Hence, investigating the global pesticide flows from supplying economies to demanding economies from the perspective of international trade networks (ITNs) would deepen our understanding of pesticides reallocation all over the world.

The evolving structure and dynamics of the international trade networks or world trade webs (WTWs) with aggregated goods have been investigated extensively at macro-, meso- and micro-scales, be they undirected or directed, unweighted or weighted \cite{Serrano-Boguna-2003-PhysRevE,Garlaschelli-Loffredo-2005-PhysicaA,Garlaschelli-DiMatteo-Aste-Caldarelli-Loffredo-2007-EurPhysJB,Fagiolo-Reyes-Schiavo-2008-PhysicaA,Fagiolo-Reyes-Schiavo-2009-PhysRevE,Fagiolo-Reyes-Schiavo-2010-JEvolEcon,Squartini-Fagiolo-Garlaschelli-2011a-PhysRevE,Squartini-Fagiolo-Garlaschelli-2011b-PhysRevE,Karpiarz-Fronczak-Fronczak-2014-PhysRevLett,Campi-Duenas-Fagiolo-2020-EnvironResLett}. 
There are quite a few studies focusing on the structure and evolution of international food trade networks of, including aggregate food \cite{Raynolds-2004-WorldDev,ErcseyRavasz-Toroczkai-Lakner-Baranyi-2012-PLoSOne,Suweis-Carr-Maritan-Rinaldo-DOdorico-2015-ProcNatlAcadSciUSA,Marchand-Carr-Dell'Angelo-Fader-Gephart-Kummu-Magliocca-Porkka-Puma-Ratajczak-Rulli-Seekell-Suweis-Tavoni-D'Odorico-2016-EnvironResLett}, individual food commodities \cite{Torreggiani-Mangioni-Puma-Fagiolo-2018-EnvironResLett,Qiang-Niu-Wang-Zhang-Liu-Cheng-2020-Sustainability,Zhang-Zhou-2021-Entropy}, cereals \cite{Dupas-Halloy-Chatzimpiros-2019-PLoSOne},
rice \cite{Zhang-Zhou-2021-Entropy},
maize \cite{Wu-Guclu-2013-RiskAnal,Zhang-Zhou-2021-Entropy},
wheat \cite{Puma-Bose-Chon-Cook-2015-EnvironResLett,Fair-Bauch-Anand-2017-SciRep,Zhang-Zhou-2021-Entropy,GutierrezMoya-AdensoDiaz-Lozano-2021-FoodSecur}, 
soybean \cite{SchafferSmith-Tomscha-Jarvis-Maguire-Treglia-Liu-2018-EcolSoc,Zhang-Zhou-2021-Entropy},
seafood \cite{Gephart-Pace-2015-EnvironResLett,Gephart-Rovenskaya-Dieckmann-Pace-Brannstrom-2016-EnvironResLett,Stoll-Crona-Fabinyi-Farr-2018-FrontMarSci}, and 
meat \cite{Chung-Kapsar-Frank-Liu-2020-SciRep}. The ITNs of agricultural inputs are relatively rare \cite{Li-Xie-Zhou-2021-FrontPhysics}. Different types of models  have been proposed to understand the formation of  the international trade networks \cite{Garlaschelli-Loffredo-2004-PhysRevLett,Bhattacharya-Mukherjee-Saramaki-Kaski-Manna-2008-JStatMech,Fagiolo-2010-JEconInteractCoord,Duenas-Fagiolo-2013-JEconInteractCoord,Mastrandrea-Squartini-Fagiolo-Garlaschelli-2014-PhysRevE,Almog-Bird-Garlaschelli-2019-FrontPhys}. Studies of risk transmission and the Vulnerability of international food trade networks to shocks are also important to food security \cite{Wu-Guclu-2013-RiskAnal,Gephart-Rovenskaya-Dieckmann-Pace-Brannstrom-2016-EnvironResLett,Marchand-Carr-Dell'Angelo-Fader-Gephart-Kummu-Magliocca-Porkka-Puma-Ratajczak-Rulli-Seekell-Suweis-Tavoni-D'Odorico-2016-EnvironResLett,Distefano-Laio-Ridolfi-Schiavo-2018-PLoSOne}.
 
At the meso-scale, community structure in the  international trade networks is of particular interest, since it often relates to the important function of node groups \cite{Fortunato-2010-PhysRep,Malliaros-Vazirgiannis-2013-PhysRep}. Tzekina, Danthi and Rockmore investigated the ITNs of 186 economies from 1948 to 2005 and unveiled  trade ``islands'' which is interpreted as mixed evidence for globalization  \cite{Tzekina-Danthi-Rockmore-2008-EurPhysJB}.
Barigozzi, Fagiolo and Mangioni identified the aggregate and commodity-specific community structures of the directed ITNs of 97 commodities traded among 162 economies from 1992 to 2003 by optimizing the modularity using taboo search and found that they are much more correlated with geographic communities than partitions induced by regional-trade agreements \cite{Barigozzi-Fagiolo-Mangioni-2011-PhysicaA,Barigozzi-Fagiolo-Mangioni-2011}. 
Estrada obtained three communities by using N-ComBa $K$-means method for the undirected international miscellaneous metal manufactures network \cite{Estrada-2011-Chaos}. Fan et al. introduced the weighted extremal optimisation algorithm for community detection in the ITN in 2010 and also identified three communities \cite{Fan-Ren-Cai-Cui-2014-EconModel}. Applying the modularity optimization method to the undirected ITN during the period from 1995 to 2011, Zhu et al. identified three main communities (the America community, the Europe community, and the Asia-Oceania community) \cite{Zhu-Cerina-Chessa-Caldarelli-Riccaboni-2014-PLoSOne}.
Reyes, Wooster and Shirrell detected the community structure of the undirected ITN for every year between 1970 and 2000 and found that regional trade agreements have a strengthening effect over time with cyclical components on the formation of the world trade network community structure \cite{Reyes-Wooster-Shirrell-2014-WorldEcon}. Bartesaghi, Clemente and Grassi used the communicability distance approach to investigate the community structure of the undirected international trade network in 2016 and the optimal partitioning gives 19 nontrivial communities \cite{Bartesaghi-Clemente-Grassi-2022-JEconInteractCoord}. 
Dong et al. investigated the undirected international copper trade networks from 2007 to 2015 and identified three to five communities in different years \cite{Dong-Gao-Sun-Liu-2018-ResourPolicy}. Torreggiani et al. investigated the undirected international food trade networks of 16 commodities from 1992 to 2011 and found about 5-6 communities in each commodity-specific network \cite{Torreggiani-Mangioni-Puma-Fagiolo-2018-EnvironResLett}. 
We note that Piccardi and Tajoli used four approaches (modularity optimization, cluster analysis, stability functions, and persistence probabilities) to analyze communities in the ITNs from 1962 to 2008 and found no agreed evidence of significant partitions but a few weak communities suggesting a truly globalizing trading system \cite{Piccardi-Tajoli-2012-PhysRevE}.

The community structure of a complex network reflects underlying higher-order interactions among different nodes \cite{Grilli-Barabas-MichalskaSmith-Allesina-2017-Nature,Battiston-Amico-Barrat-Bianconi-deArruda-Franceschiello-Iacopini-Kefi-Latora-Moreno-Murray-Peixoto-Vaccarino-Petri-2021-NatPhys}, which shape the dynamics of complex networks. In this work, we aim at identify the evolving community structure of five categories of iPTNs. There are numerous community detection methods \cite{Fortunato-2010-PhysRep,Malliaros-Vazirgiannis-2013-PhysRep}, which often provide distinct partitioning results. Since the iPTNs are directed and weighted, we utilize the Infomap algorithm \cite{Rosvall-Bergstrom-2008-ProcNatlAcadSciUSA}. Certainly, there are other algorithms capable of coping with such graphs, for example, the Order Statistics Local Optimization Method (OSLOM) \cite{Lancichinetti-Radicchi-Ramasco-Fortunato-2011-PLoSOne}. However, there are stable clusters identified in temporal networks that we call intrinsic community blocks. The detected communities are found to have evident regional patterns and vary over different pesticide categories. The largest and stablest intrinsic community block contains a few developed economies in Europe.












\section{Construction of iPTNs}
\label{S1:NetworkConstruction}

\subsection{Data description}

The data sets we analyze were extracted from the UN Comtrade Database (https://comtrade.un.org), under Heading 3808. We consider the international trade flows of five categories of pesticides, including insecticides (380891), fungicides (380892), herbicides (380893),  disinfectants (380894), and rodenticides and other similar products (380899). Since the trade data earlier than 2007 are not available for at least one pesticide category and the data for 2019 are still incomplete \cite{Li-Xie-Zhou-2021-FrontPhysics}, the sample period considered in this work is from 2007 to 2018.

\subsection{Network construction}

Denoting the category of pesticide by $cmd \in$ \{380891, 380892, 380893, 380894, 380899\} and the trade value (in units of US\$) of pesticide $cmd$ exported from economy $i$ to economy $j$ in year $t$ by $w_{ij}^{cmd}(t)$, we express the temporary international pesticide trade network as follows,
\begin{equation}
    {\mathbf{W}}^{cmd}(t) = \left[w_{ij}^{cmd}(t)\right].
    \label{Eq:PesticideNet:cmd:W:t}
\end{equation}
Since national pesticide trade is not consider, we have $w_{ii}^{cmd}(t)=0$ for all pesticide categories in all years and there are no self-loops in the iPTNs. Simple preprocessing of the data applies as in Ref.~\cite{Li-Xie-Zhou-2021-FrontPhysics} that we do not repeat here. The main statistical properties have also been investigated in Ref.~\cite{Li-Xie-Zhou-2021-FrontPhysics}.






\section{Empirical results}
\label{S1:Results}

\subsection{Community structure of the undirected iPTNs}
\label{S2:iPTN:Graph:Community}

For the five undirected and weighted iPTNs, we adopt the two-level Infomap algorithm to identify the community partitioning \cite{Rosvall-Bergstrom-2008-ProcNatlAcadSciUSA}. The Infomap algorithm obtains the community partitioning $\mathbf{C}$ when the expected description length of a random walk reaches its minimum and the per step average description length is given by a map equation:
\begin{equation}
    L({\mathbf{C}}) = L_{\mathcal{Q}}({\mathbf{C}}) + L_{\mathcal{P}}({\mathbf{C}}),
\end{equation}
where $L_{\mathcal{Q}}({\mathbf{C}})$ is the per level codelength for communities, describing the inter-community movements of random walkers, and $L_{\mathcal{P}}({\mathbf{C}})$ is the per level codelength for leaf nodes, describing the intra-community movements of random walkers.




\begin{figure}[!ht]
    \centering
    \includegraphics[width=0.483\linewidth]{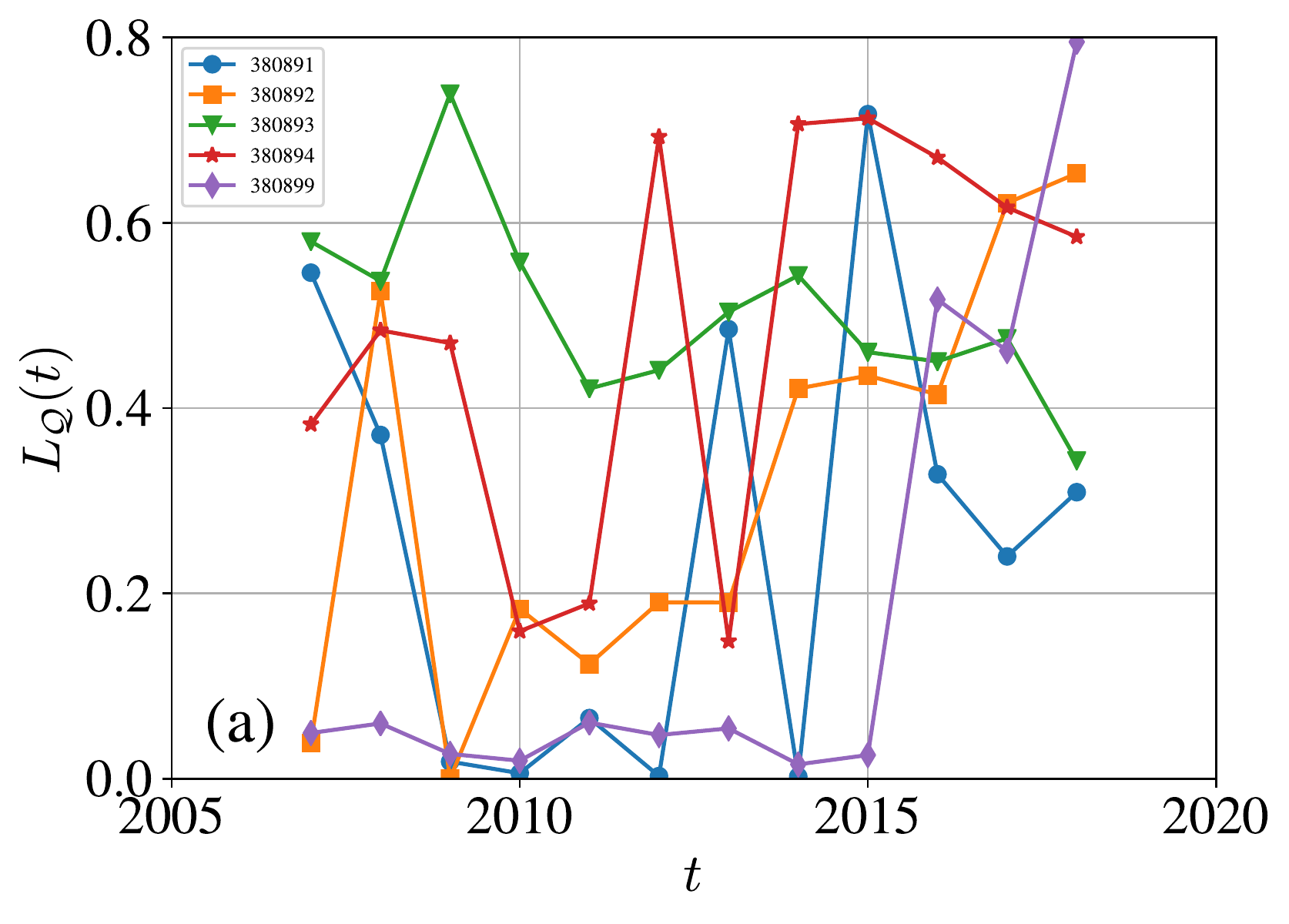}
    \includegraphics[width=0.483\linewidth]{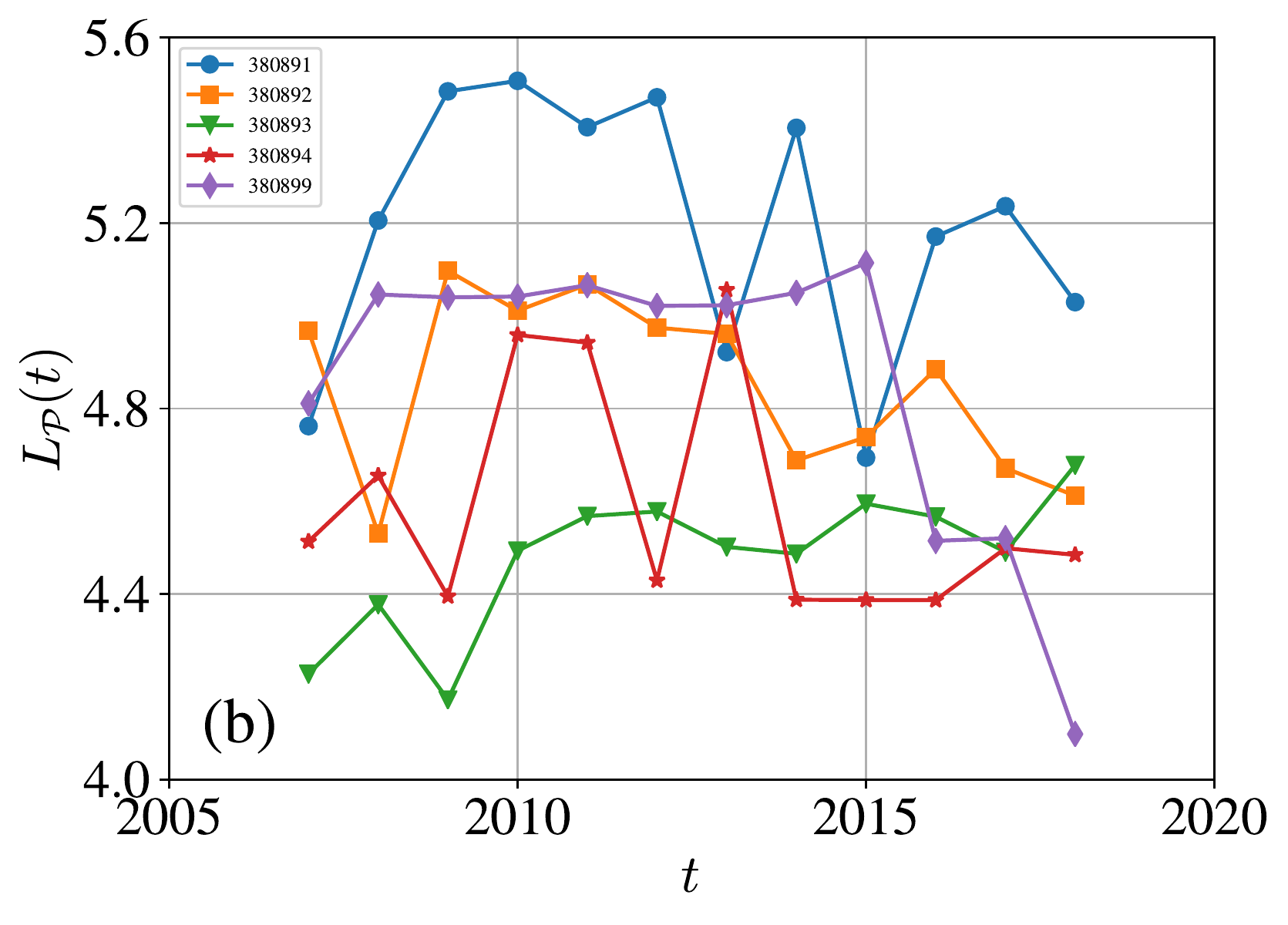}
    \caption{Yearly evolution of the per level codelengths for modules (a) and leaf nodes (b) for the undirected international trade networks of insecticides (380891), fungicides (380892), herbicides (380893),  disinfectants (380894), and rodenticides and other similar products (380899) from 2007 to 2018.}
    \label{Fig:iPTN:undirected:Community:CodeLen:t}
\end{figure}

For each undirected and weighted international pesticide trade network, we obtain the community partitioning ${\mathbf{C}}^{cmd}(t)$. We illustrate the evolution of the per level codelength for communities ($L_{\mathcal{Q}}$) in Fig.~\ref{Fig:iPTN:undirected:Community:CodeLen:t}(a) and the per level codelength for leaf nodes ($L_{\mathcal{P}}$) in Fig.~\ref{Fig:iPTN:undirected:Community:CodeLen:t}(b). Most of the per level codelength curves fluctuate remarkably. It is interesting to see that, for a given category of pesticides, when $L_{\mathcal{Q}}$ increases, $L_{\mathcal{P}}$ decreases, and vice versa. Actually, the $L(t)$ curves fluctuate much less and do not show evident trends.

\begin{figure}[!ht]
    \centering
    \includegraphics[width=0.783\linewidth]{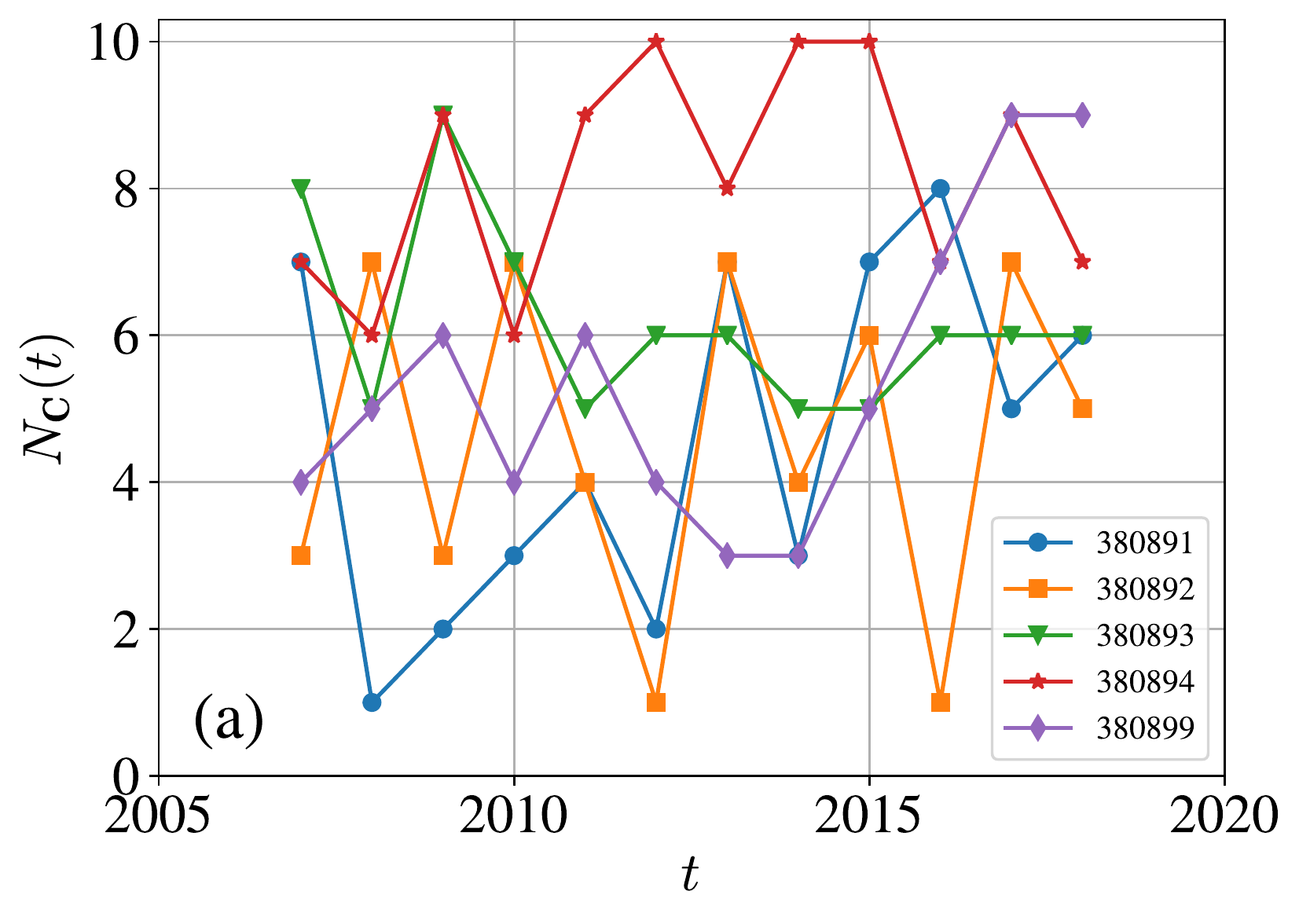}
    \caption{Evolution of the number of communities for the undirected international trade networks of insecticides (380891),  fungicides (380892), herbicides (380893), disinfectants (380894), and rodenticides and other similar products (380899) from 2007 to 2018.}
    \label{Fig:iPTN:undirected:Community:num:t}
\end{figure}

In Fig.~\ref{Fig:iPTN:undirected:Community:num:t}, we show the evolution of the number $N_{\mathbf{C}}$ of communities for the five categories of iPTNs. It is found that the number of identified communities changes over different years. There are three networks with only one community, that is, insecticides in 2008, fungicides in 2012, and fungicides in 2016. There are also two networks (insecticides in 2009 and 2012) having two communities. All other networks have at least three communities. The international trade networks of disinfectants have relatively more communities than other pesticides, with the largest number 10 of communities. We also find that the curves of the networks of herbicides and disinfectants are relatively stable compared to other pesticides.

The most popular measure for the goodness of community partitioning of a network is the modularity \cite{Newman-Girvan-2004-PhysRevE,Fortunato-2010-PhysRep}. 
The modularity of an undirected but weighted network is 
\begin{equation}
    Q = \frac{1}{2w}\sum_i\sum_j\left(w_{ij}-\frac{w_iw_j}{2w}\right)\delta(C_i,C_j)
\end{equation}
where $\delta(C_i, C_j)$ is the Kronecker delta function such that
\begin{equation}
    \delta(C_i, C_j) = 
    \begin{cases}
      1, & \mathrm{if} ~ C_i=C_j\\
      0, & \mathrm{otherwise}
    \end{cases},
\end{equation}
$w_i$ is the strength of node $i$
\begin{equation}
    w_i = \sum_j w_{ij},
\end{equation}
and $2w$ is the total strength
\begin{equation}
    W = \sum_iw_i = \sum_i\sum_j w_{ij}.
\end{equation}

\begin{figure}[!ht]
    \centering
    \includegraphics[width=0.783\linewidth]{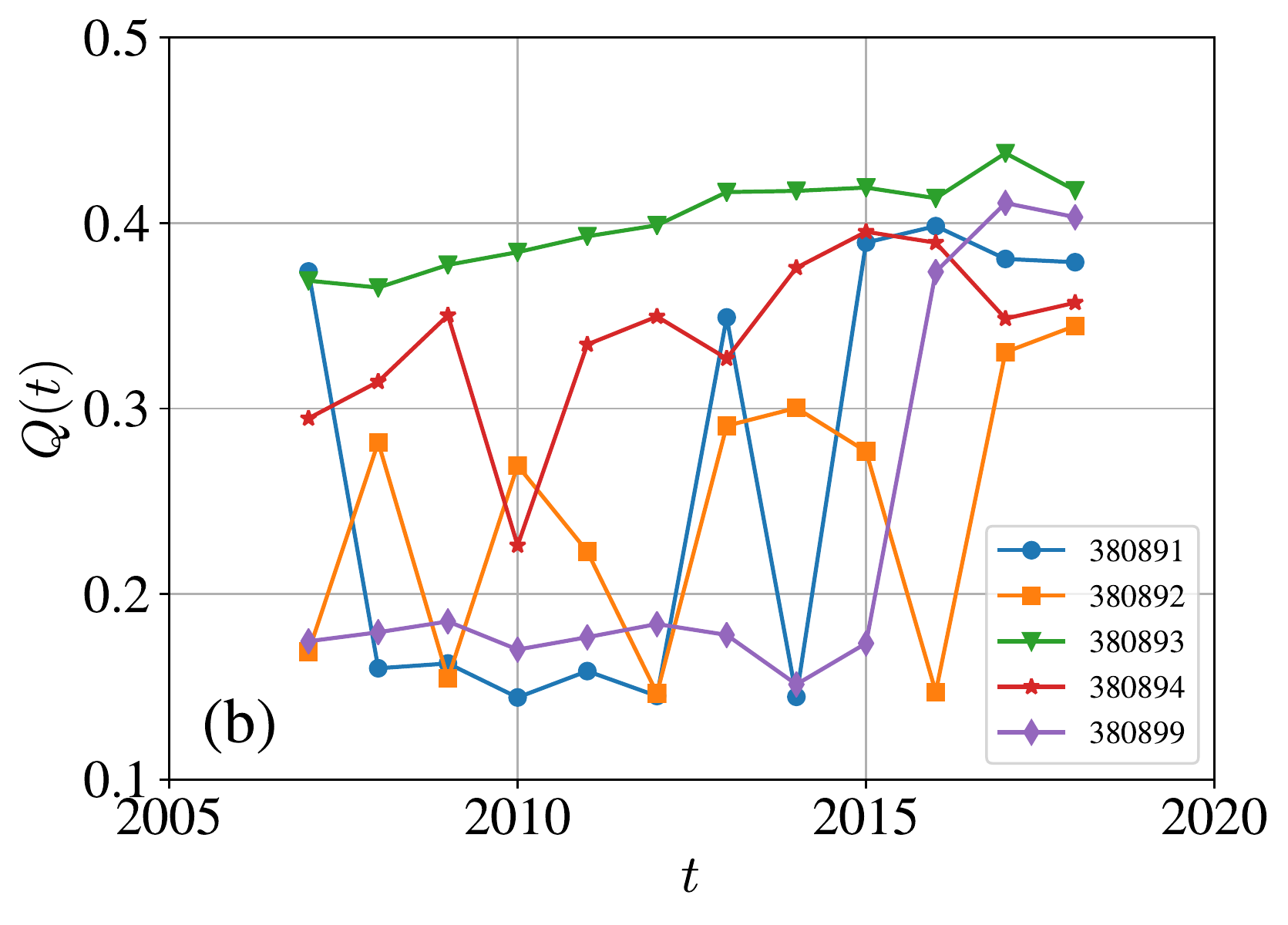}
    \caption{Evolution of the modularity for the undirected international trade networks of insecticides (380891), fungicides (380892), herbicides (380893),  disinfectants (380894), and rodenticides and other similar products (380899) from 2007 to 2018.}
    \label{Fig:iPTN:undirected:Community:modularity:t}
\end{figure}

Figure~\ref{Fig:iPTN:undirected:Community:modularity:t} illustrates the evolution of modularity for the undirected iPTNs from 2007 to 2018. We find that the modularity curve of herbicides (380893) has the least fluctuations and largest values and shows a stably increasing trend. The modularity curve of disinfectants (380894) has also relatively large values and mild fluctuations and shows roughly an increasing trend. Correspondingly, their curves of community number are relatively stable. The modularity curve of rodenticides and other similar products (380899) is stable before 2015 and increases sharply since 2016. The modularities of the three network with only one community are all very low.

\begin{figure}[!ht]
    \centering
    \includegraphics[width=0.321\linewidth]{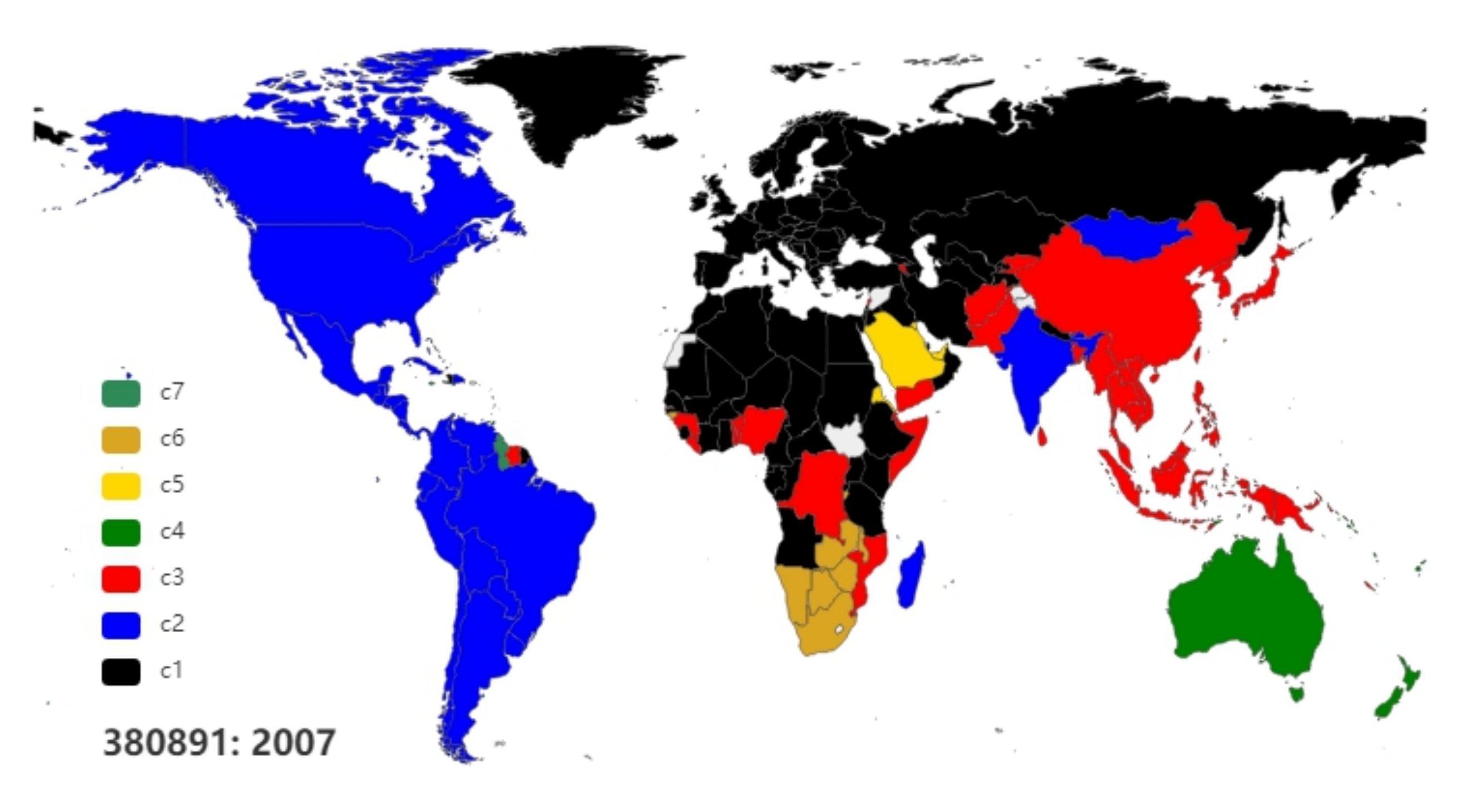}
    \includegraphics[width=0.321\linewidth]{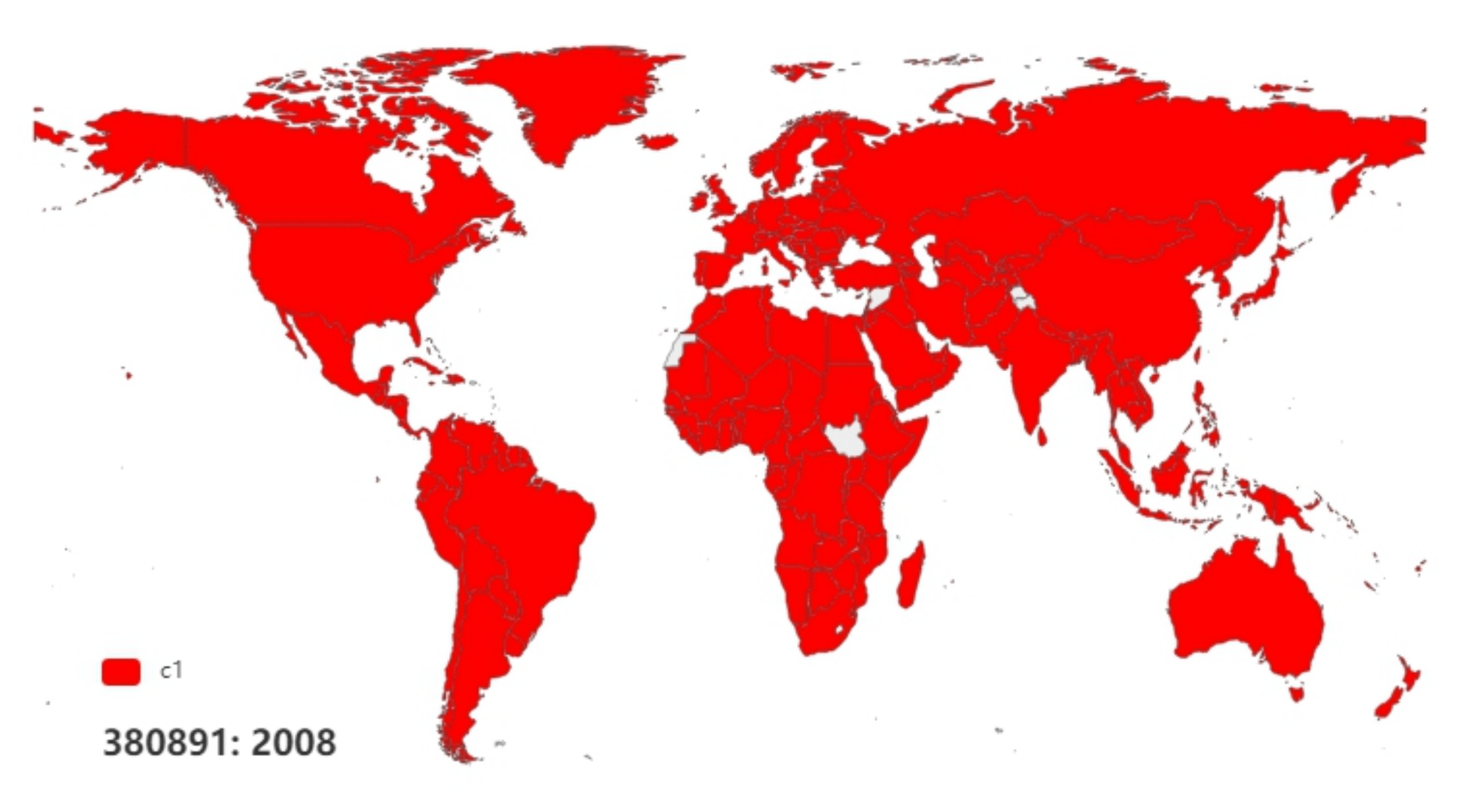}
    \includegraphics[width=0.321\linewidth]{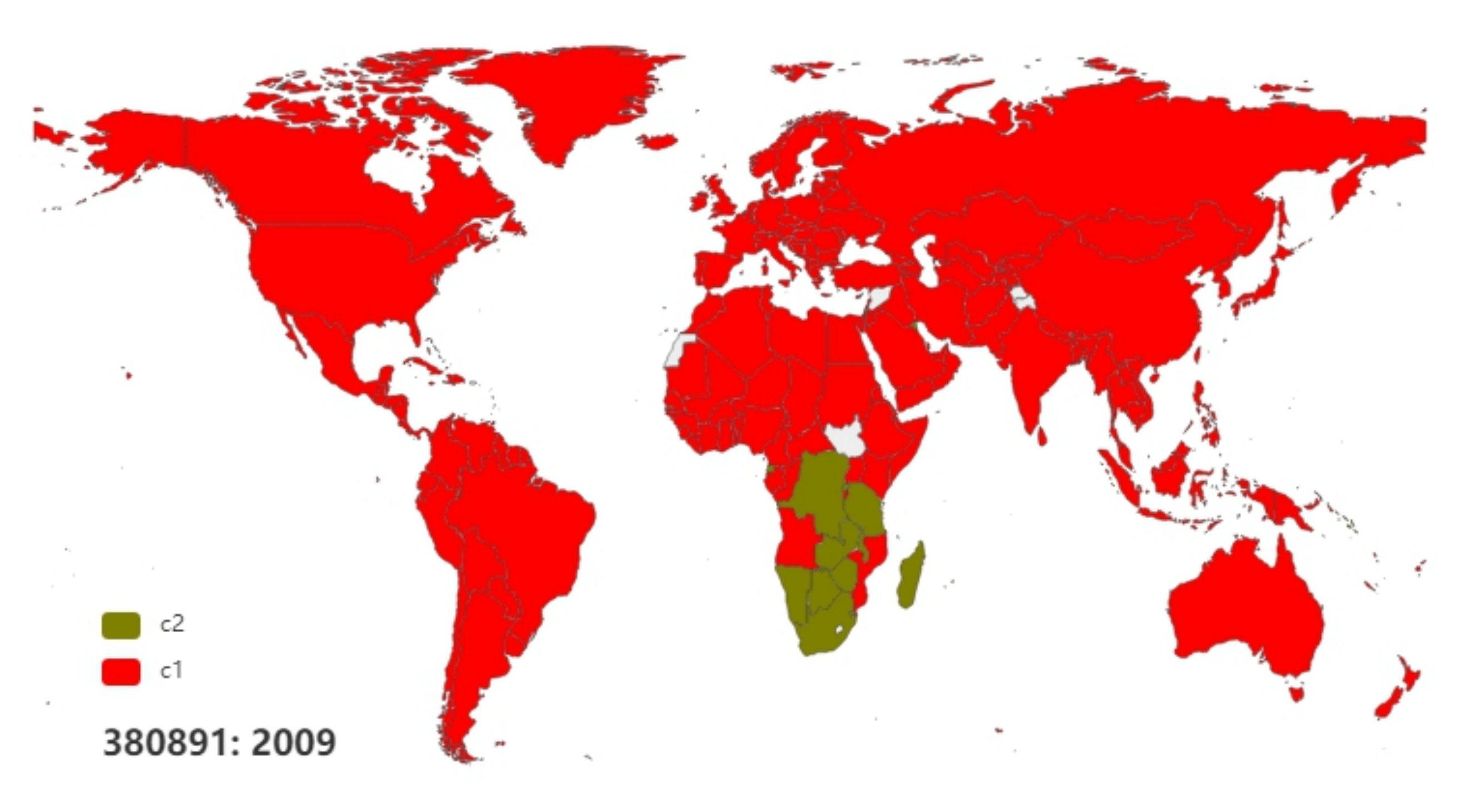}
    \includegraphics[width=0.321\linewidth]{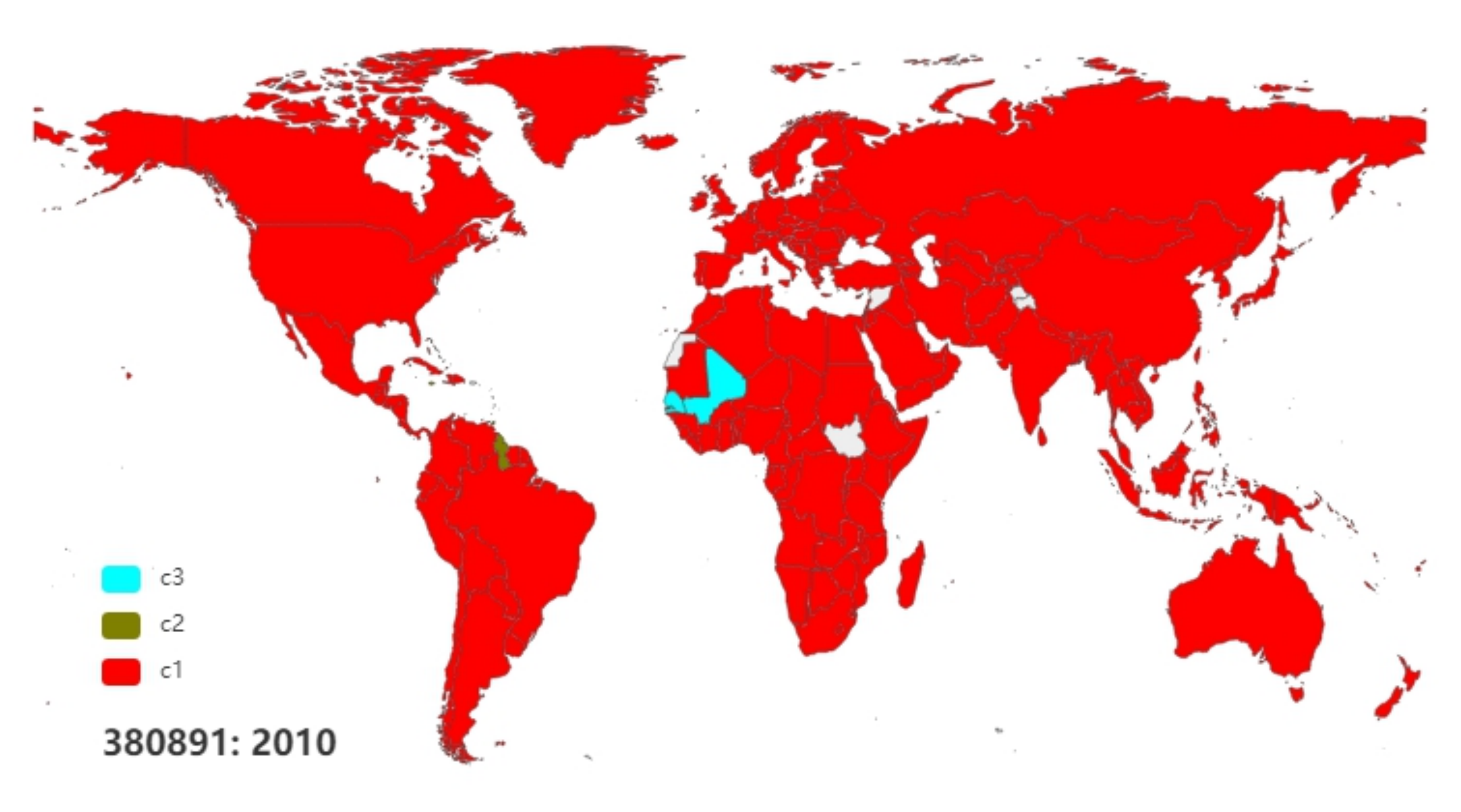}
    \includegraphics[width=0.321\linewidth]{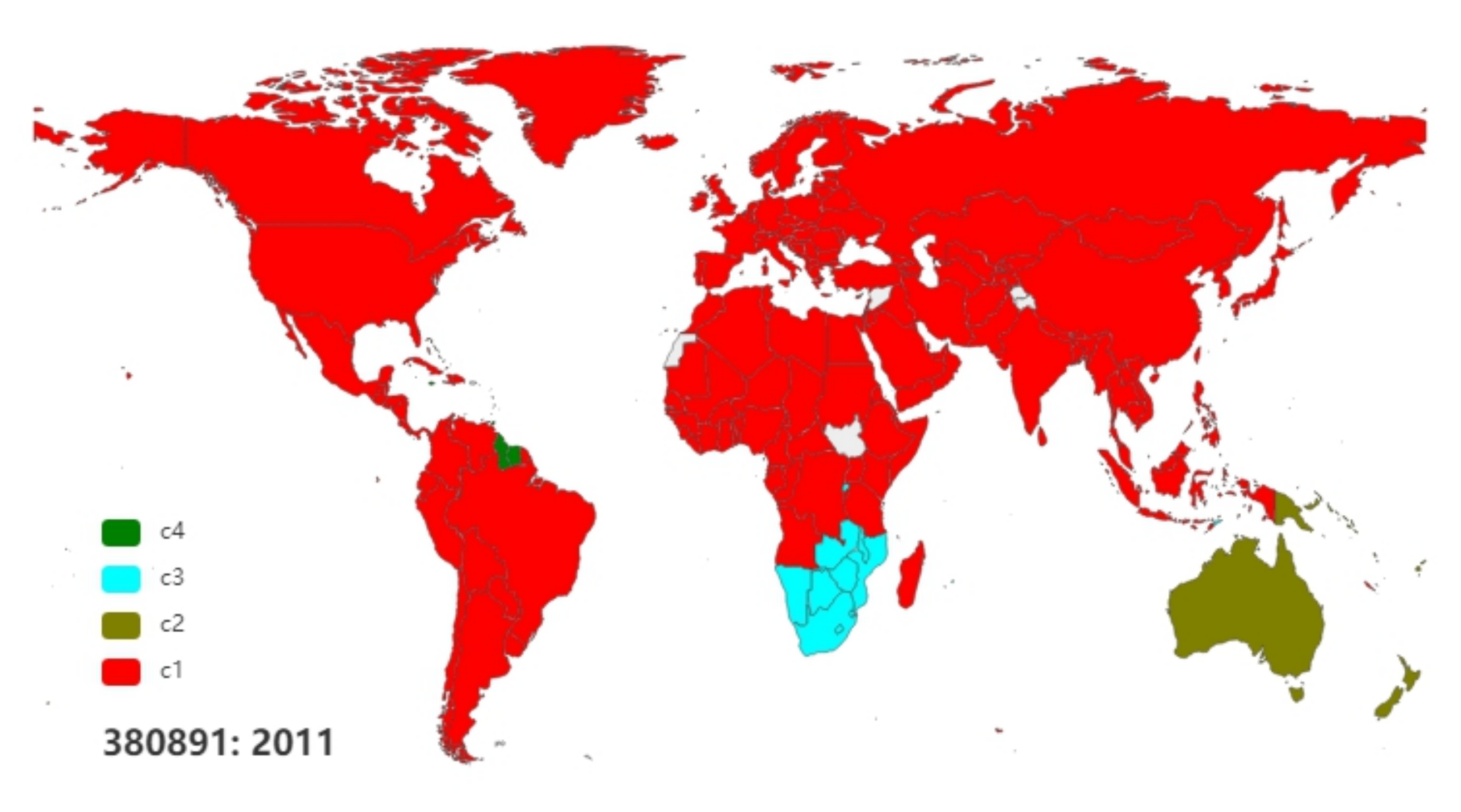}
    \includegraphics[width=0.321\linewidth]{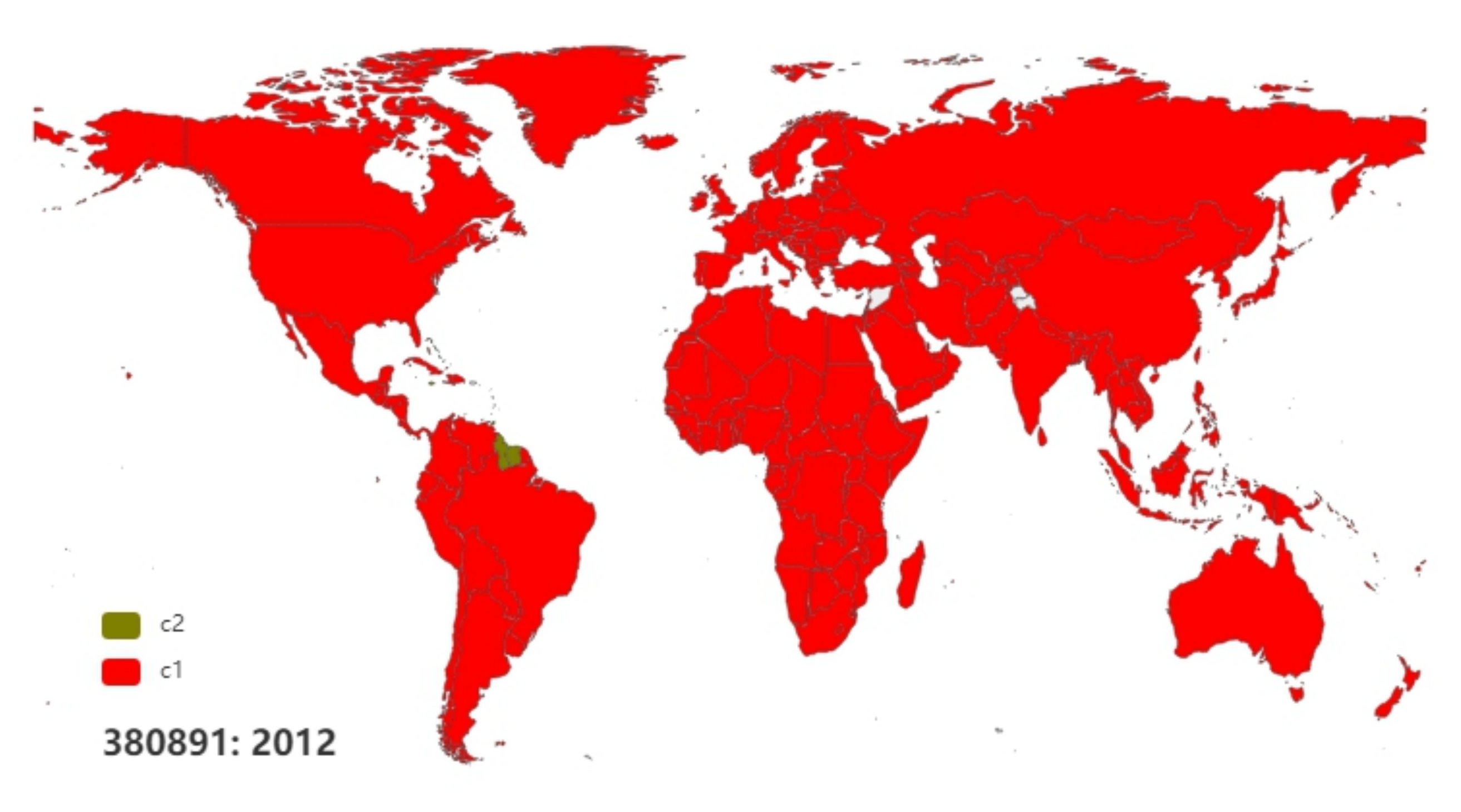}
    \includegraphics[width=0.321\linewidth]{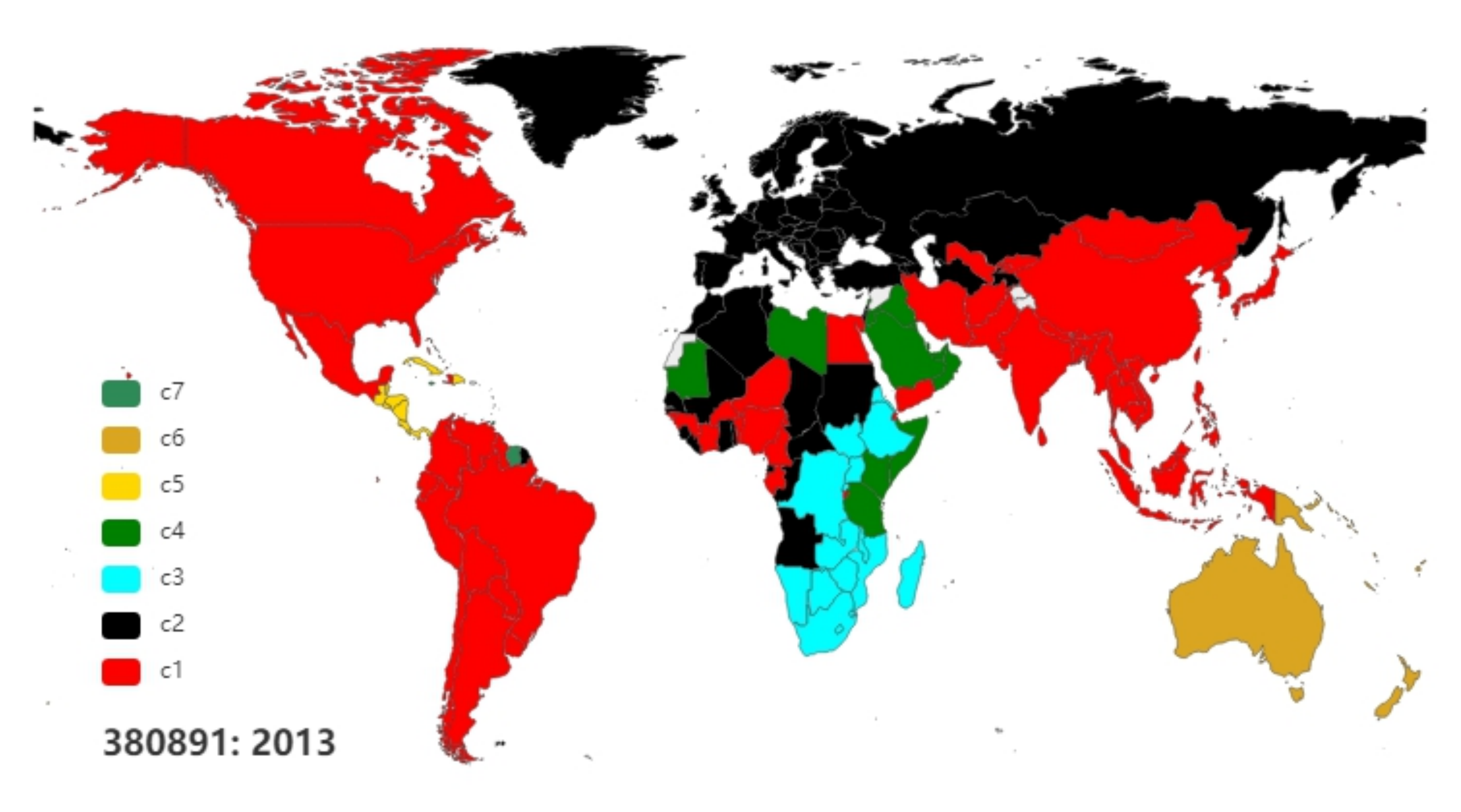}
    \includegraphics[width=0.321\linewidth]{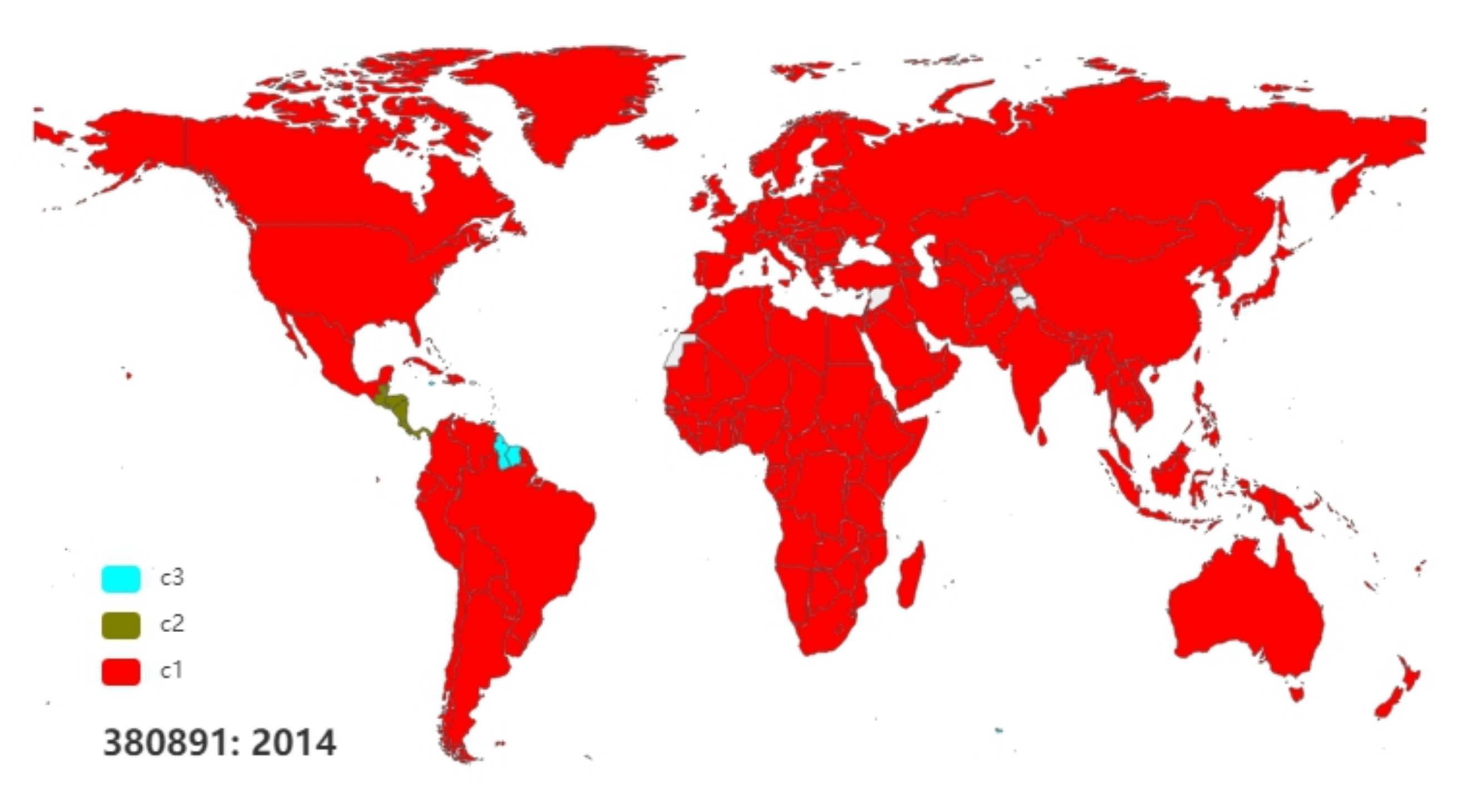}
    \includegraphics[width=0.321\linewidth]{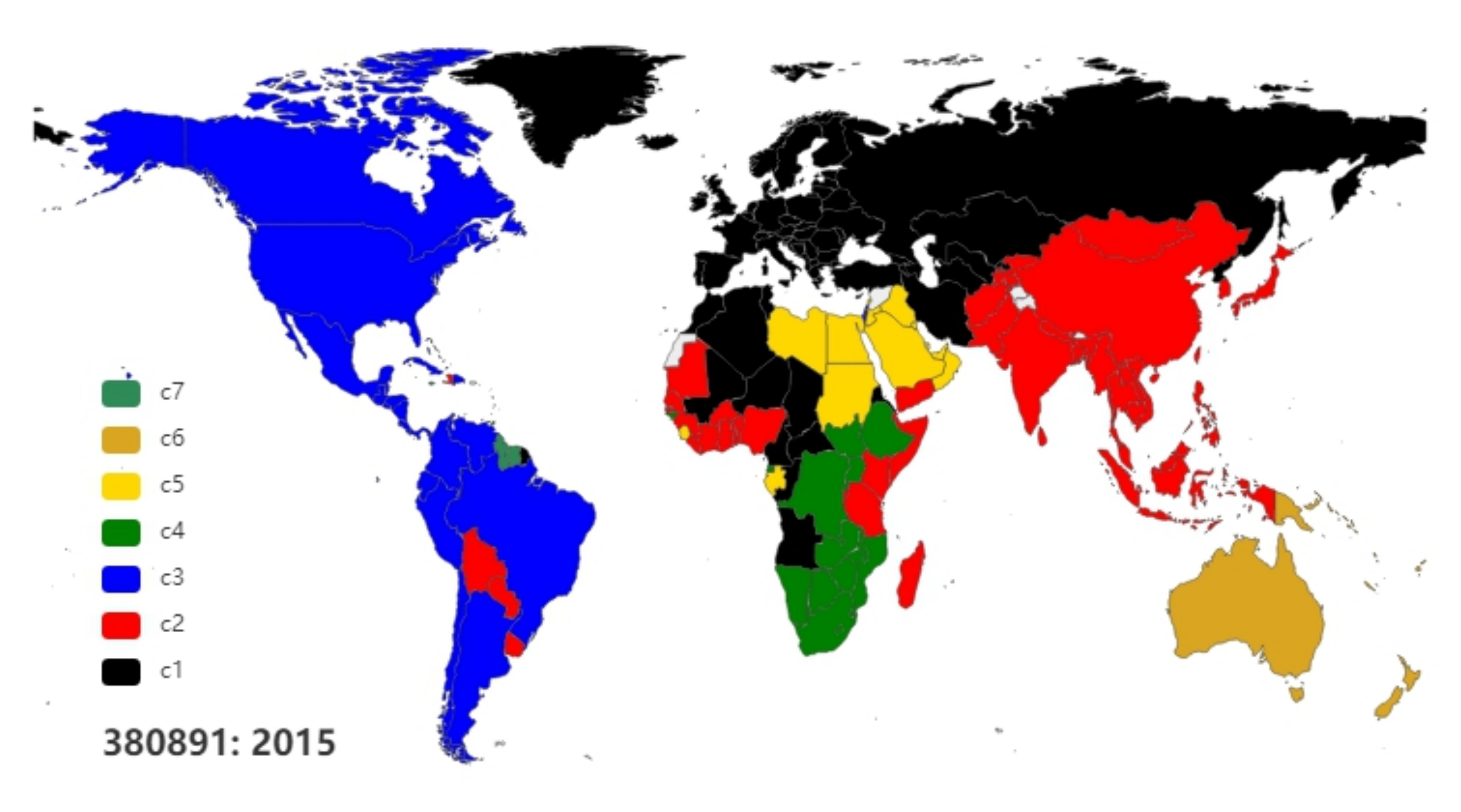}
    \includegraphics[width=0.321\linewidth]{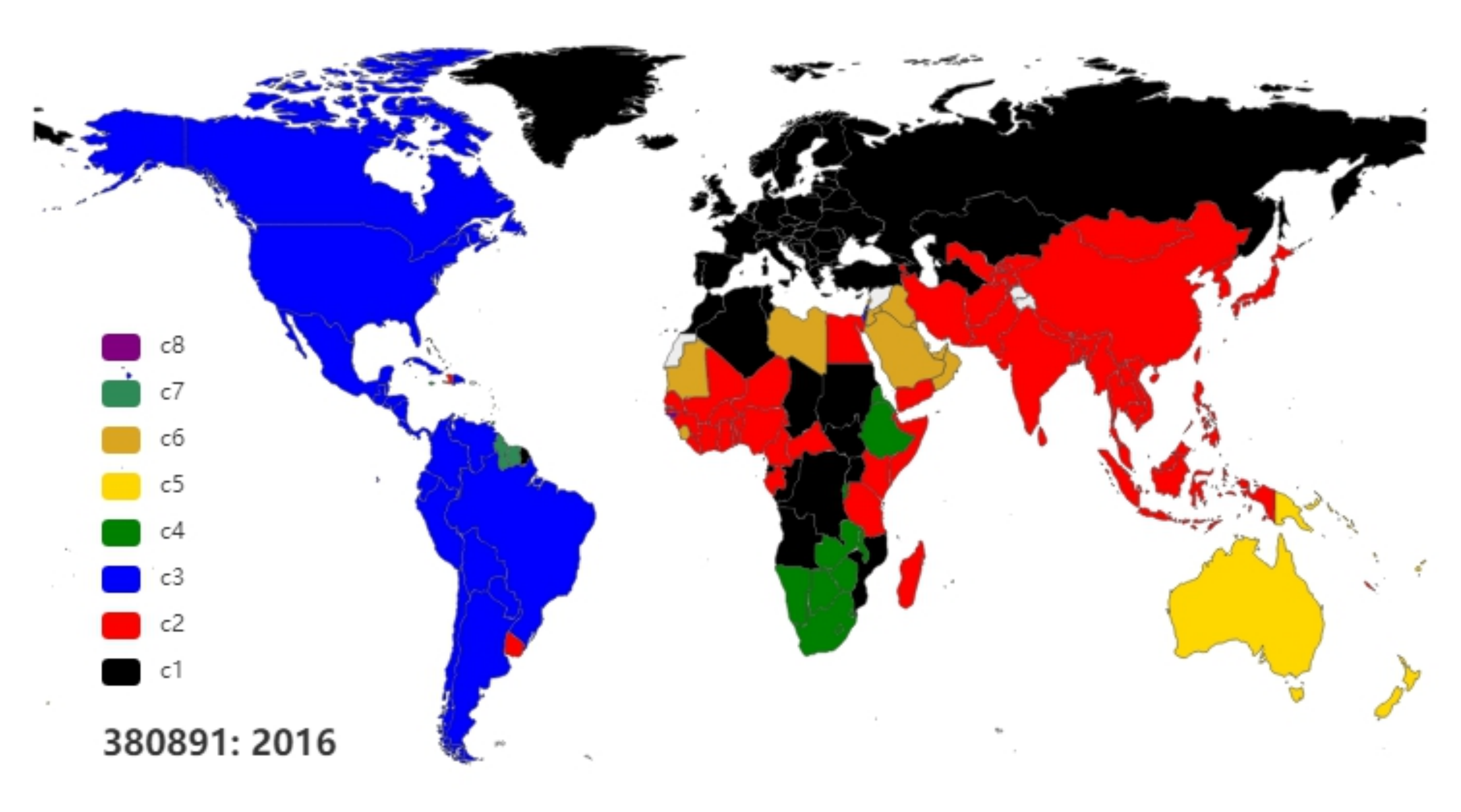}
    \includegraphics[width=0.321\linewidth]{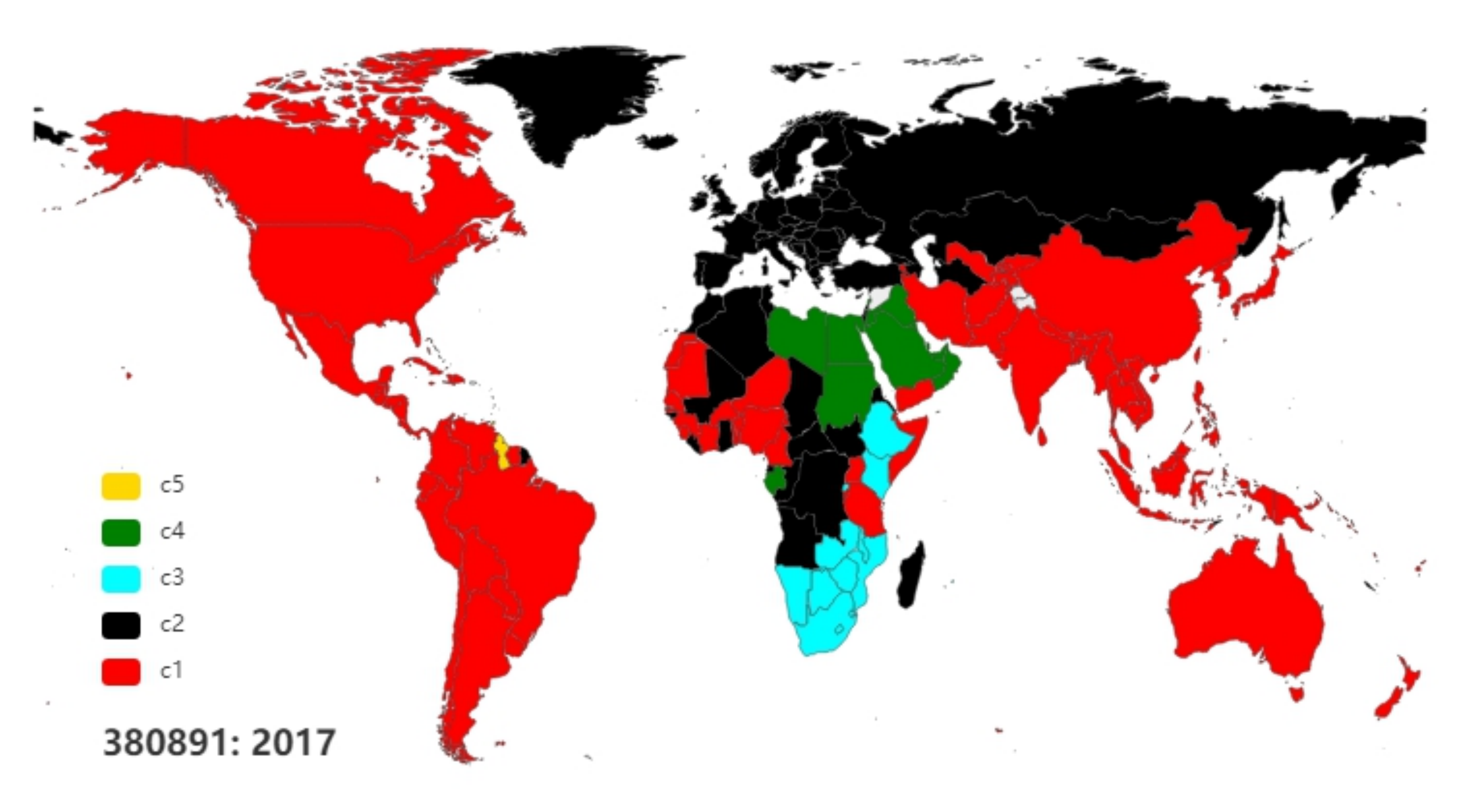}
    \includegraphics[width=0.321\linewidth]{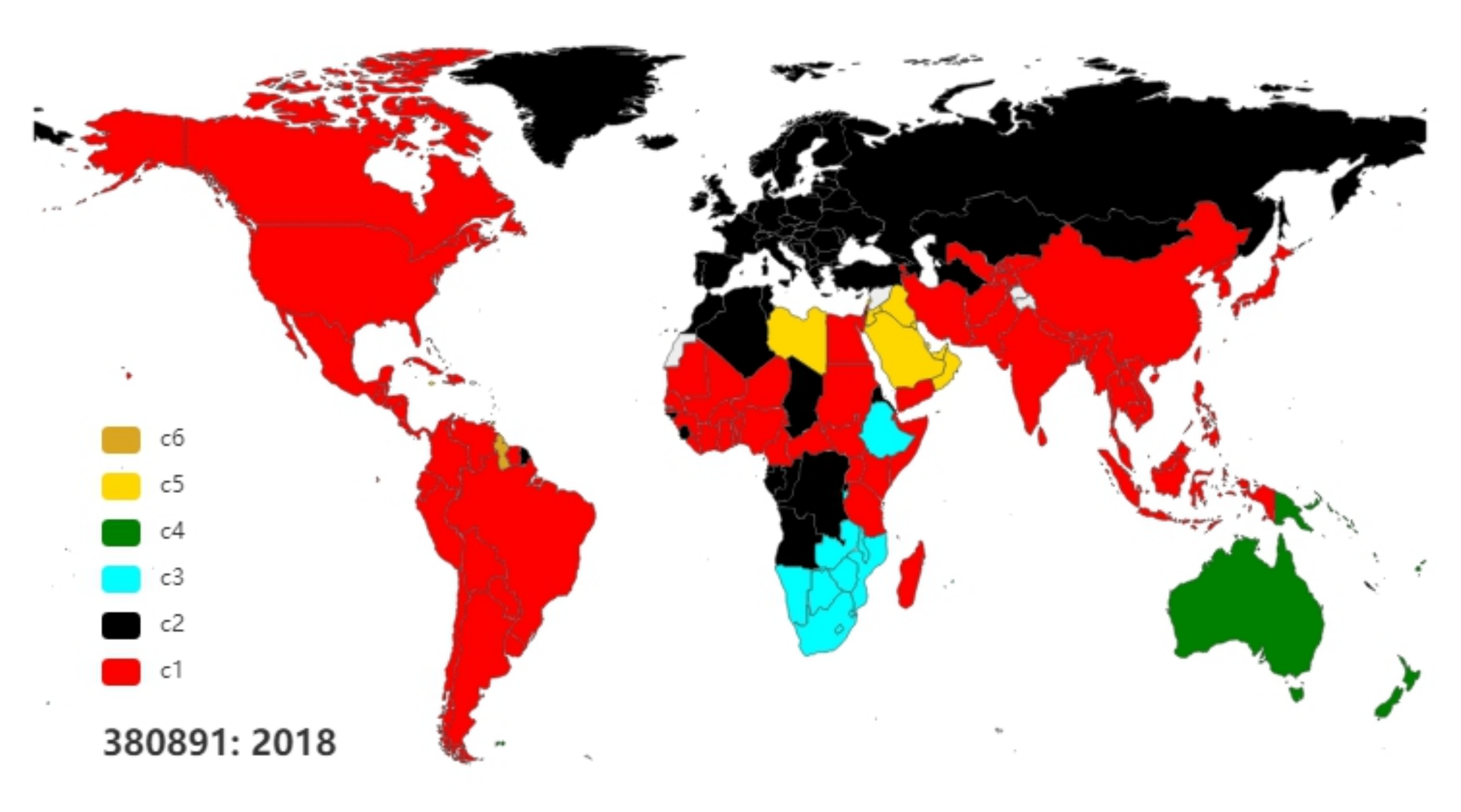}
    \caption{Community evolution of the undirected iPTNs of insecticides (380891) from 2007 to 2018.}
    \label{Fig:iPTN:undirected:CommunityMap:380891}
\end{figure}

Figure~\ref{Fig:iPTN:undirected:CommunityMap:380891} illustrates the evolution of communities of the undirected iPTNs of insecticides (380891) from 2007 to 2018. It is striking to observe that there are 6 years when one community dominates, that are 2008, 2009, 2010, 2011, 2012, and 2014. 
In 2008, all the economies form a community. 
In 2009, we obtain two communities. The smaller community $C_2$ contains 18 economies, 16 in southern Africa (Botswana, Dem. Rep. Congo, Eq. Guinea, Madagascar, Malawi, Mauritius, Namibia, Rwanda, Saint Helena, Seychelles, Solomon Is., South Africa, Swazilande, Tanzania, Zambia, and Zimbabw, where 11 economies belong to the Southern African Development Community) and 2 in middle east (Palestine and Kuwait).
In 2010, we have three communities. The second community $C_2$ contains 12 economies, including Barbados, Dominica, Grenada, Guyana, Jamaica, Montserrat, Saint Kitts and Nevis, Anguilla, Saint Lucia, St. Vin. and Gren., and Trinidad and Tobago, which locate around the eastern Caribbean Sea. The third community $C_3$ contains two economies, Mali and Senegal in northwest Africa.
In 2011, we idenfy four communities. The second community $C_2$ contains 15 economies (Australia, Solomon Is., Cook Islands, Fiji, Kiribati, Nauru, Vanuatu, New Zealand, Niue, Norfolk Island, Papua New Guinea, Tonga, Tuvalu, Wallis and Futuna, and Samoa) located in the Oceania and around. The third community $C_3$ contains 14 economies (Botswana, Comoros, Lesotho, Malawi, Mauritius, Mozambique, Namibia, Timor-Leste, Rwanda, Saint Helena, South Africa, Zimbabwe, Swaziland, and Zambia) in southen Africa. The fourth community $C_4$ contains 11 economies (Barbados, Dominica, Grenada, Guyana, Jamaica, Montserrat, Saint Kitts and Nevis, Saint Lucia, St. Vin. and Gren., Suriname, and Trinidad and Tobago) in Latin-America.
In 2012, we find two communities. The second community $C_2$ contains 12 economies (Antigua and Barb., Barbados, Dominica, Grenada, Guyana, Jamaica, Montserrat, Saint Kitts and Nevis, Saint Lucia, St. Vin. and Gren., Suriname, and Trinidad and Tobago), which locate around the eastern Caribbean Sea.
In 2014, there are three communities. The second community $C_2$ contains seven economies (Belize, Costa Rica, El Salvador, Guatemala, Honduras, Nicaragua, and Panama) located along the western Caribbean Sea, while the third community $C_3$ contains 13 economies (Antigua and Barb., Barbados, Dominica, Fr. S. Antarctic Lands, Grenada, Guyana, Jamaica, Montserrat, Saint Kitts and Nevis, Saint Lucia, St. Vin. and Gren., Suriname, and Trinidad and Tobago) located along the eastern Caribbean Sea.

The other six networks in 2007, 2013, 2015, 2016, 2017, and 2018 have more communities than networks in other years and the communities occupy relatively comparable areas. Comparing these community maps in different years, we see four main regions: American Continent, European Continent, Asia, and Oceania. In contrast, economies in Africa and Middle East are clustered into smaller communities or belong to other big communities. It shows the importance of geographical distance in the international pesticide trade.

\begin{figure}[!ht]
    \centering
    \includegraphics[width=0.321\linewidth]{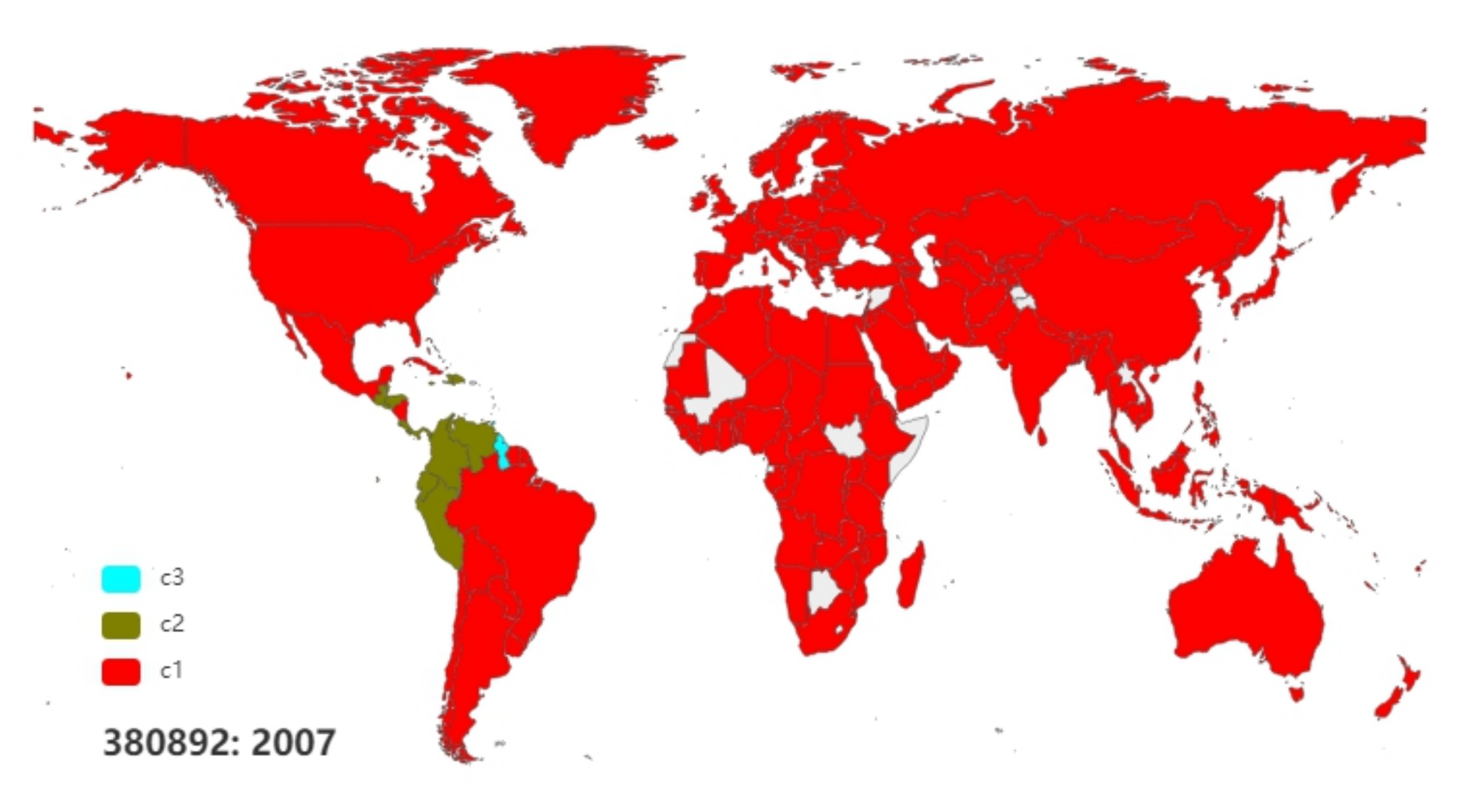}
    \includegraphics[width=0.321\linewidth]{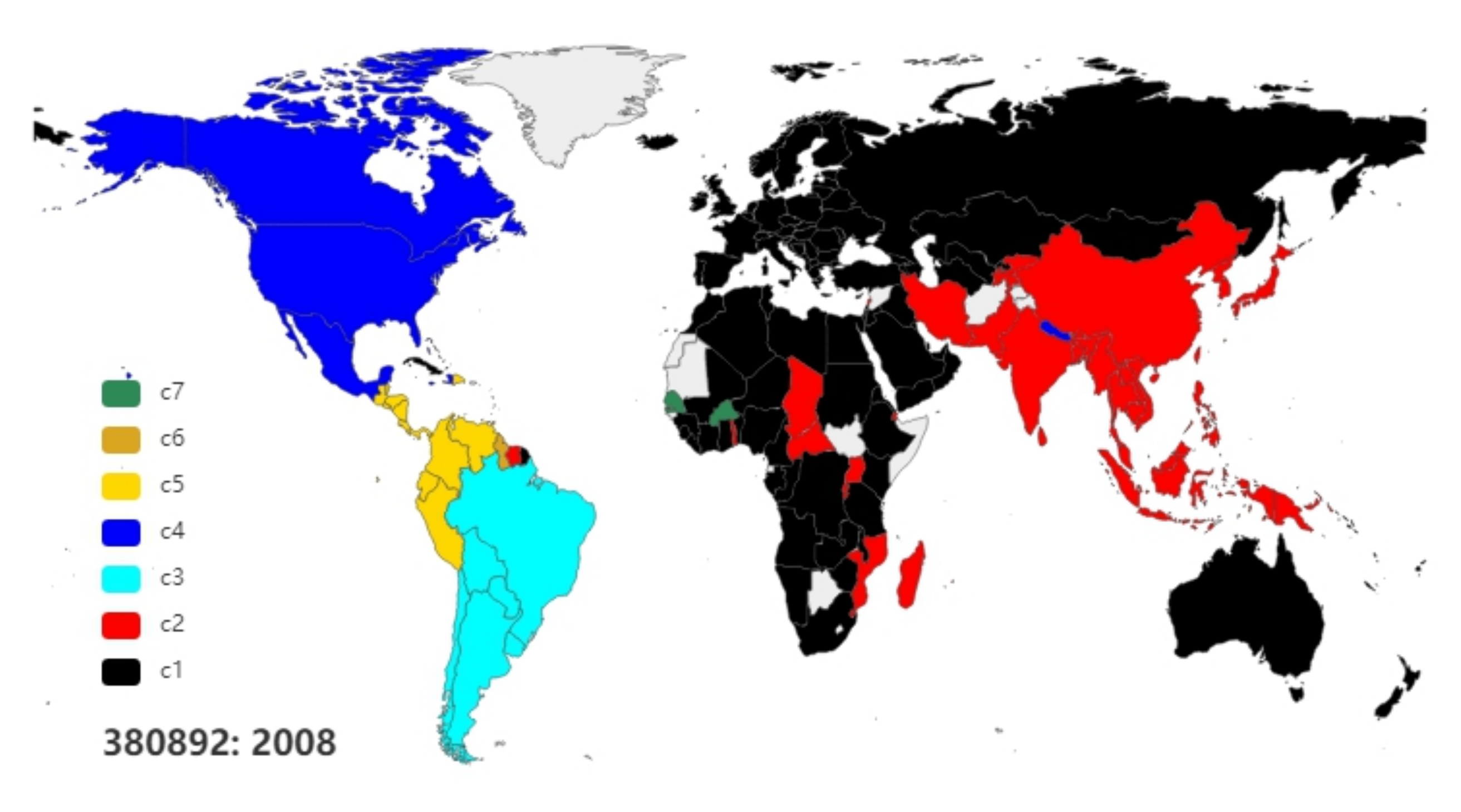}
    \includegraphics[width=0.321\linewidth]{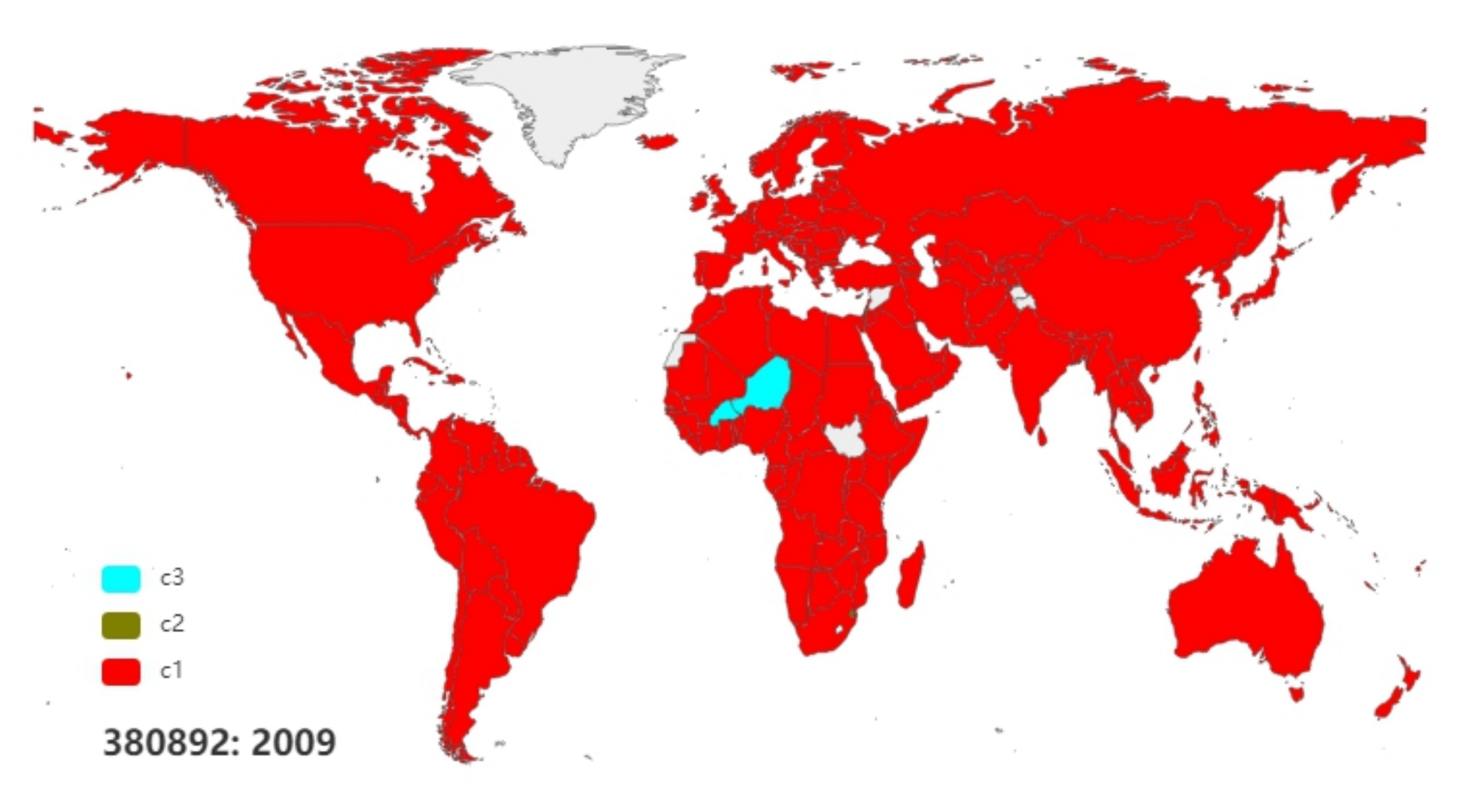}
    \includegraphics[width=0.321\linewidth]{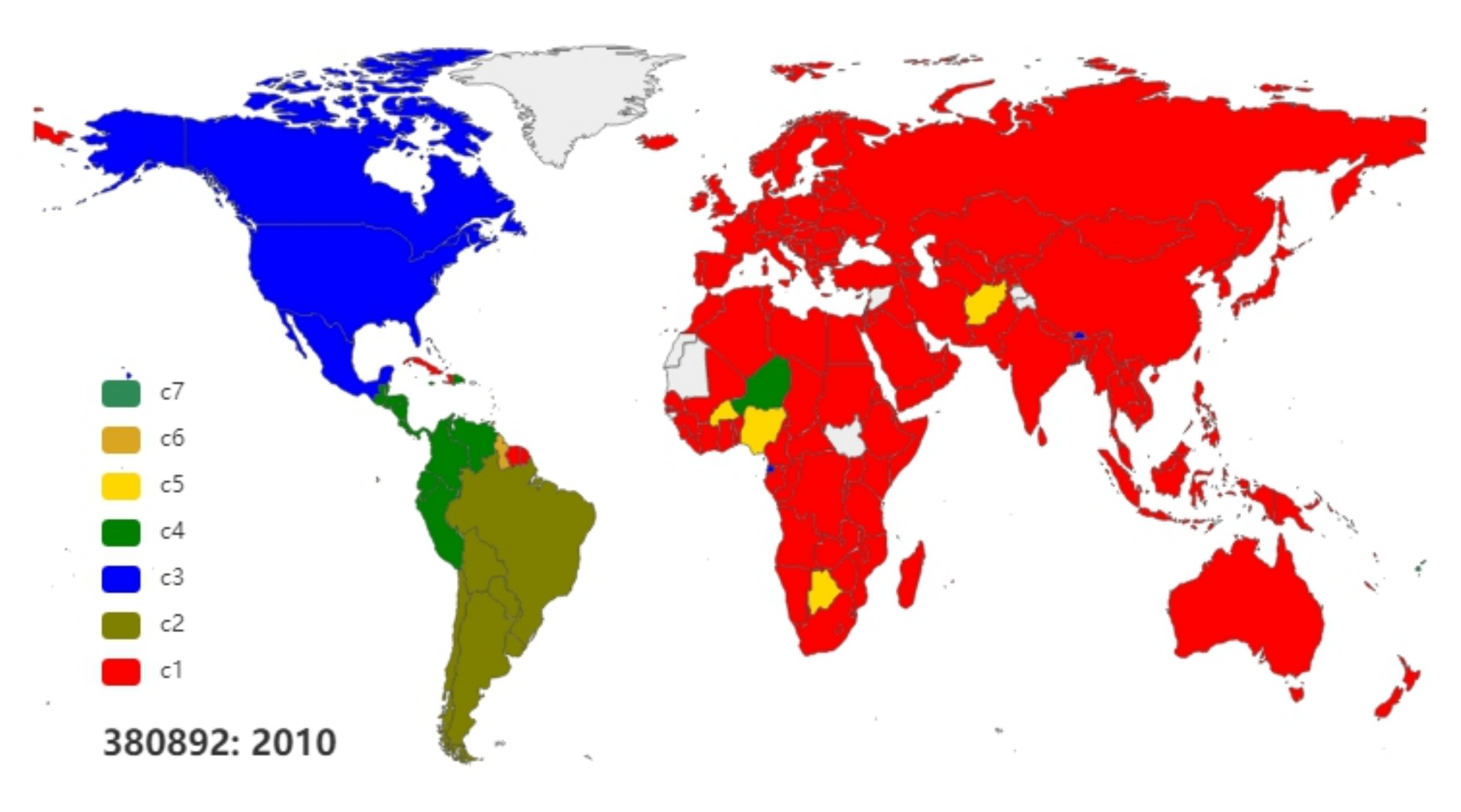}
    \includegraphics[width=0.321\linewidth]{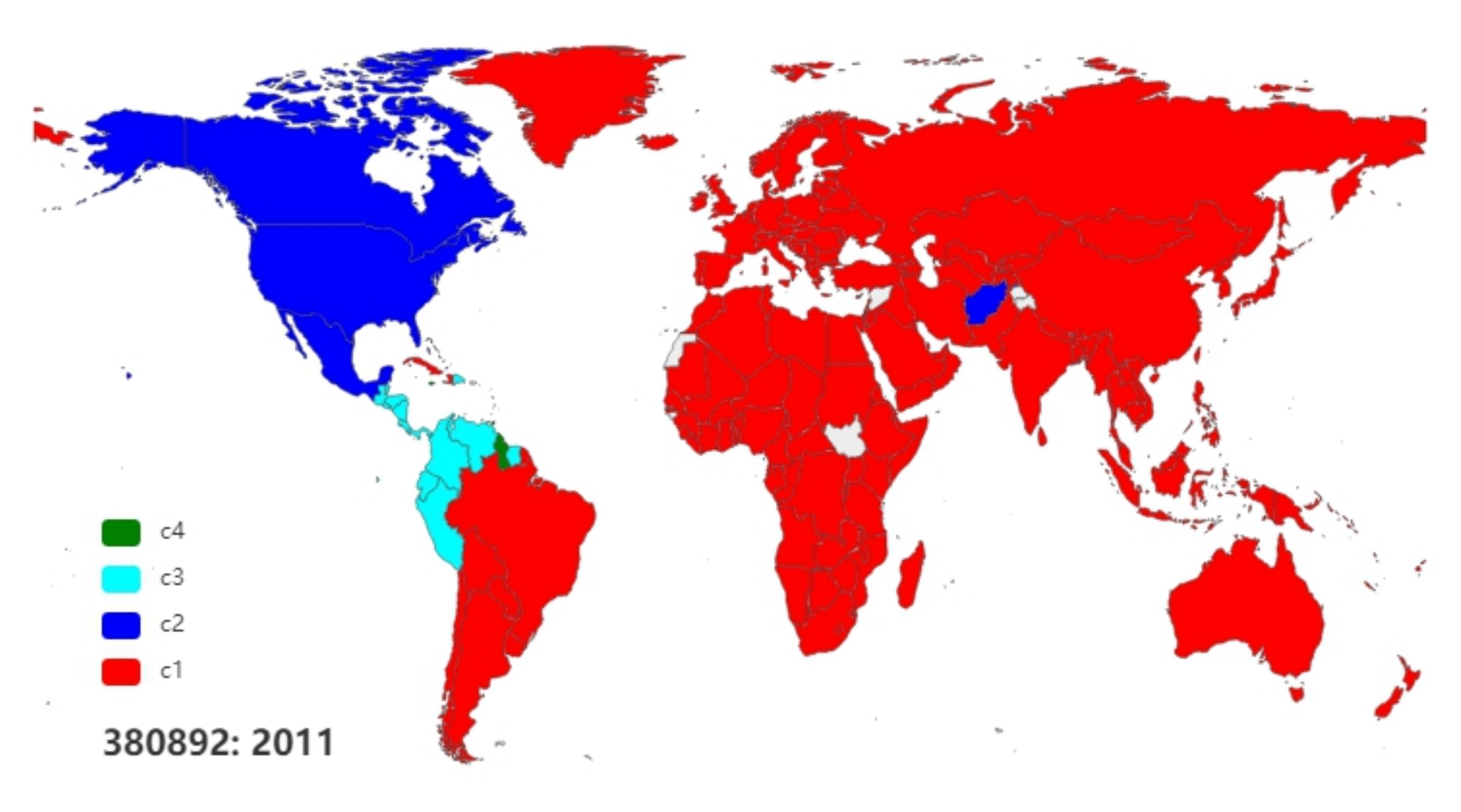}
    \includegraphics[width=0.321\linewidth]{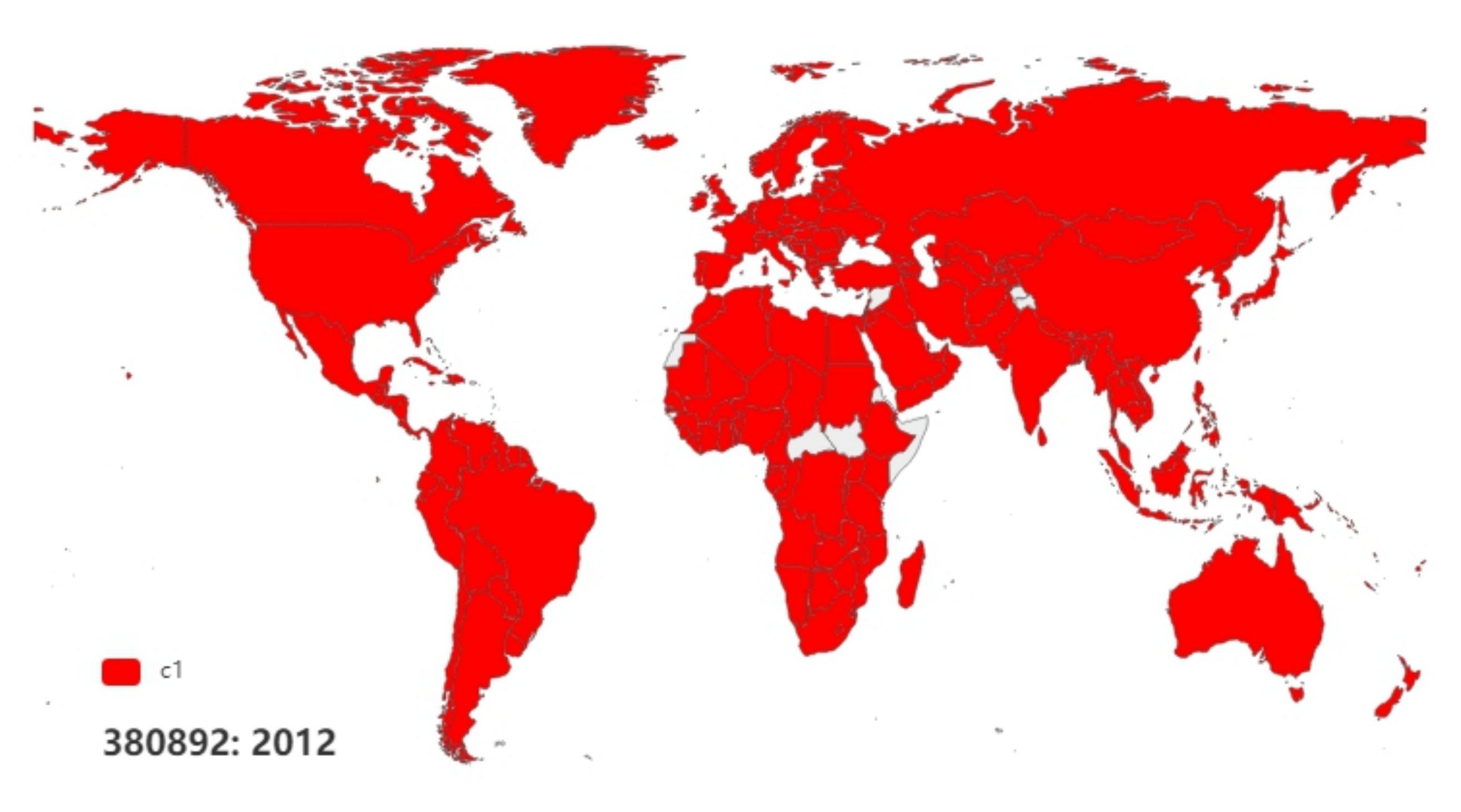}
    \includegraphics[width=0.321\linewidth]{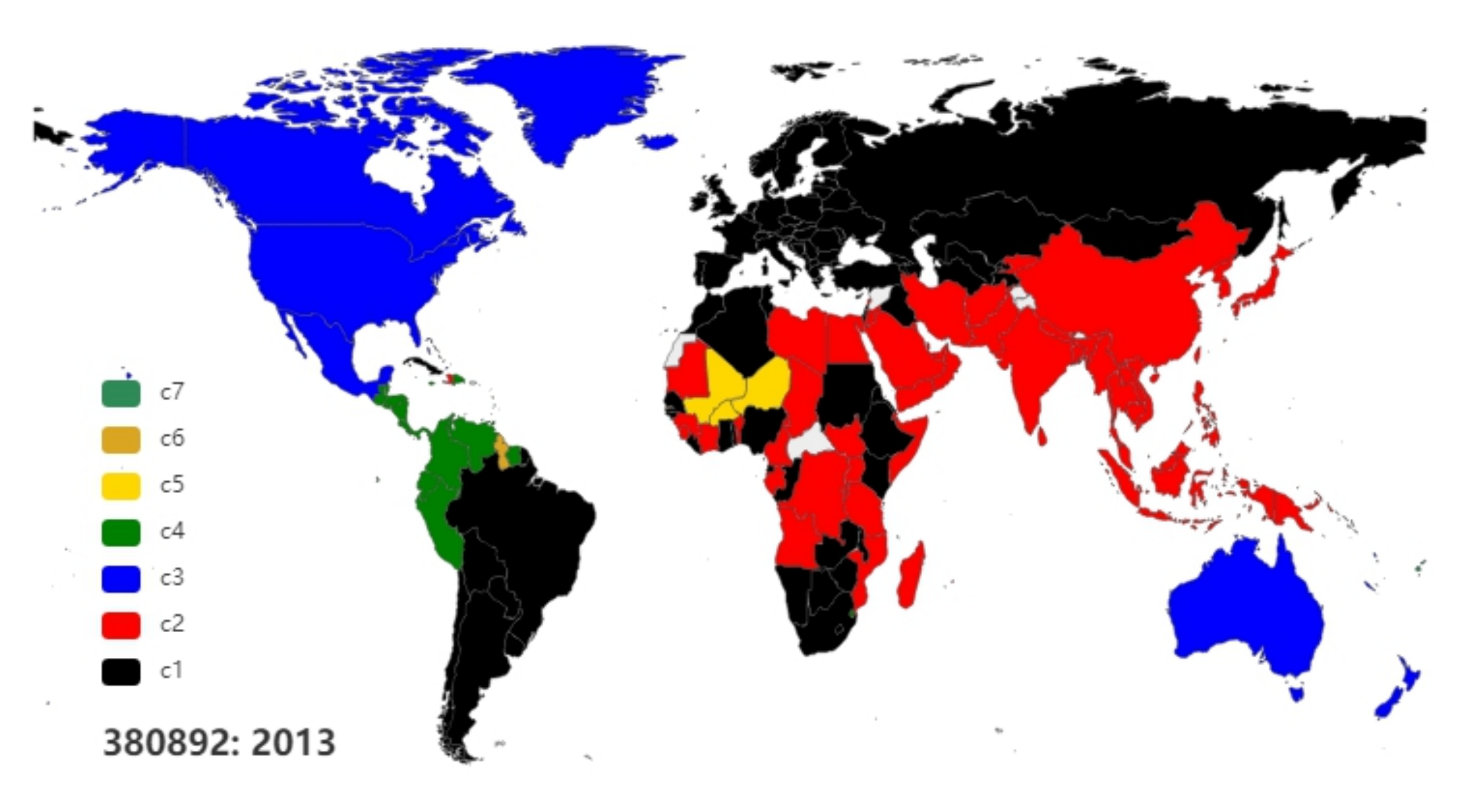}
    \includegraphics[width=0.321\linewidth]{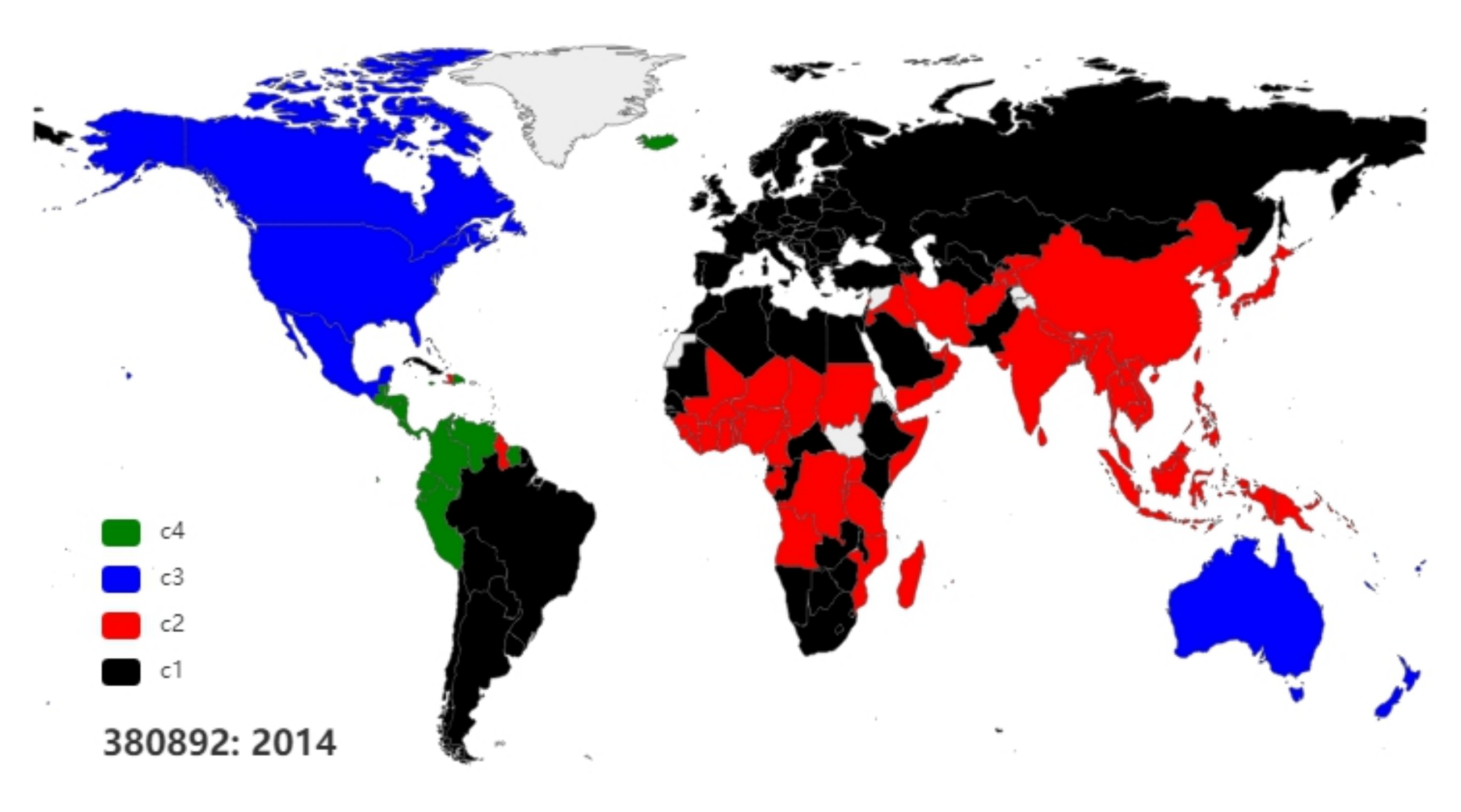}
    \includegraphics[width=0.321\linewidth]{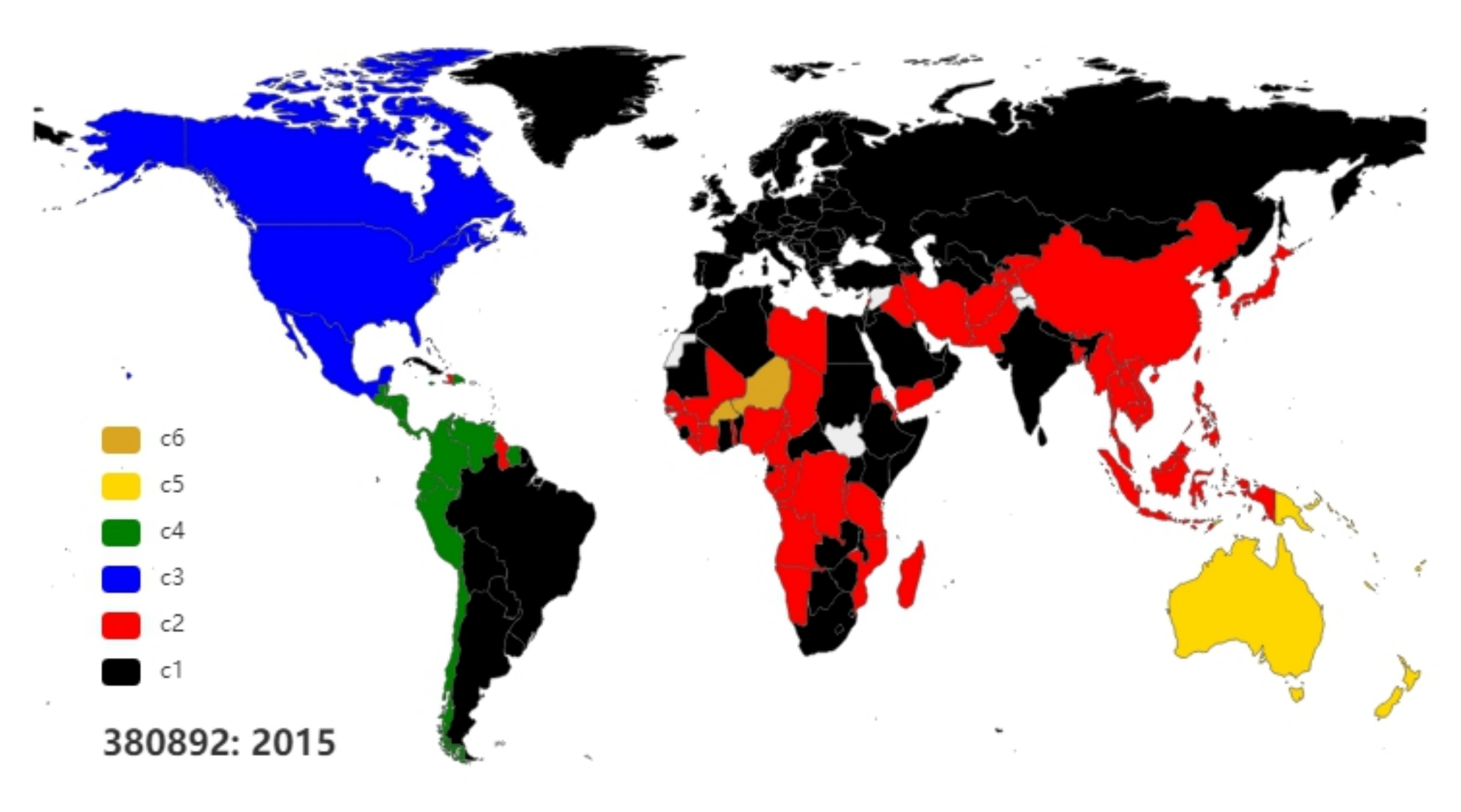}
    \includegraphics[width=0.321\linewidth]{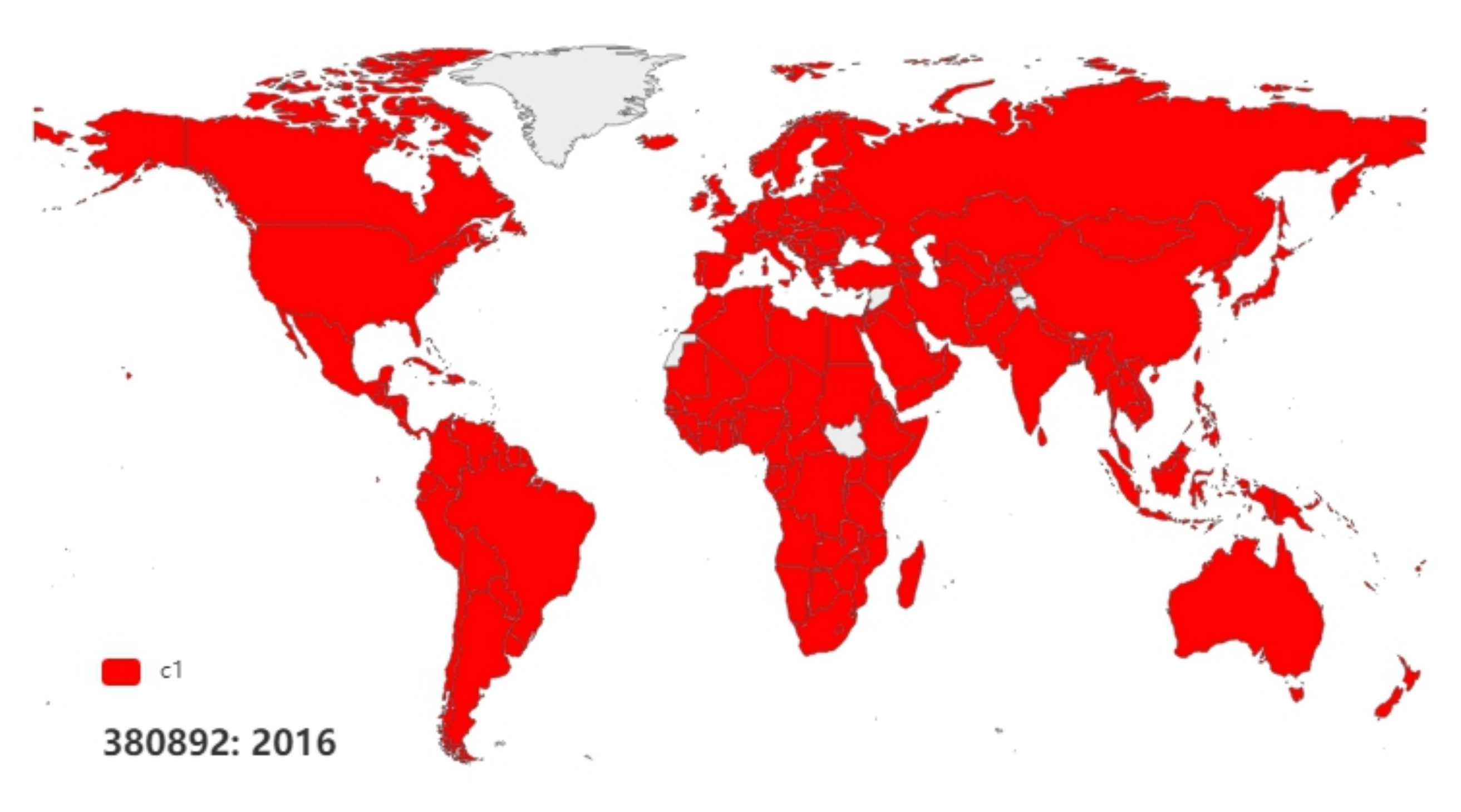}
    \includegraphics[width=0.321\linewidth]{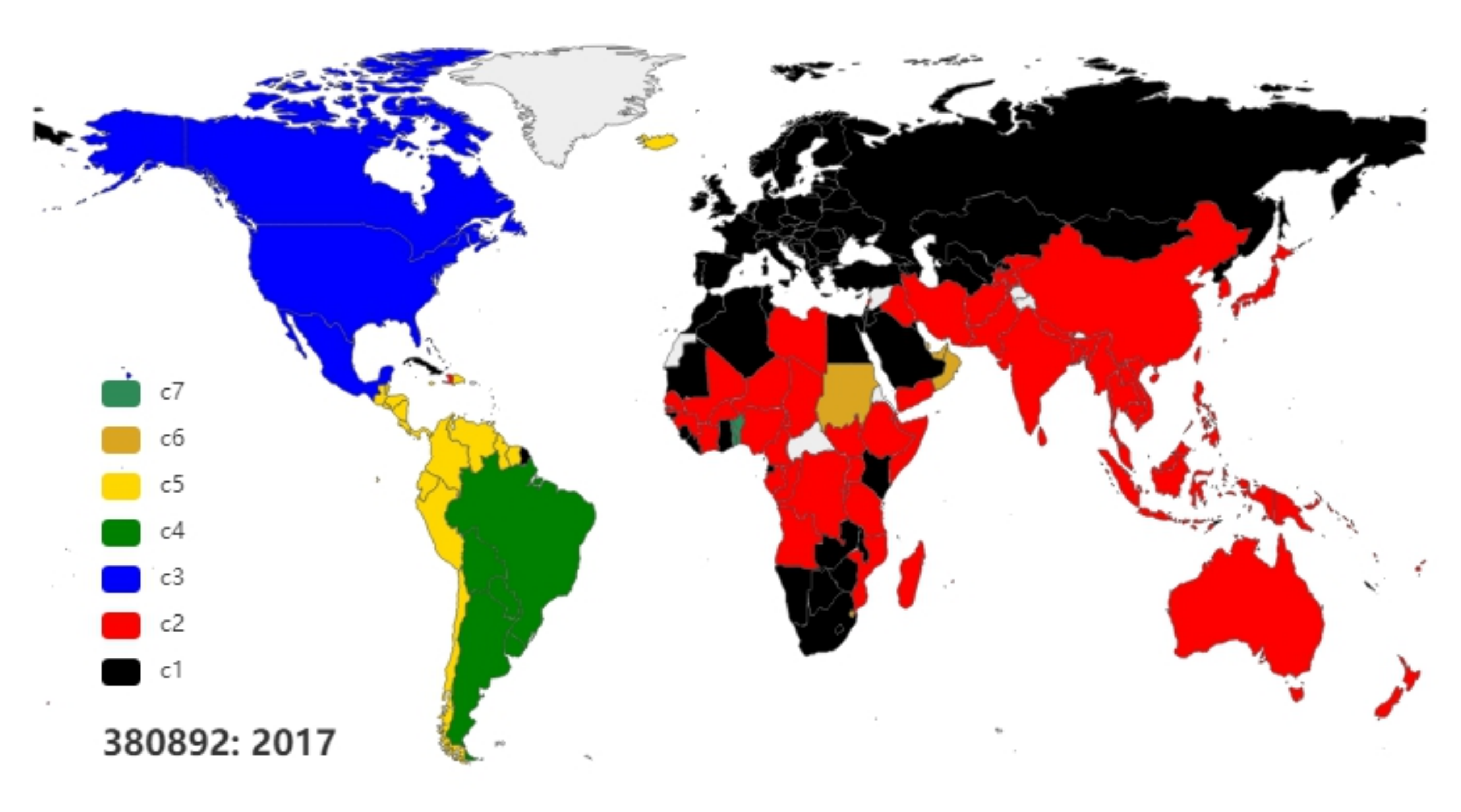}
    \includegraphics[width=0.321\linewidth]{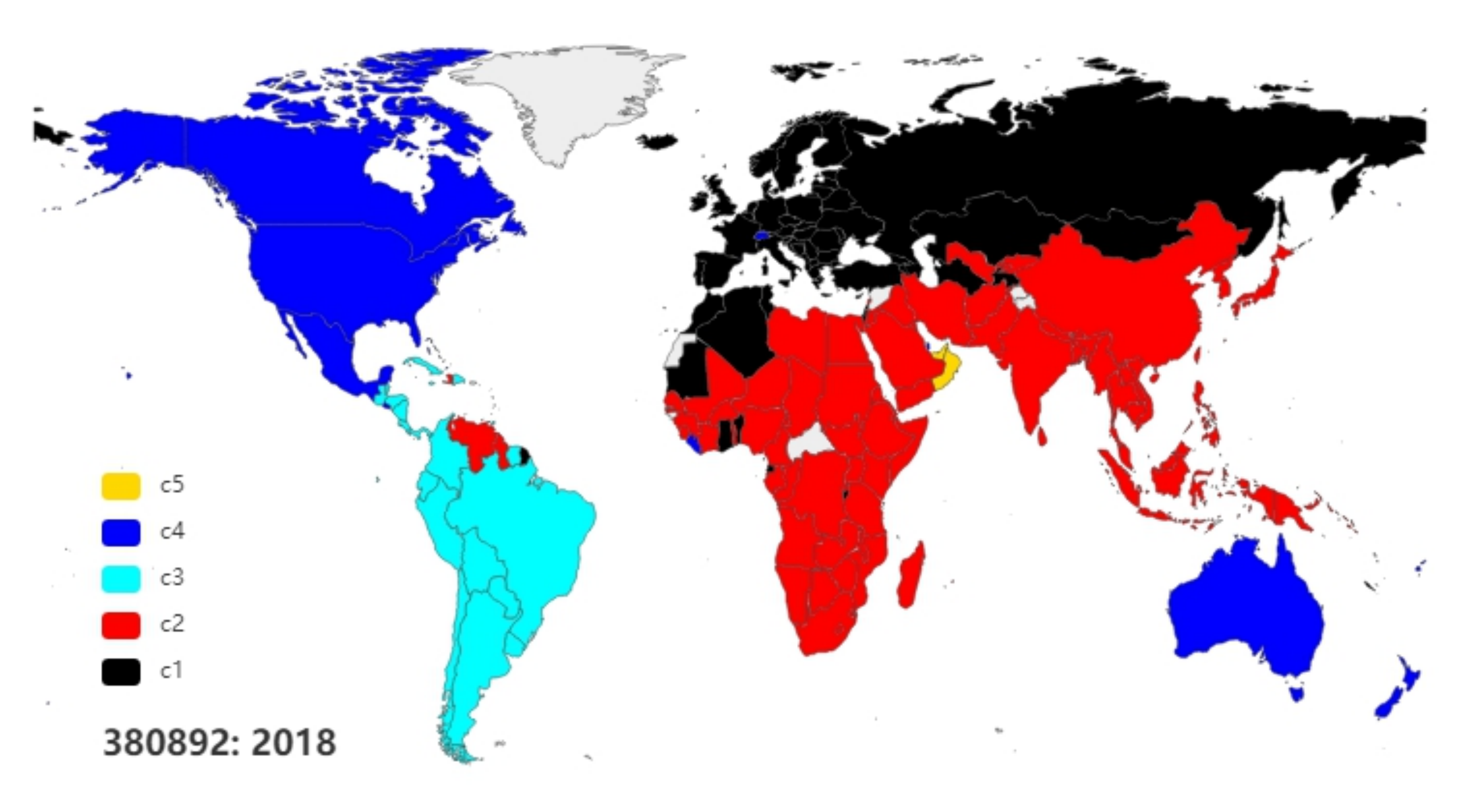}
    \caption{Community evolution of the undirected iPTNs of fungicides (380892) from 2007 to 2018.}
    \label{Fig:iPTN:undirected:CommunityMap:380892}
\end{figure}

Figure~\ref{Fig:iPTN:undirected:CommunityMap:380892} illustrates the evolution  of communities of the undirected iPTNs of fungicides (380892) from 2007 to 2018. There are two networks, respectively in 2012 and 2016, that have only one community. The two networks in 2007 and 2009 have three communities. However, the first community $C_1$ contains most of the economies. 
In 2007, the second community $C_2$ contains 13 economies (Belize, Colombia, Costa Rica, Dominican Rep., Ecuador, El Salvador, Guatemala, Haiti, Honduras, Jamaica, Panama, Peru, and Venezuela), while the third community $C_3$ contains 10 economies (Antigua and Barb., Barbados, Grenada, Guyana, Montserrat, Aruba, Saint Kitts and Nevis, Saint Lucia, St. Vin. and Gren., and Trinidad and Tobago). The economies in these two communities locate around the Caribean Sea.
In 2009, the second community $C_2$ contains two economies (Palestine in Middle East and Swaziland in southern Africa), while the third community $C_3$ also contains two economies (Niger and Burkina Faso), both in northern Africa.

The other eight networks in 2008, 2010, 2011, 2013, 2014, 2015, 2017, and 2018 have at least four communities. We observe several clustered regions, including the three economies of the North American Free Trade Area (NAFTA) together with a few small economies in the east Caribbean, Guyana and smaller economies in the east Caribbean, economies around the Caribbean, Asia, Oceania, Europe, and fragmented regions in Africa. These regions are identified communities in some years or a few regions form a community in other years. For the economies in the Economic Commission for Latin America and the Caribbean, we mainly observe one, two or three communities in different years and some small economies may belong to the NAFTA-based community. Economies in the Oceania may clustered with Europe (2008, 2010, 2011), the NAFTA (2013, 2014, 2018), or Asia (2010, 2011, 2017). It suggests that geographic distance and trade organizations play important roles in the formation of international fungicide trade communities.

\begin{figure}[!ht]
    \centering
    \includegraphics[width=0.321\linewidth]{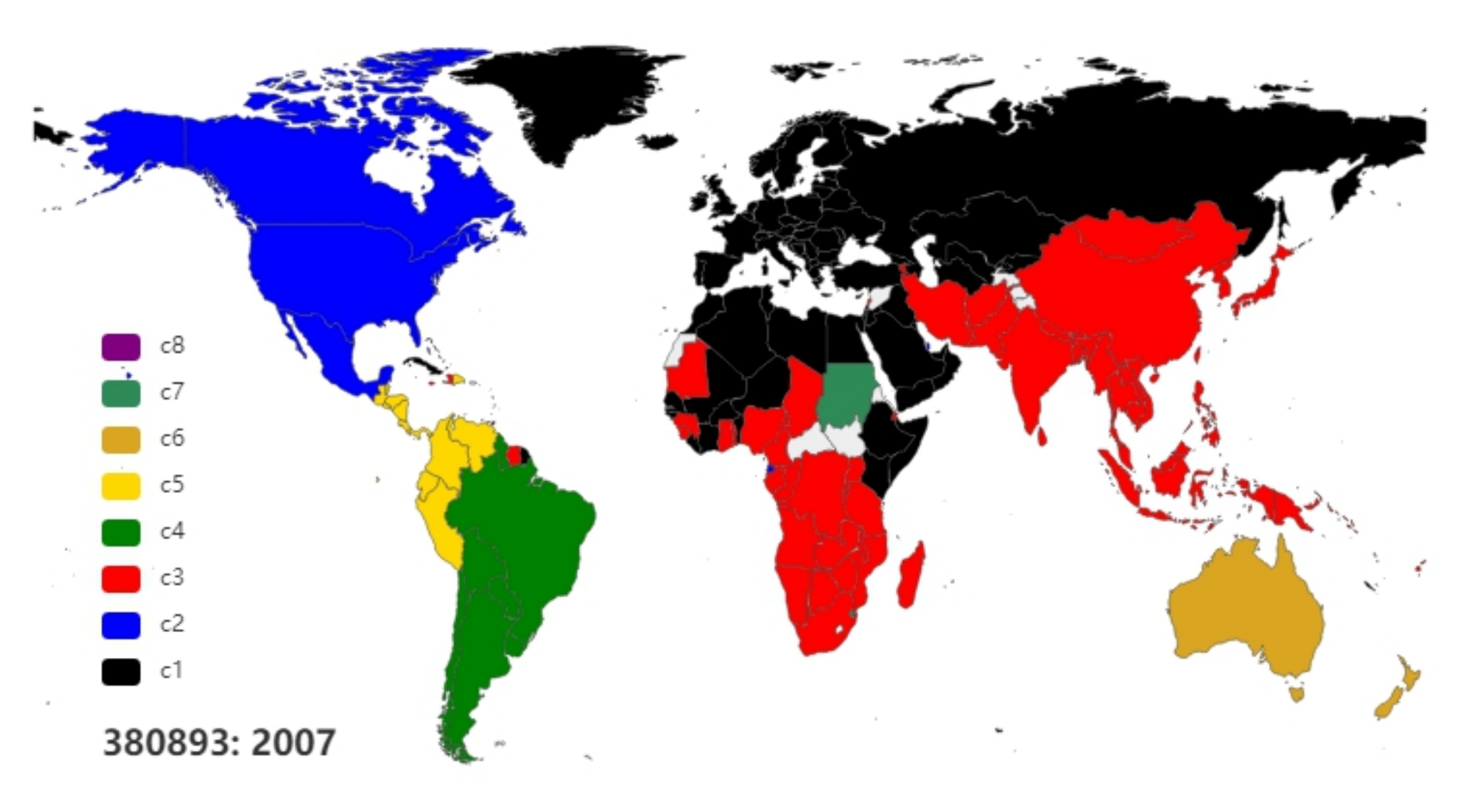}
    \includegraphics[width=0.321\linewidth]{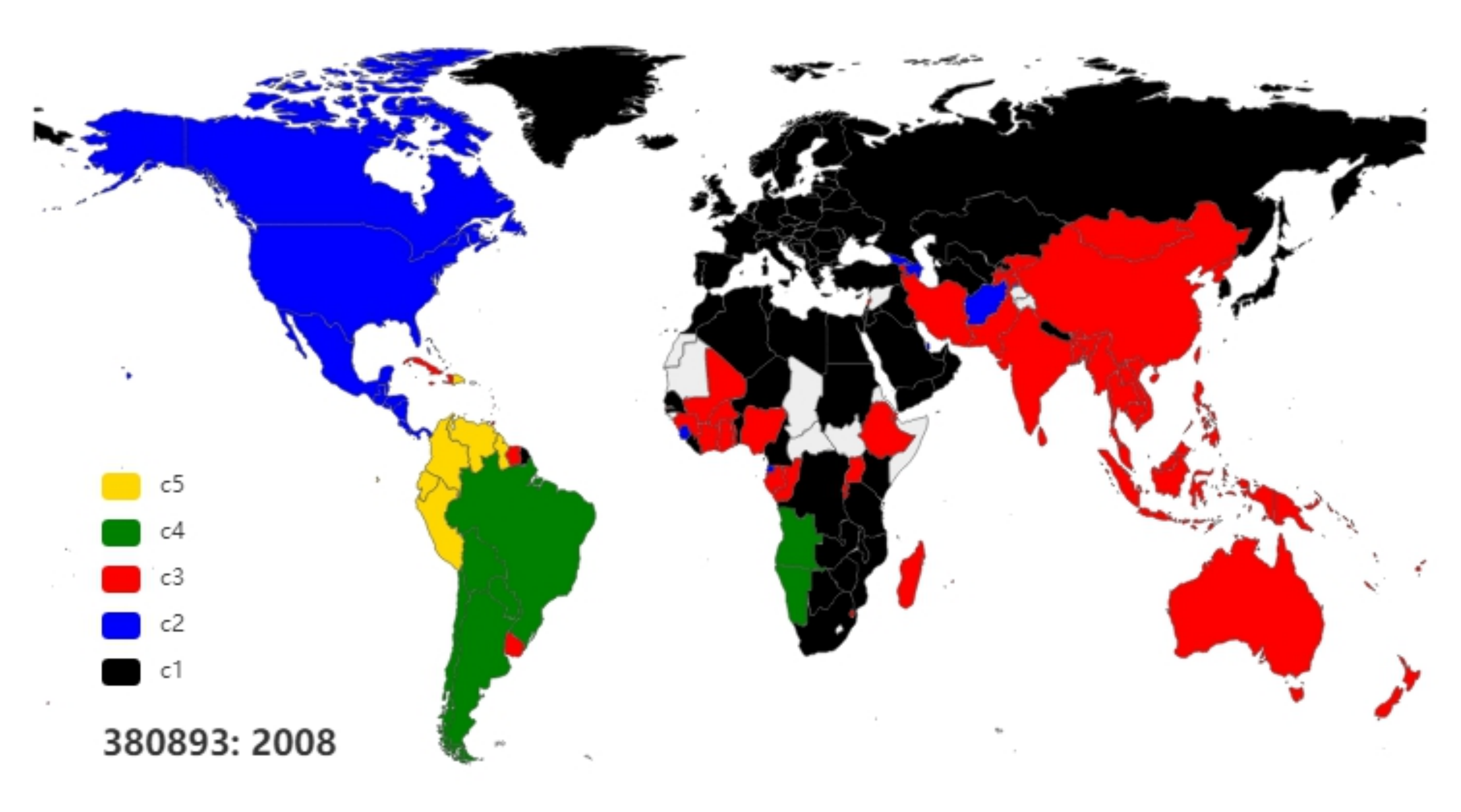}
    \includegraphics[width=0.321\linewidth]{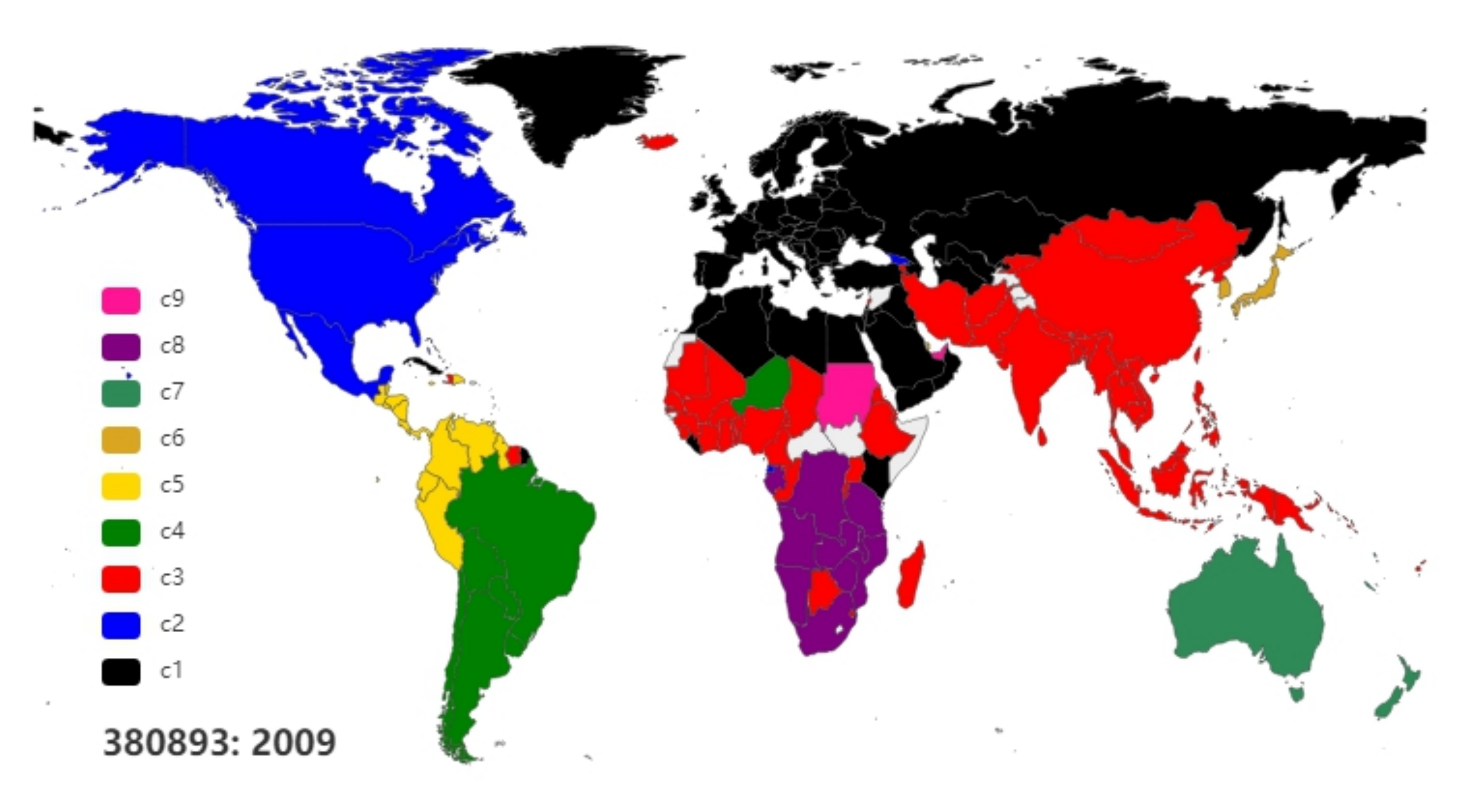}
    \includegraphics[width=0.321\linewidth]{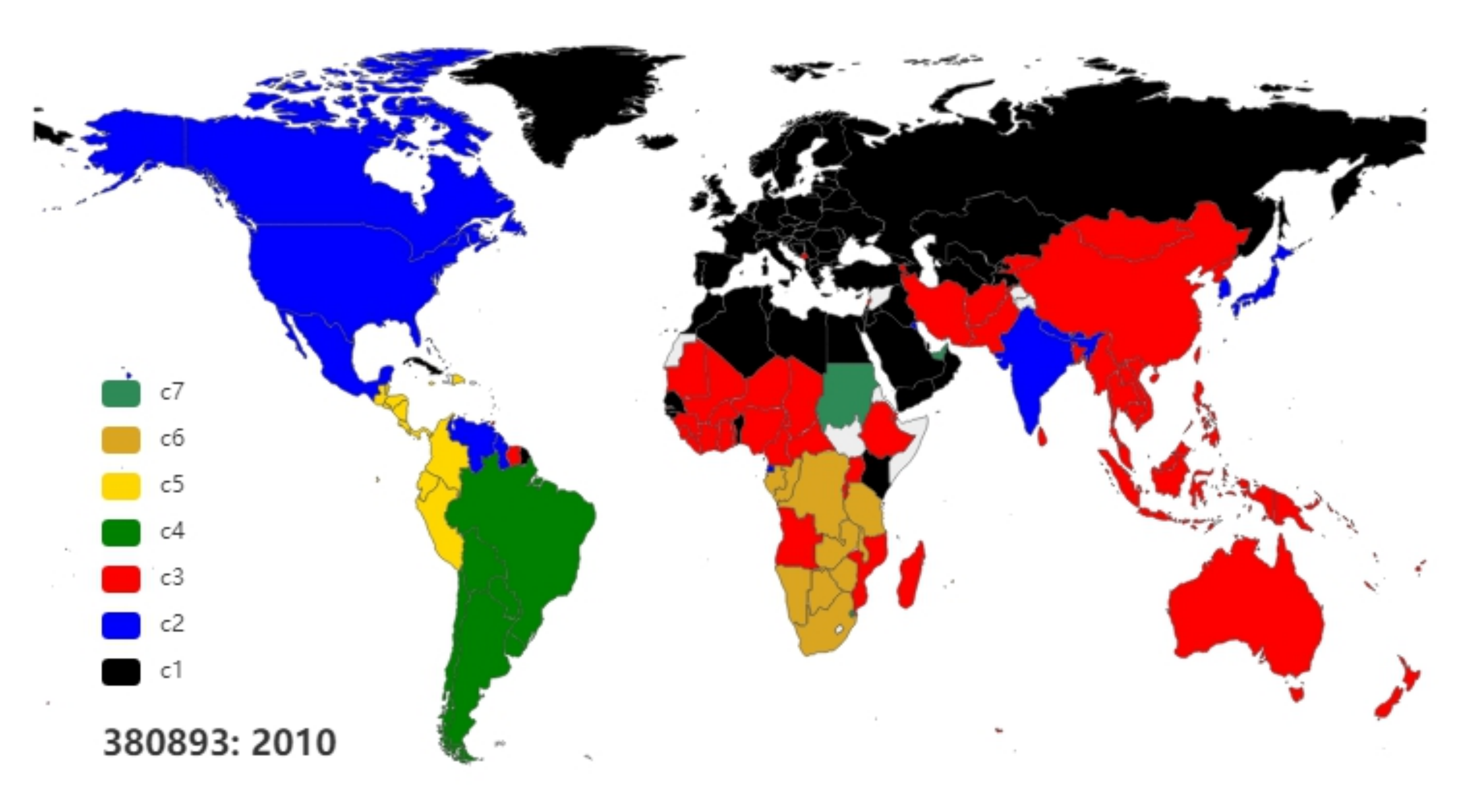}
    \includegraphics[width=0.321\linewidth]{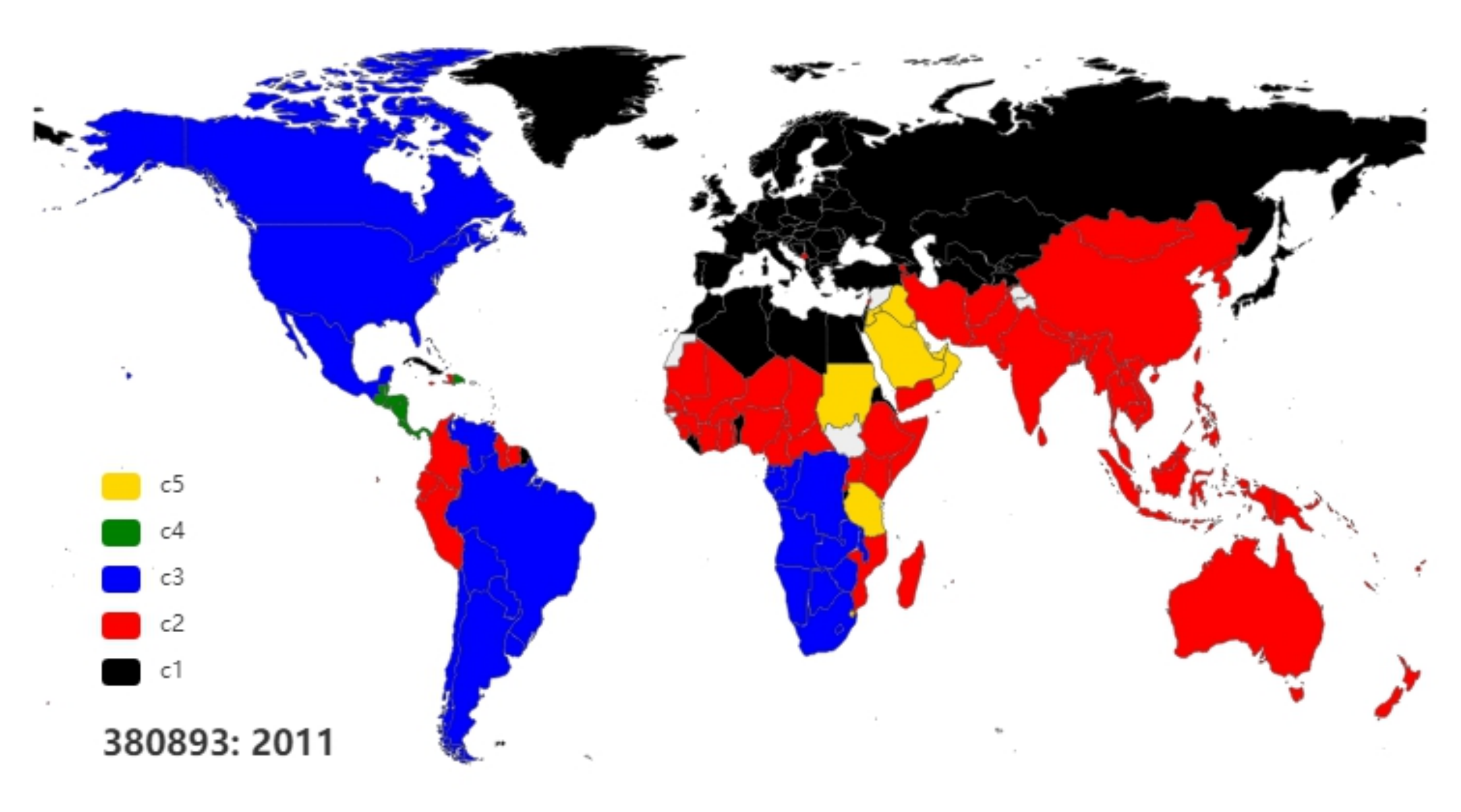}
    \includegraphics[width=0.321\linewidth]{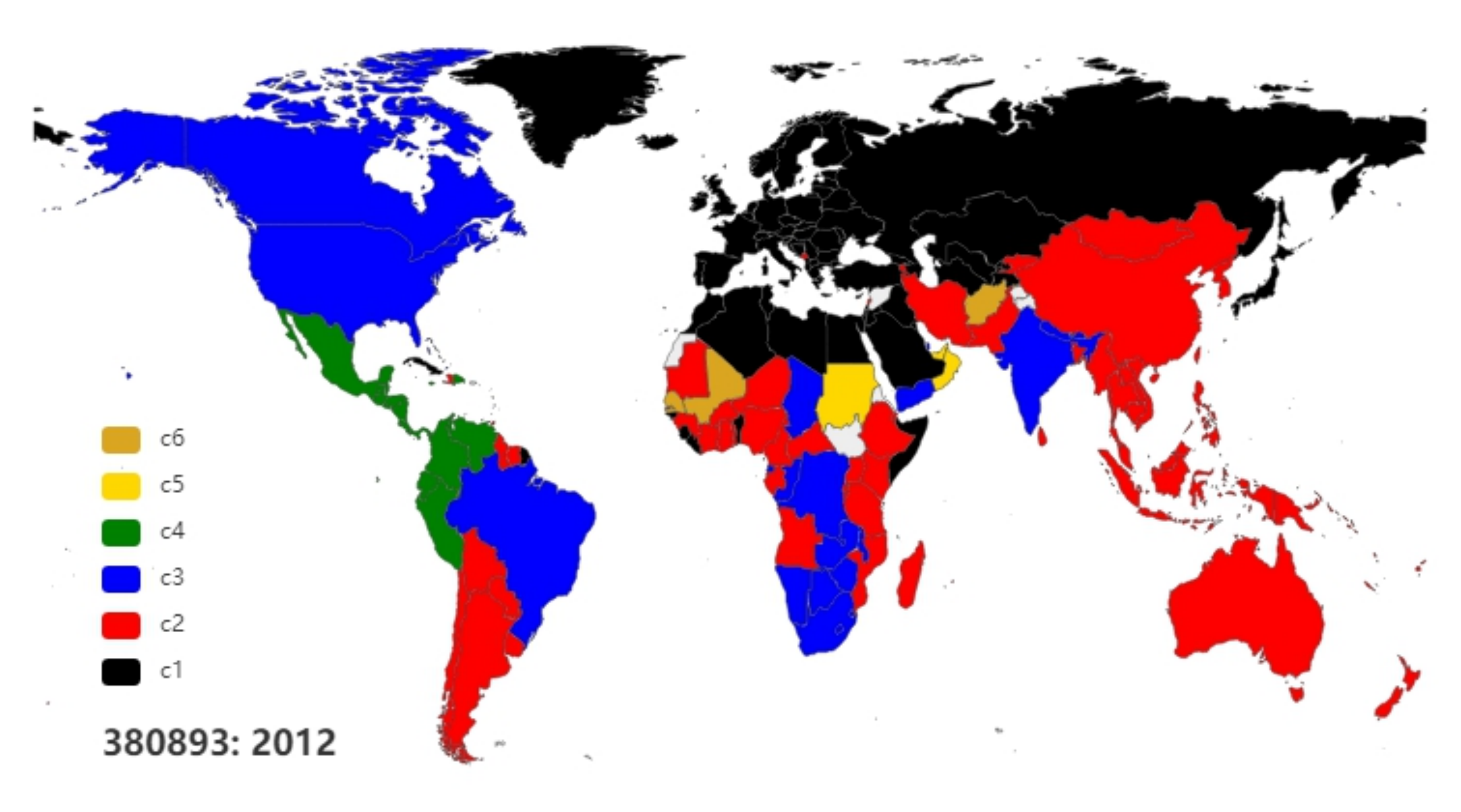}
    \includegraphics[width=0.321\linewidth]{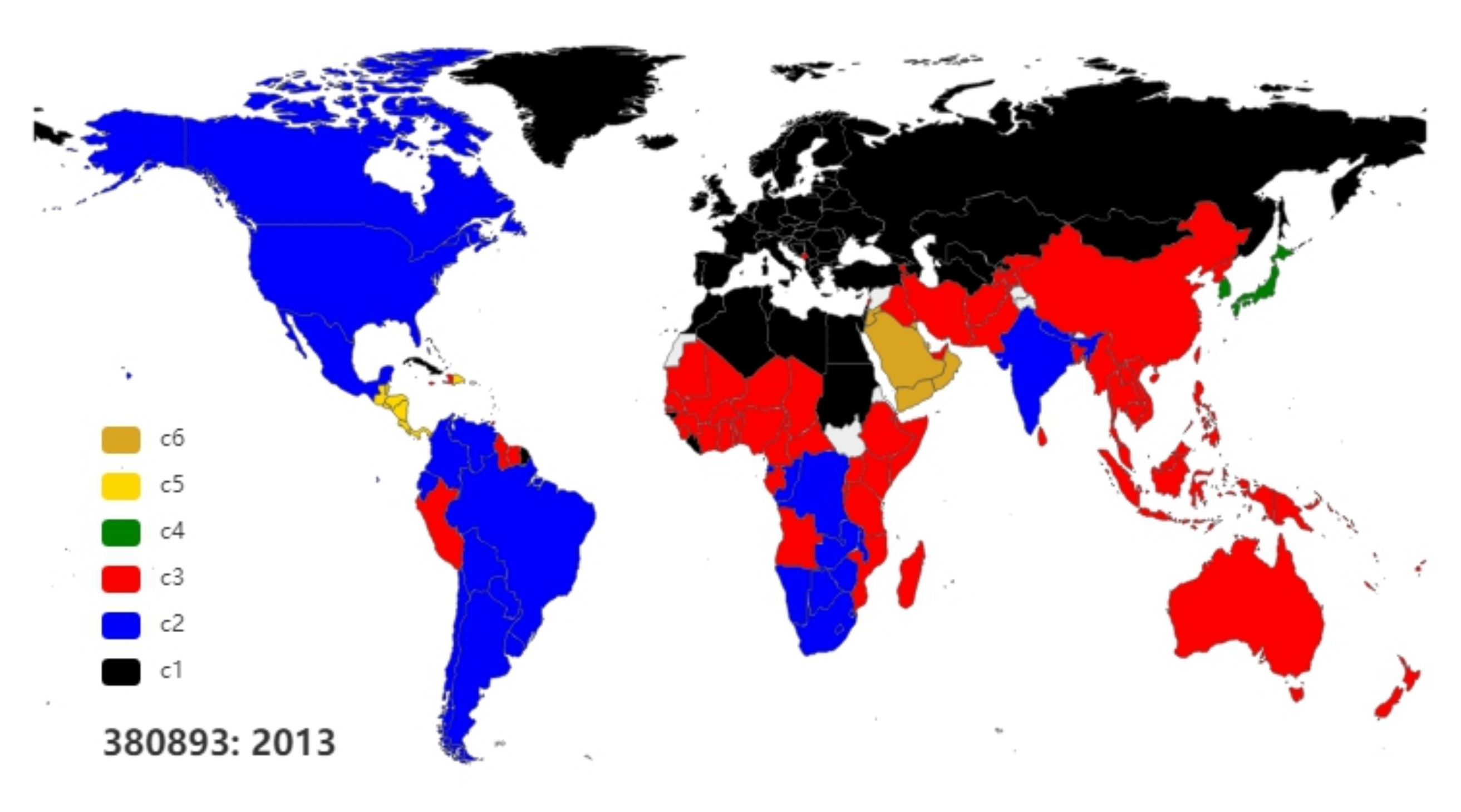}
    \includegraphics[width=0.321\linewidth]{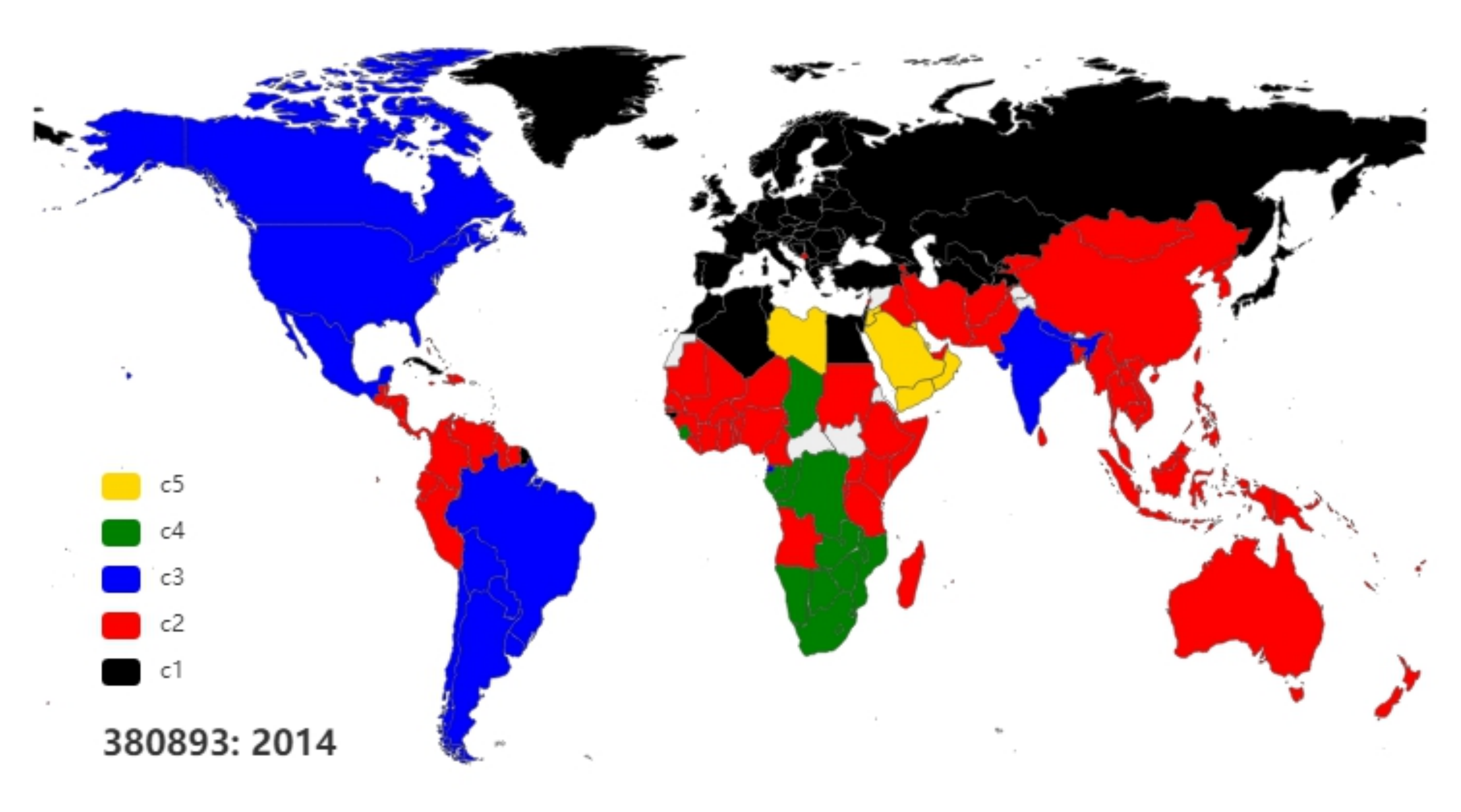}
    \includegraphics[width=0.321\linewidth]{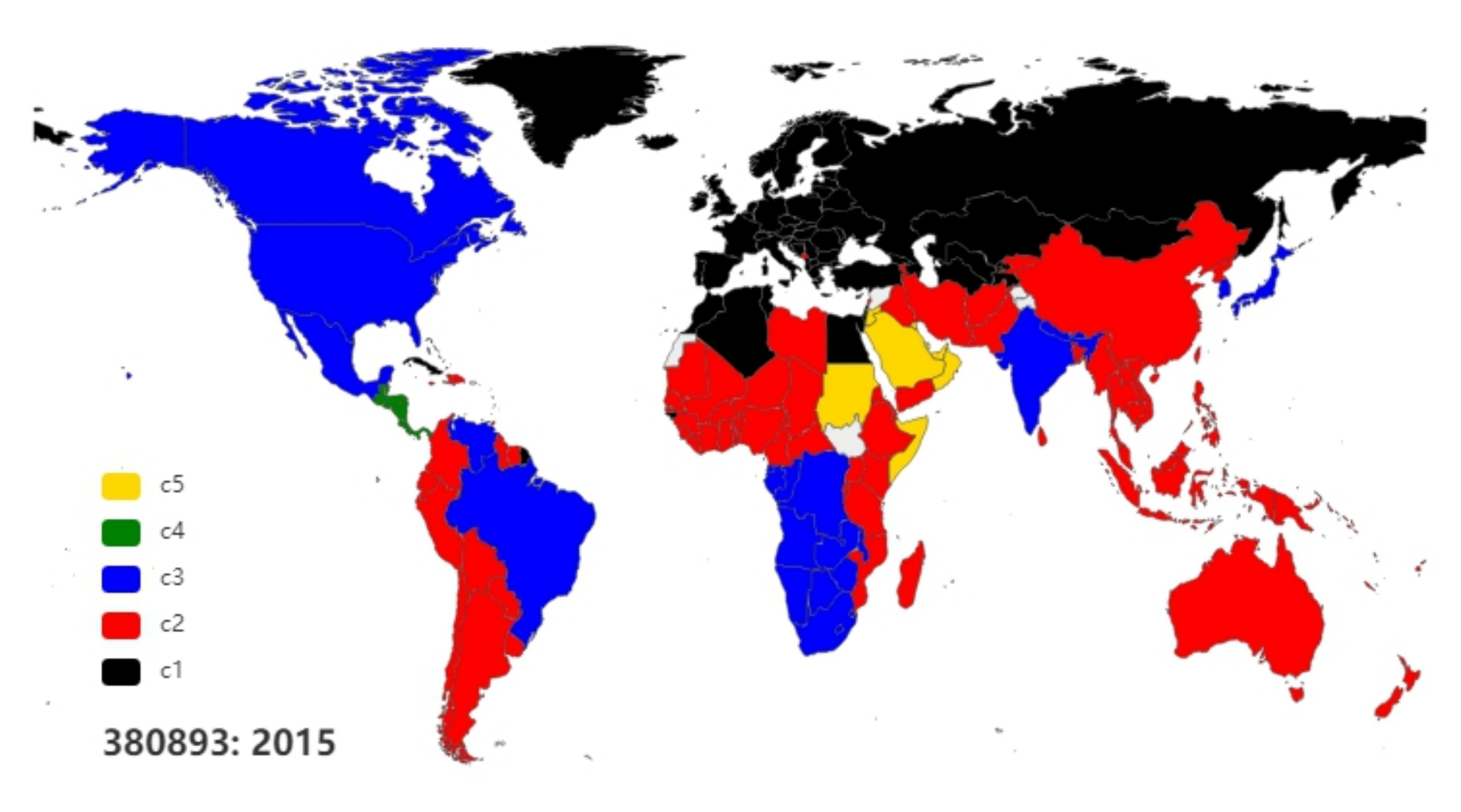}
    \includegraphics[width=0.321\linewidth]{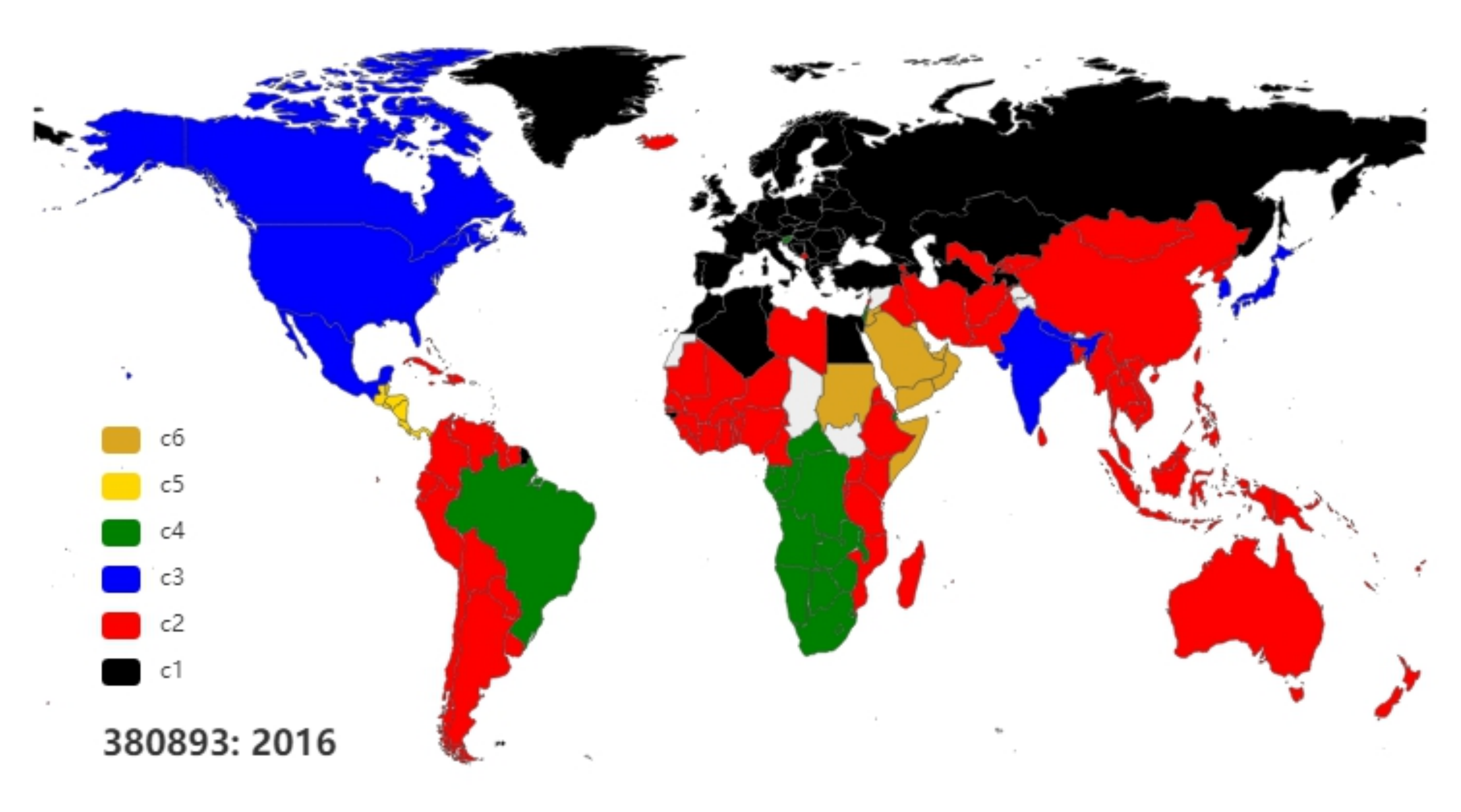}
    \includegraphics[width=0.321\linewidth]{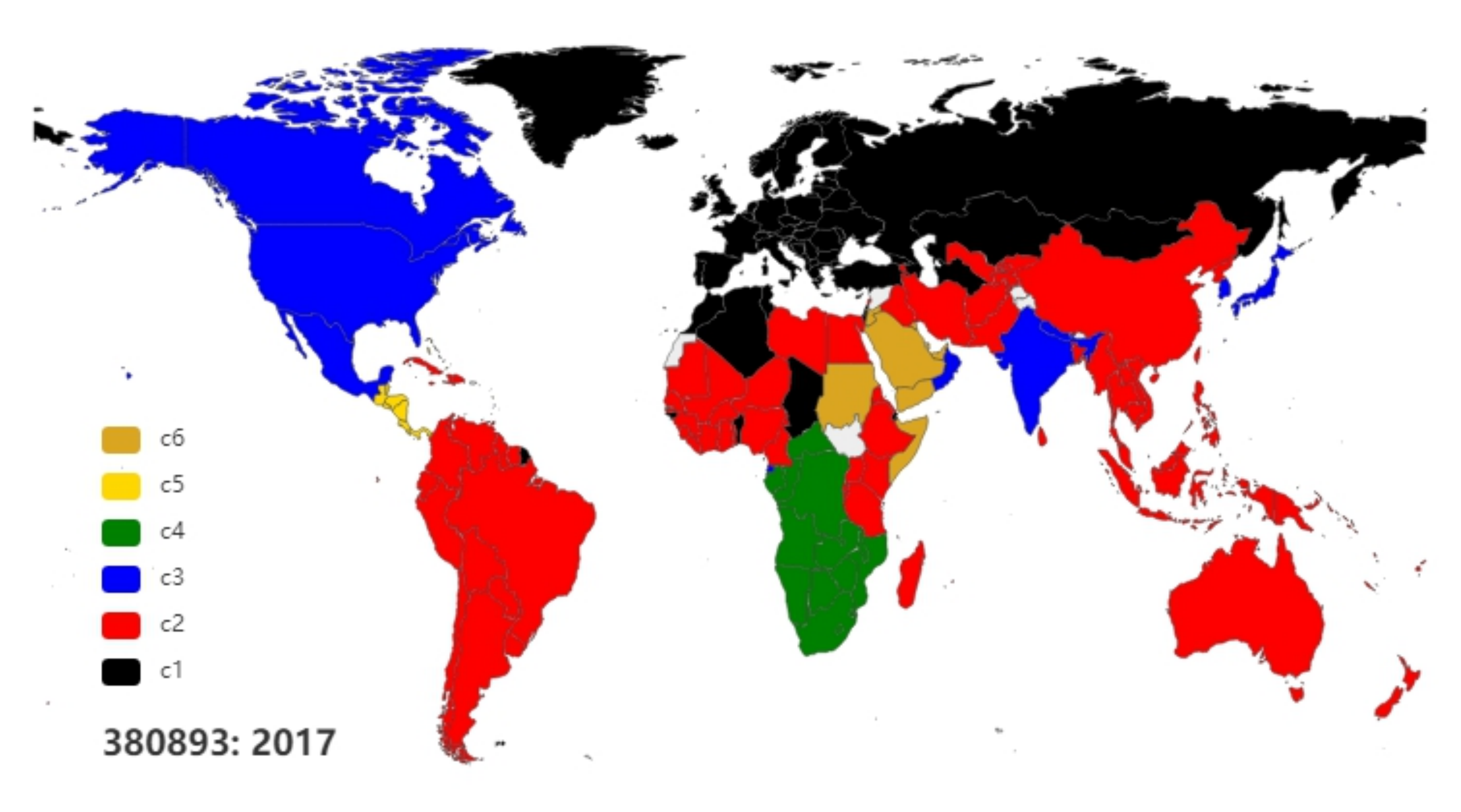}
    \includegraphics[width=0.321\linewidth]{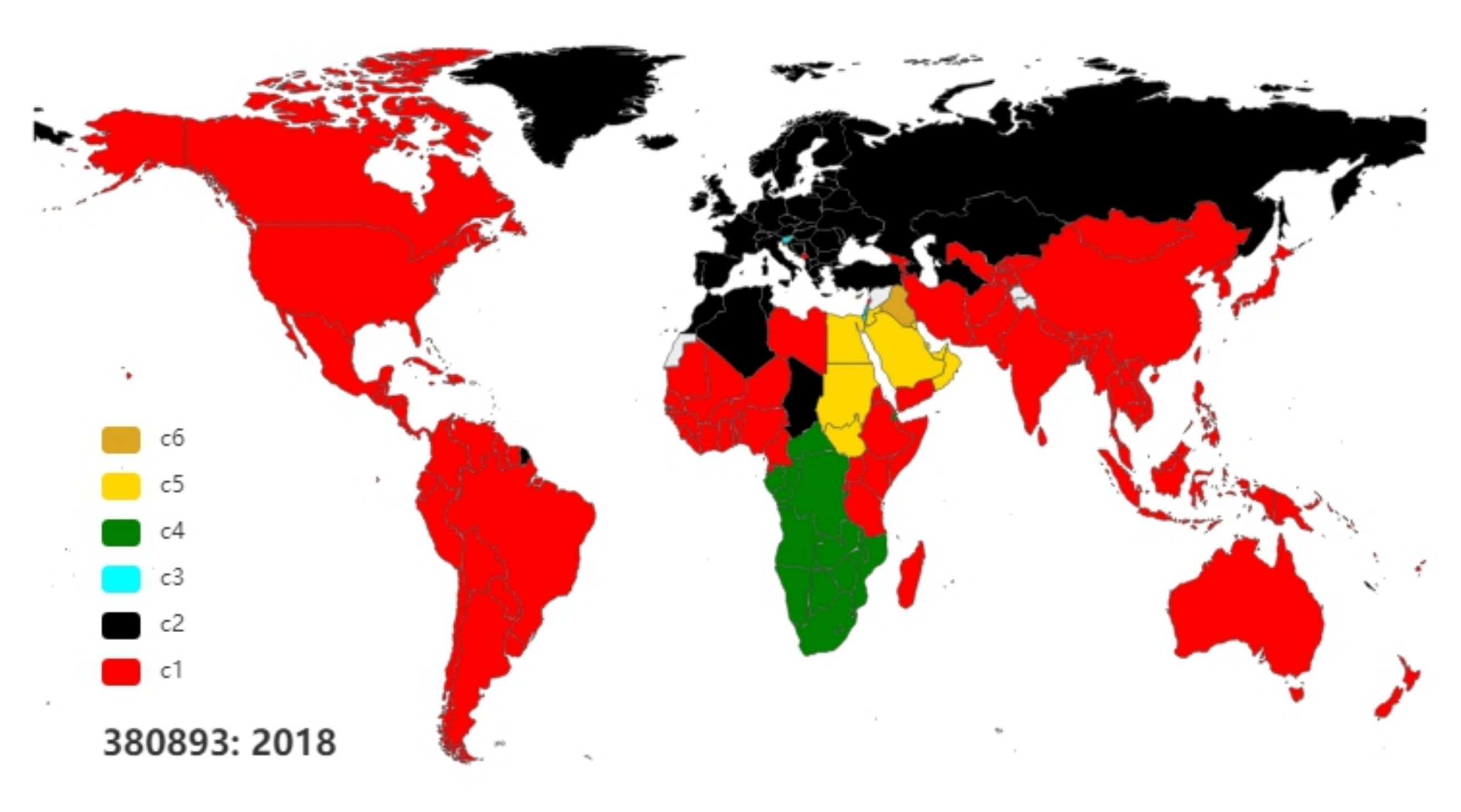}
    \caption{Community evolution of the undirected iPTNs of herbicides (380893) from 2007 to 2018.}
    \label{Fig:iPTN:undirected:CommunityMap:380893}
\end{figure}

Figure~\ref{Fig:iPTN:undirected:CommunityMap:380893} illustrates the evolution of communities of the undirected iPTNs of herbicides (380893) from 2007 to 2018. It is found that there are at least five communities in each year. 
The three economies of the NAFTA always belong to the same community, which often contains a few small economies in the Caribbean. In contrast, other economies in Latin America and the Caribbean usually form or belong to other communities, showing the similar regional pattern as in the international fungicide trade networks. In 2018, most of the economies on the American Continent form a unique community. 
In all the years, most economies in Europe belong to a community, and most economies in Asia also belong to a community. Japan and South Korea sometimes belong to the north America community or the Europe community. Economies in the Middle East more likely belong to the Europe community before 2012 and form their own community afterwards. 
Economies in Oceania belong to the Asia community in most years, except that they form a separate community in 2007 and belong to the South America community in 2009.

There are about 60 economies in Africa, which belong to three or more communities. The economies in each of these communities are not randomly distributed in Africa; rather, they are clustered in a localized way, suggesting that geographical closeness is the dominating deriving force in the formation of trade communities. Some economies are more likely to belong to the Asia community, while some other economies tend to join the Europe community. This pattern reflects their corresponding stable economic relationship. The Southern African Development Community and other geographically close economies often belong to a same, separate community in most years.

\begin{figure}[!ht]
    \centering
    \includegraphics[width=0.321\linewidth]{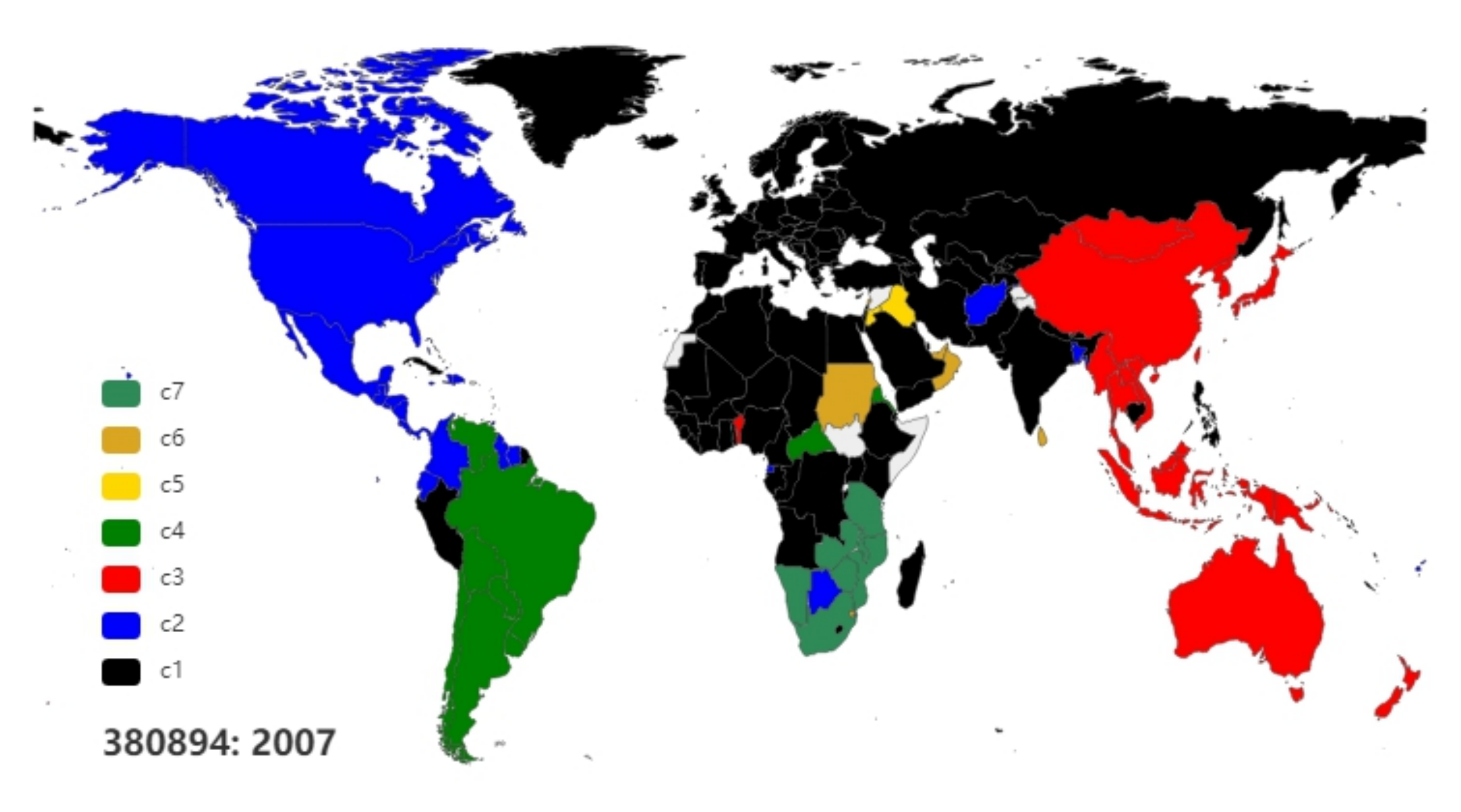}
    \includegraphics[width=0.321\linewidth]{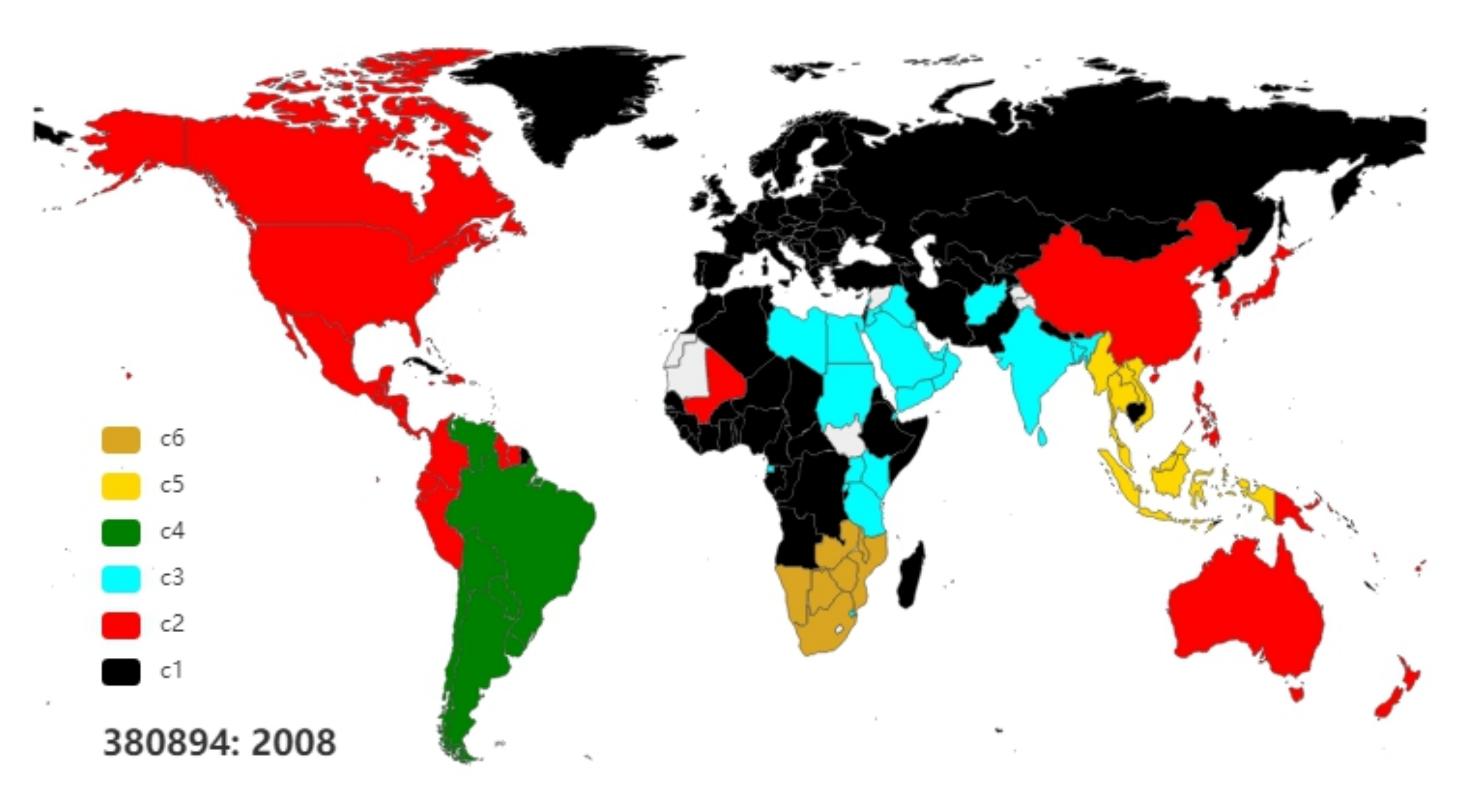}
    \includegraphics[width=0.321\linewidth]{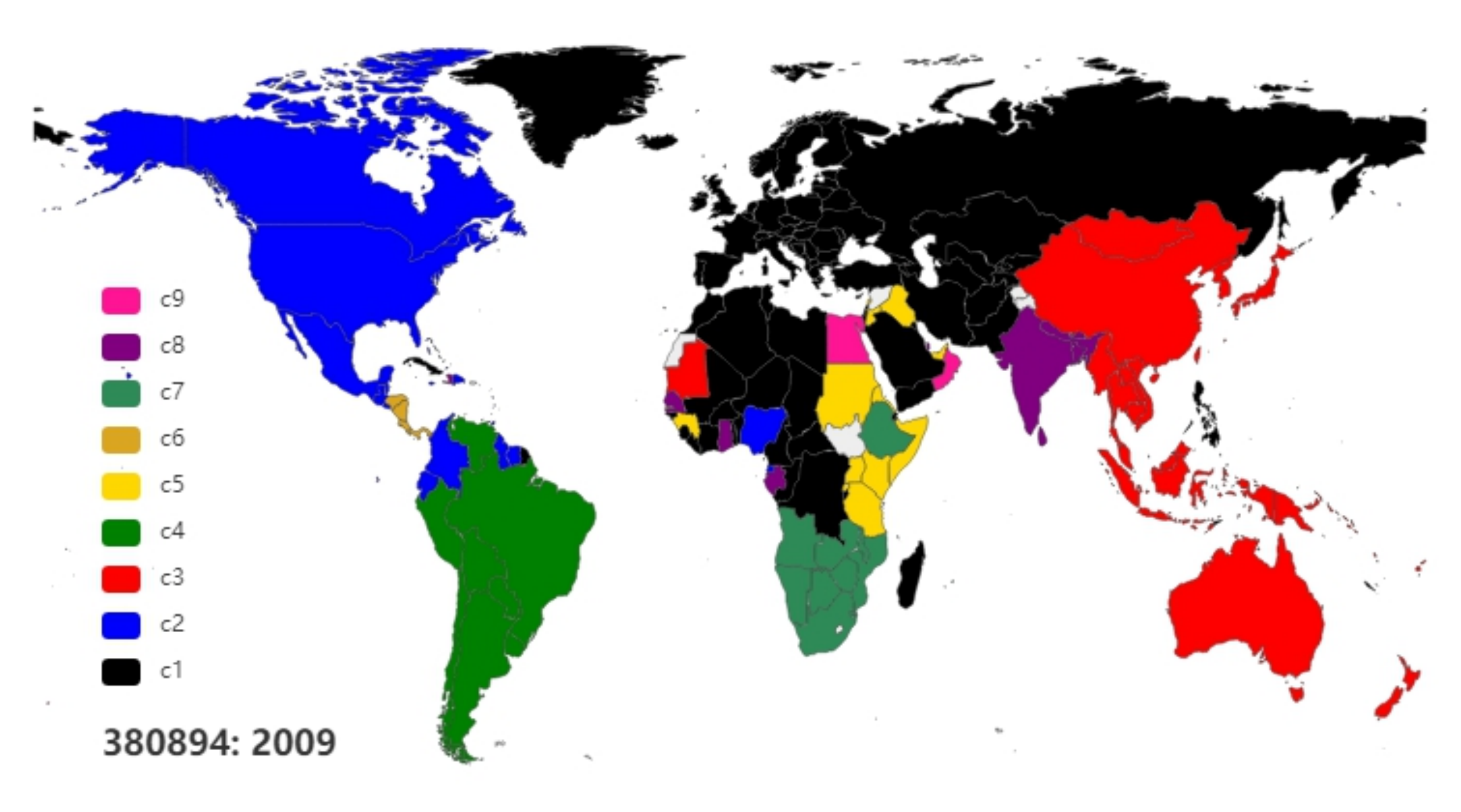}
    \includegraphics[width=0.321\linewidth]{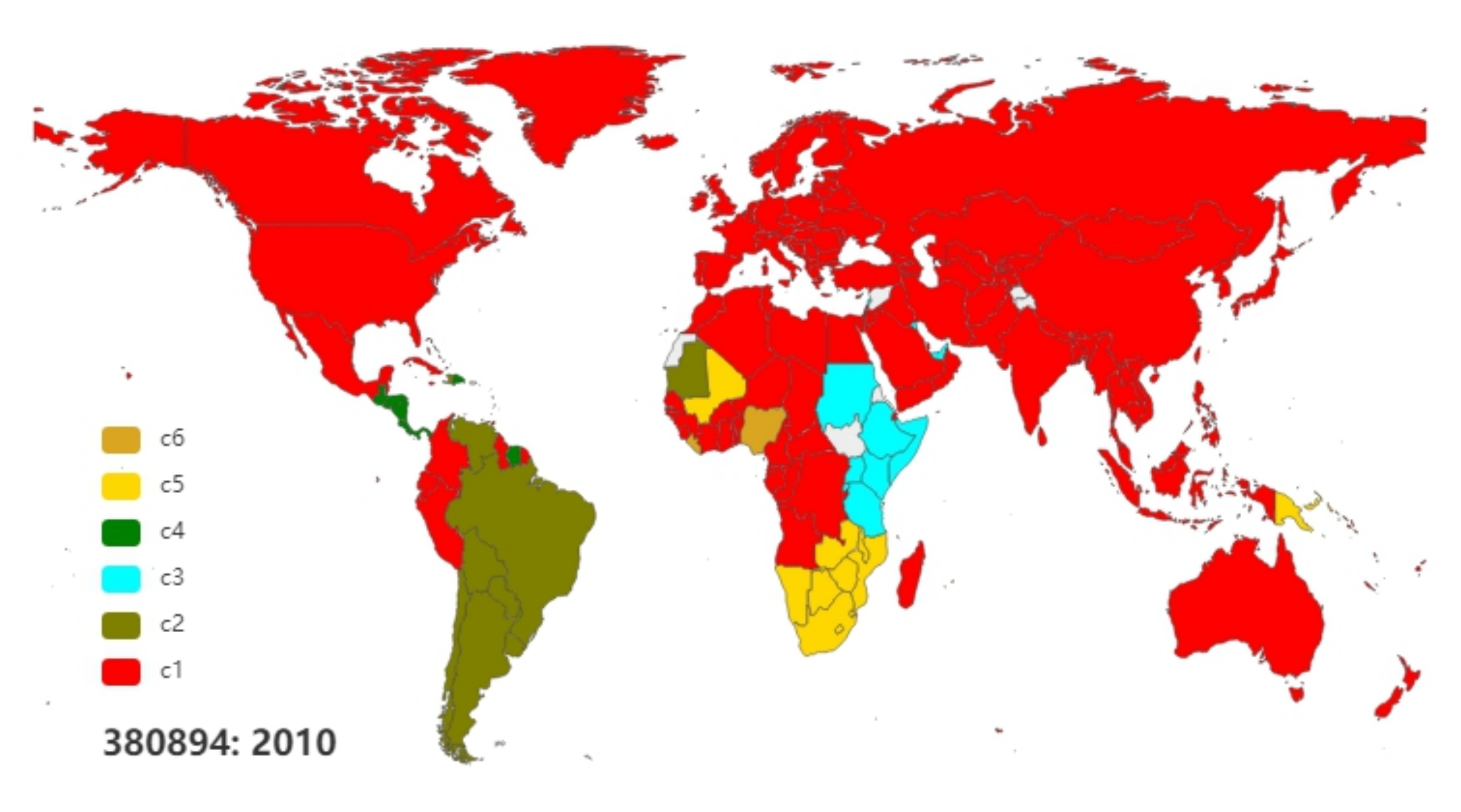}
    \includegraphics[width=0.321\linewidth]{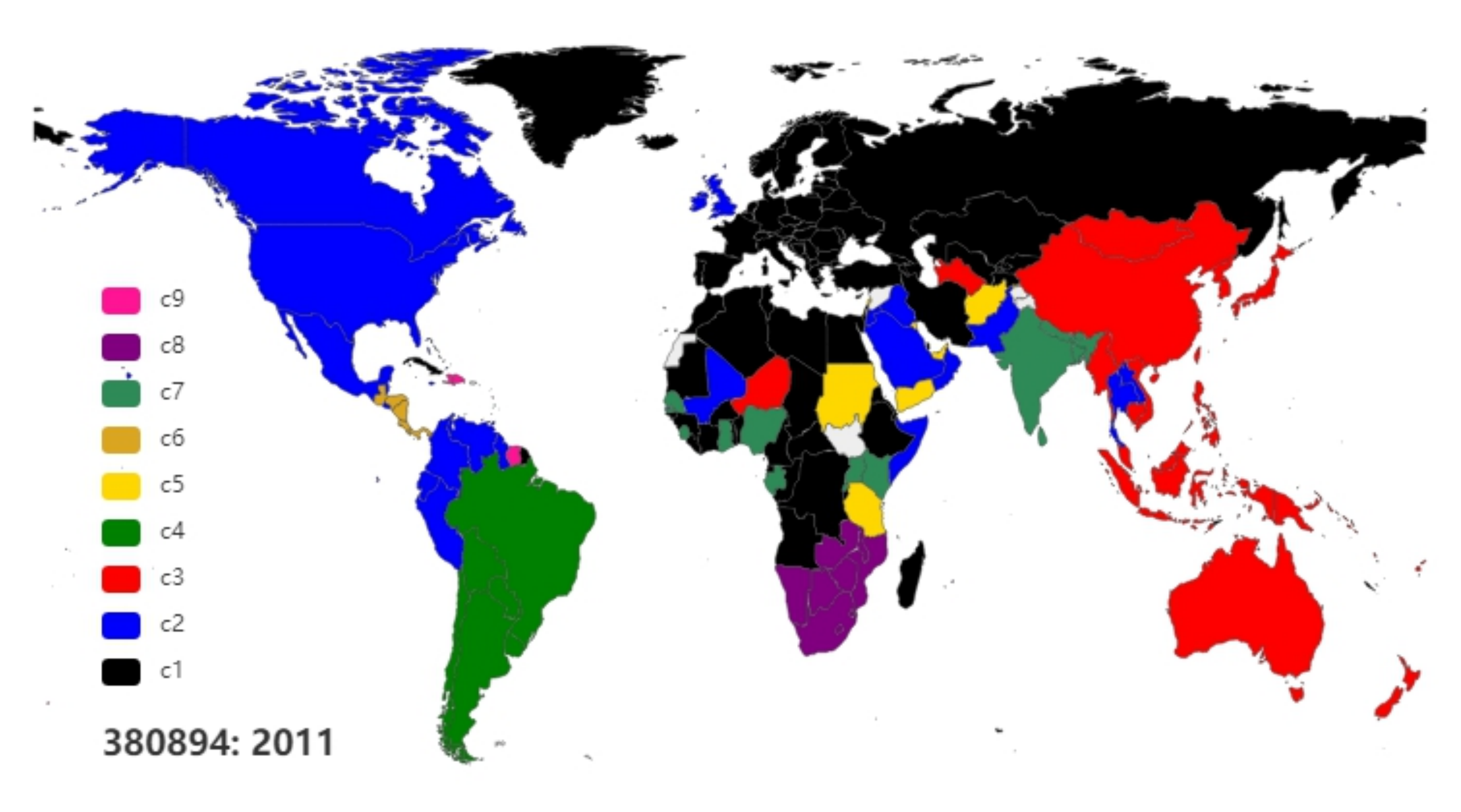}
    \includegraphics[width=0.321\linewidth]{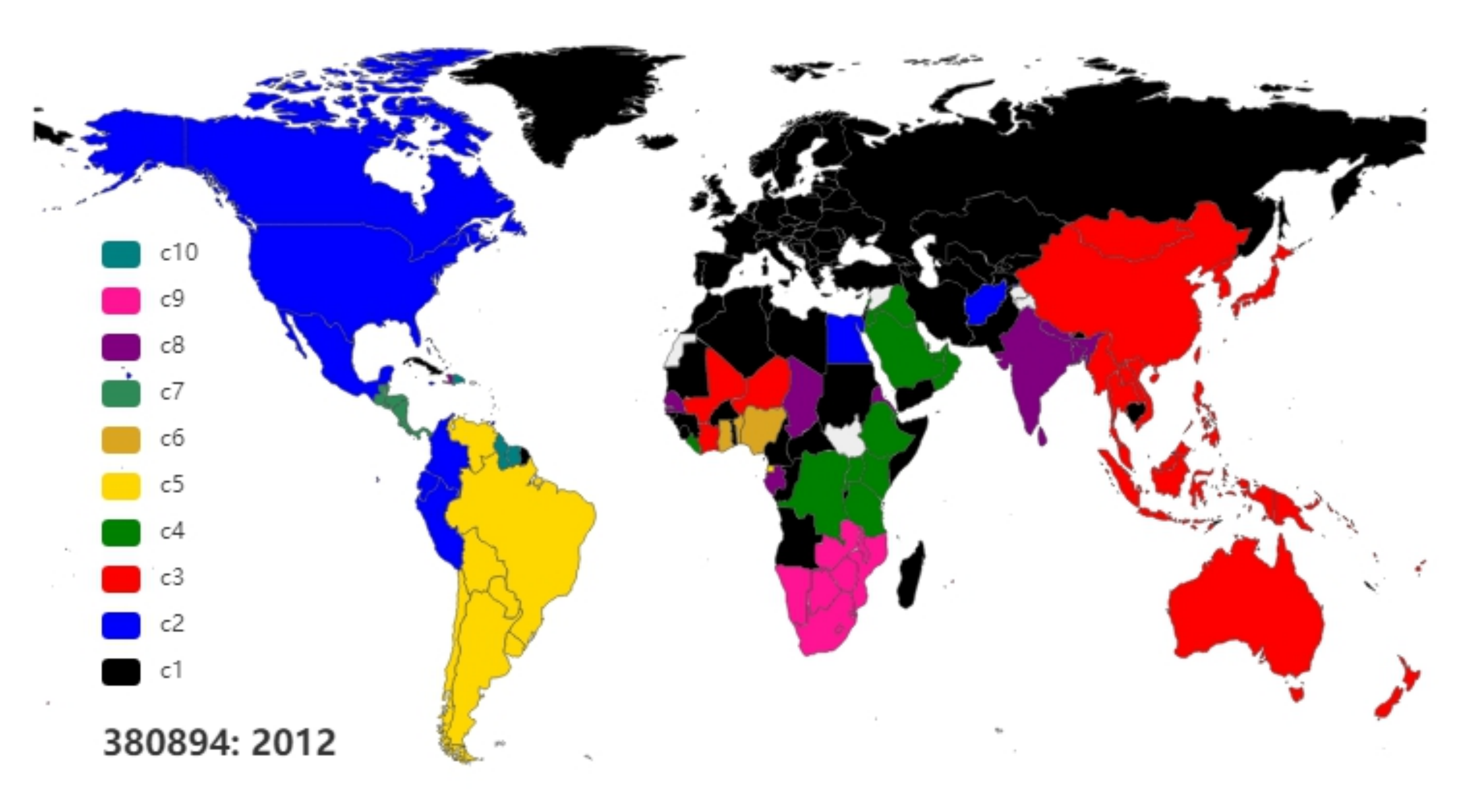}
    \includegraphics[width=0.321\linewidth]{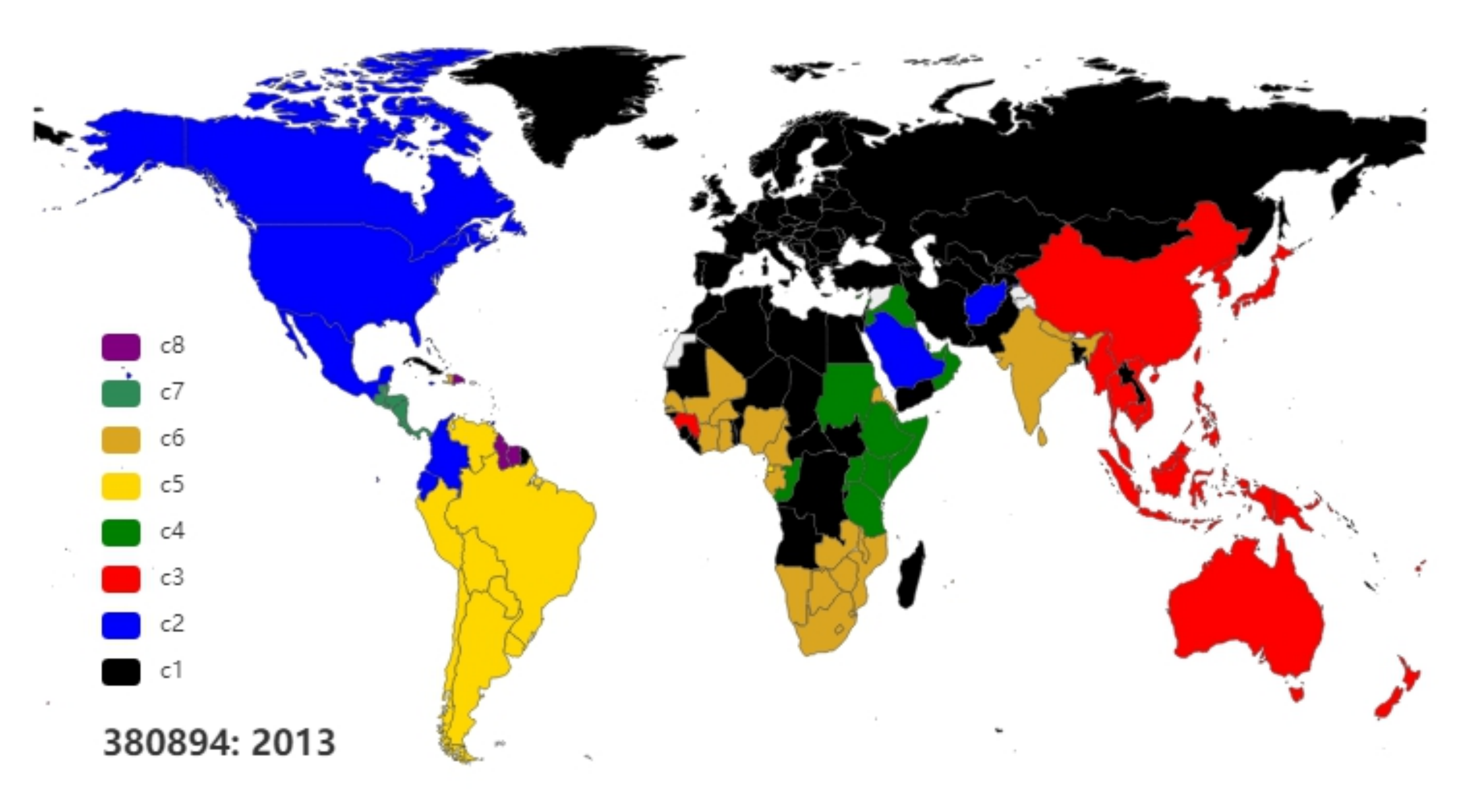}
    \includegraphics[width=0.321\linewidth]{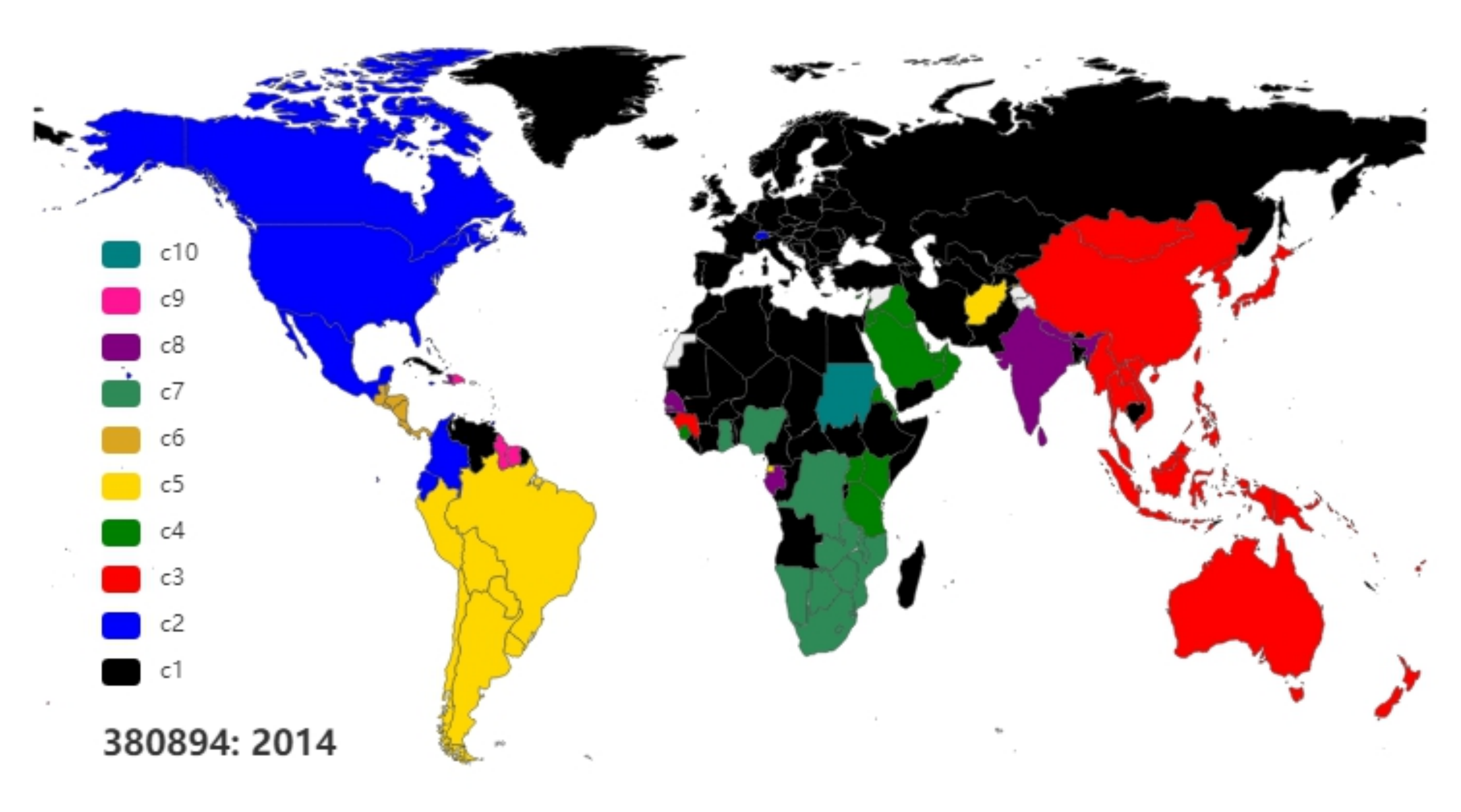}
    \includegraphics[width=0.321\linewidth]{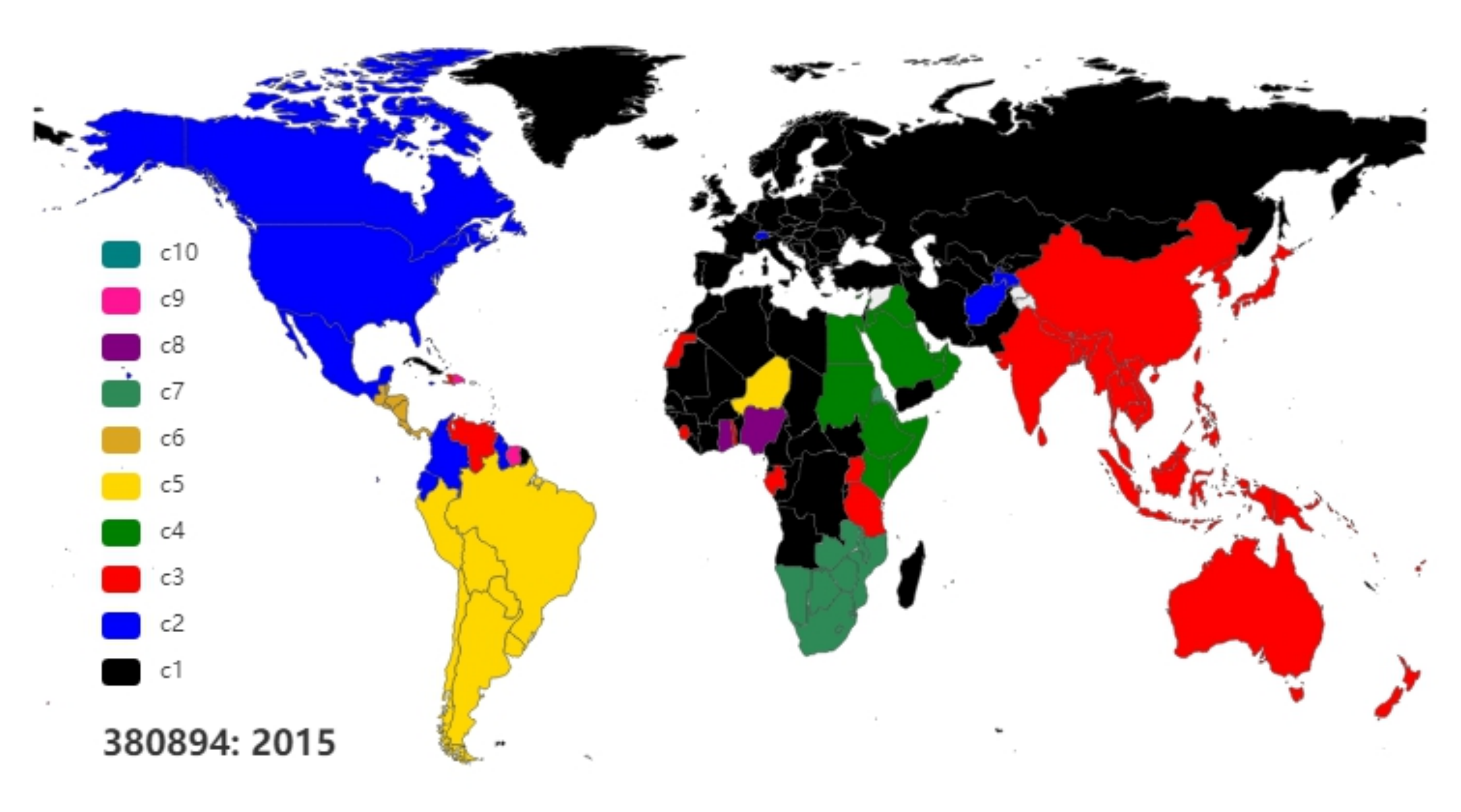}
    \includegraphics[width=0.321\linewidth]{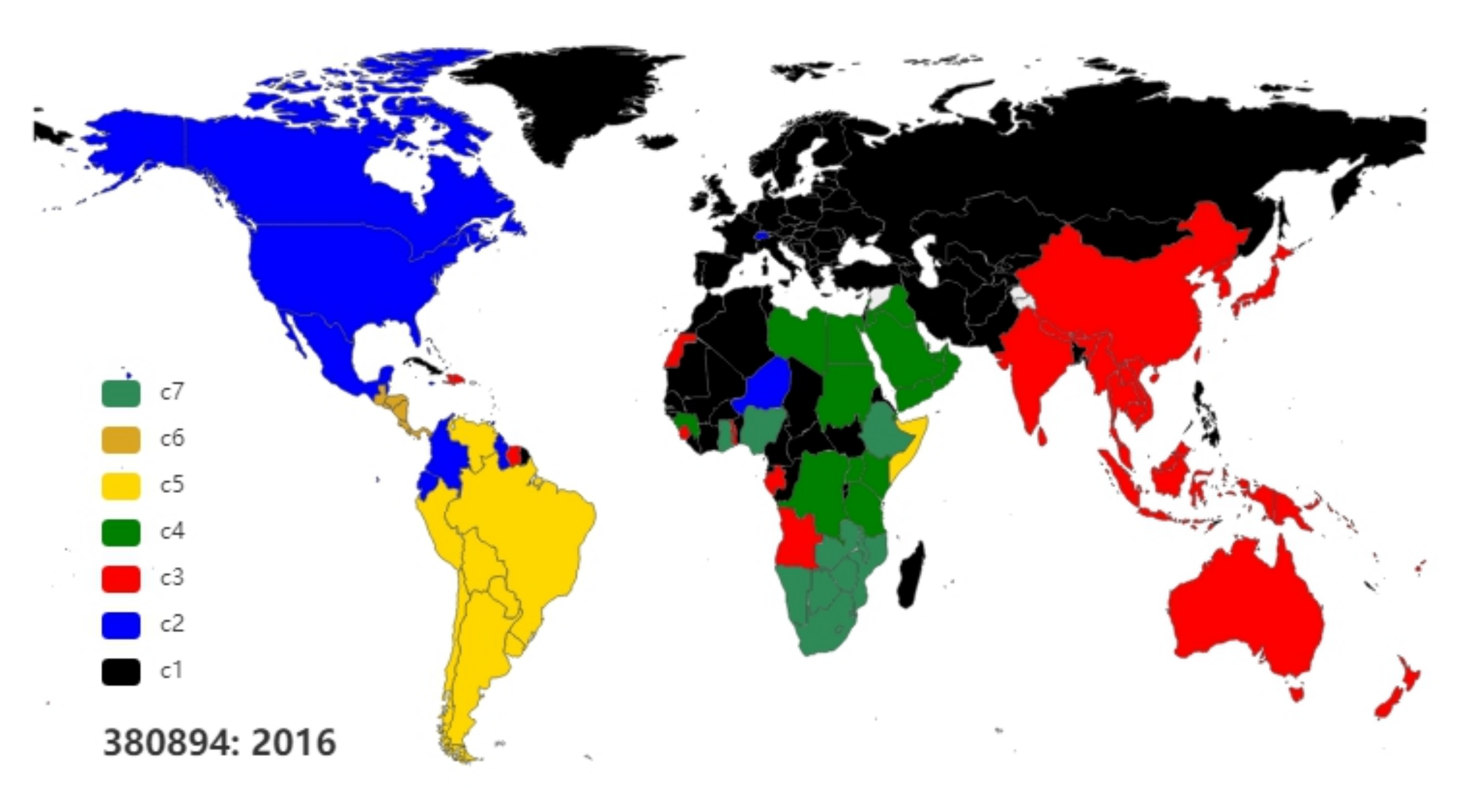}
    \includegraphics[width=0.321\linewidth]{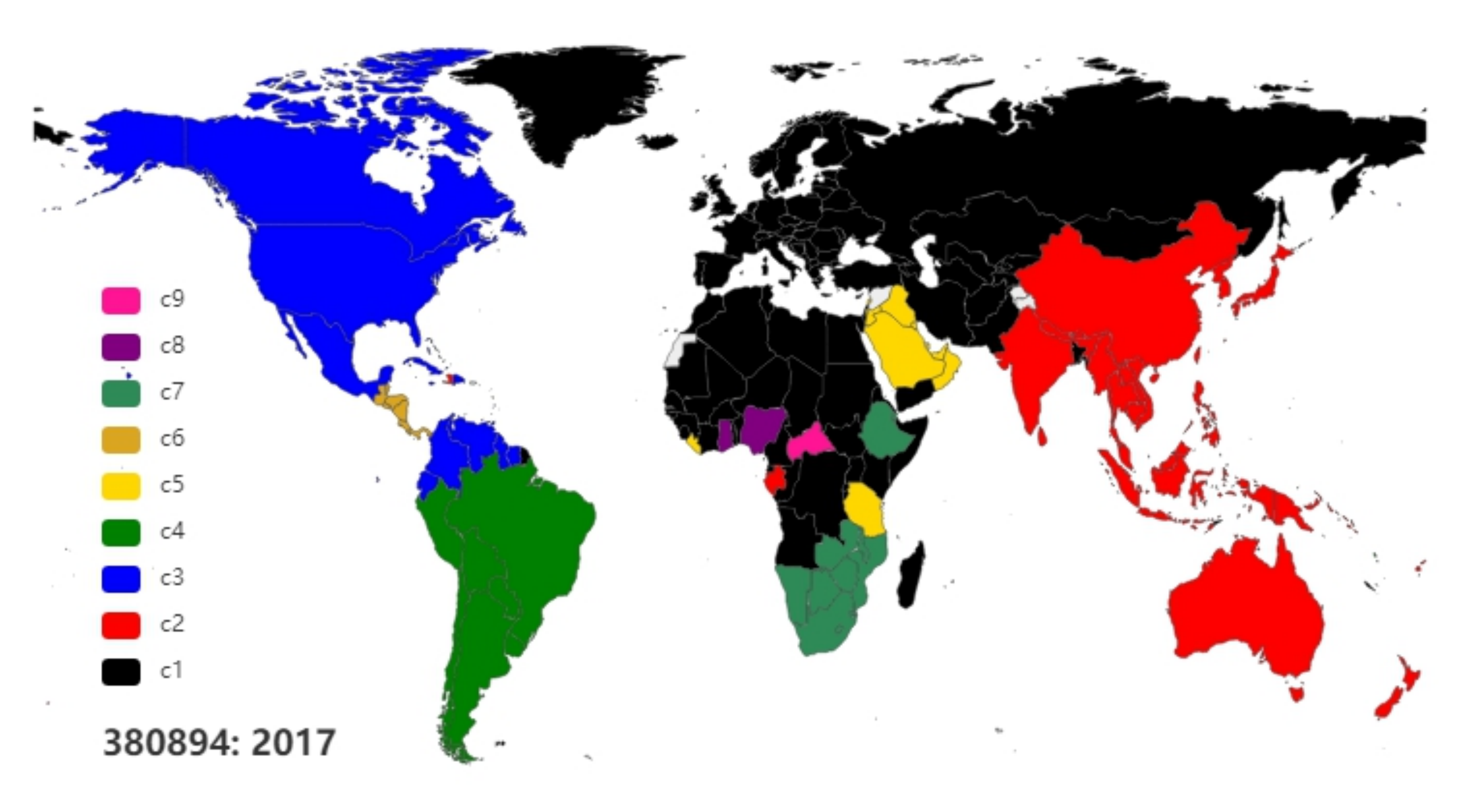}
    \includegraphics[width=0.321\linewidth]{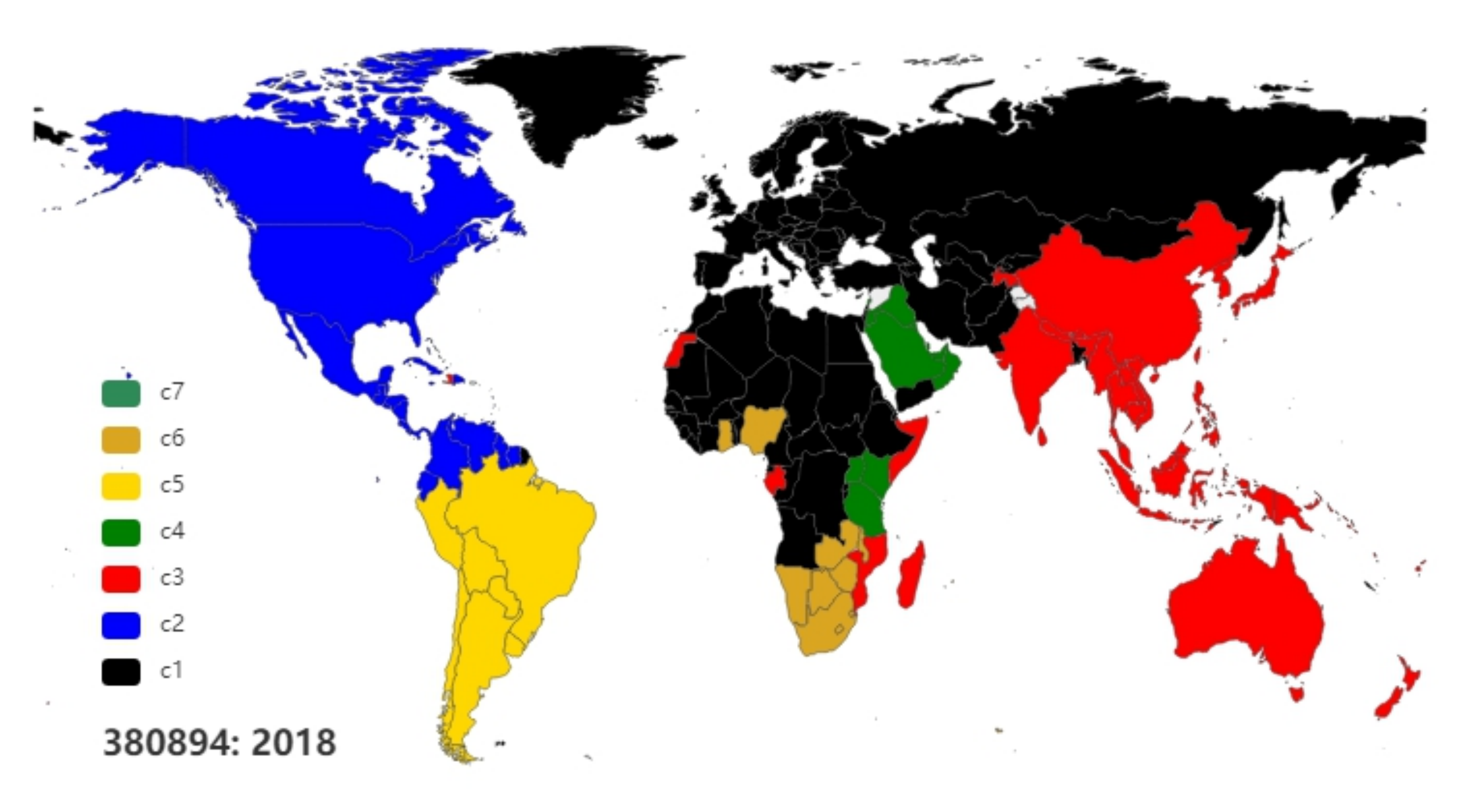}
    \caption{Community evolution of the undirected iPTNs of disinfectants (380894) from 2007 to 2018.}
    \label{Fig:iPTN:undirected:CommunityMap:380894}
\end{figure}

Figure~\ref{Fig:iPTN:undirected:CommunityMap:380894} illustrates the evolution of communities of the undirected iPTNs of disinfectants (380894) from 2007 to 2018. These networks contain 6 to 10 communities and the division of communities in different years is quite similar except for the network in 2010. The first community $C_1$ has 166 economies and contains most economies in North America, Europe, Asia and Oceania. The only exception is Papua New Guinea, the second largest country in Oceania, which belongs to the fifth community $C_5$. The second community $C_2$ has 9 economies (Argentina, Bolivia, Brazil, Chile, Haiti, Paraguay, Uruguay, and Venezuela in South America and Mauritania in Northwest Africa). The third community $C_3$ has 13 economies (Ethiopia, Kenya, Rwanda, Somalia, Sudan, Uganda, and Tanzania in Eastern Africa, Bahrain, Kuwait, Lebanon, Syrian, and United Arab Emirates within the Middle East region, and Anguilla in Eastern Caribbean). The fourth community $C_4$ has 8 economies (Costa Rica, El Salvador, Guatemala, Honduras, Nicaragua, and Panama in Western Caribbean, Dominican Rep. in Eastern Caribbean, and Suriname in South America). The fifth community has 13 economies (Botswana, Lesotho, Malawi, Mali, Mauritius, Mozambique, Namibia, Saint Helena, South Africa, Zimbabwe, Swaziland, Zambia in Africa and Papua New Guinea in Oceania). The sixth community $C_6$ has three economies (Antigua and Barb. in Eastern Caribbean and Liberia and Nigeria in Western Africa). The African economies distribute in five communities (except the fourth community $C_4$), showing that they are less integrated in the economic development.

The patterns of community maps of the Other 12 networks look quite similar. It is found that the economies in North America and South America never unite to form a big community in all years including 2010. Economies in the Caribbean and around may belong to different communities rather than the main part of South America. Most economies in Oceania, East Asia, South Asia and Southeast Asia fall in the same community in most years, except that in 2008 most economies in Southeast Asia form a separate community $C_5$. The relatively large numbers of identified communities in the international disinfectant trade networks are mainly attributed to African economies, which belong to several communities in all years.

\begin{figure}[!ht]
    \centering
    \includegraphics[width=0.321\linewidth]{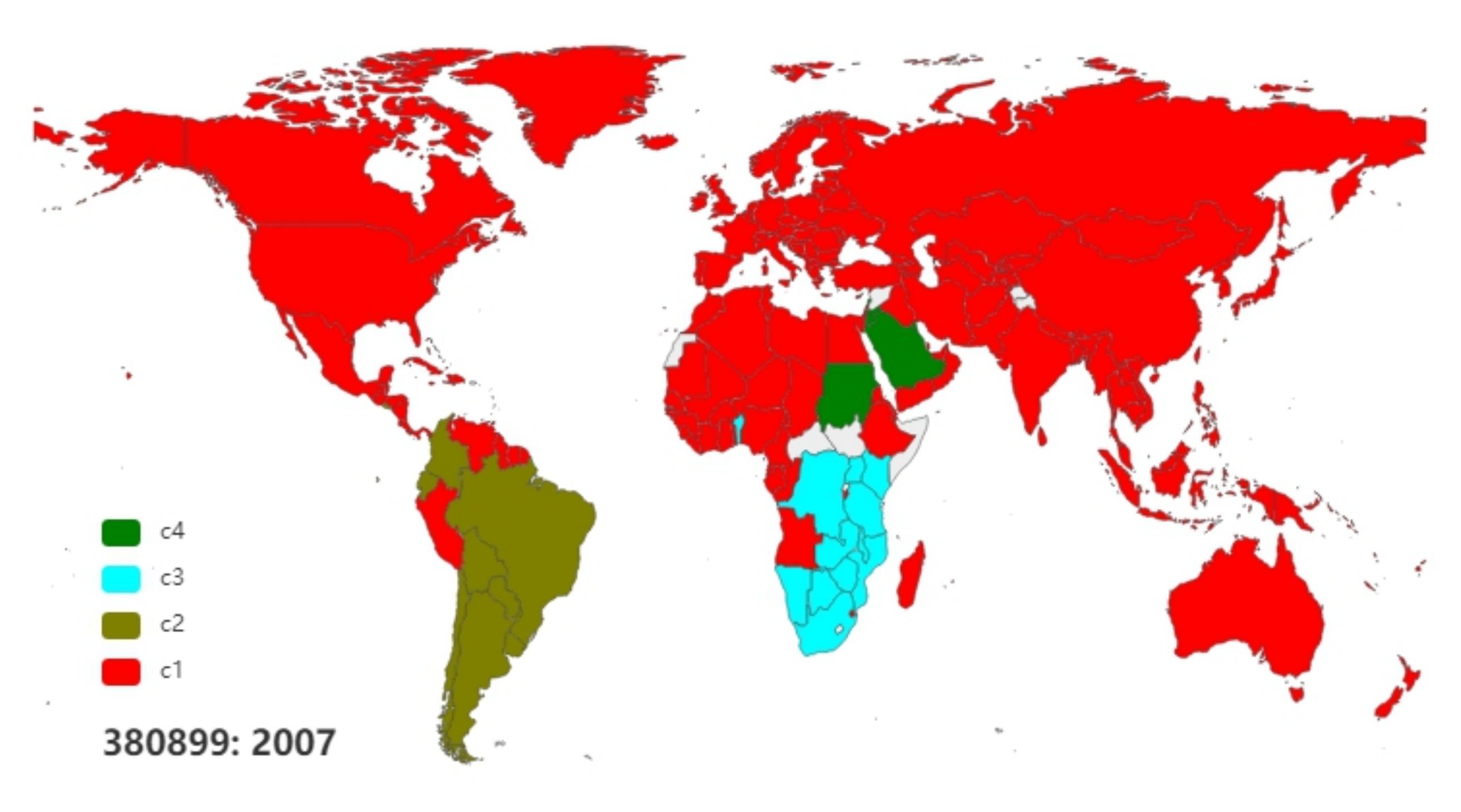}
    \includegraphics[width=0.321\linewidth]{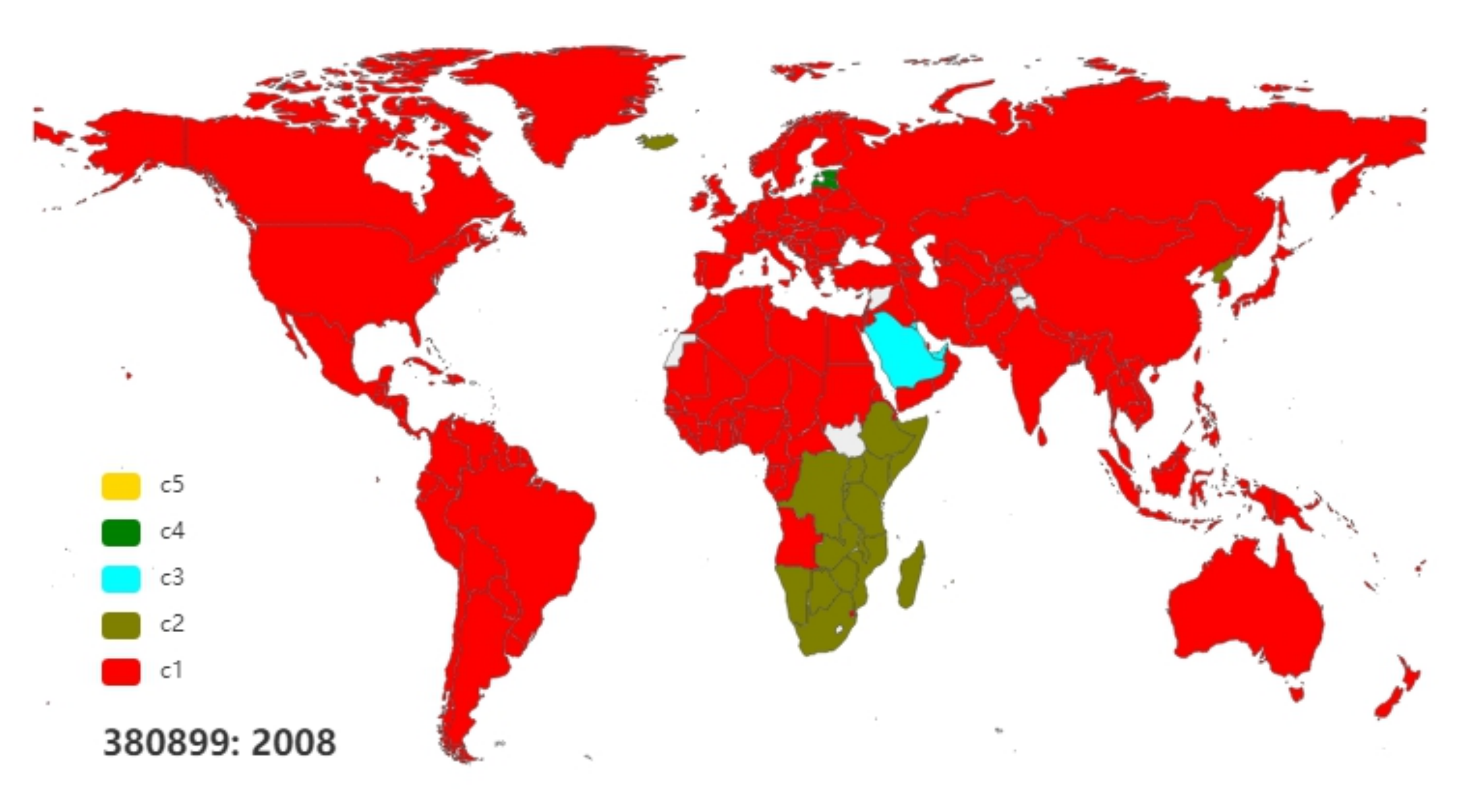}
    \includegraphics[width=0.321\linewidth]{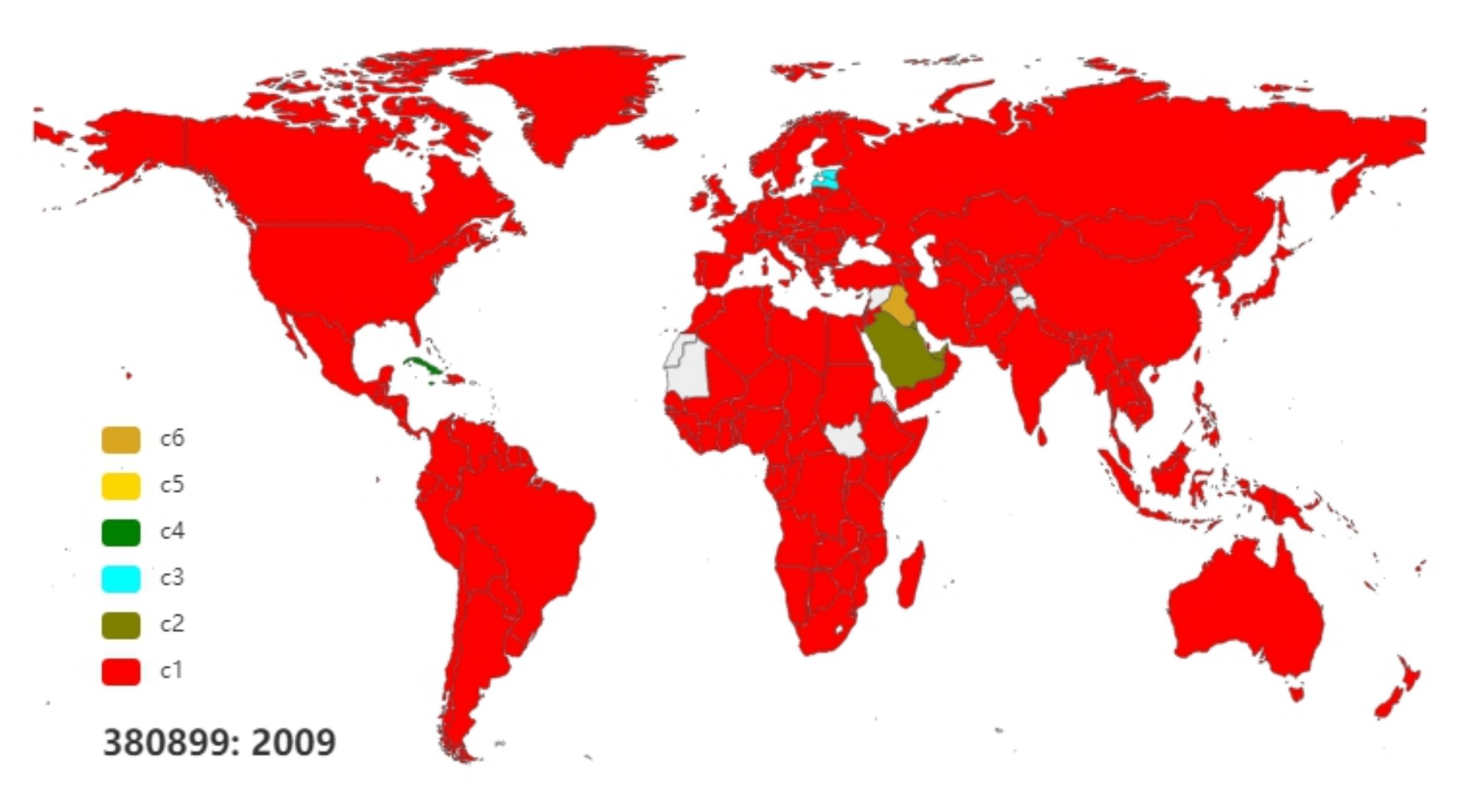}
    \includegraphics[width=0.321\linewidth]{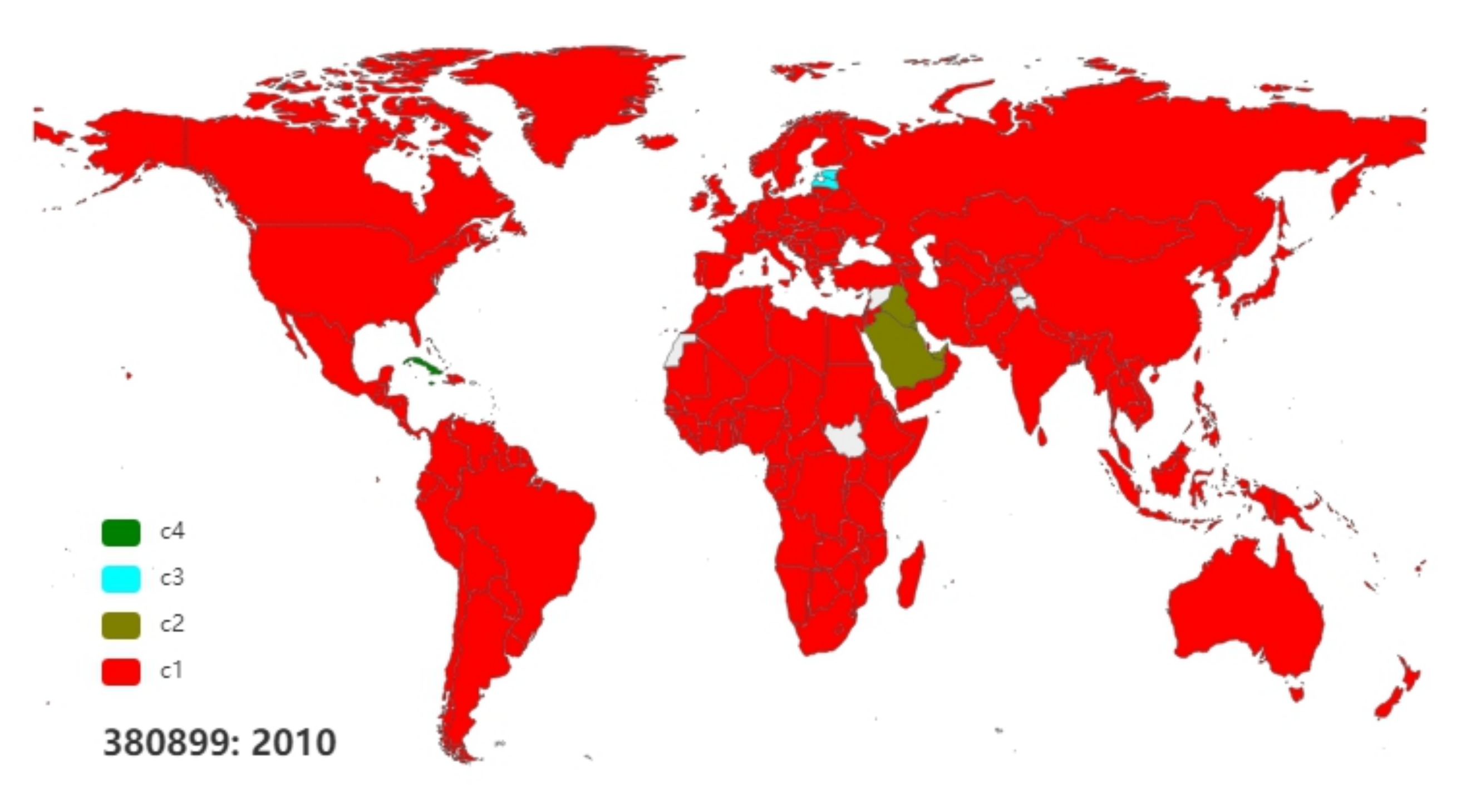}
    \includegraphics[width=0.321\linewidth]{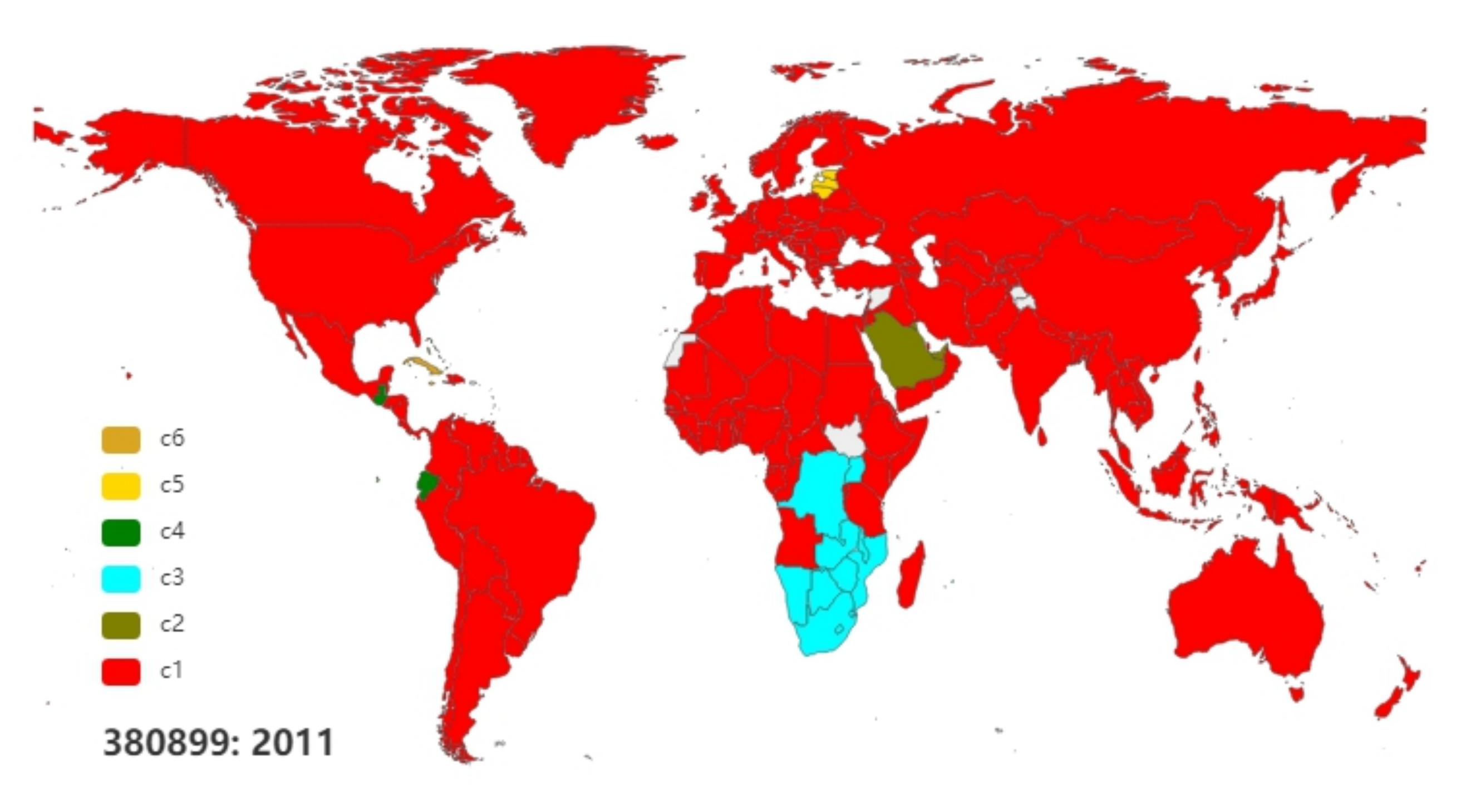}
    \includegraphics[width=0.321\linewidth]{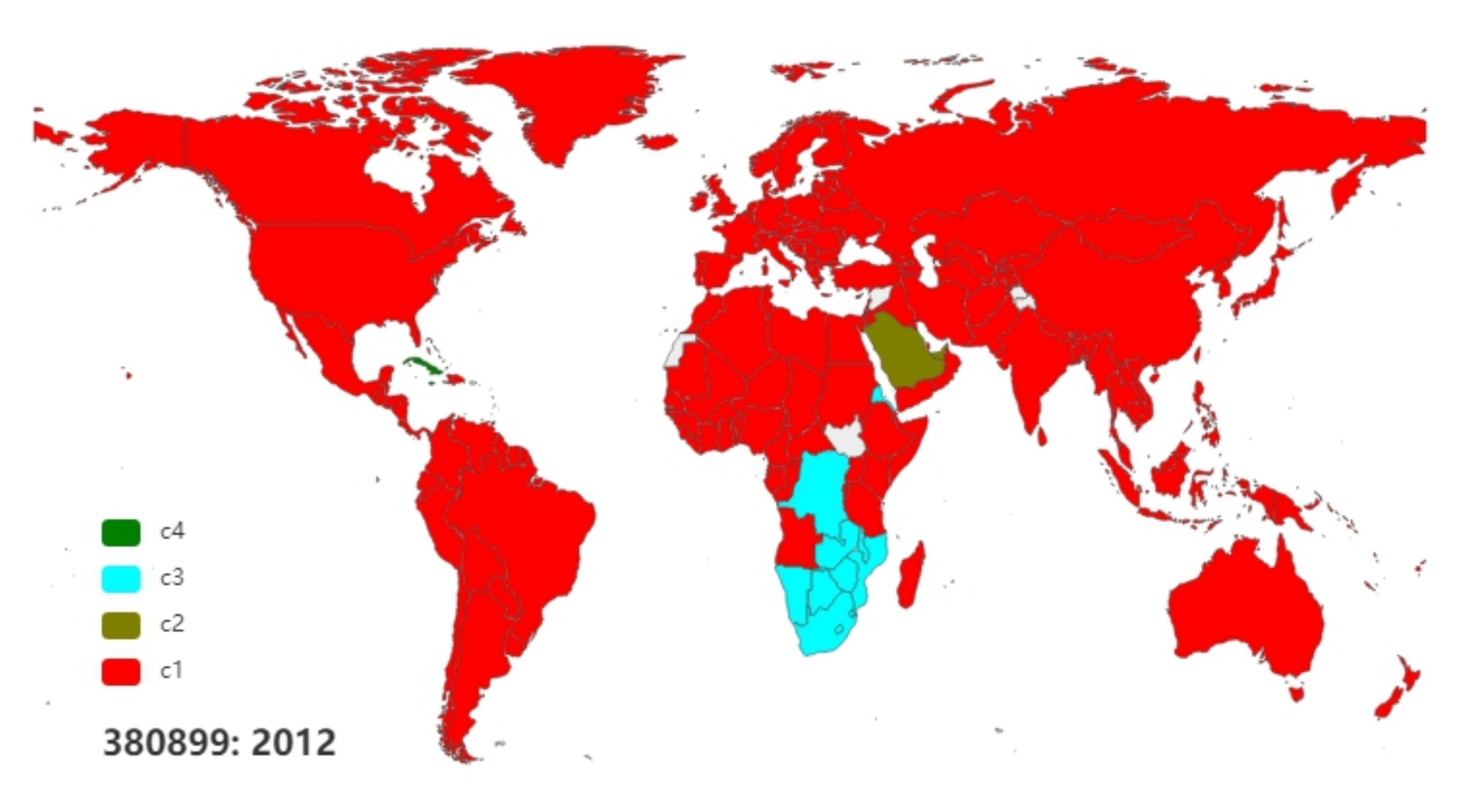}
    \includegraphics[width=0.321\linewidth]{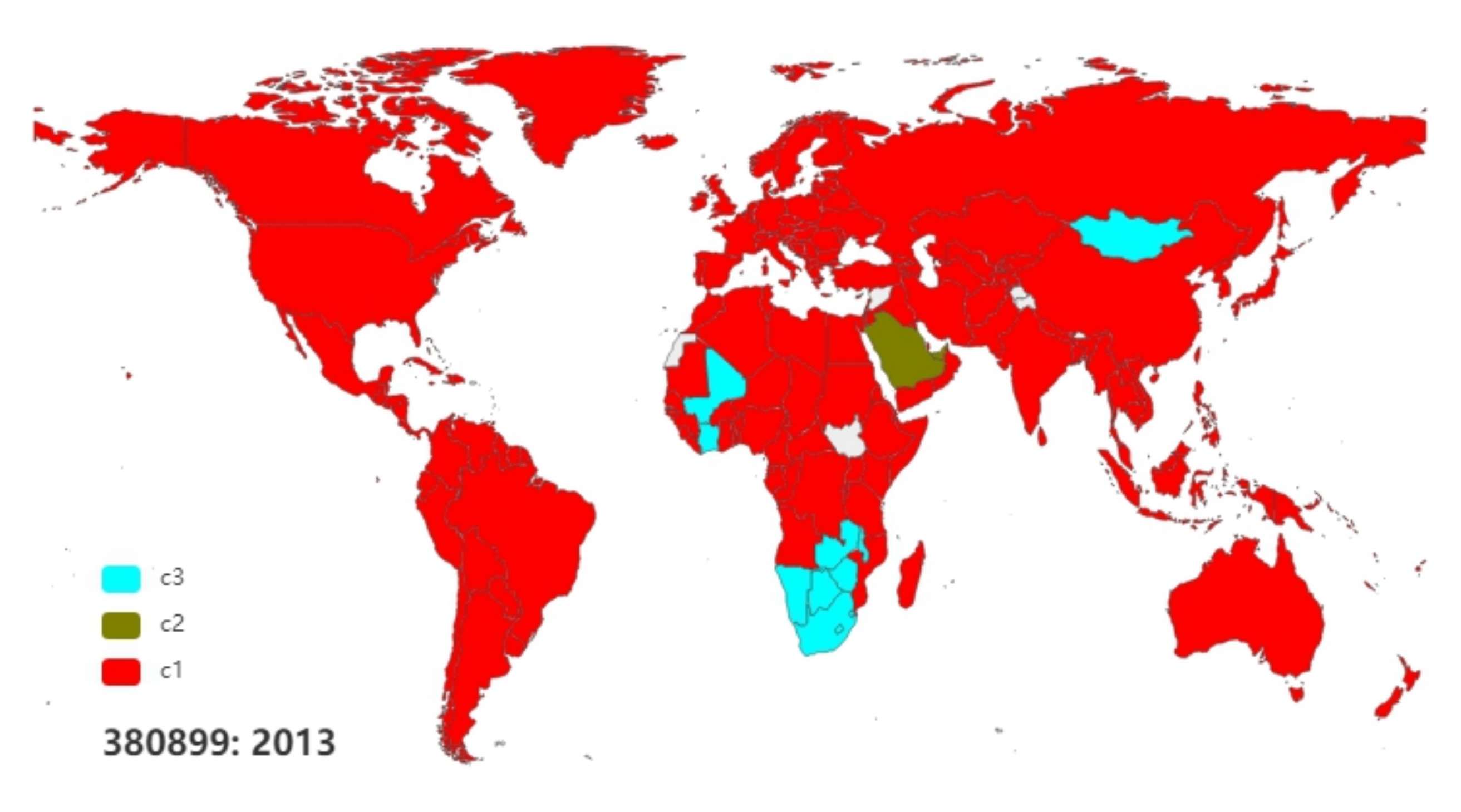}
    \includegraphics[width=0.321\linewidth]{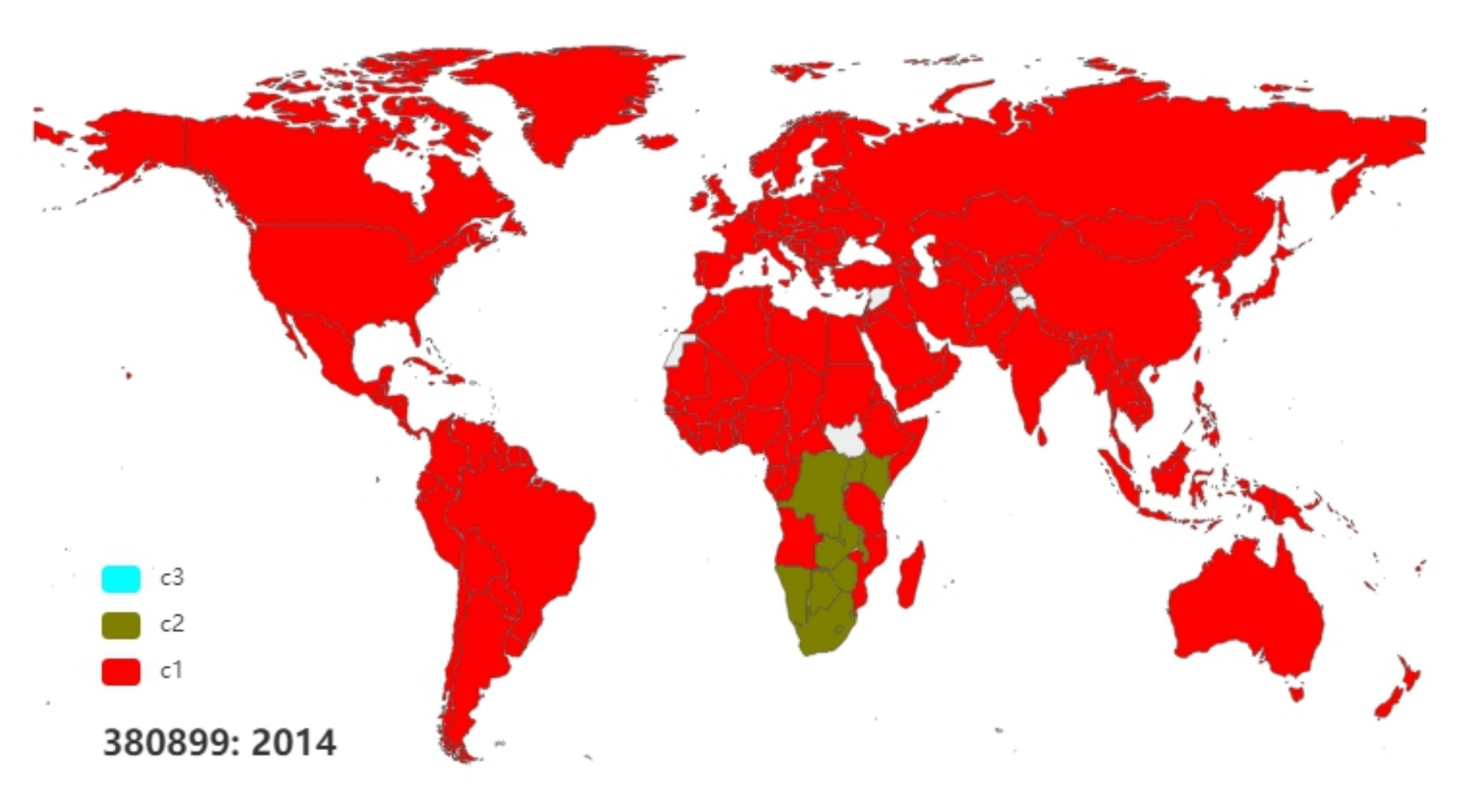}
    \includegraphics[width=0.321\linewidth]{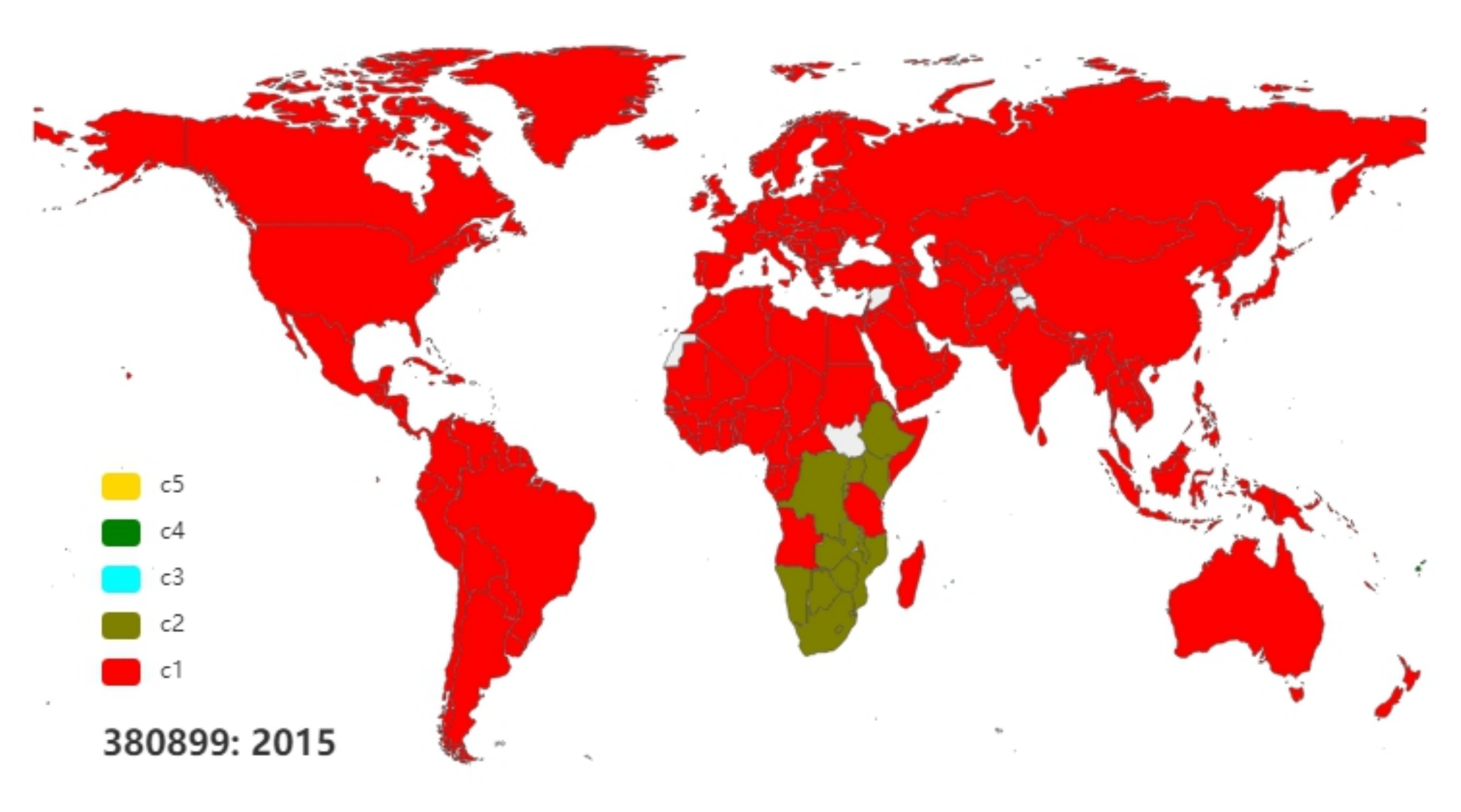}
    \includegraphics[width=0.321\linewidth]{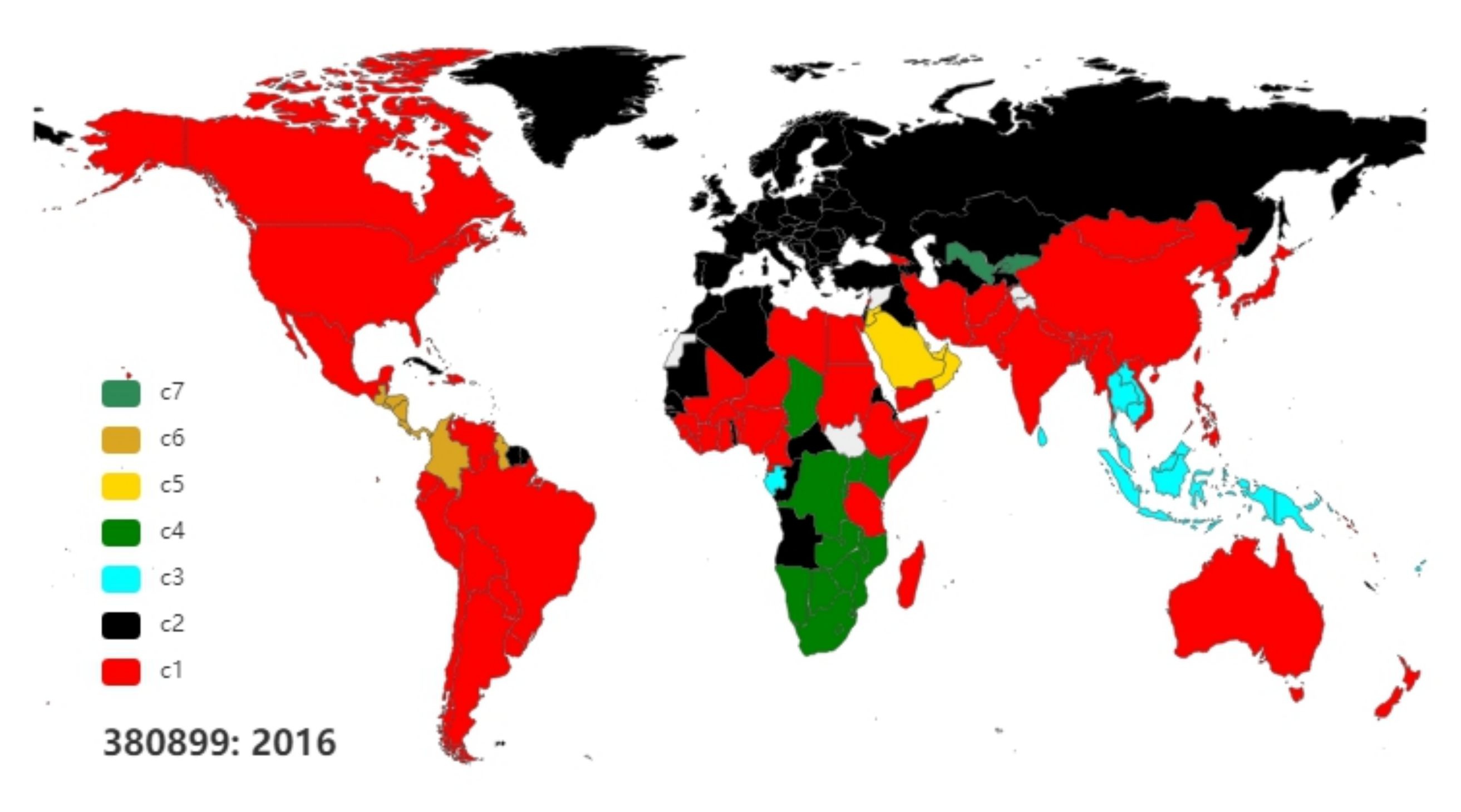}
    \includegraphics[width=0.321\linewidth]{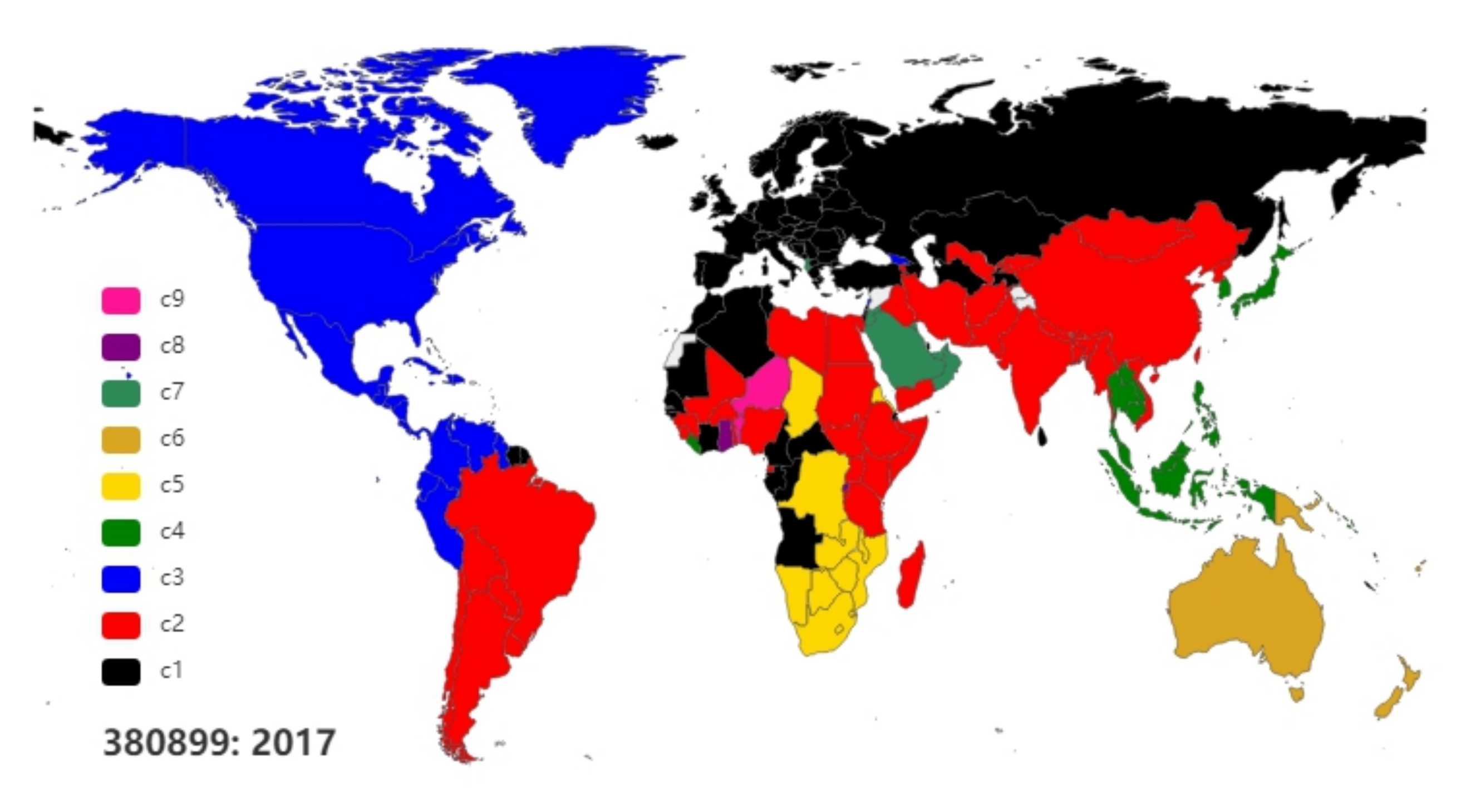}
    \includegraphics[width=0.321\linewidth]{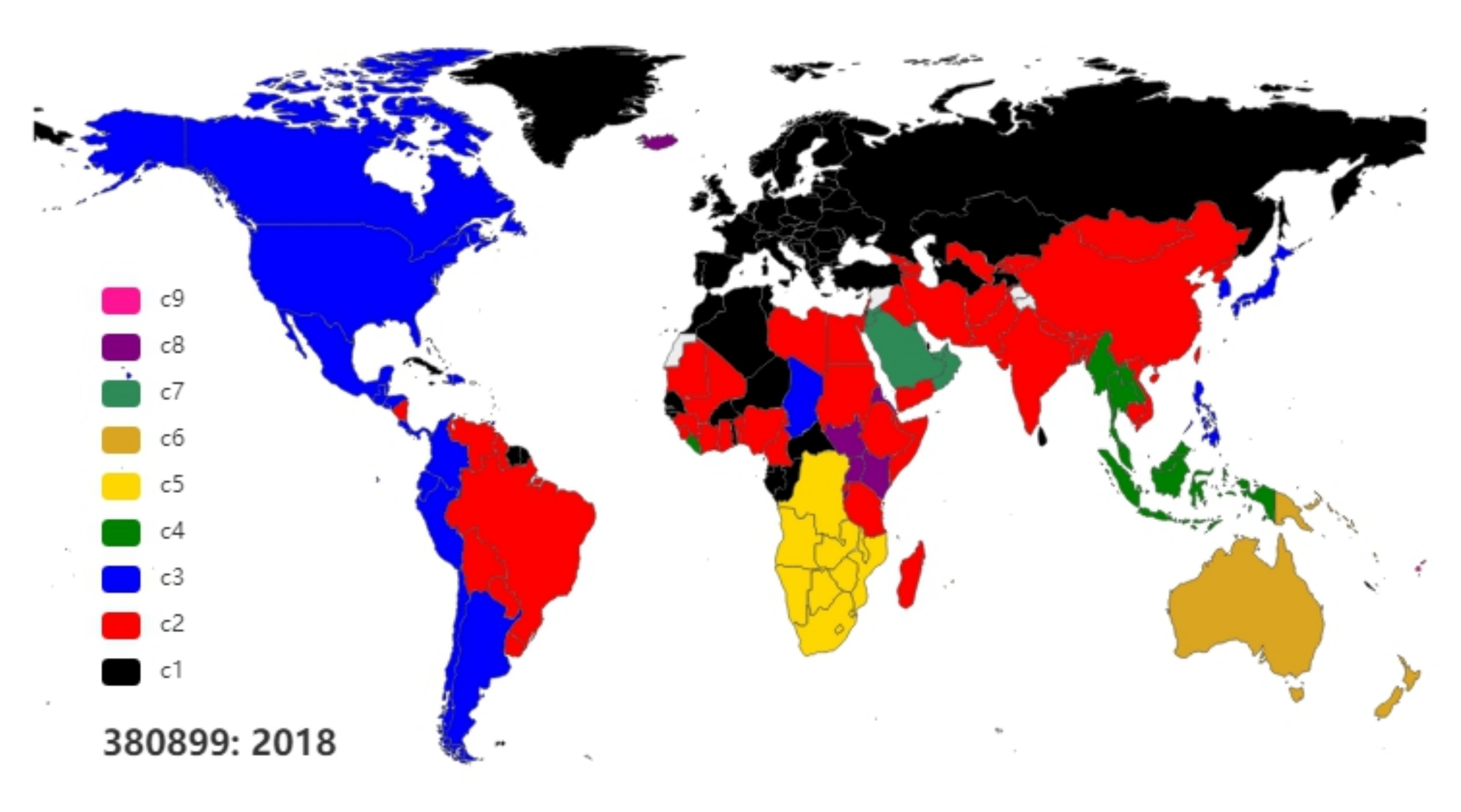}
    \caption{Community evolution of the undirected iPTNs of rodenticides and other similar products (380899) from 2007 to 2018.}
    \label{Fig:iPTN:undirected:CommunityMap:380899}
\end{figure}

Figure~\ref{Fig:iPTN:undirected:CommunityMap:380899} illustrates the evolution of communities of the undirected iPTNs of rodenticides and other similar products (380899) from 2007 to 2018. In the nine networks from 2007 to 2015, we obtain three to five communities and each map is dominated by a biggest community decorated with other communities. South America belongs to the biggest community in all year but 2007. In addition, quite a few economies in Middle East and Africa are not belonging to the biggest community. This observation is consistent with other iPTNs.

The three recent networks in 2016, 2017 and 2018 exhibit different community patterns. These networks have seven or nine communities. The majority of economies in Southern Africa and Europe become two separate communities. In 2016, we observe three separate communities in Southeast Asia, Caribbean and Middle East. In 2017 and 2018, we find a united community in Southeast Asia and Middle East, and a new community in Oceania.

\subsection{Community structure of the directed iPTNs}
\label{S2:iPTN:DiGraph:Community}

We adopt the multi-level Infomap algorithm \cite{Rosvall-Bergstrom-2008-ProcNatlAcadSciUSA} to identify the community partitioning of the five categories of iPTNs that are directed and weighted. We show below the community maps of the iPTNs on the yearly basis.

\begin{figure}[!ht]
    \centering
    \includegraphics[width=0.483\linewidth]{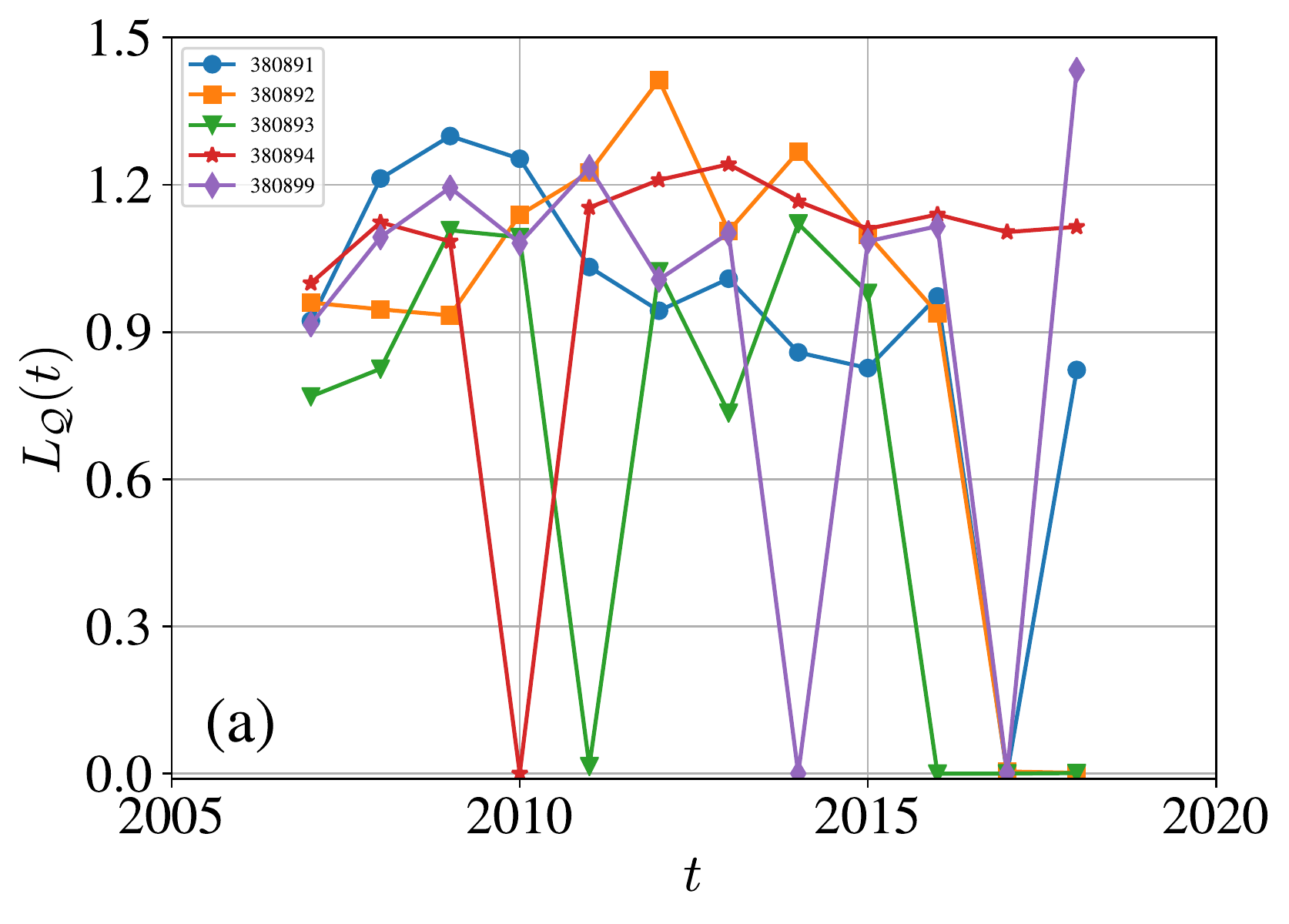}
    \includegraphics[width=0.483\linewidth]{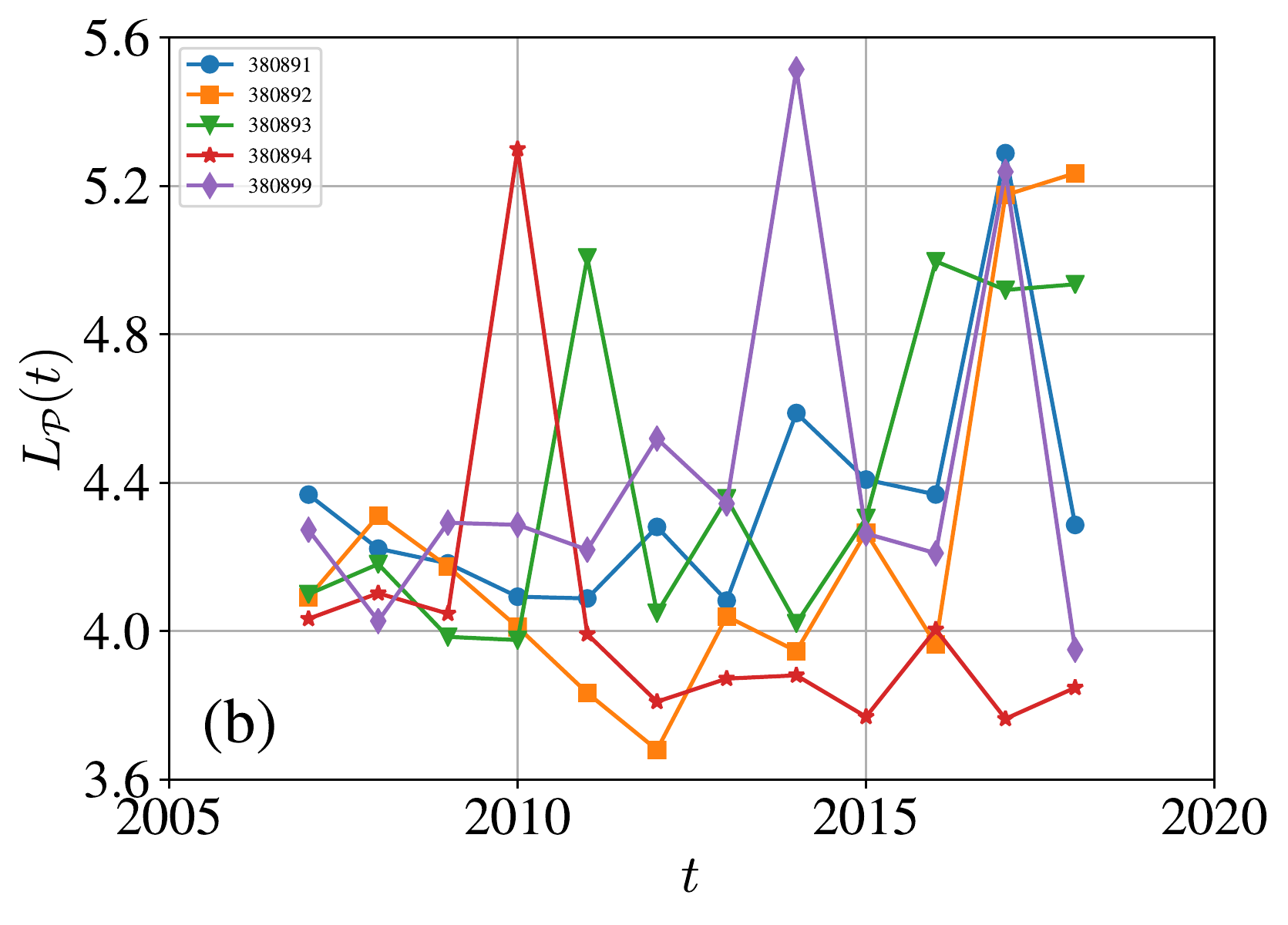}
    \caption{Yearly evolution of the per level codelengths for modules (a) and leaf nodes (b) in the directed international trade networks of insecticides (380891),  fungicides (380892), herbicides (380893),  disinfectants (380894), and rodenticides and other similar products (380899) from 2007 to 2018.}
    \label{Fig:iPTN:directed:Community:CodeLen:t}
\end{figure}

For each directed international pesticide trade network, we obtain the community partitioning ${\mathbf{C}}^{cmd}(t)$. We illustrate the evolution of the per level codelength for communities ($L_{\mathcal{Q}}$) in Fig.~\ref{Fig:iPTN:directed:Community:CodeLen:t}(a). While most of the $L_{\mathcal{Q}}$ values fluctuate around 1, there are 10 points close to 0. Since $L_{\mathcal{Q}}$ represents the codelength of inter-community walks, a small value shows that there are very few identified communities. The evolution of the per level codelength for leaf nodes ($L_{\mathcal{P}}$) is shown in Fig.~\ref{Fig:iPTN:directed:Community:CodeLen:t}(b). Despite that most of the $L_{\mathcal{P}}$ values fluctuate around 4.2, we again observe 10 points above $L_{\mathcal{P}}=4.9$, which correspond to the 10 outlier networks identified in Fig.~\ref{Fig:iPTN:directed:Community:CodeLen:t}(a). Since $L_{\mathcal{P}}$ represents the codelength of intra-community walks, larger communities usually have greater $L_{\mathcal{P}}$ values. Comparing Fig.~\ref{Fig:iPTN:undirected:Community:CodeLen:t} and Fig.~\ref{Fig:iPTN:directed:Community:CodeLen:t}, we find that
\begin{equation}
    L_{\mathcal{P}}^{\rm{dir}}>L_{\mathcal{P}}^{\rm{undir}} ~~{\rm{and}}~~ L_{\mathcal{Q}}^{\rm{dir}}<L_{\mathcal{Q}}^{\rm{undir}}
    \label{Eq:LP:LQ:dir:undir}
\end{equation}
for most of the corresponding pairs of direct iPTNs and undirected iPTNs, where the superscripts ``dir'' and ``undir'' stand respectively for the directed and undirected international trade networks for the same category of pesticides in the same year.

\begin{figure}[!ht]
    \centering
    \includegraphics[width=0.483\linewidth]{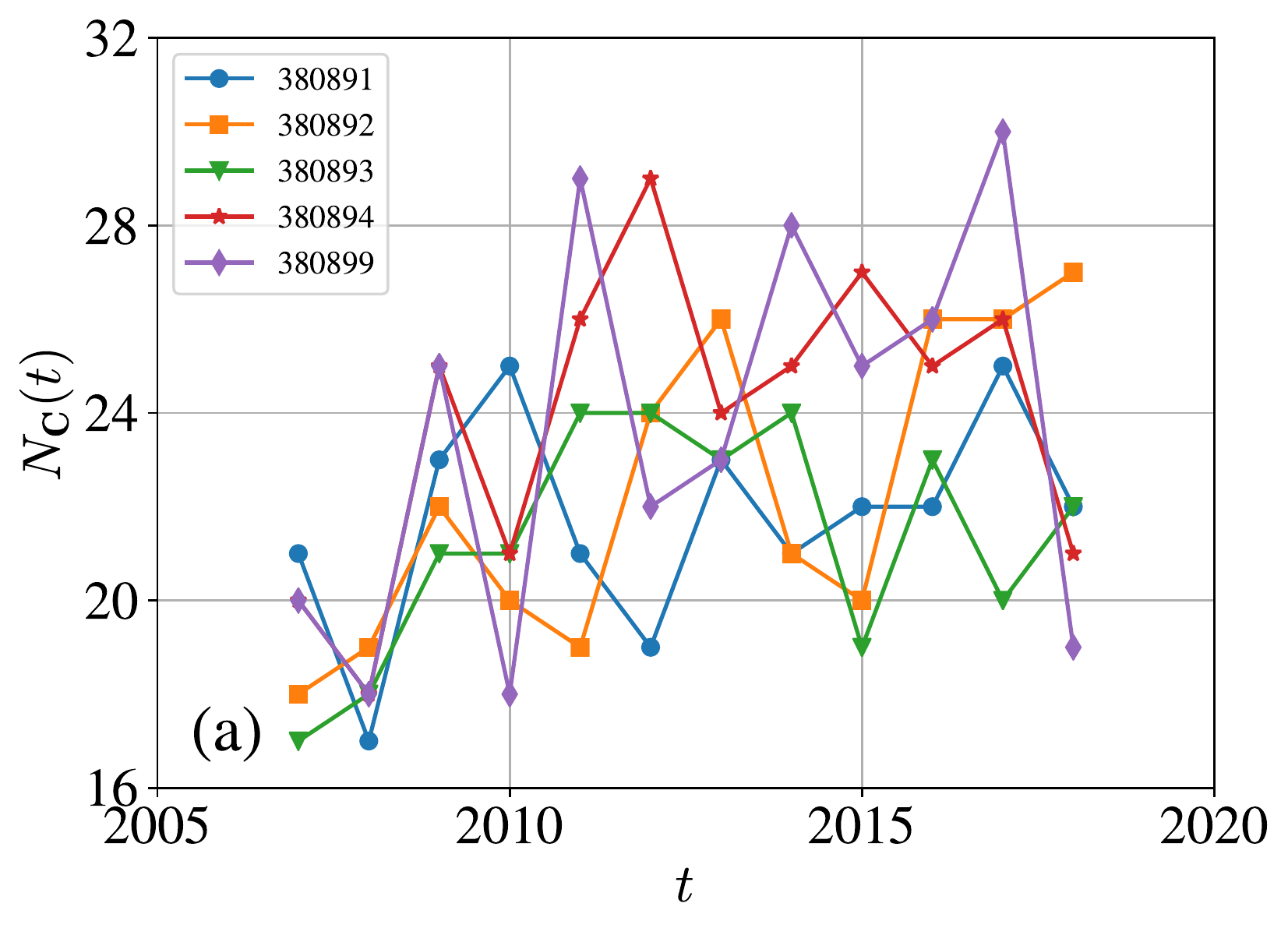}
    \includegraphics[width=0.483\linewidth]{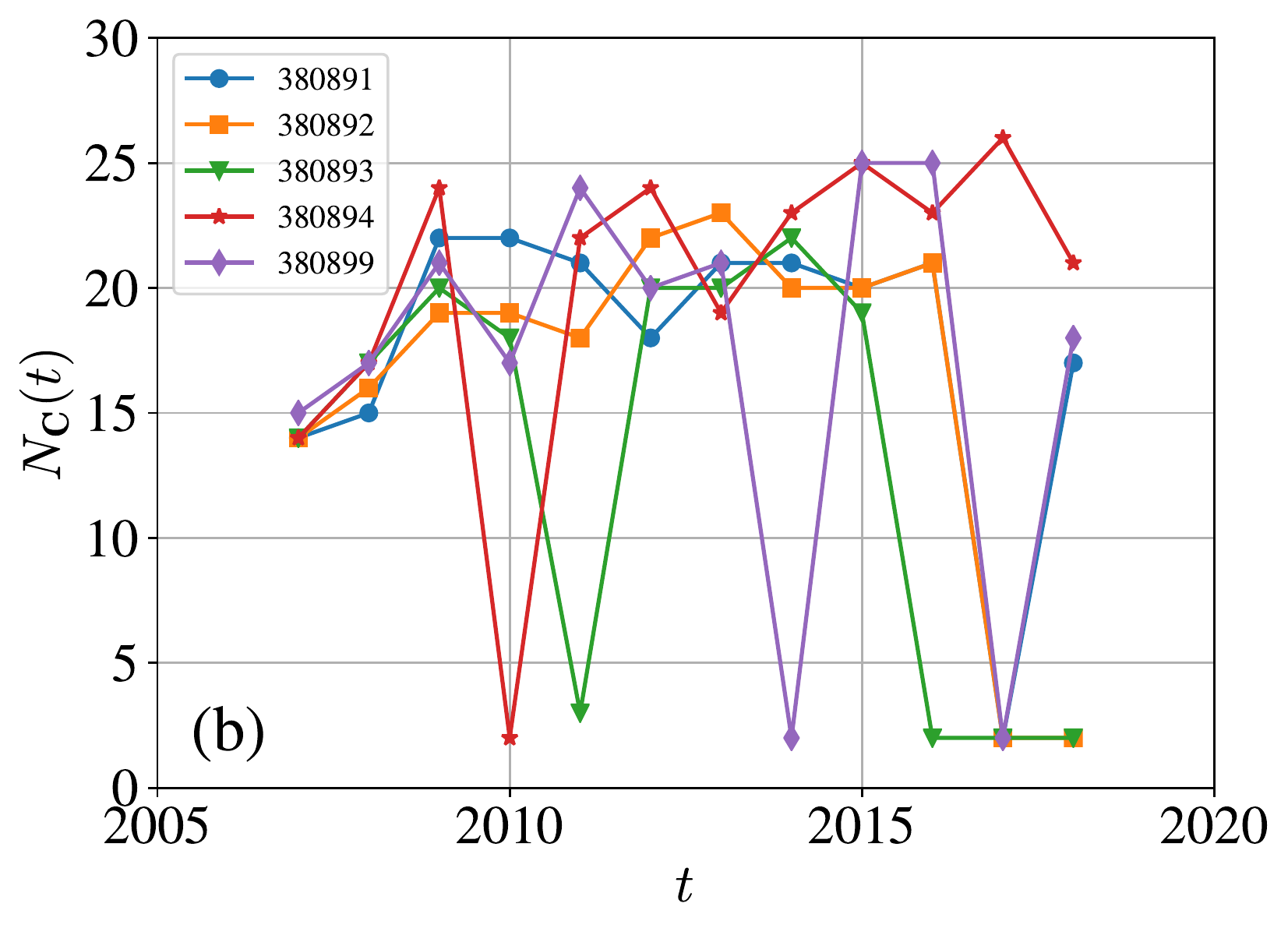}
    \caption{Evolution of the number of communities (a) and non-trivial top communities (b) for the directed international trade networks of insecticides (380891), fungicides (380892), herbicides (380893), disinfectants (380894), and rodenticides and other similar products (380899) from 2007 to 2018.}
    \label{Fig:iPTN:directed:Community:num:t}
\end{figure}

In Fig.~\ref{Fig:iPTN:directed:Community:num:t}(a), we show the evolution of the number $N^{\mathrm{dir}}_{\mathbf{C}}$ of communities for the five categories of directed iPTNs. We observe a mild increasing trend in the first years. In Fig.~\ref{Fig:iPTN:directed:Community:num:t}(b), we show the evolution of the number $N^{\mathrm{dir}}_{\mathbf{C^{>1}}}= N^{\mathrm{dir}}_{\mathbf{C}} - N^{\mathrm{dir}}_{\mathbf{C^1}}$ of non-trivial top communities for the five categories of directed iPTNs, where $N^{\mathrm{dir}}_{\mathbf{C^1}}$ is the number of trivial communities. A trivial community is a module with either one or all nodes within it. In the networks we analyzed, there are one-node communities, but not all-nodes communities.

As shown in Fig.~\ref{Fig:iPTN:directed:Community:num:t}(b), after removing trivial communities, we obtain nine networks with $N^{\mathrm{dir}}_{\mathbf{C^{>1}}}=2$ and one network with $N^{\mathrm{dir}}_{\mathbf{C^{>1}}}=3$, which are the same outlier networks recognized from Fig.~\ref{Fig:iPTN:directed:Community:CodeLen:t} and hard to be recognized from Fig.~\ref{Fig:iPTN:directed:Community:num:t}(a). Excluding these 10 networks, other networks each has at least 14 communities. Comparing Fig.~\ref{Fig:iPTN:directed:Community:num:t} and Fig.~\ref{Fig:iPTN:undirected:Community:num:t}, we find that there are more communities identified in the directed networks than in the undirected networks such that
\begin{equation}
    N^{\mathrm{dir}}_{\mathbf{C}} 
    \geq N^{\mathrm{dir}}_{\mathbf{C^{>1}}} 
    > N^{\mathrm{undir}}_{\mathbf{C}}
\end{equation}
for most cases (year and pesticide).



The modularity of a directed and weighted network is defined as follows \cite{Arenas-Duch-Fernandez-Gomez-2007-NewJPhys}:
\begin{equation}
    Q = \frac{1}{2W}\sum_i\sum_j\left(w_{ij}-\frac{s_i^{\mathrm{out}}s_j^{\mathrm{in}}}{2W}\right)\delta(C_i, C_j),
\end{equation}
where $s_i^{\mathrm{out}}$ is the total export of economy $i$
\begin{equation}
    s_i^{\mathrm{out}} = \sum_j w_{ij},
\end{equation}
$s_j^{\mathrm{in}}$ is the total import of economy $j$
\begin{equation}
    s_j^{\mathrm{in}} = \sum_i w_{ij},
\end{equation}
and $W$ is the total trade volume all over the world
\begin{equation}
    W = \sum_i s_i^{\mathrm{out}} = \sum_js_j^{\mathrm{in}}  = \sum_i\sum_j w_{ij}.
\end{equation}

\begin{figure}[!ht]
    \centering
    \includegraphics[width=0.783\linewidth]{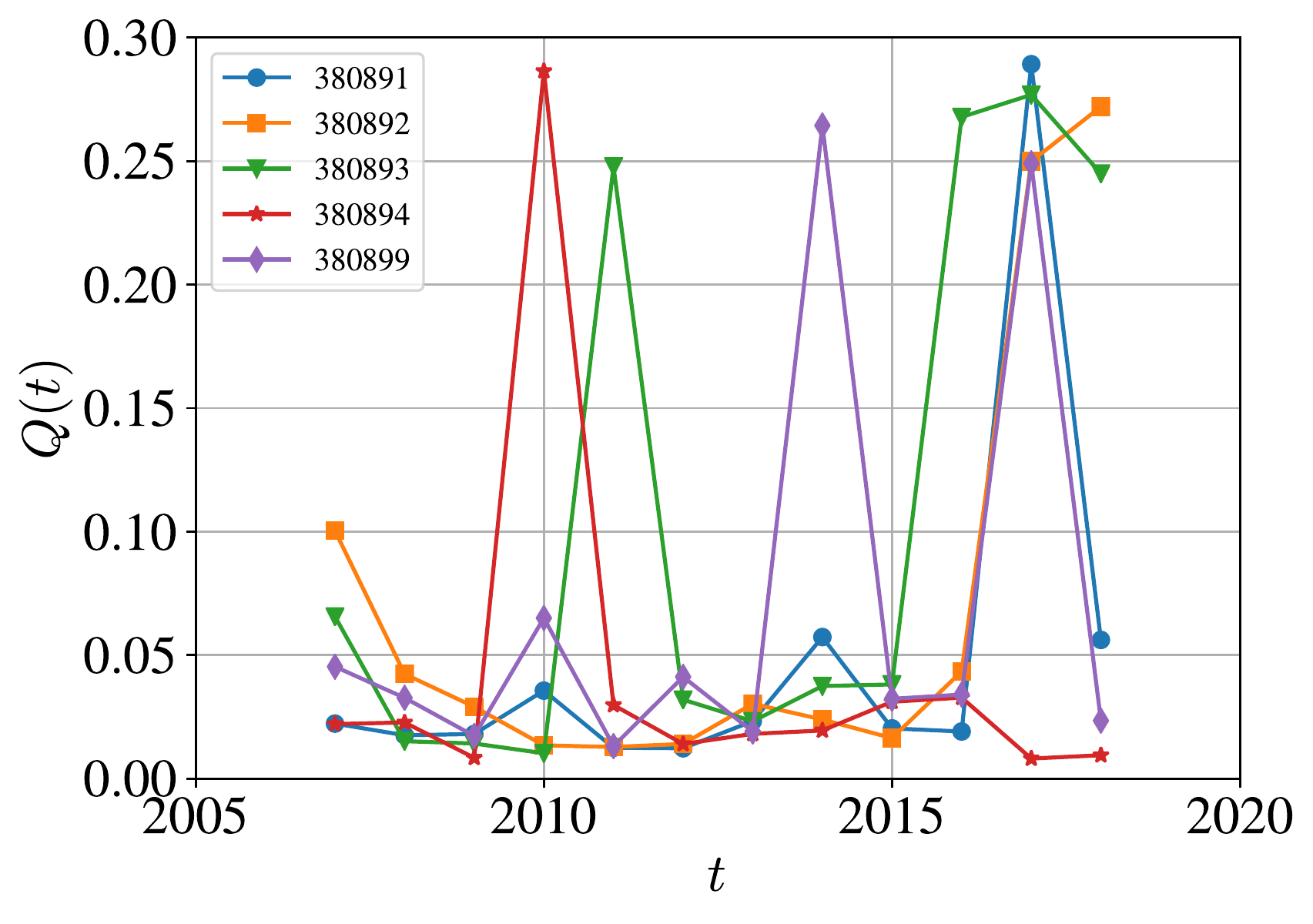}
    \caption{Evolution of the modularity for the directed international trade networks of insecticides (380891), fungicides (380892), herbicides (380893),  disinfectants (380894), and rodenticides and other similar products (380899) from 2007 to 2018.}
    \label{Fig:iPTN:directed:Community:modularity:t}
\end{figure}

Figure~\ref{Fig:iPTN:directed:Community:modularity:t} illustrates the evolution of the modularity for the directed iPTNs from 2007 to 2018. We find that the 10 outlier networks have much larger modularity than the rest networks. The rest networks have very small modularities, which are much less than the modularities of the undirected networks. Hence, for most cases (year and pesticide), a directed network usually have more communities but lower modulairy than its counterpart undirected network.

\begin{figure}[!ht]
    \centering
    \includegraphics[width=0.321\linewidth]{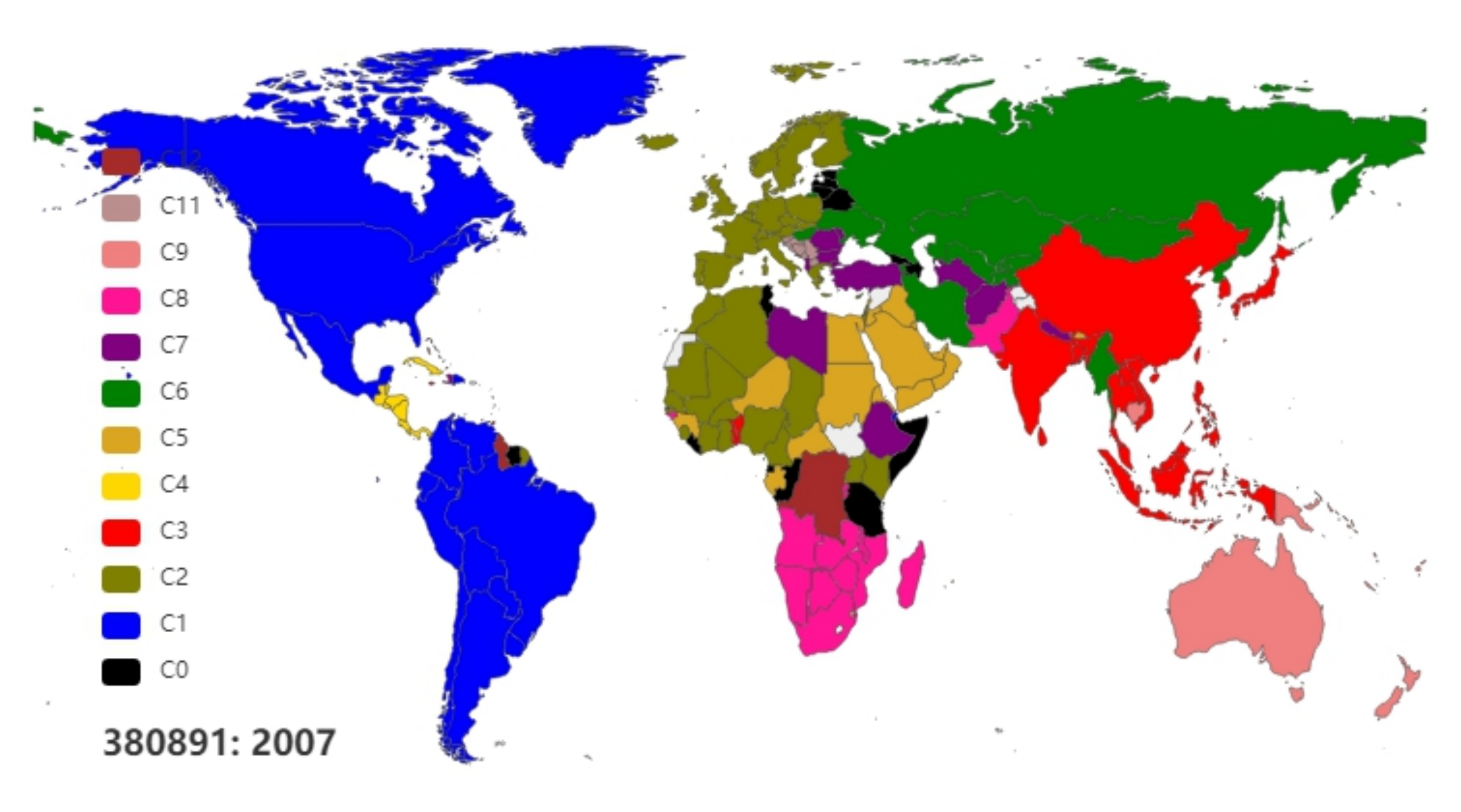}
    \includegraphics[width=0.321\linewidth]{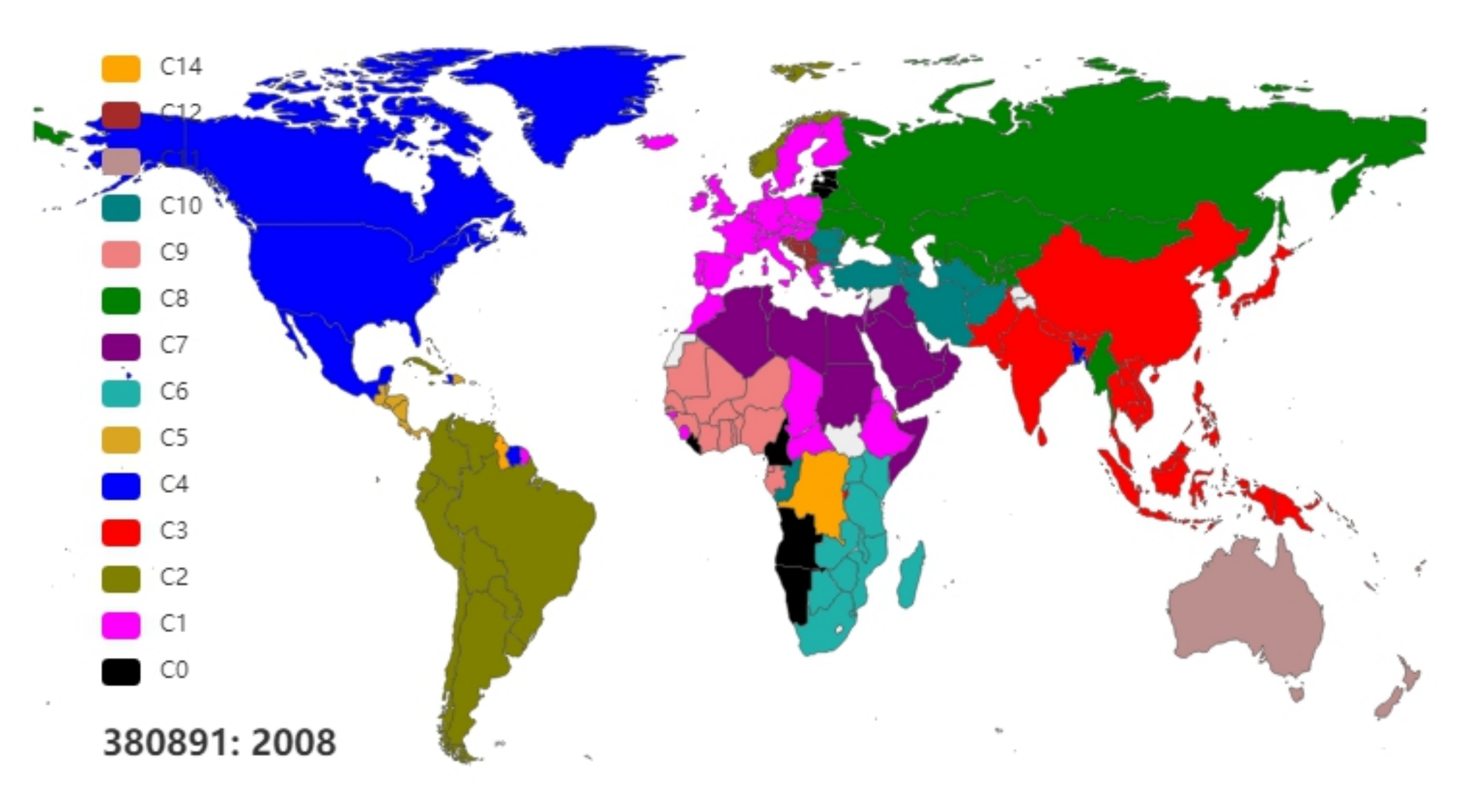}
    \includegraphics[width=0.321\linewidth]{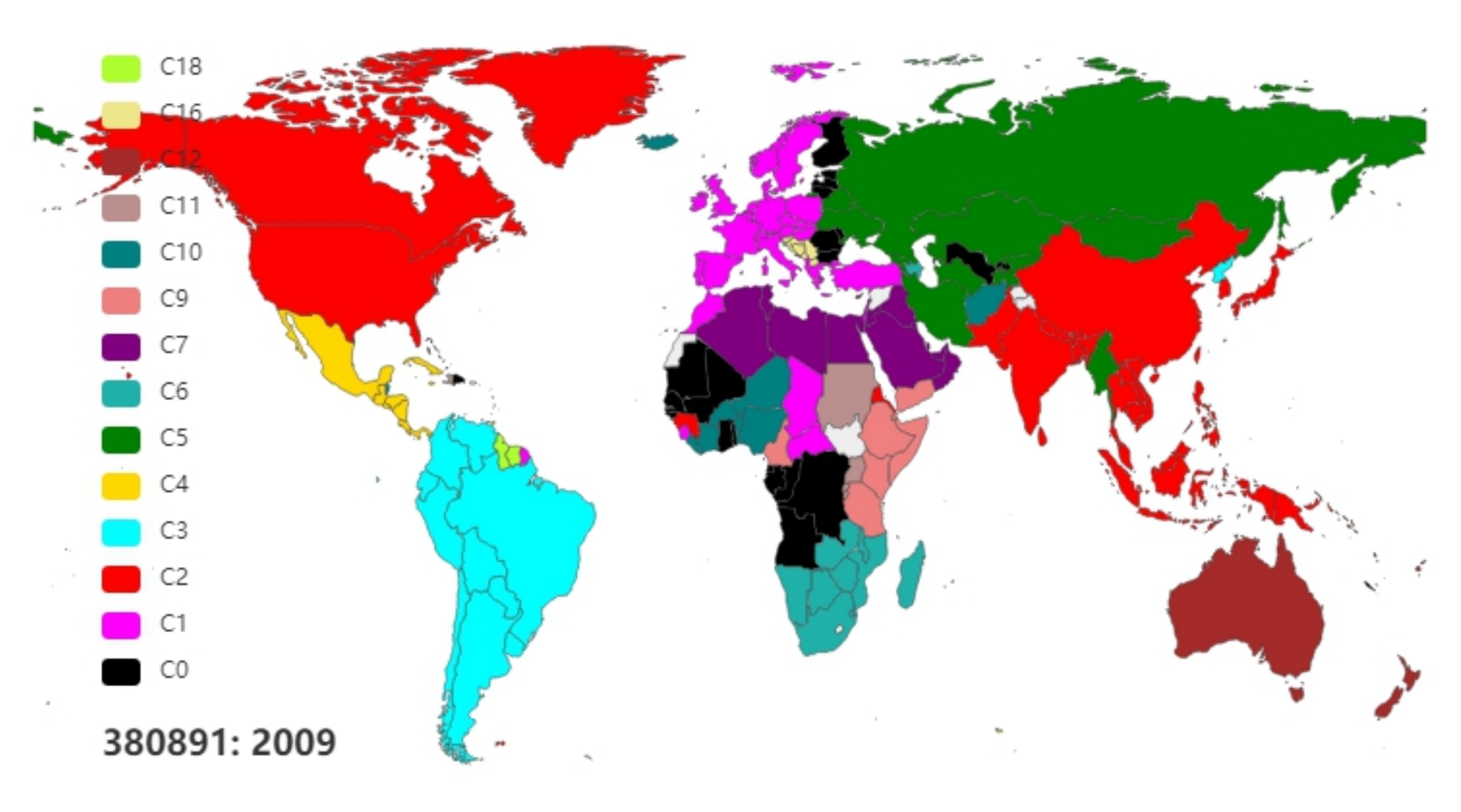}
    \includegraphics[width=0.321\linewidth]{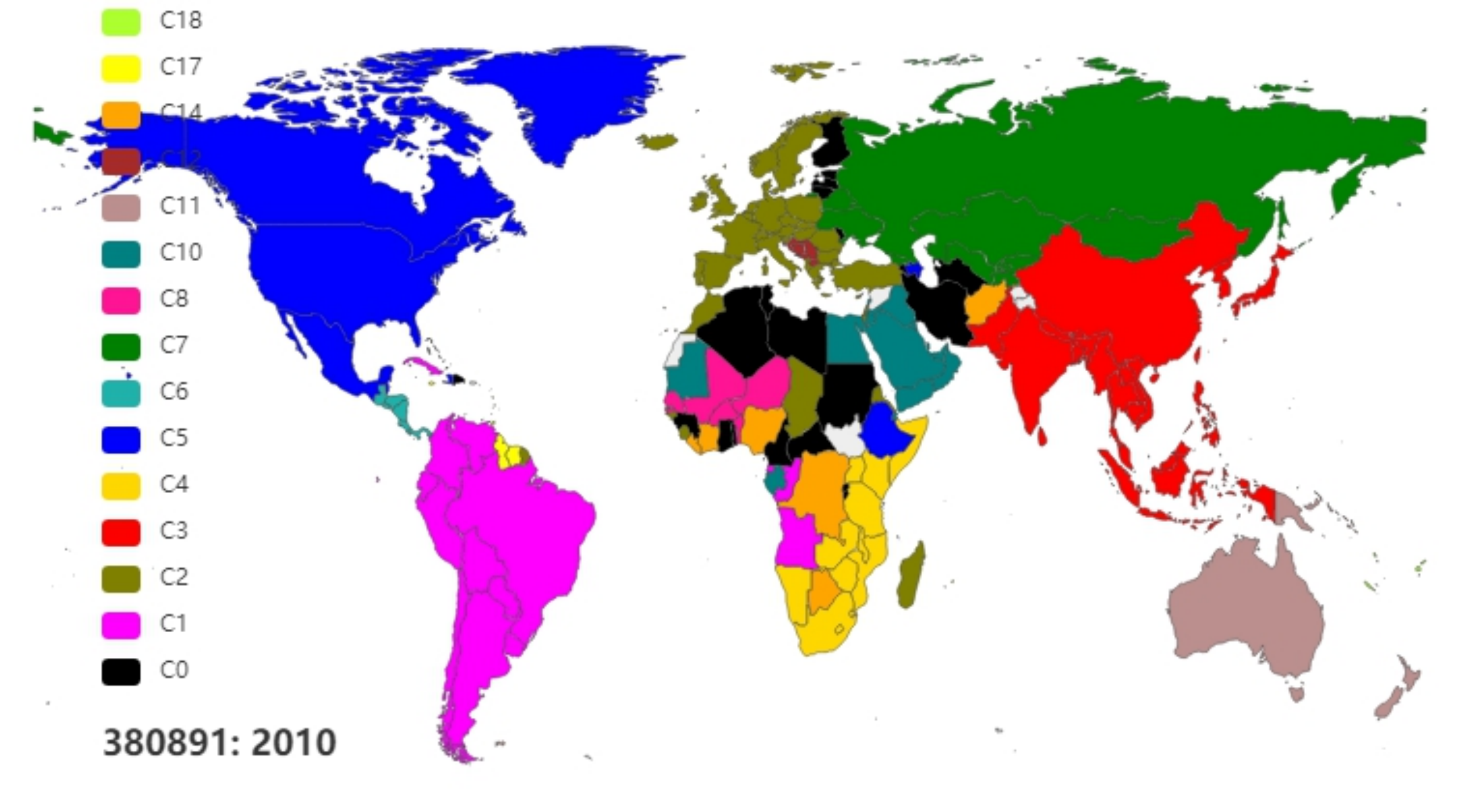}
    \includegraphics[width=0.321\linewidth]{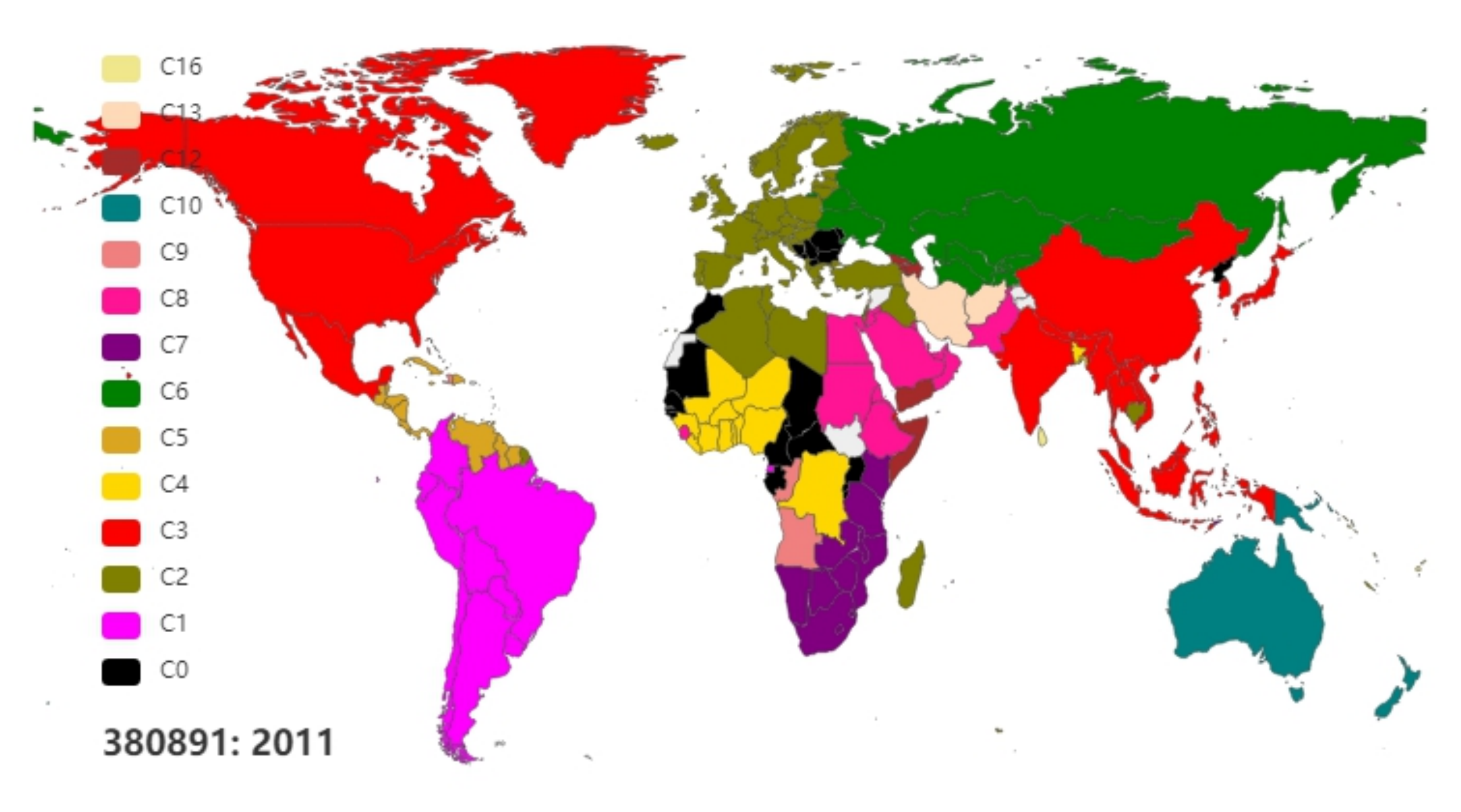}
    \includegraphics[width=0.321\linewidth]{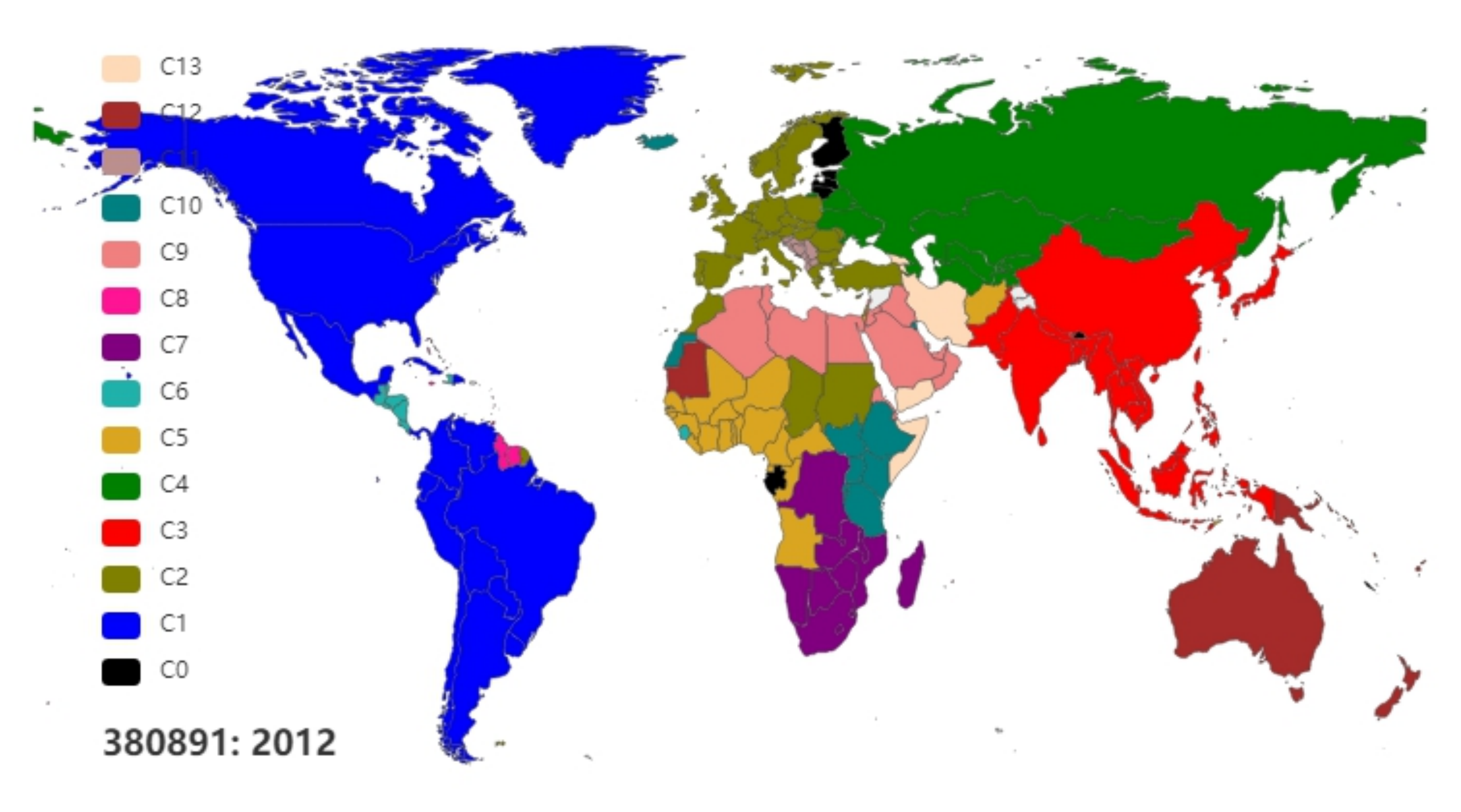}
    \includegraphics[width=0.321\linewidth]{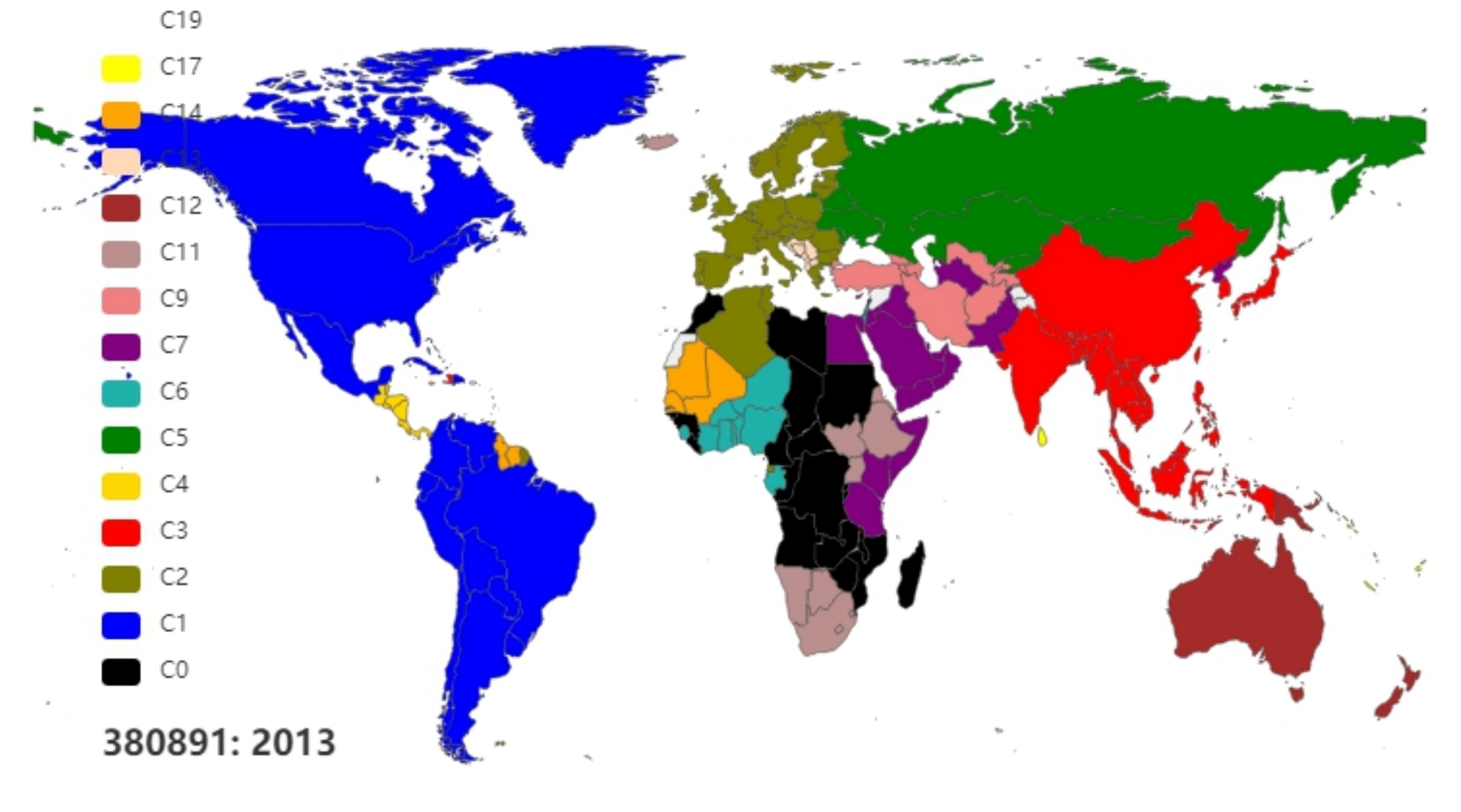}
    \includegraphics[width=0.321\linewidth]{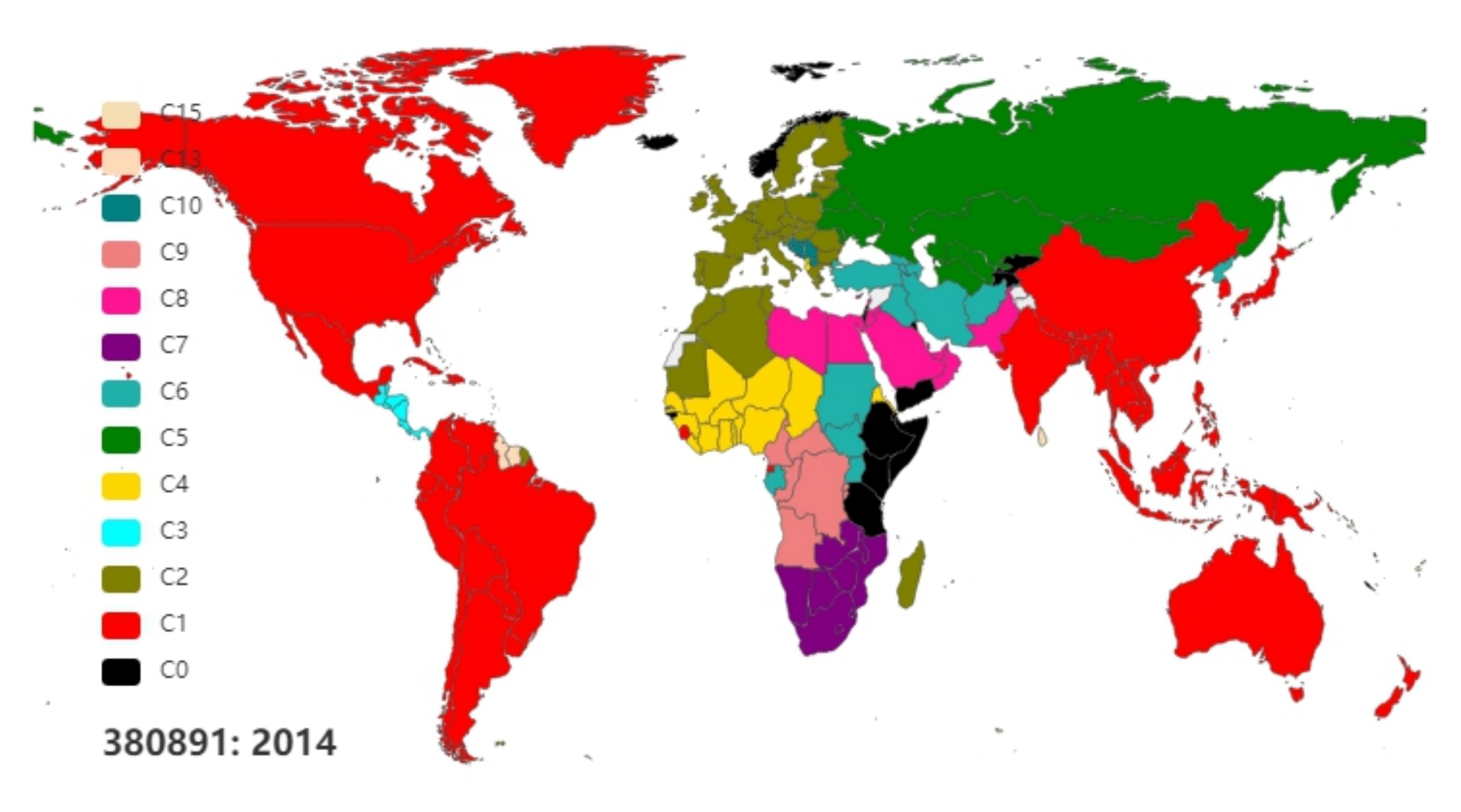}
    \includegraphics[width=0.321\linewidth]{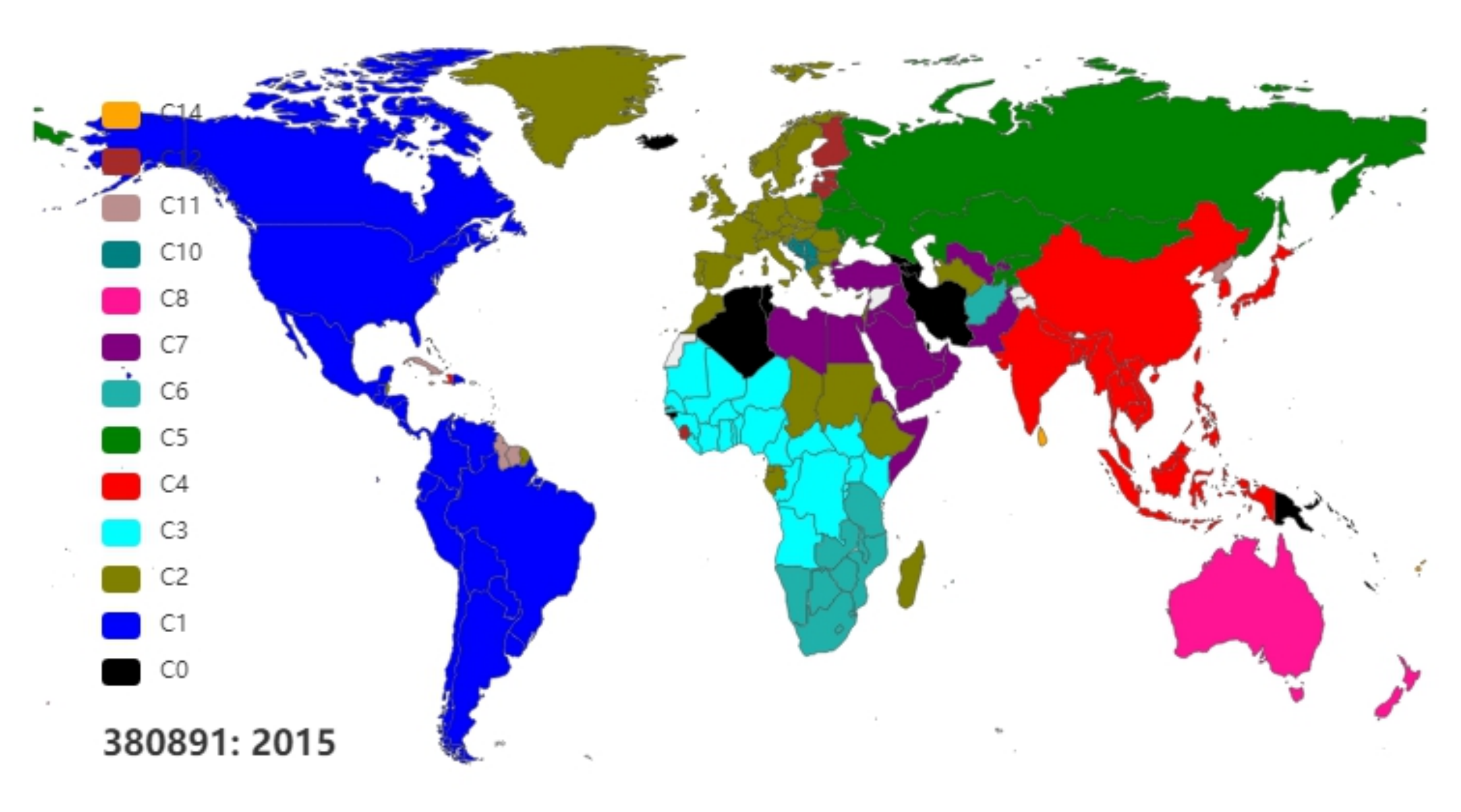}
    \includegraphics[width=0.321\linewidth]{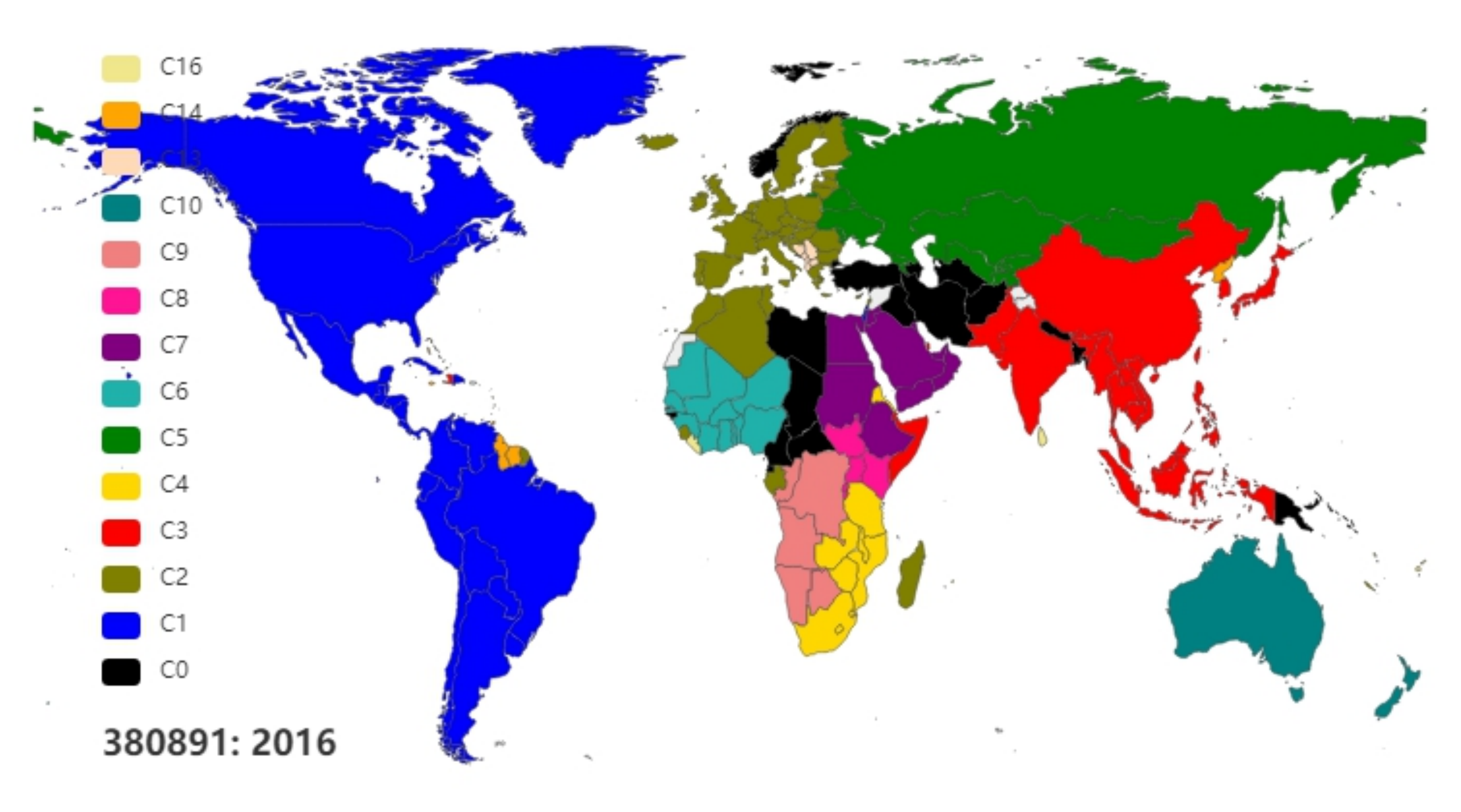}
    \includegraphics[width=0.321\linewidth]{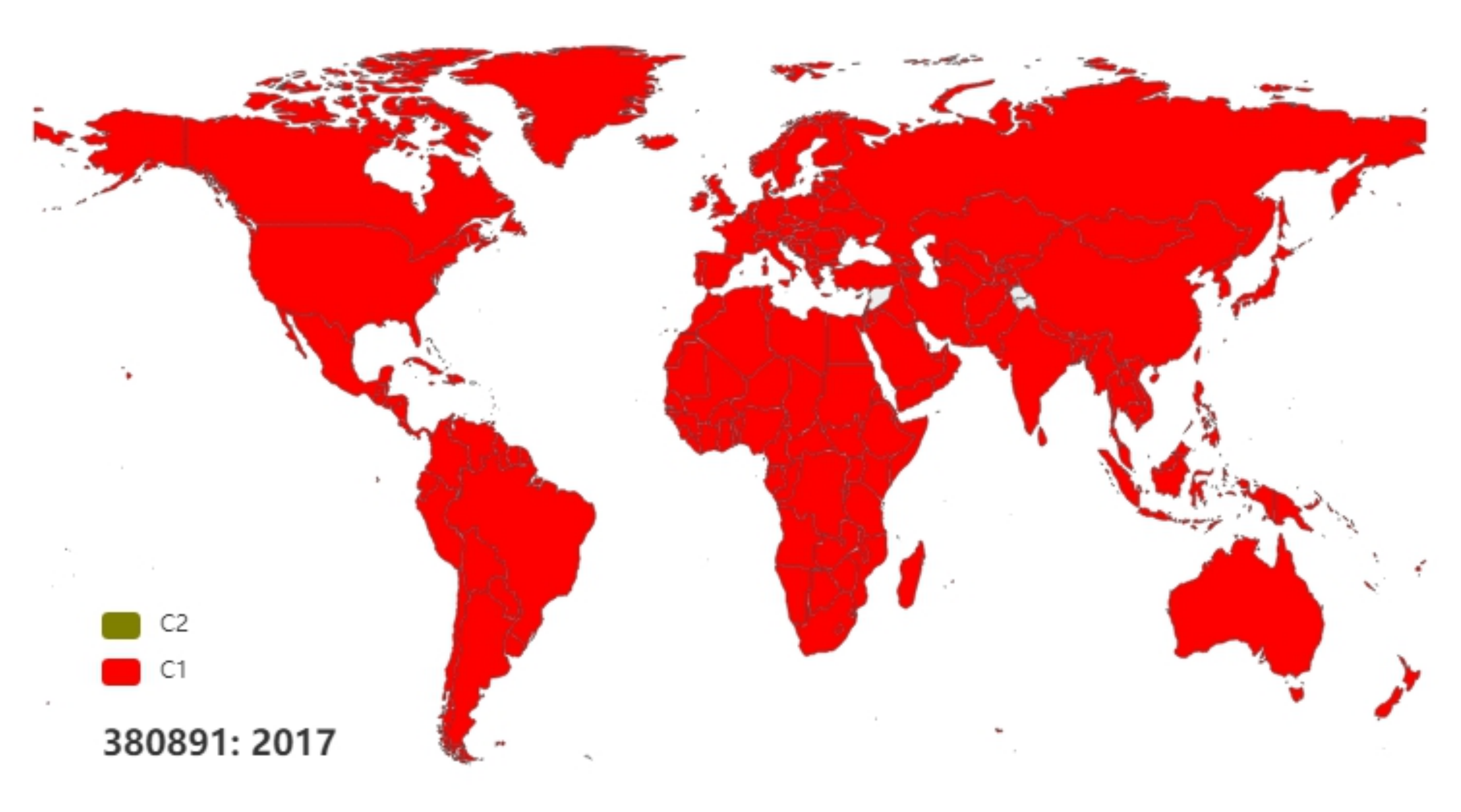}
    \includegraphics[width=0.321\linewidth]{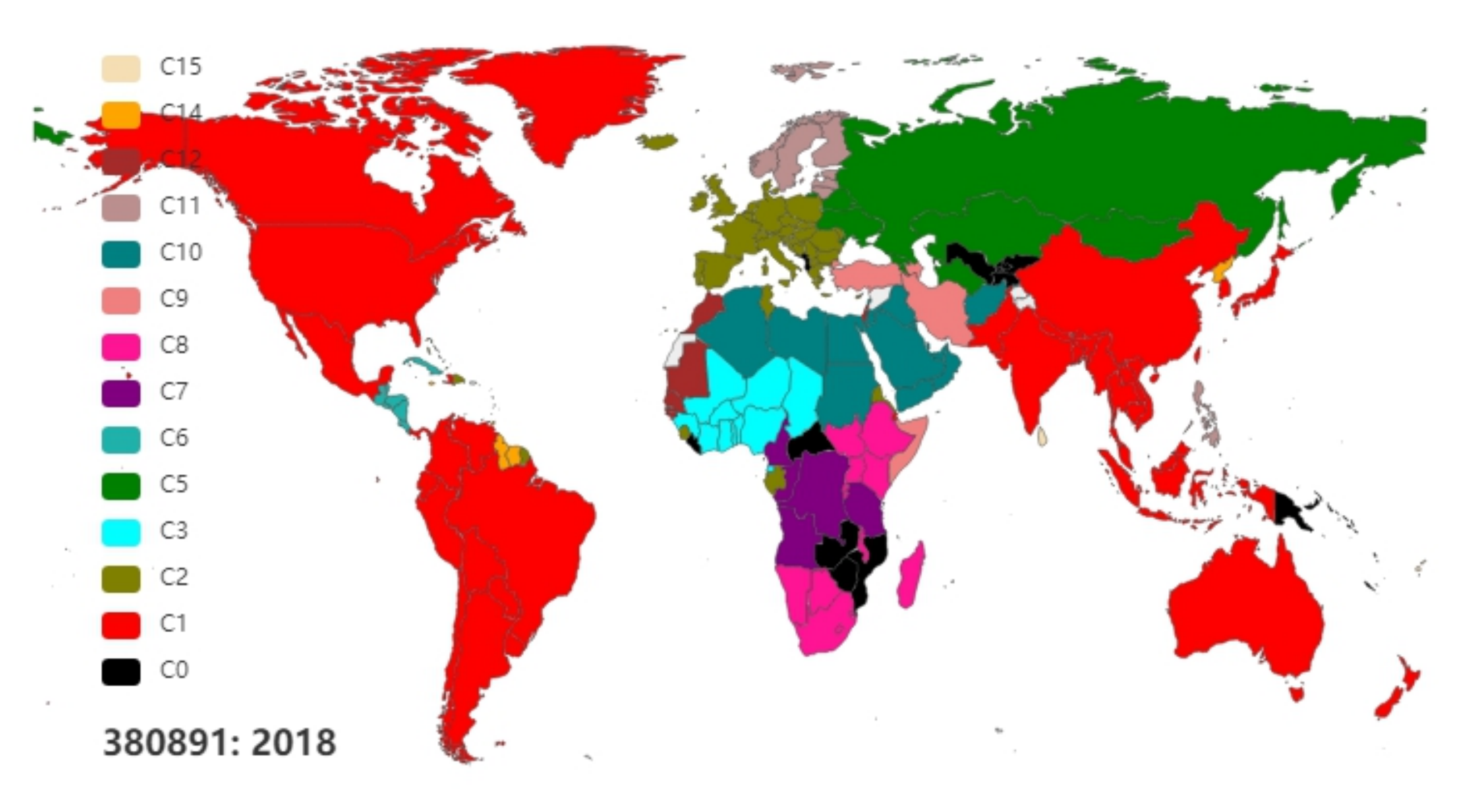}
    \caption{Community evolution of the directed iPTNs of insecticides (380891) from 2007 to 2018. For better visibility, the non-trivial communities containing less than 5 economies and trivial communities are merged to $C_0$.}
    \label{Fig:iPTN:directed:CommunityMap:380891}
\end{figure}

Figure~\ref{Fig:iPTN:directed:CommunityMap:380891} illustrates the evolution of communities of the directed iPTNs of insecticides (380891) from 2007 to 2018. For better visibility, the non-trivial communities containing less than 5 economies and trivial communities are merged to $C_0$. For instance, in the community map of 2007, the merged cluster $C_0$ contains 3 non-trivial communities ($C_{10}$ with Belarus, Estonia, Latvia, and Lithuania, $C_{13}$ with Azerbaijan and Georgia, and $C_{14}$ with Tunisia and Mayotte) and 7 trivial communities ($C_{15}$ Tanzania, $C_{16}$ Gambia, $C_{17}$ Liberia, $C_{18}$, Suriname, $C_{19}$ Congo, $C_{20}$ Somalia, and $C_{21}$ Eq. Guinea). 
In Figure~\ref{Fig:iPTN:directed:CommunityMap:380891}, only the network in 2017 does not have cluster $C_0$, which contains 2 communities. The second community $C_{2}$ contains 6 economies (Virgin Islands (British), Saint Kitts and Nevis, St. Vin. and Gren., Cura{\c{c}}ao, Aruba, and Turks and Caicos Is.), all located in the eastern Caribbean Sea.

We now investigate the community structure of the 11 networks (except 2017). Most of the communities exhibit regional clustering or are combinations of regionally clustered areas. The community in green appears in all 11 networks, which mainly contains Russia and several economies geographically close to Russia. China, India, Japan, Korea and main economies in South Asia and Southeast Asia usually belong to a community. For instance, in 2007, this Asia community contains 24 economies including Indonesia, Japan, Philippines, Thailand, Togo, China, Hong Kong, Korea, India, Singapore, Bangladesh, Sri Lanka, Benin, Macao, Malaysia, Vietnam, Kiribati, Maldives, Micronesia, Marshall Islands, Guam, N. Mariana Is., Lao PDR, and Brunei. Australia, New Zealand and other economies in Oceania forms a community. In 2007, the Oceania community contains 17 economies, including Australia, New Zealand, Solomon Is., Cook Islands, Fiji, F. Polynesia, New Caledonia, Vanuatu, Palau, Papua New Guinea, Timor-Leste, Tonga, Samoa, Cambodia, Niue, Norfolk Island, and Swaziland. The Asia community and the Oceania community merge to one community in 2014 and 2018.

Economies in West Europe and sometimes in North Africa and Middle East form a community. This West Europe community is offen the largest community in the sense of the number of economies it contains. In 2007, it contains 47 economies, including Algeria, France, Germany, Denmark, Ghana, Greece, Israel, Nigeria, Senegal, Spain, Italy, Kenya, Mali, Mauritius, Netherlands, Poland, Slovakia, Switzerland, United Kingdom, Austria, Belgium, Cameroon, Cyprus, Czech Rep., Ireland, C\^{o}te d'Ivoire, Luxembourg, Norway, Portugal, Sweden, Chad, Comoros, Finland, Iceland, Malta, Morocco, Sierra Leone, Uganda, Burkina Faso, Mauritania, Faroe Is., Andorra, Cabo Verde, S{\~{a}}o Tom{\'{e}} and Principe, Wallis and Futuna, Palestine, and Gibraltar. 

In the North and South American continents, Canada, Greenland (except 2015), Mexico, and the USA always belong to a community, most economies in South America also forms a community, and other economies (especially others in the Caribbean) may belong to the North America community or the South America community, or form separate communities. From 2008 to 2011, the North America community and the South America community are separate; Otherwise, they form the biggest community in the sense of area. In 2009 and 2011, the North America community and the Asia community unite to one community. In 2014 and 2018, the North America community, the Asia community and the Oceania community united belong to the same community. The community structure of the economies in Africa is quite complex and changes frequently.

\begin{figure}[!ht]
    \centering
    \includegraphics[width=0.321\linewidth]{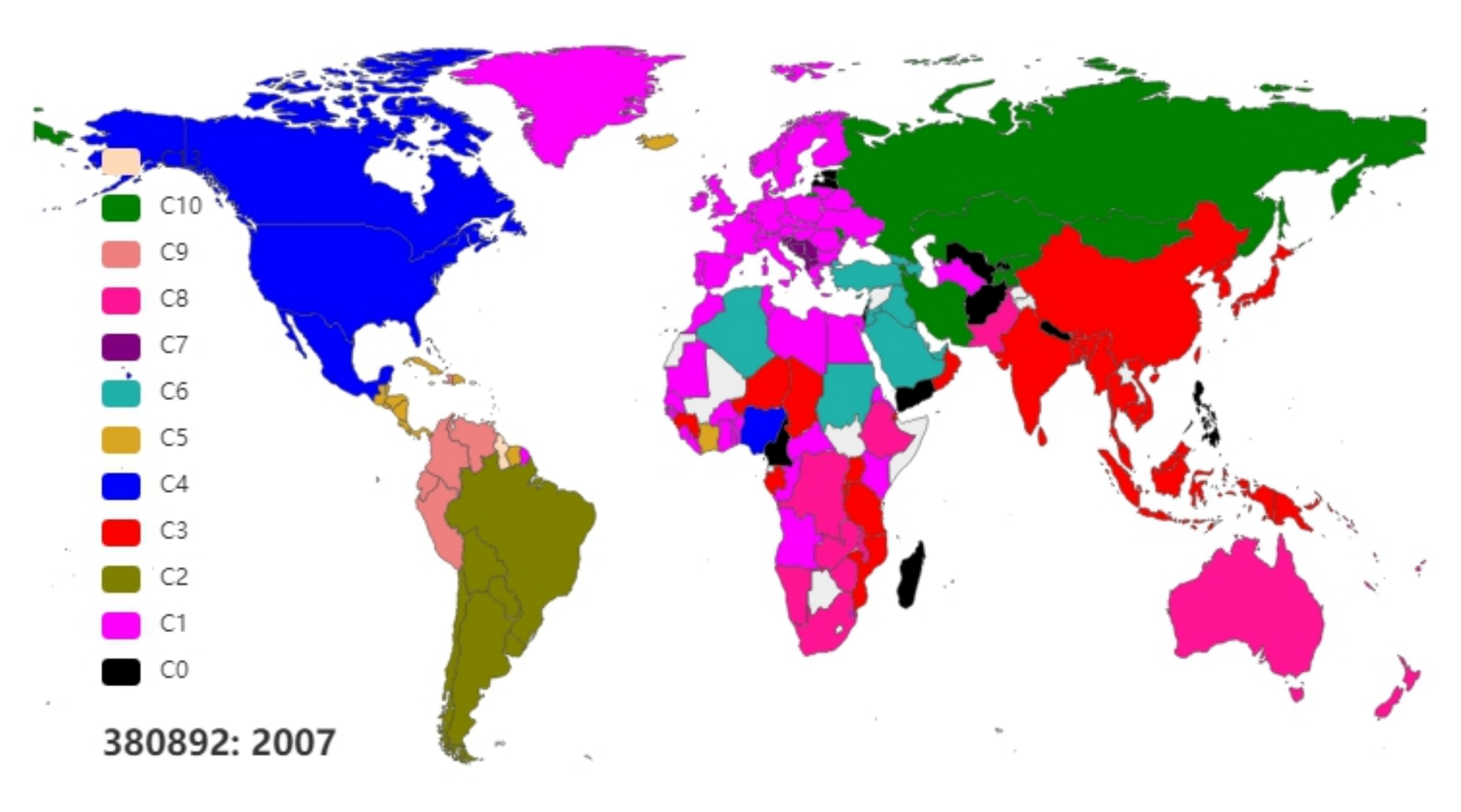}
    \includegraphics[width=0.321\linewidth]{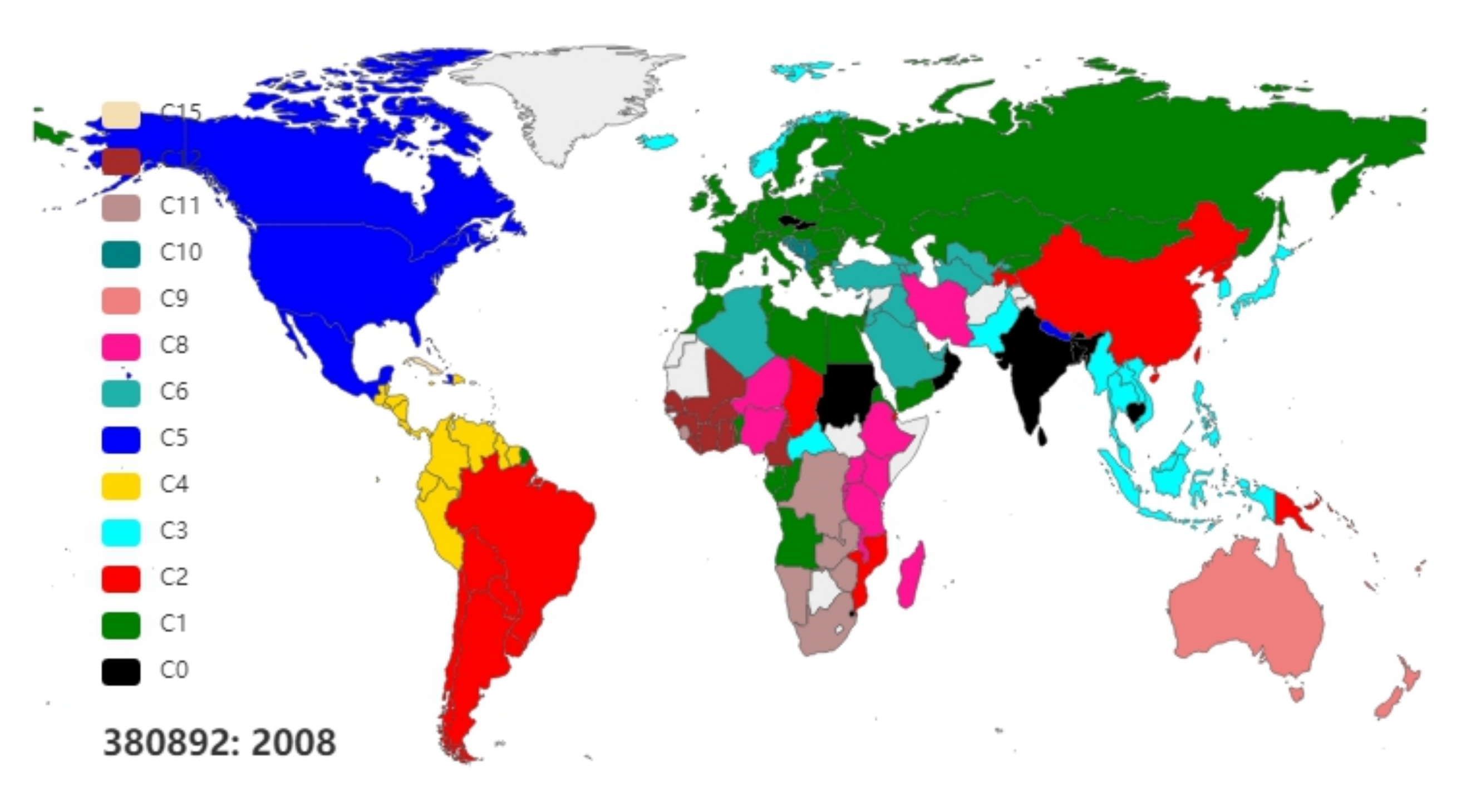}
    \includegraphics[width=0.321\linewidth]{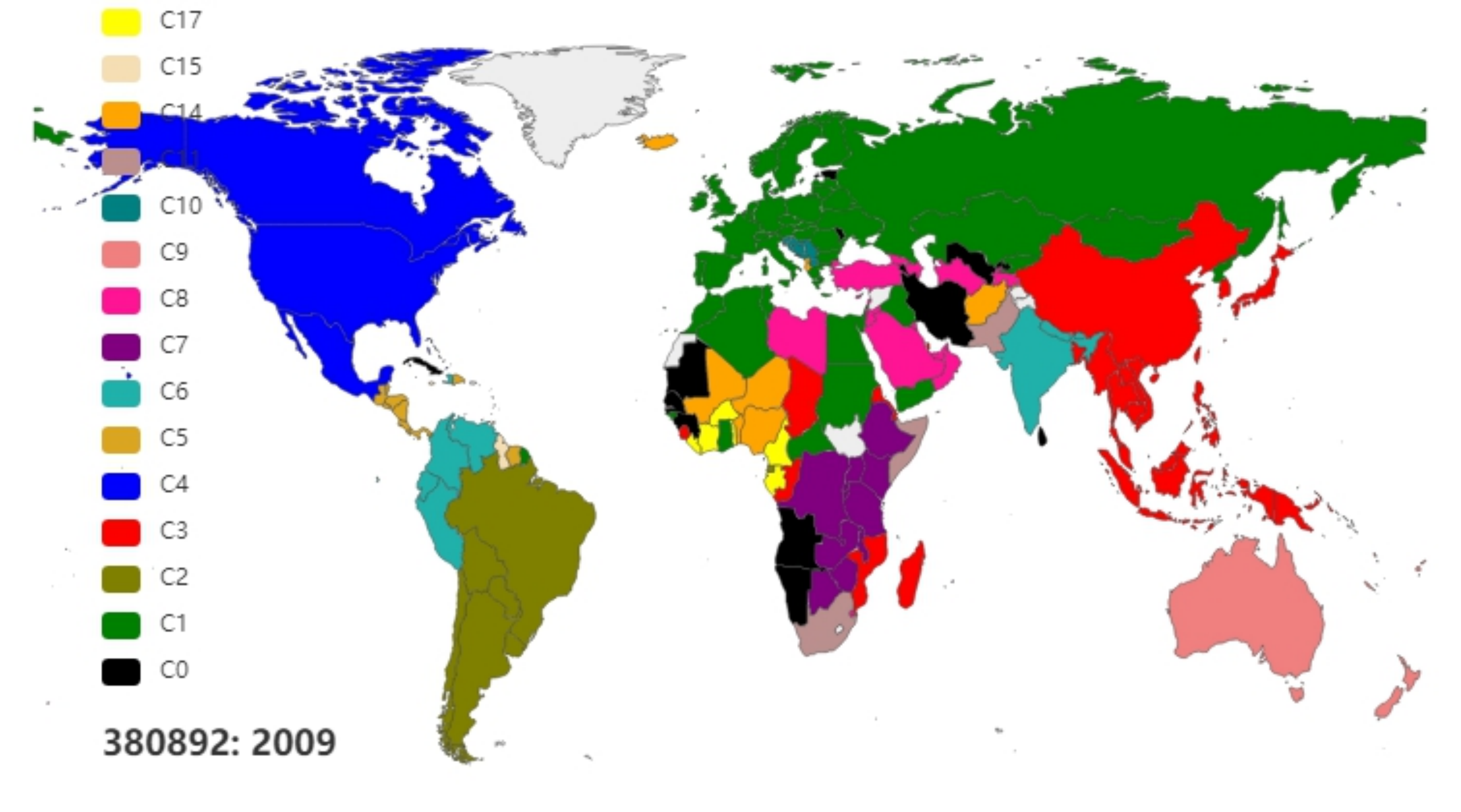}
    \includegraphics[width=0.321\linewidth]{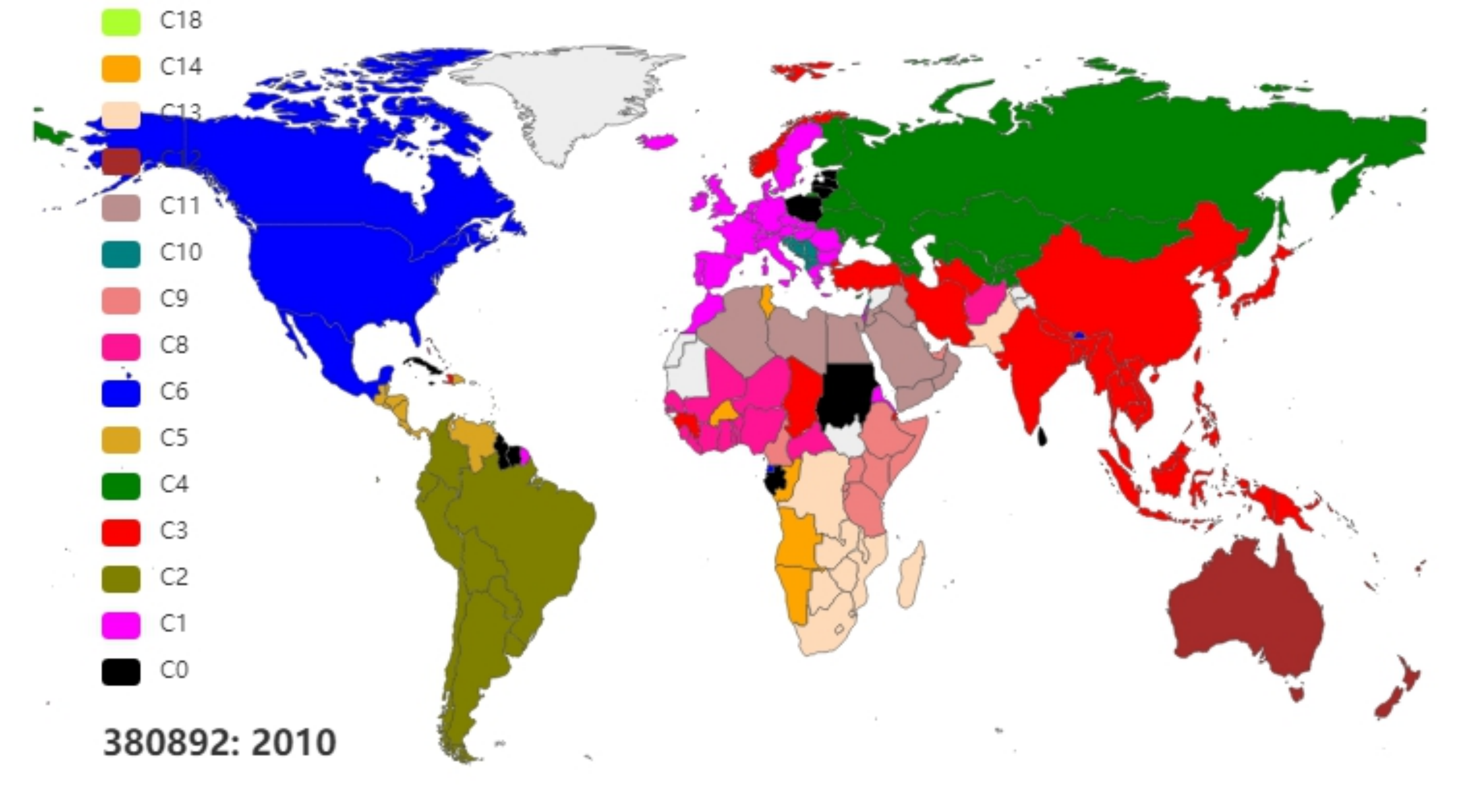}
    \includegraphics[width=0.321\linewidth]{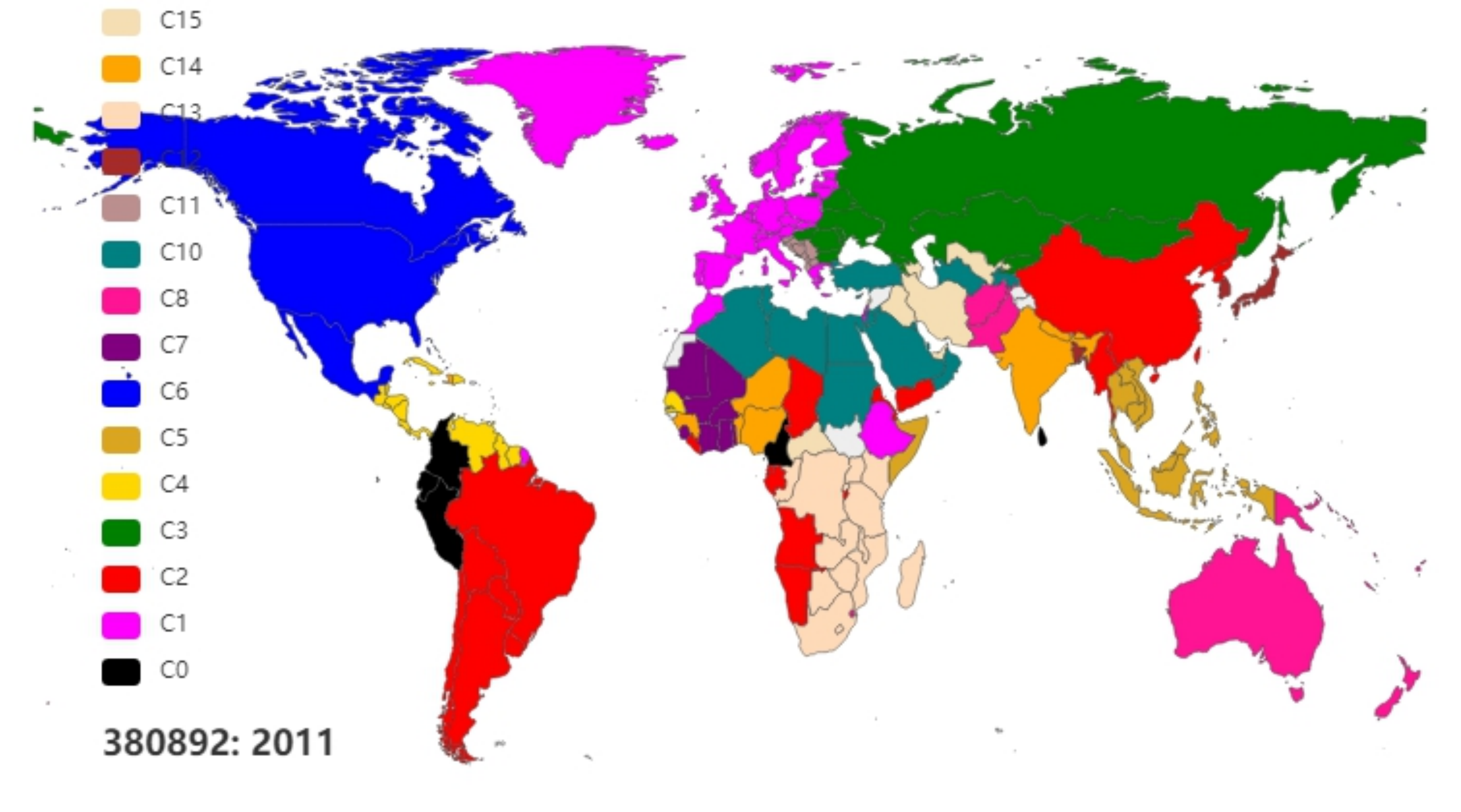}
    \includegraphics[width=0.321\linewidth]{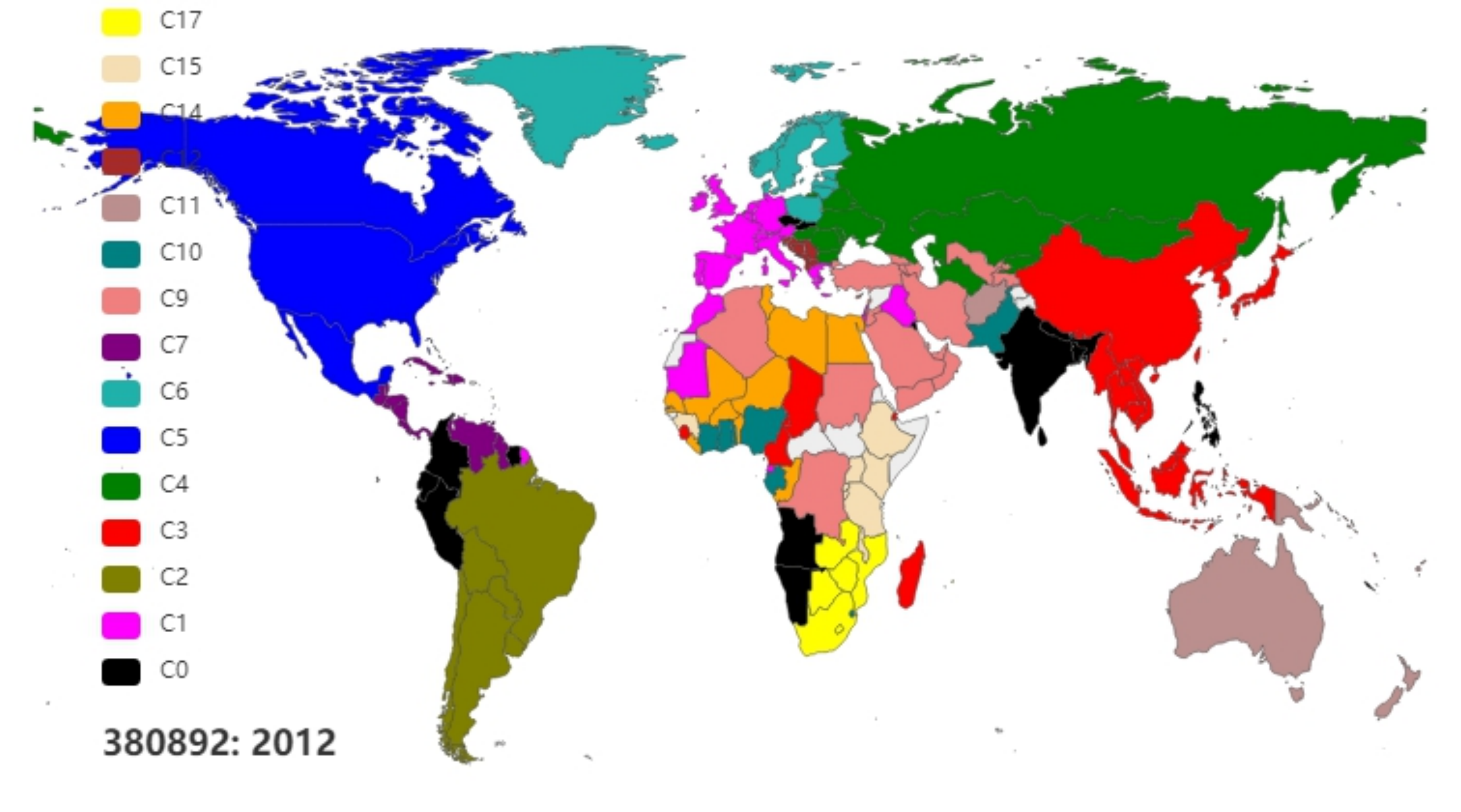}
    \includegraphics[width=0.321\linewidth]{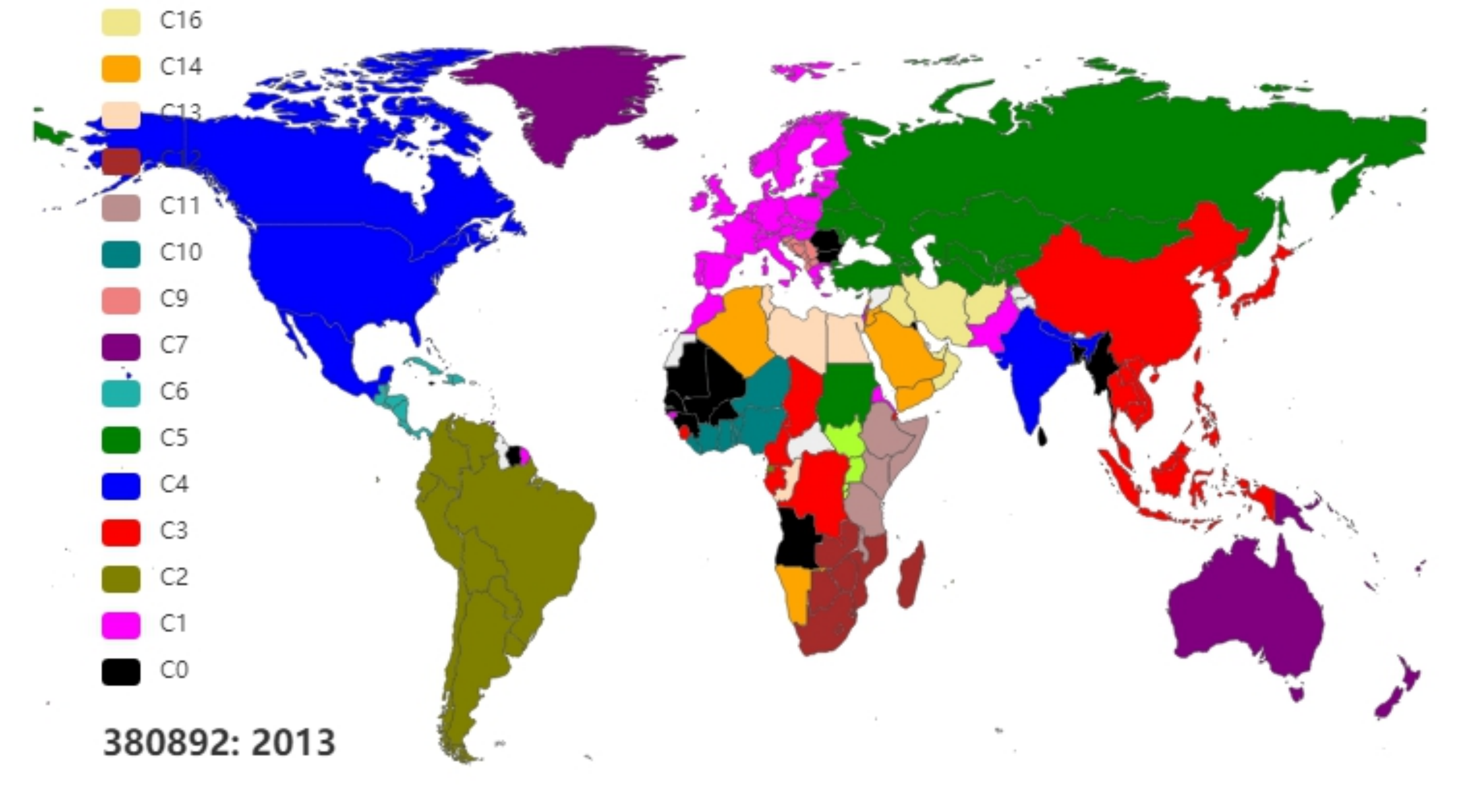}
    \includegraphics[width=0.321\linewidth]{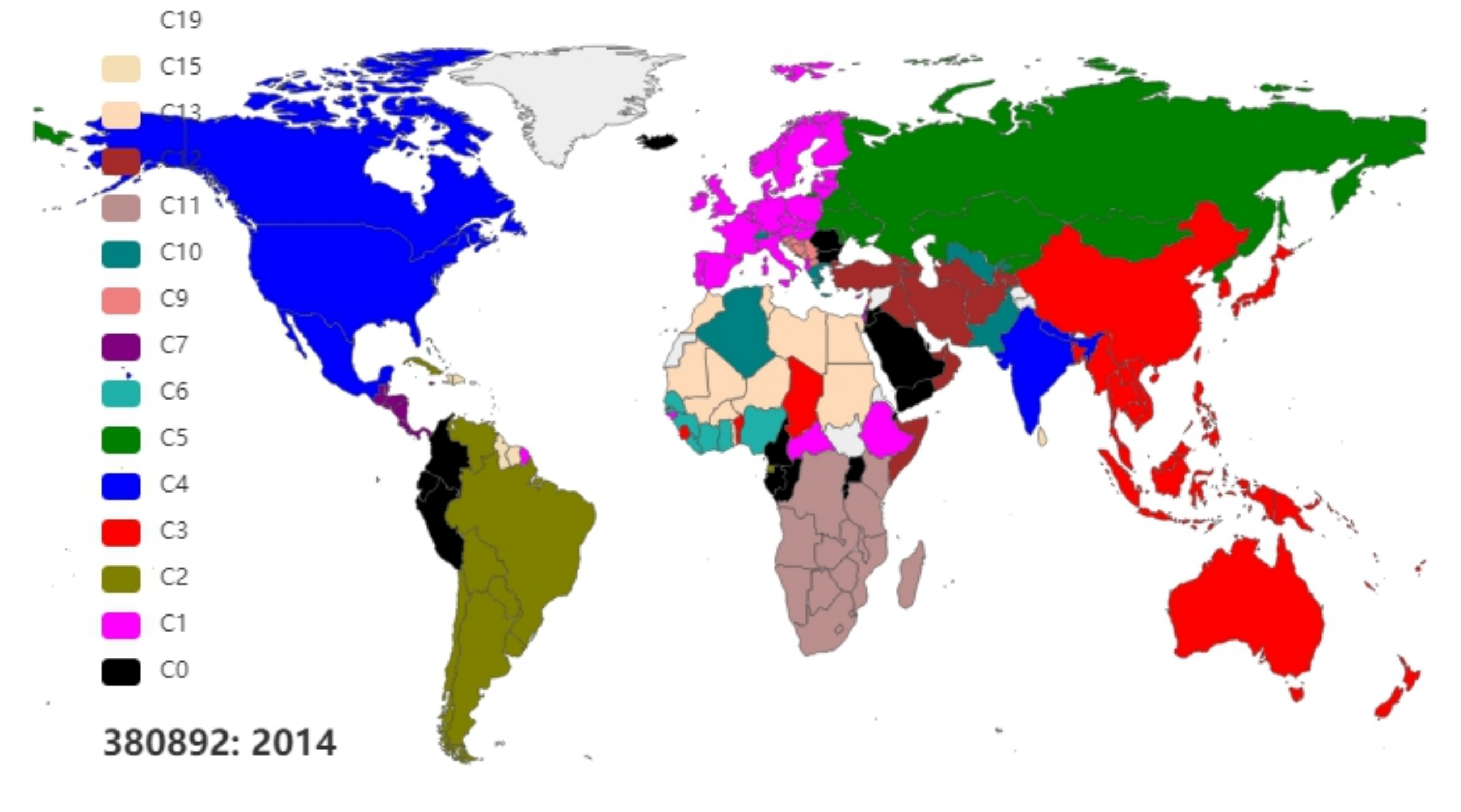}
    \includegraphics[width=0.321\linewidth]{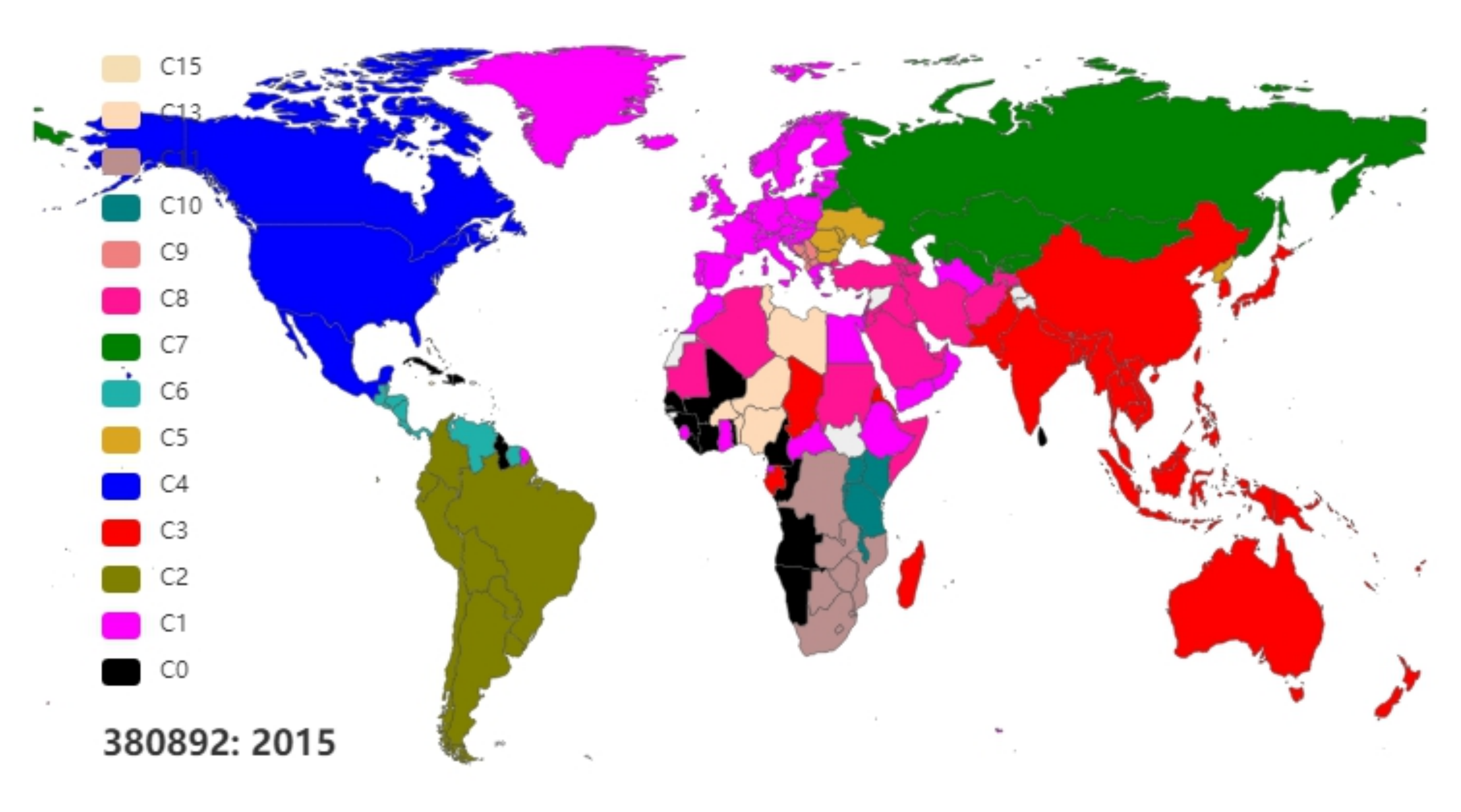}
    \includegraphics[width=0.321\linewidth]{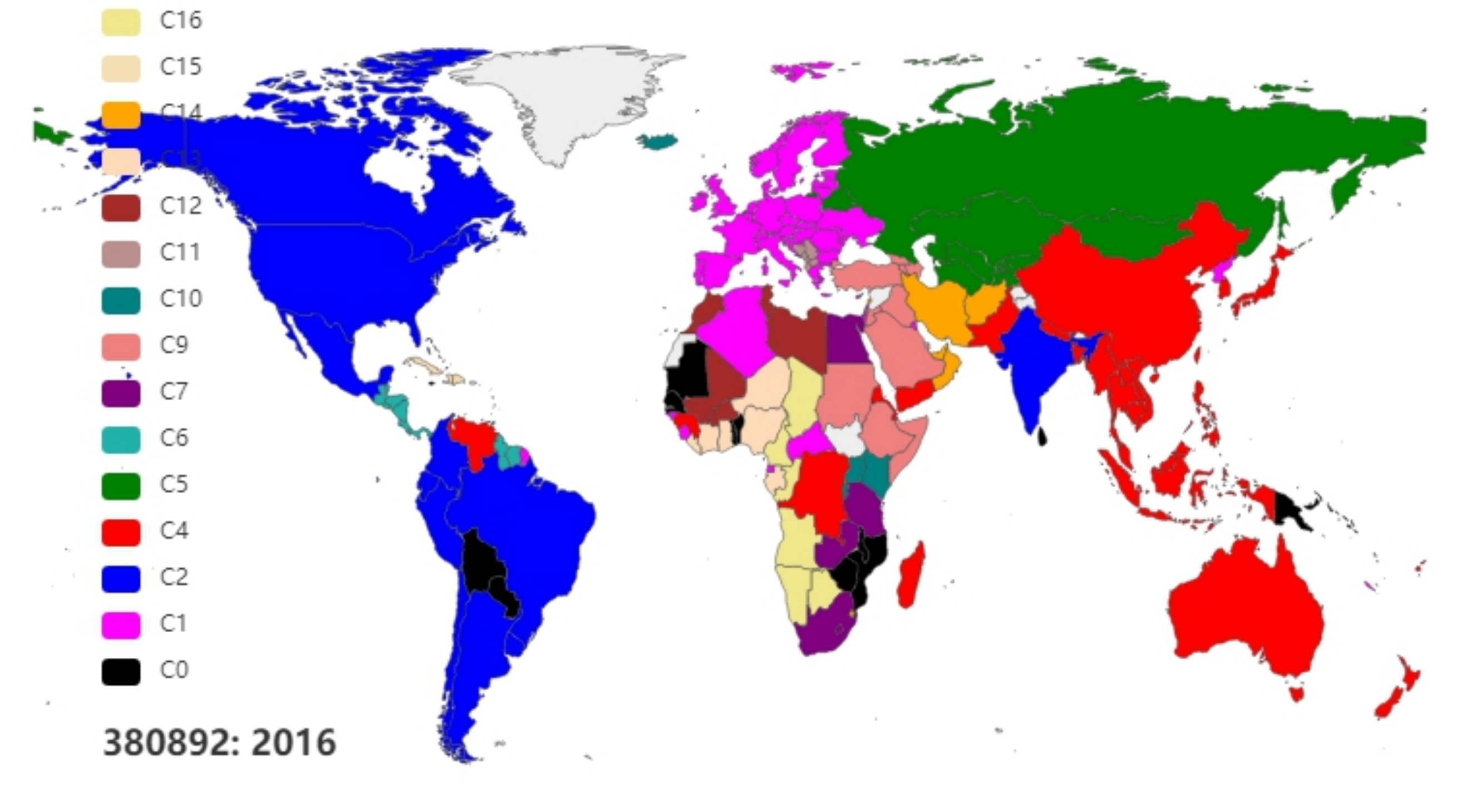}
    \includegraphics[width=0.321\linewidth]{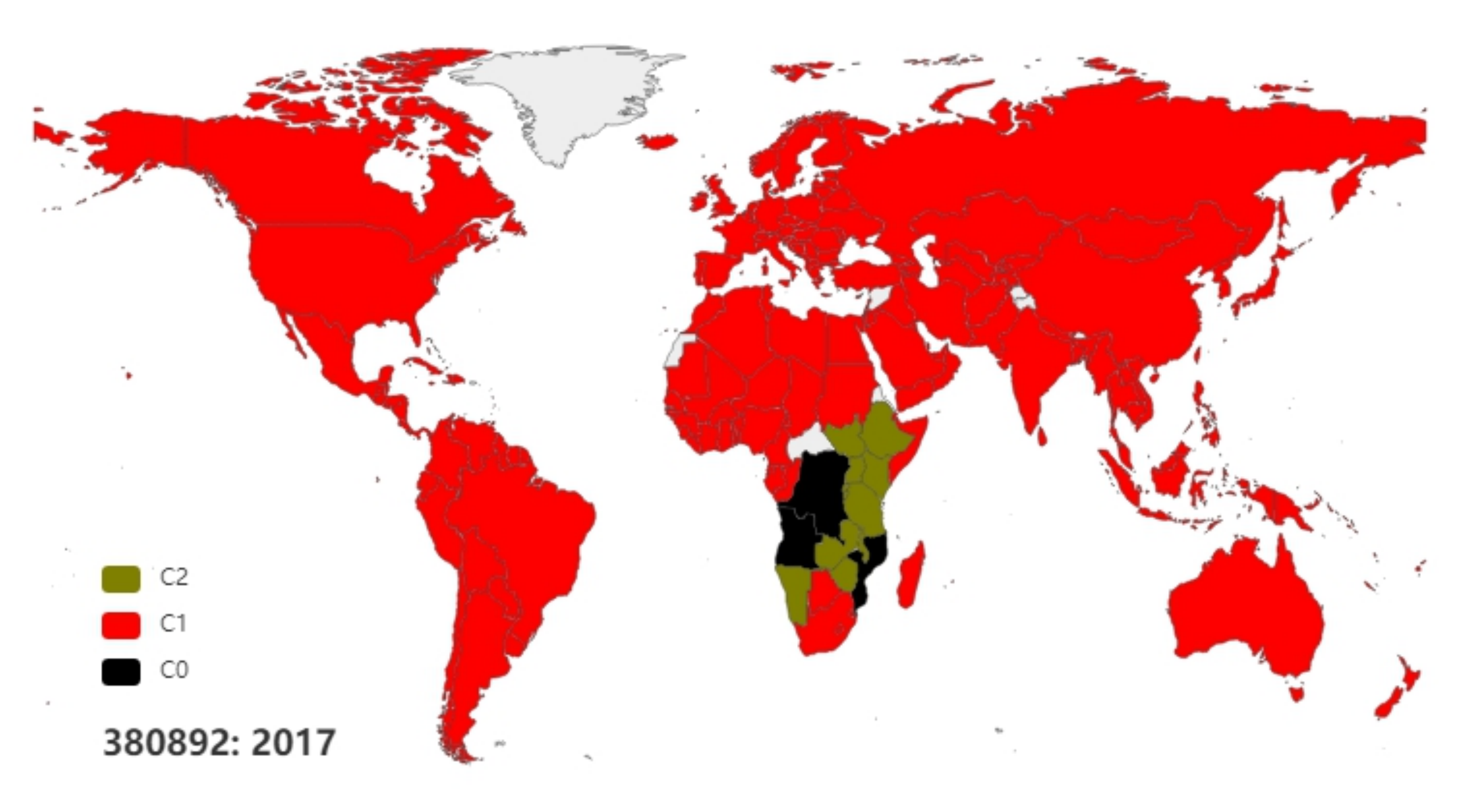}
    \includegraphics[width=0.321\linewidth]{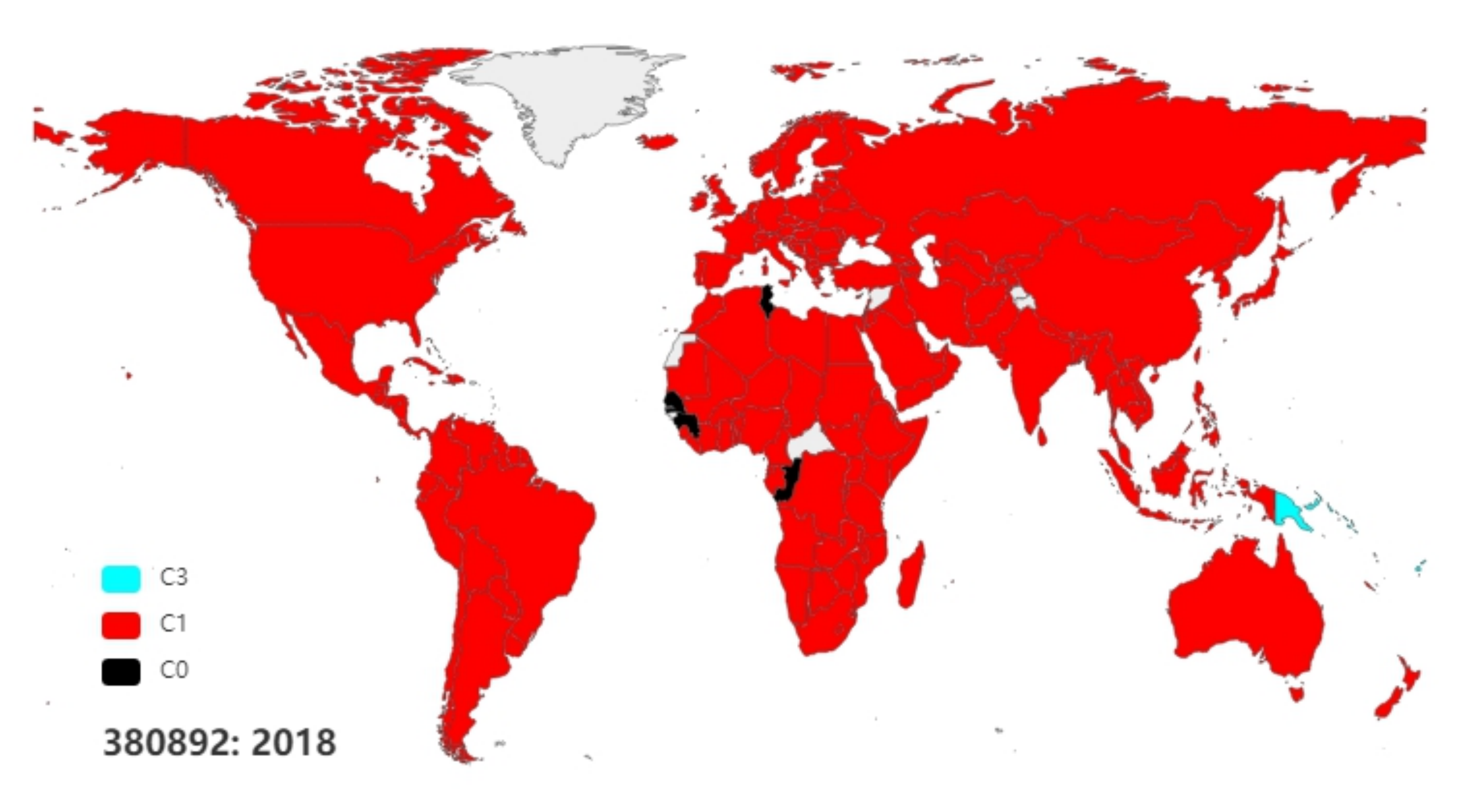}
    \caption{Community evolution of the directed iPTNs of fungicides (380892) from 2007 to 2018. For better visibility, the non-trivial communities containing less than 5 economies and trivial communities are merged to $C_0$.}
    \label{Fig:iPTN:directed:CommunityMap:380892}
\end{figure}

Figure~\ref{Fig:iPTN:directed:CommunityMap:380892} illustrates the evolution  of communities of the directed iPTNs of fungicides (380892) from 2007 to 2018. For better visibility, the non-trivial communities containing less than 5 economies and trivial communities are merged to $C_0$.

We find that the community structure of the two networks in 2017 and 2018 is dominated by a big community and very different from other networks. In 2017, there are 3 communities, in which $C_{2}$ contains 13 economies (Tanzania, Zambia, Kenya, Ethiopia, Uganda, Namibia, Burundi, Rwanda, Malawi, Zimbabwe, S. Sudan, Comoros, and Turks and Caicos Is.) and $C_{3}$ (or $C_0$) contains 3 economies (Angola, Dem. Rep. Congo, and Mozambique), mainly in Africa. In 2018, $C_{3}$ contains Solomon Is., Fiji, Papua New Guinea, Samoa, Kiribati, and Tuvalu in the South Pacific Ocean, while $C_0$ contains two communities $C_{2}$ (Tunisia and Congo) and $C_{4}$ (Gambia, Guinea, and Senegal), all in Africa.

\begin{figure}[!ht]
    \centering
    \includegraphics[width=0.321\linewidth]{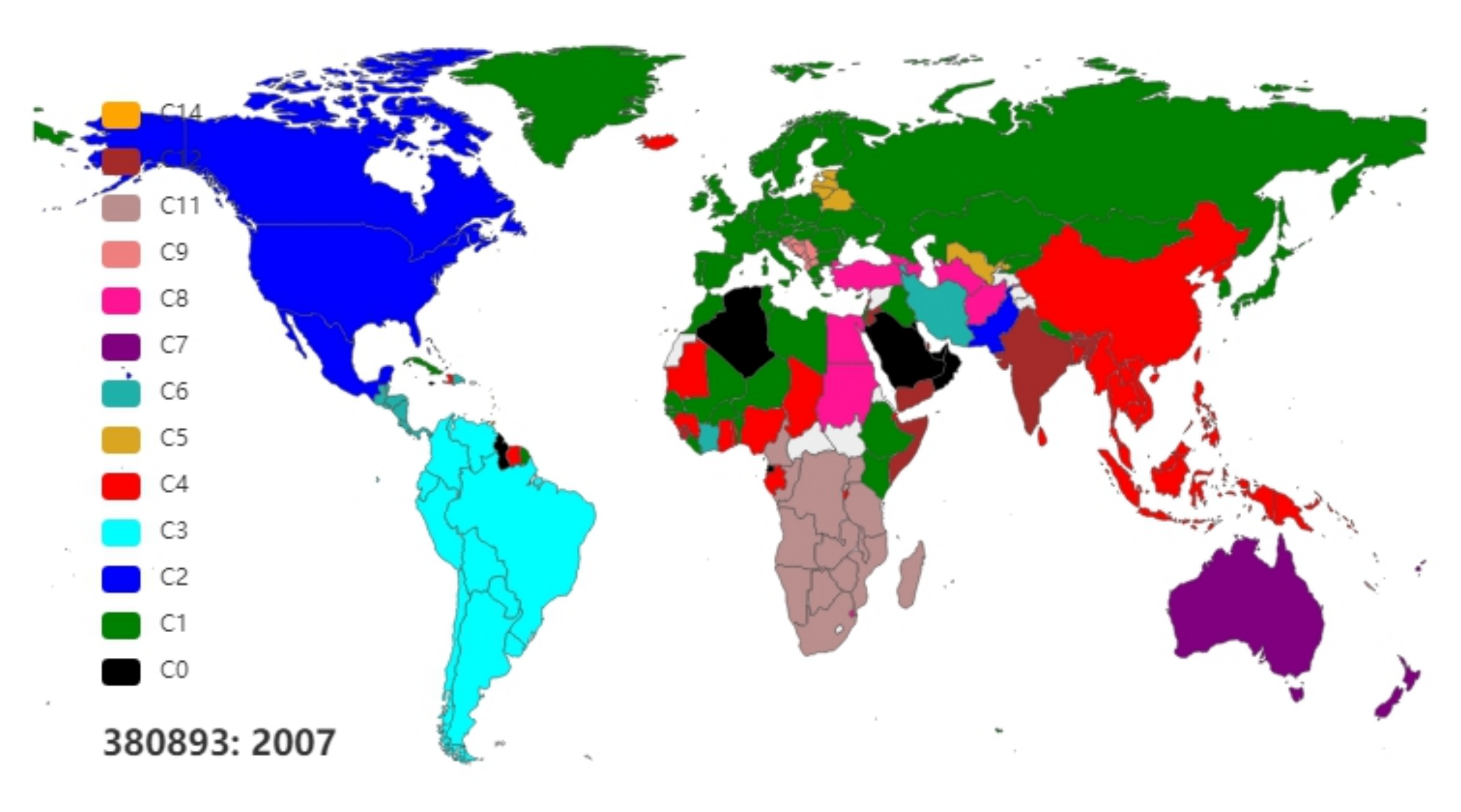}
    \includegraphics[width=0.321\linewidth]{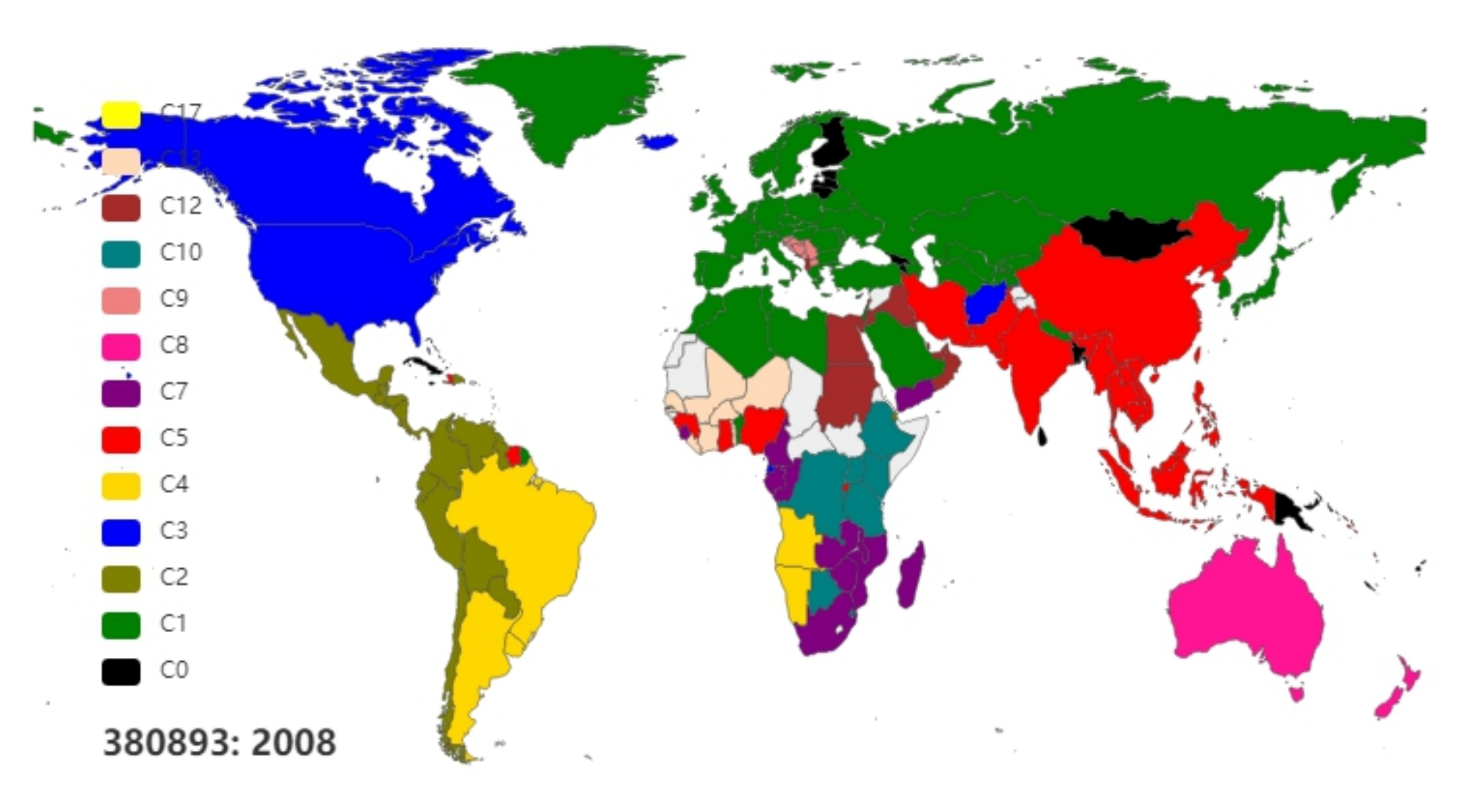}
    \includegraphics[width=0.321\linewidth]{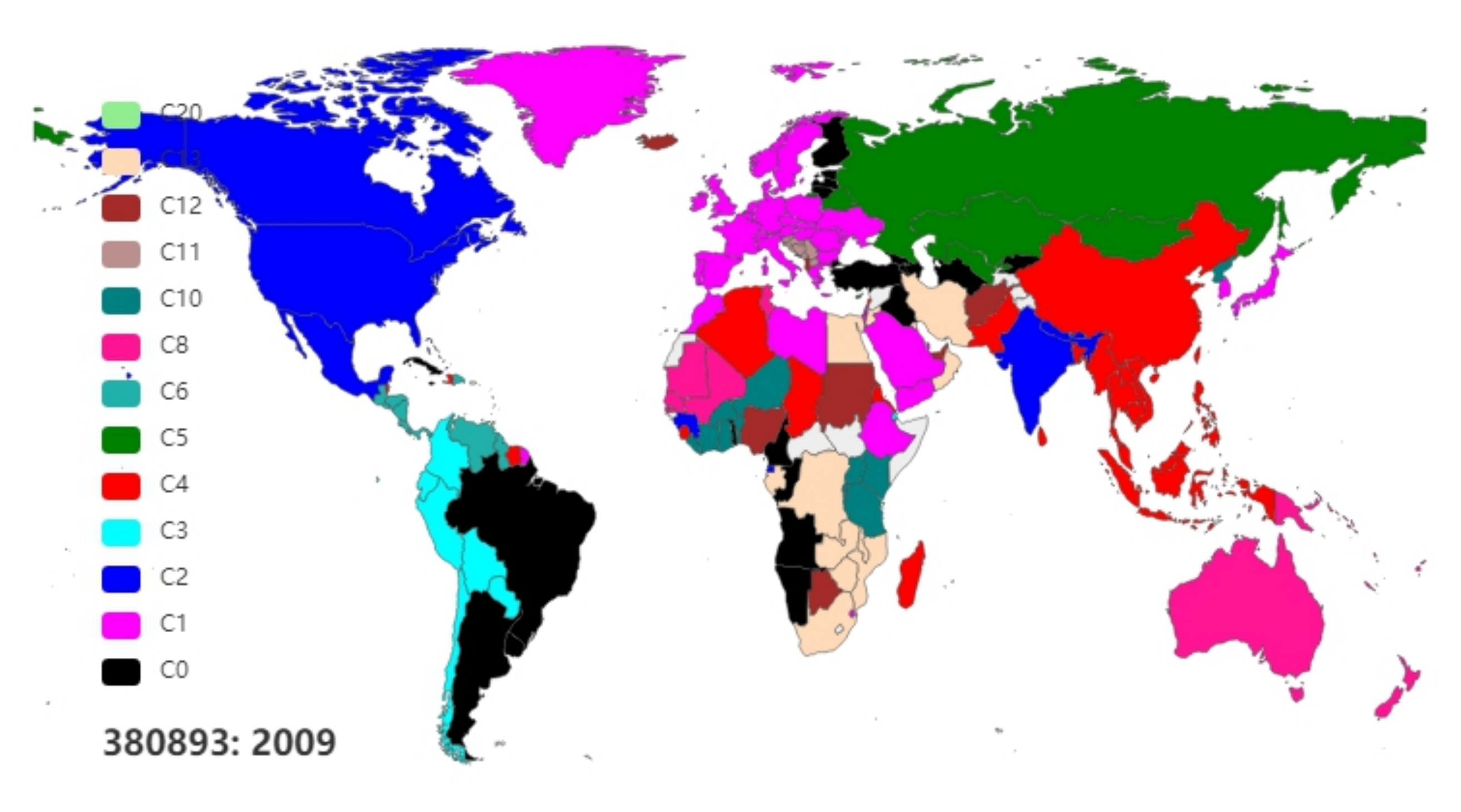}
    \includegraphics[width=0.321\linewidth]{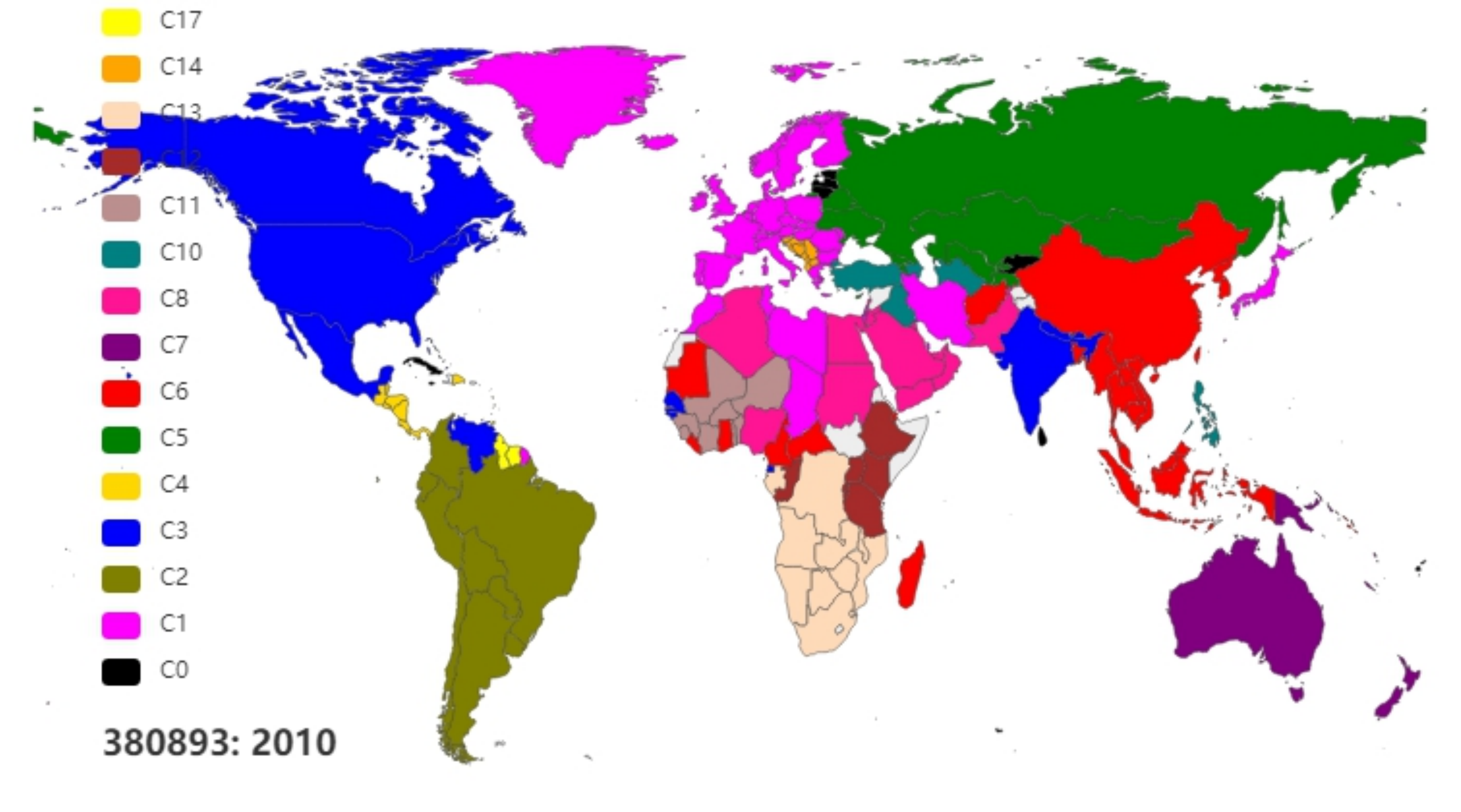}
    \includegraphics[width=0.321\linewidth]{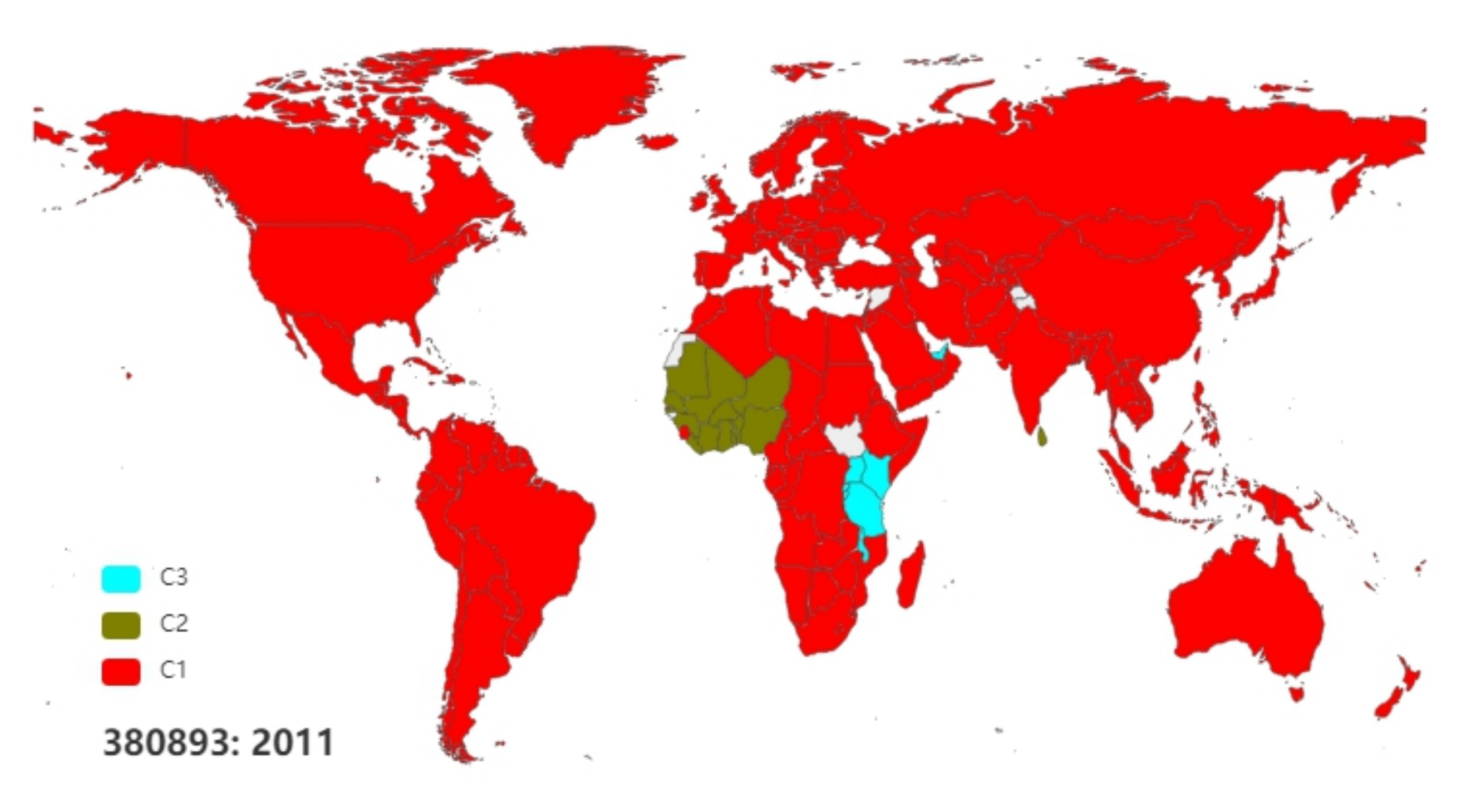}
    \includegraphics[width=0.321\linewidth]{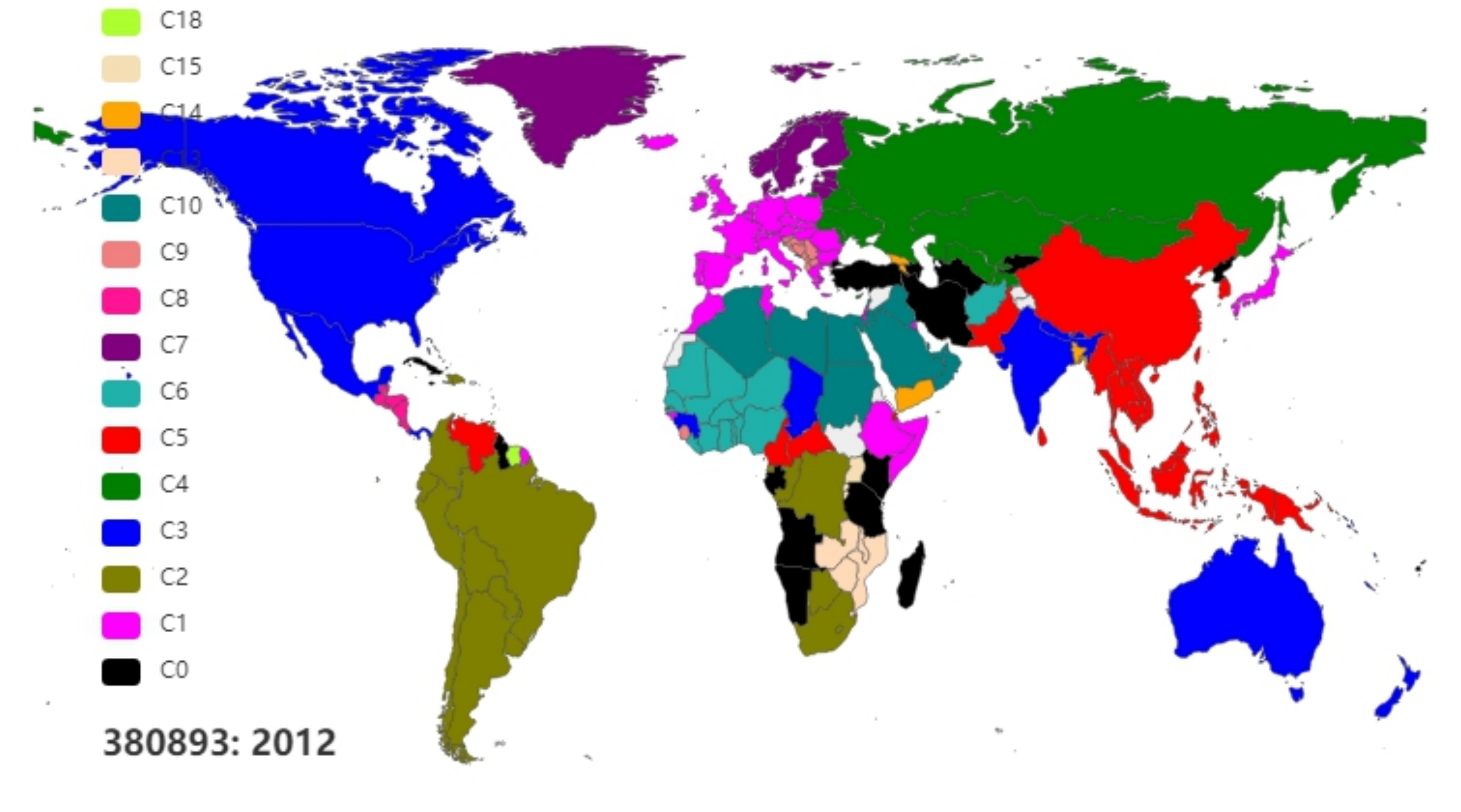}
    \includegraphics[width=0.321\linewidth]{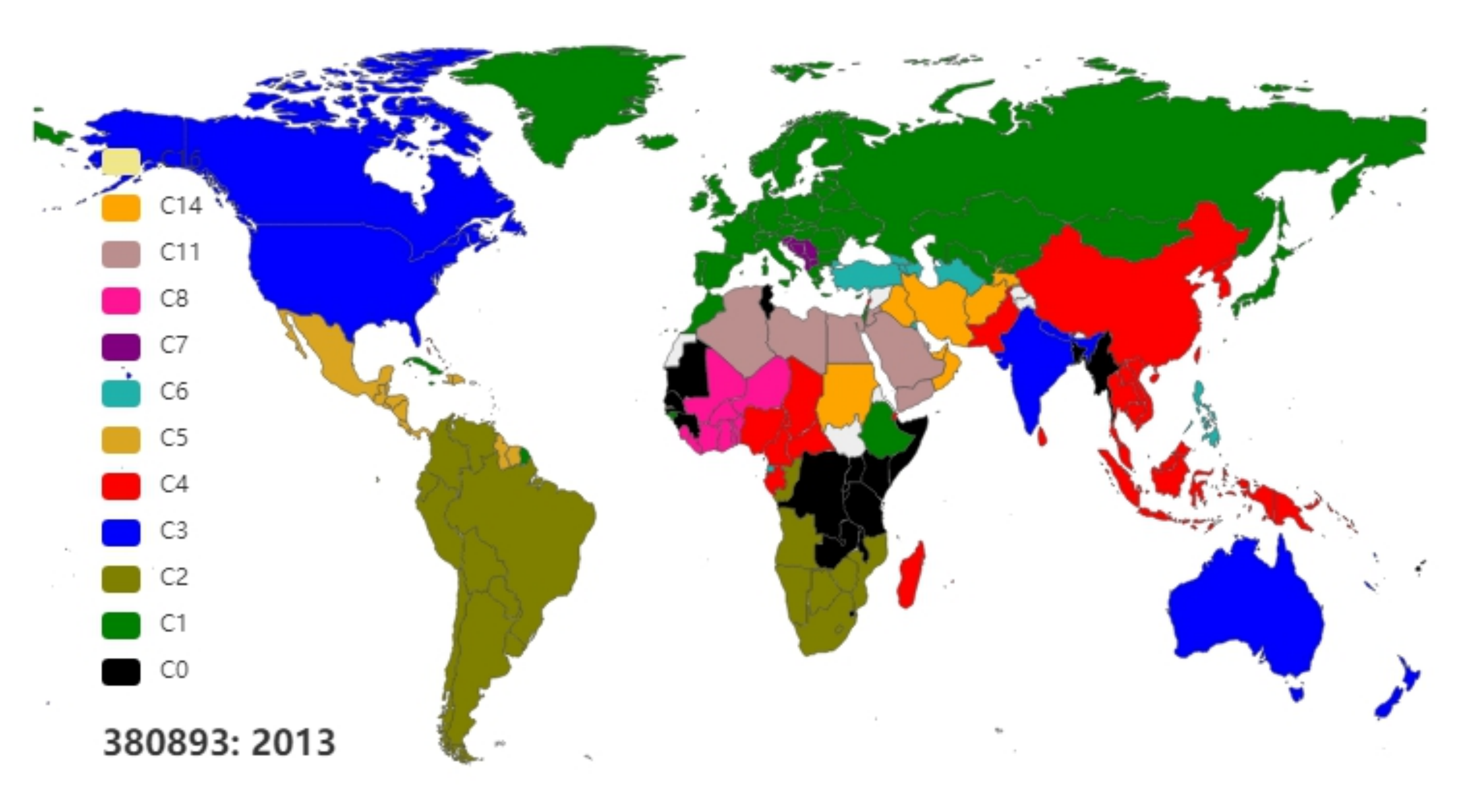}
    \includegraphics[width=0.321\linewidth]{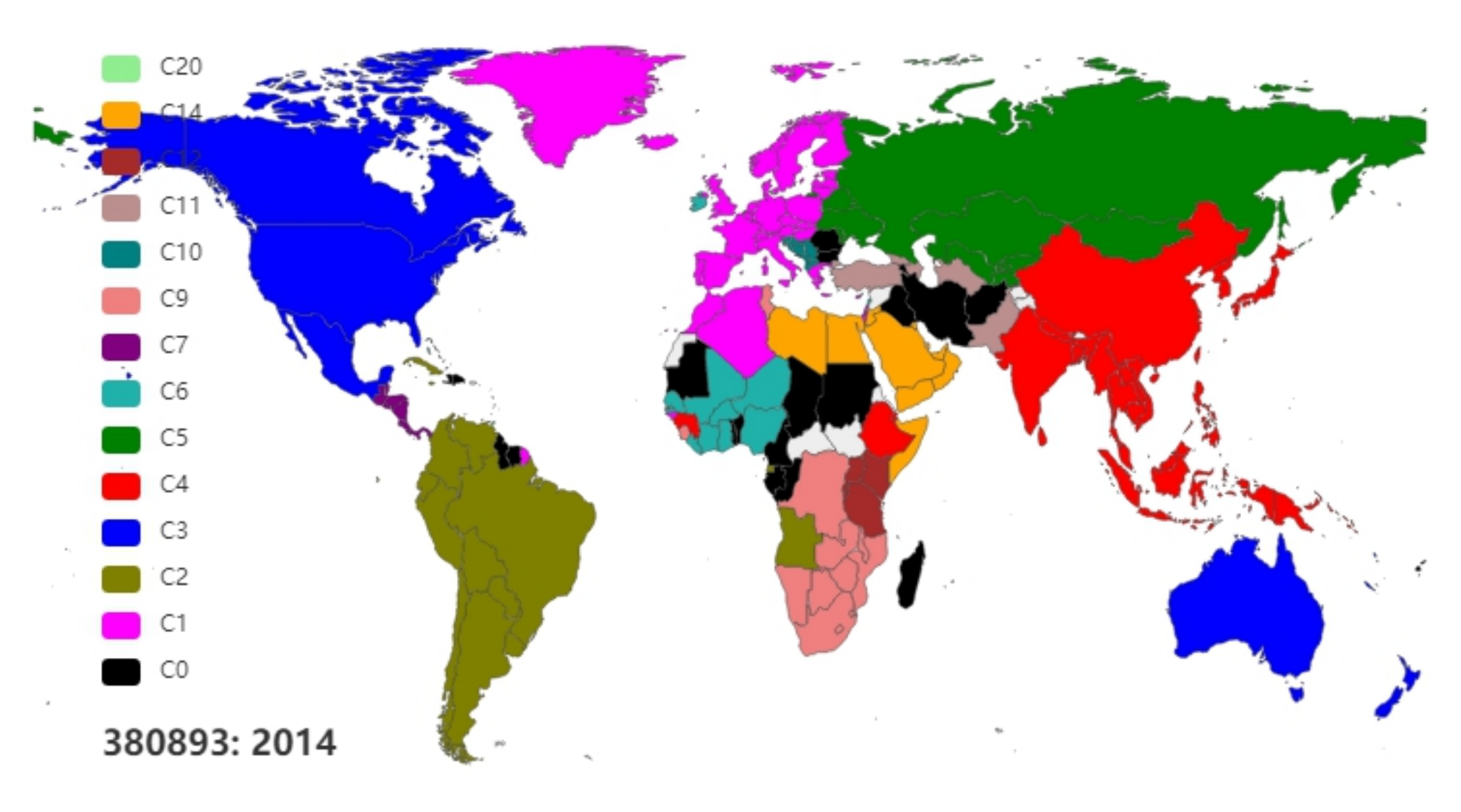}
    \includegraphics[width=0.321\linewidth]{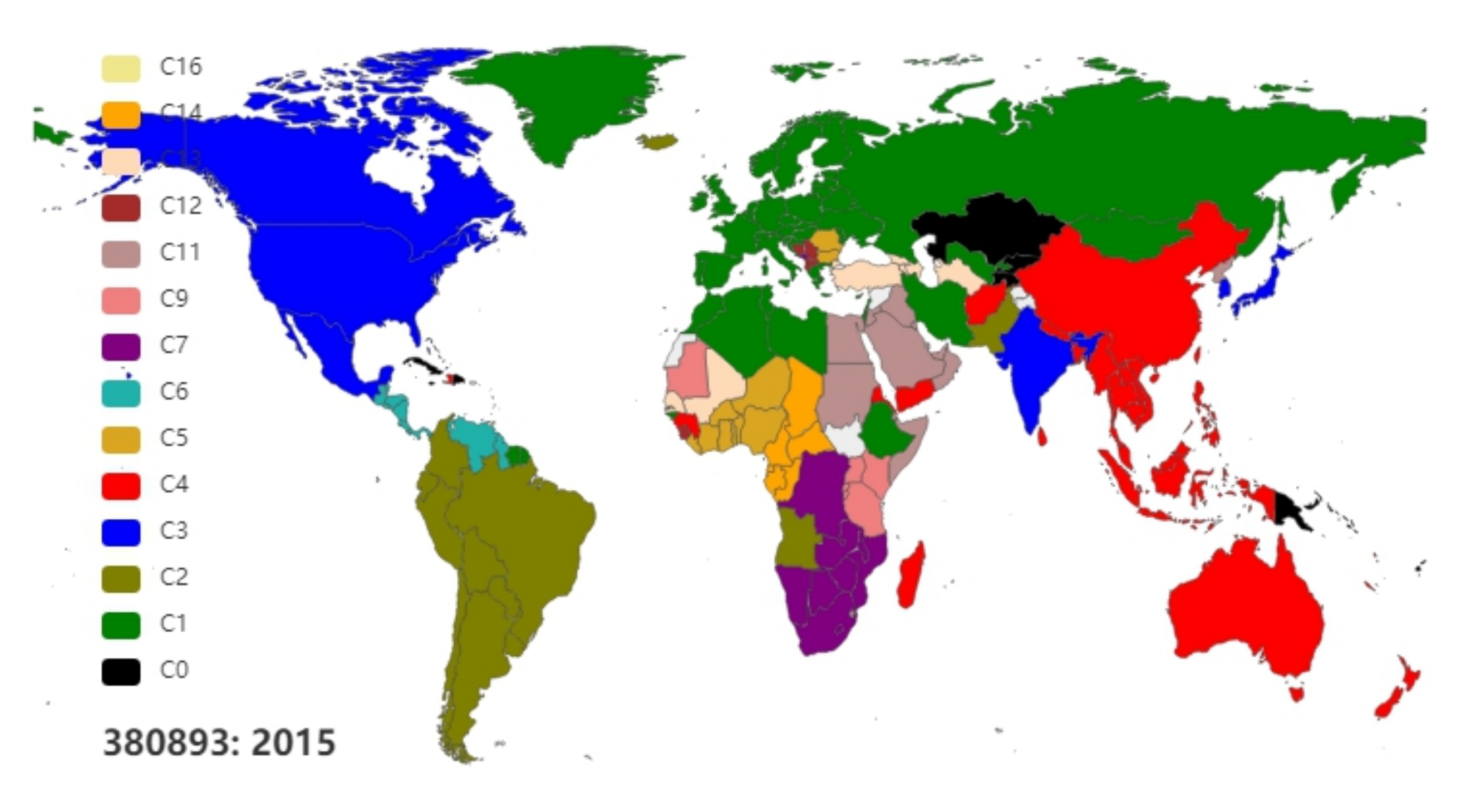}
    \includegraphics[width=0.321\linewidth]{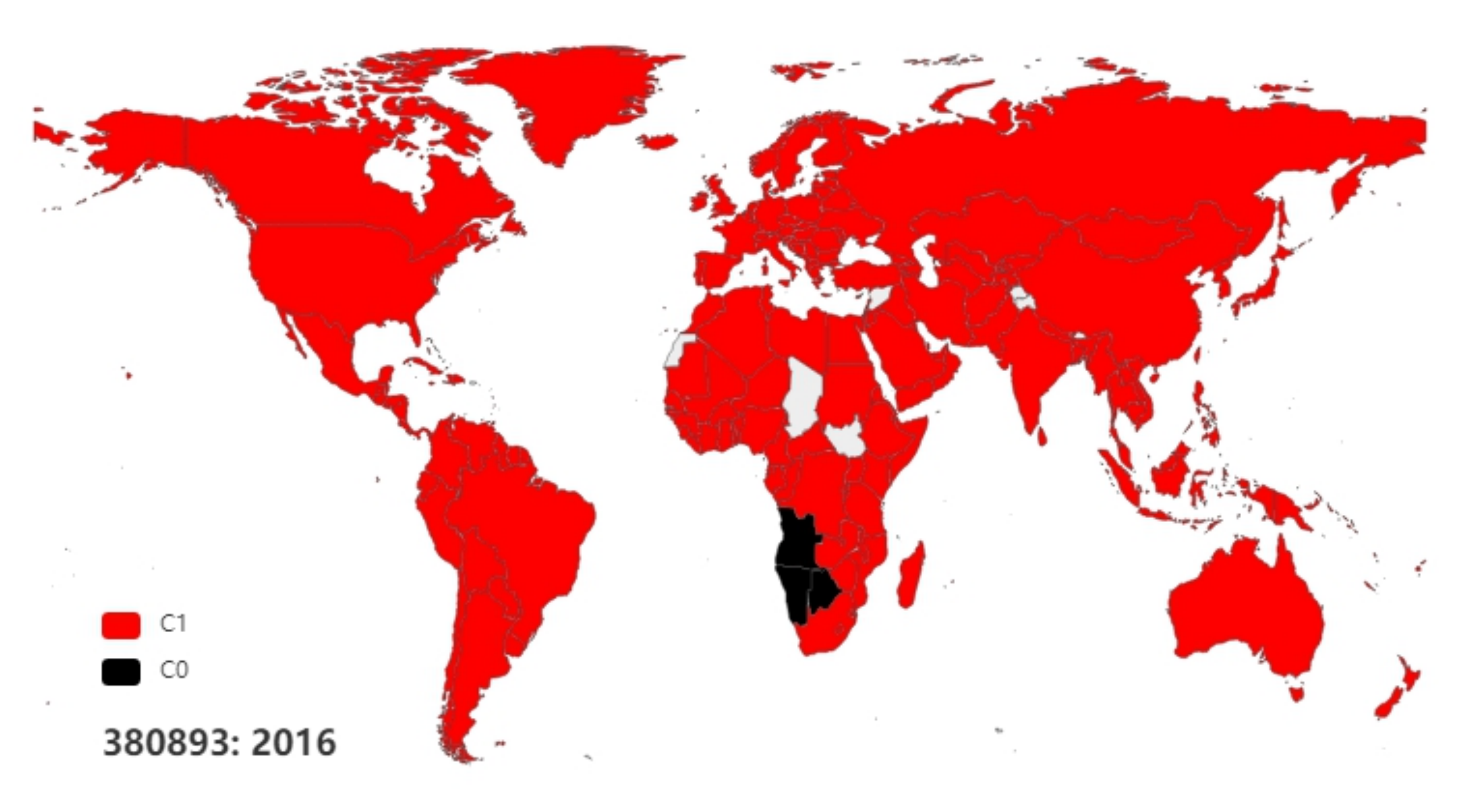}
    \includegraphics[width=0.321\linewidth]{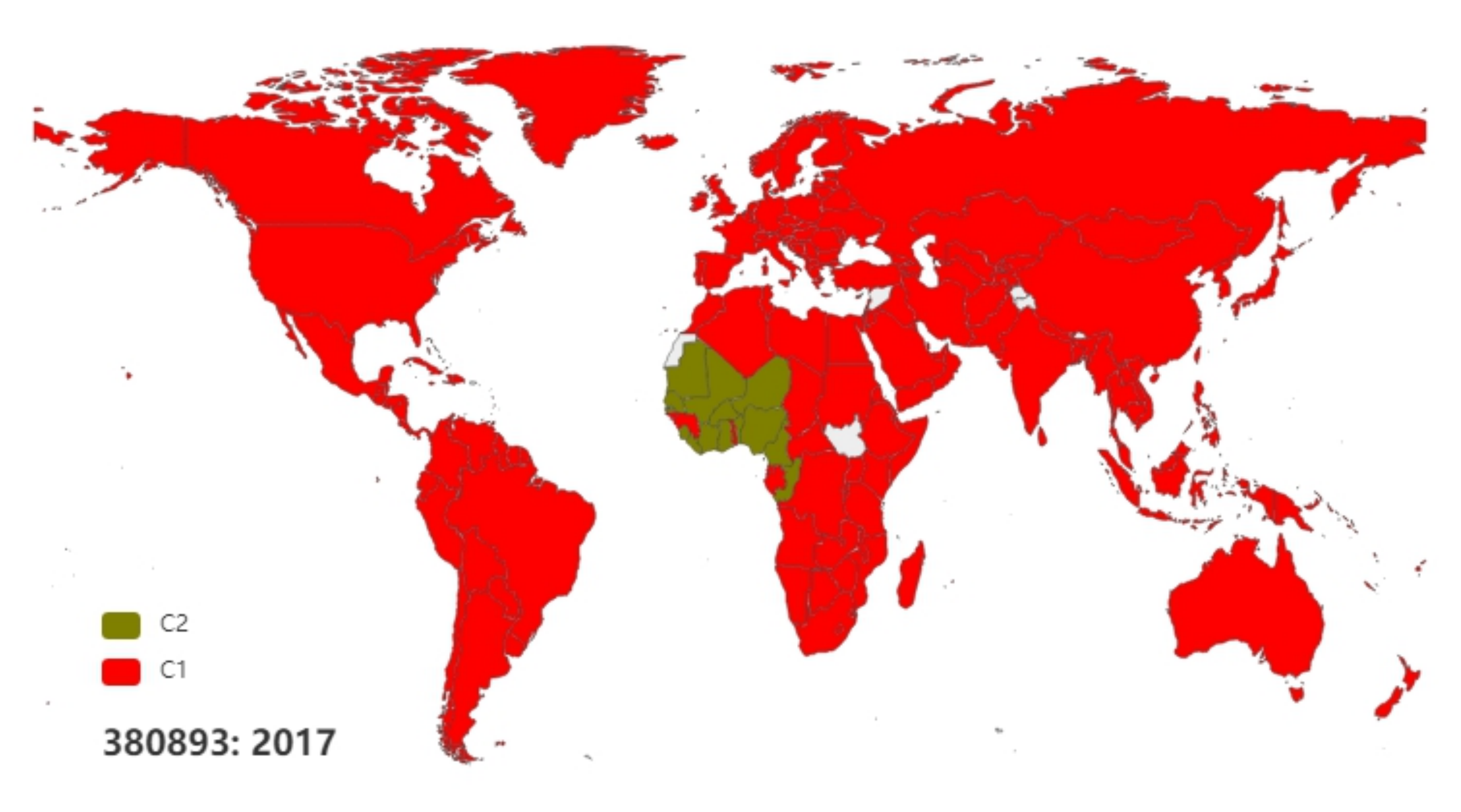}
    \includegraphics[width=0.321\linewidth]{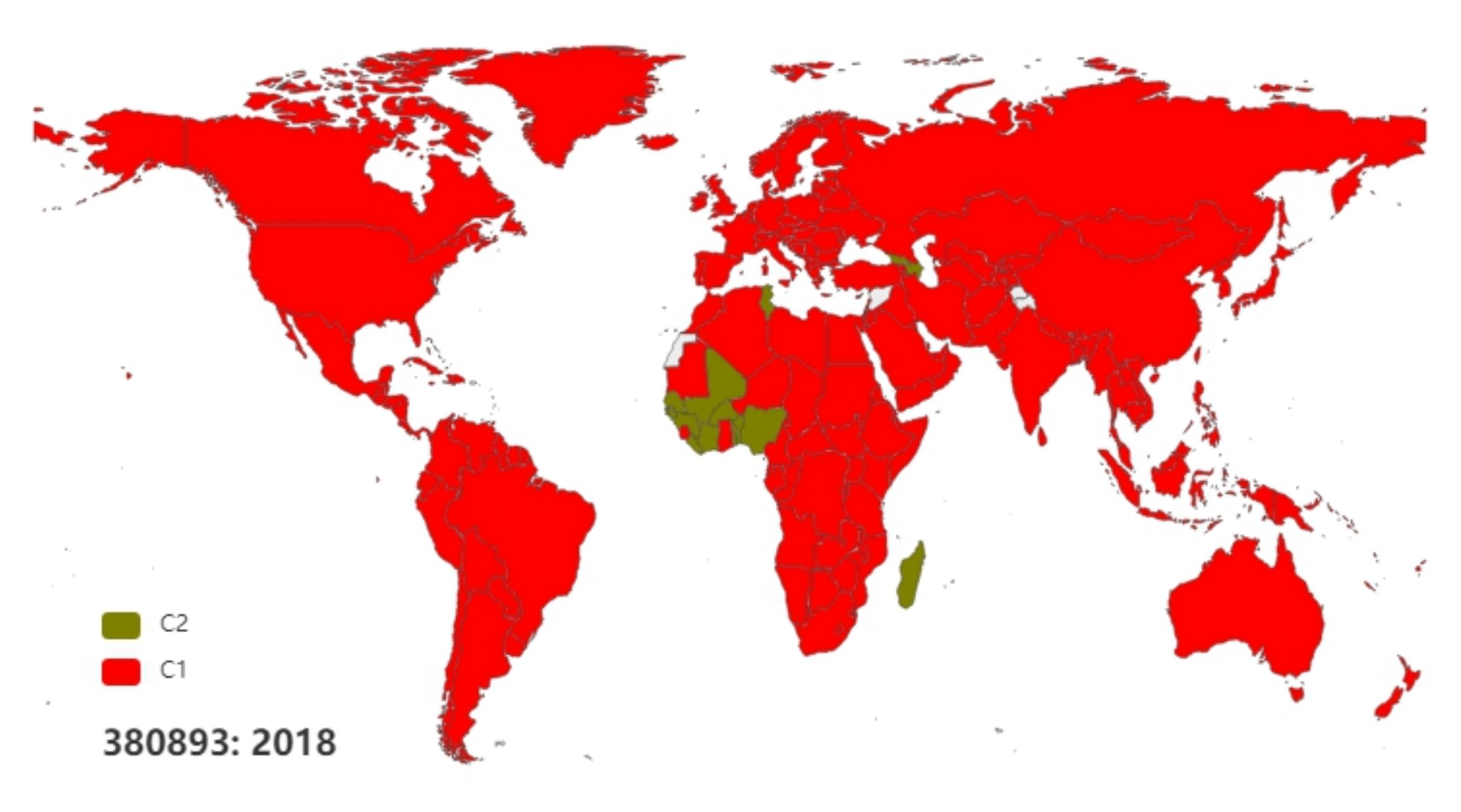}
    \caption{Community evolution of the directed iPTNs of herbicides (380893) from 2007 to 2018. For better visibility, the non-trivial communities containing less than 5 economies and trivial communities are merged to $C_0$.}
    \label{Fig:iPTN:directed:CommunityMap:380893}
\end{figure}

Figure~\ref{Fig:iPTN:directed:CommunityMap:380893} illustrates the evolution of communities of the directed iPTNs of herbicides (380893) from 2007 to 2018. For better visibility, the non-trivial communities containing less than 5 economies and trivial communities are merged to $C_0$. 
We observe that the most evident feature of the community maps is that the four networks in 2011, 2016, 2017 and 2018 are dominated by a big community. In 2011, we identify three communities, in which $C_{2}$ contains 16 economies (Antigua and Barb., Nigeria, C{\^{o}}te d'Ivoire, Senegal, Sri Lanka, Ghana, Liberia, Mali, Burkina Faso, Bouvet Island, Maldives, Benin, Guinea, Mauritania, Niger, and Togo) and $C_{3}$ contains 7 economies ('United Arab Emirates, Kenya, Burundi, Rwanda, Uganda, Tanzania, and Malawi). 
In 2016, we find two economies, in which $C_{2}$ (or $C_0$) contains Angola, S{\~{a}}o Tom{\'{e}} and Principe, Namibia, and Botswana. 
In 2017, we find two economies, in which $C_{2}$ contains Andorra, Nigeria, Cameroon, Ghana, C{\^{o}}te d'Ivoire, Liberia, Mali, Senegal, Congo, Benin, Gambia, Mauritania, Niger, Sierra Leone, and Burkina Faso. 
In 2018, we also find two economies, in which $C_{2}$:  Azerbaijan, Georgia, Senegal, Tunisia, C{\^{o}}te d'Ivoire, Mali, Nigeria, Benin, Gambia, Guinea, Liberia, Madagascar, Togo, Burkina Faso, and Guinea-Bissau. 
A common feature is that almost all the economies of the 5 small communities in the 4 networks locate in Africa.

We now turn to investigate the rest 8 networks. Firstly, the main economies in North America and South America do not form a united community. Secondly, most European economies belong to the same community in 2007, 2008, 2013, and 2015, and divide into two communities in other years. Thirdly, India does not belong to the Asia community, except in 2008 and 2012. Fourthly, the Oceania cluster forms a separate community from 2007 to 2010, falls in the same community as North America in 2012, 2013 and 2014, and joins the Asia community in 2015. Fifth, China, Dem. Rep. Korea, and most economies in South Asia and Southeat Asia form the Asia community; In contrast, Japan and Korea belong to the Asia community in 2009, 2010, 2012, and 2014, and to the Russia community in 2007, 2008, and 2013.



\begin{figure}[!ht]
    \centering
    \includegraphics[width=0.321\linewidth]{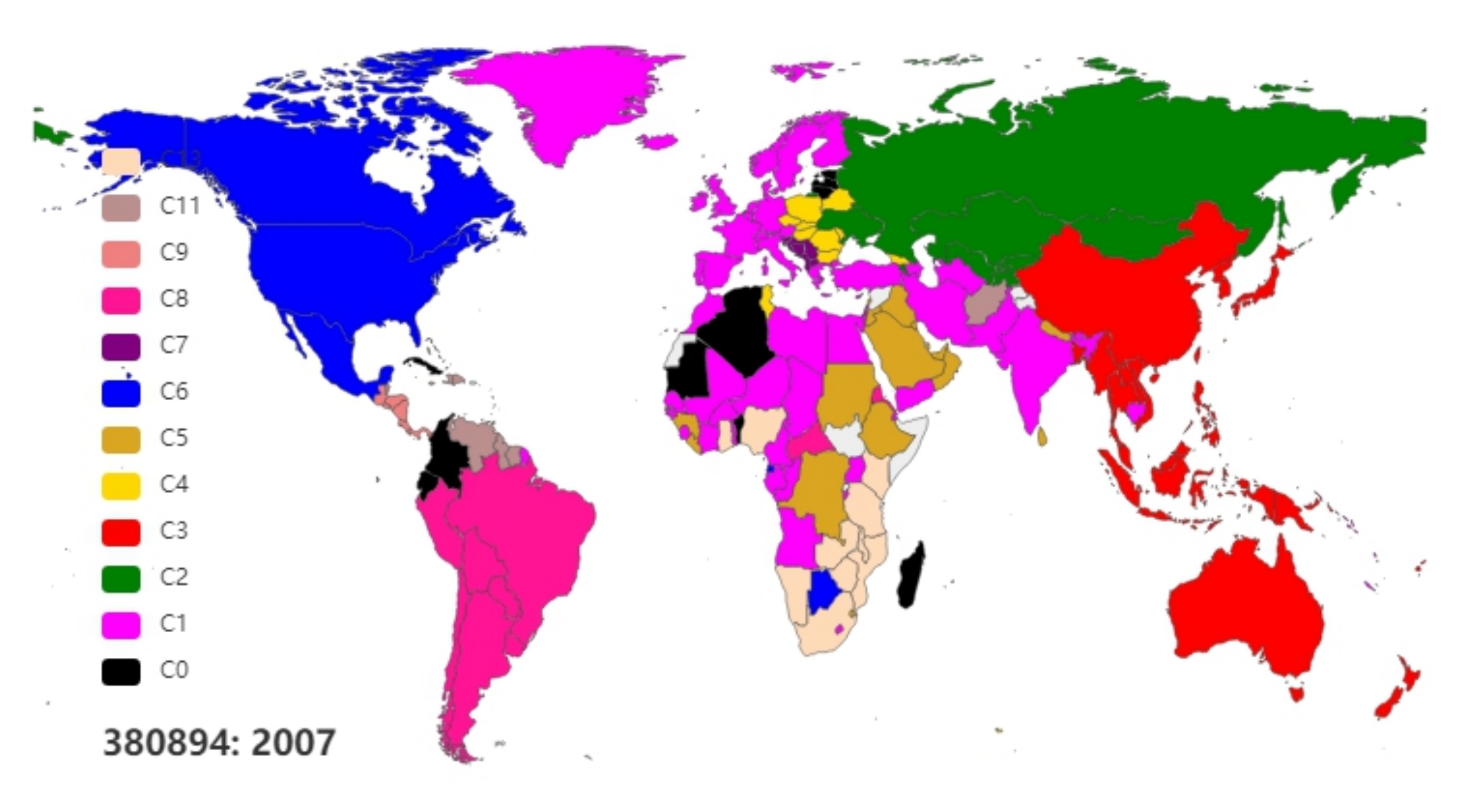}
    \includegraphics[width=0.321\linewidth]{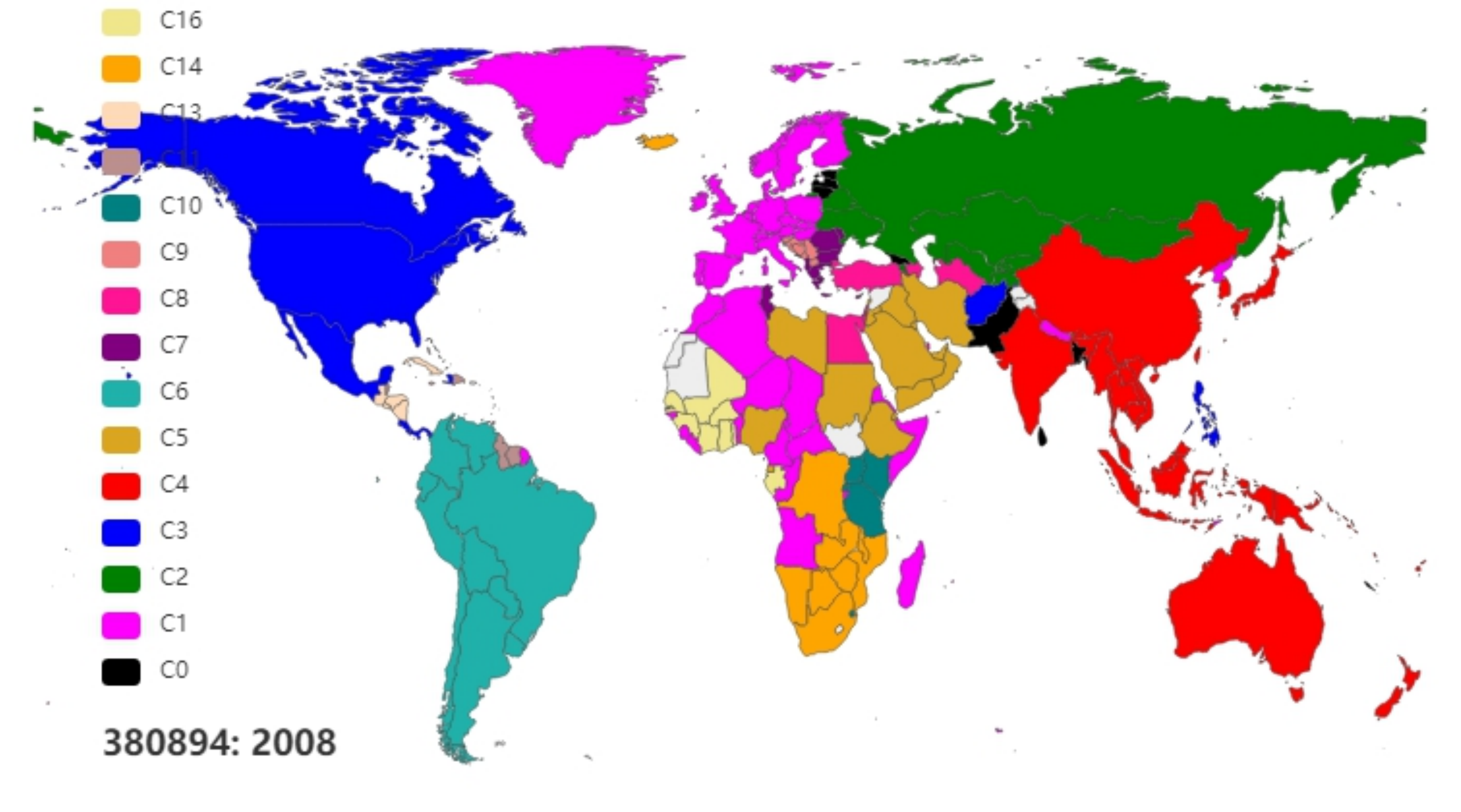}
    \includegraphics[width=0.321\linewidth]{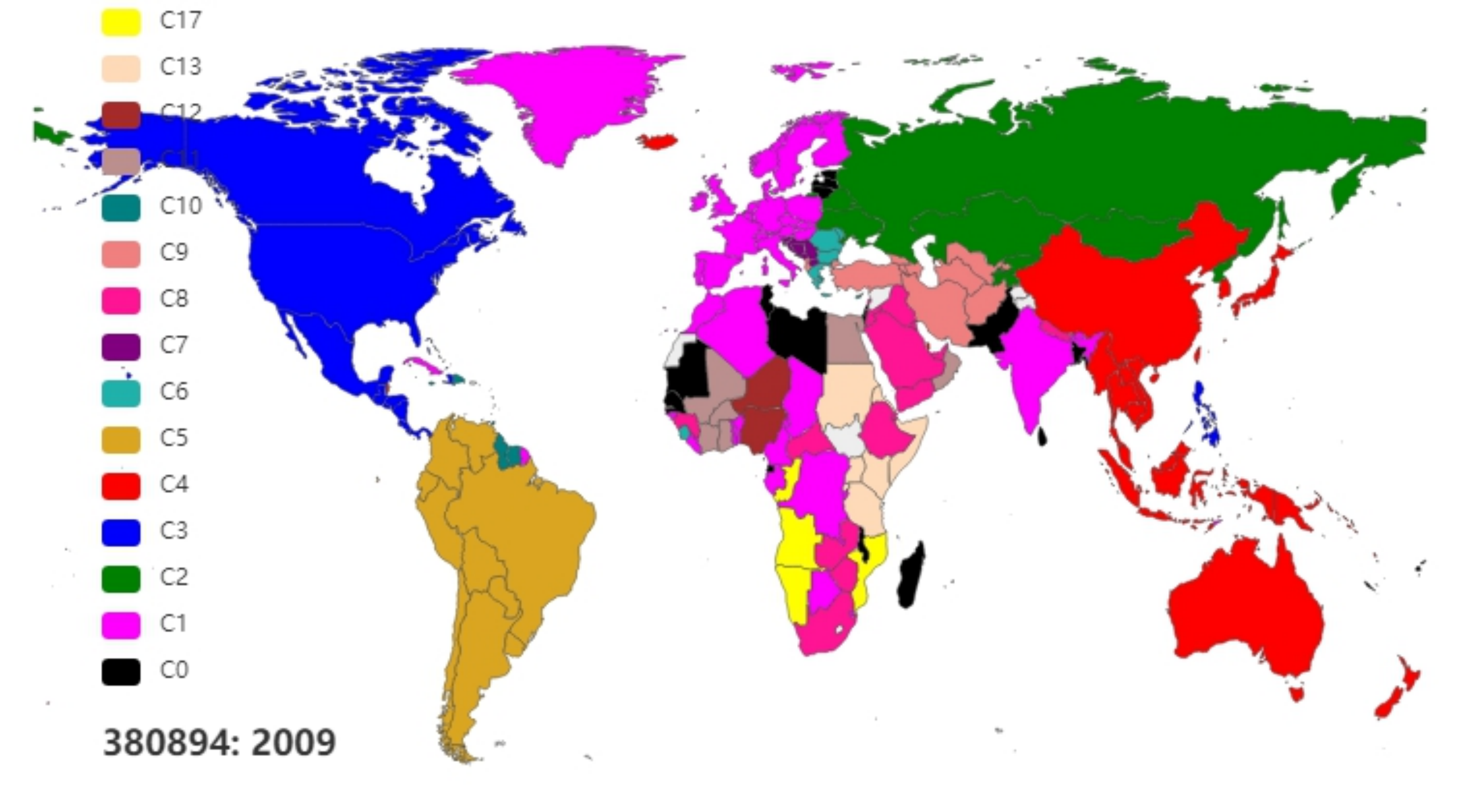}
    \includegraphics[width=0.321\linewidth]{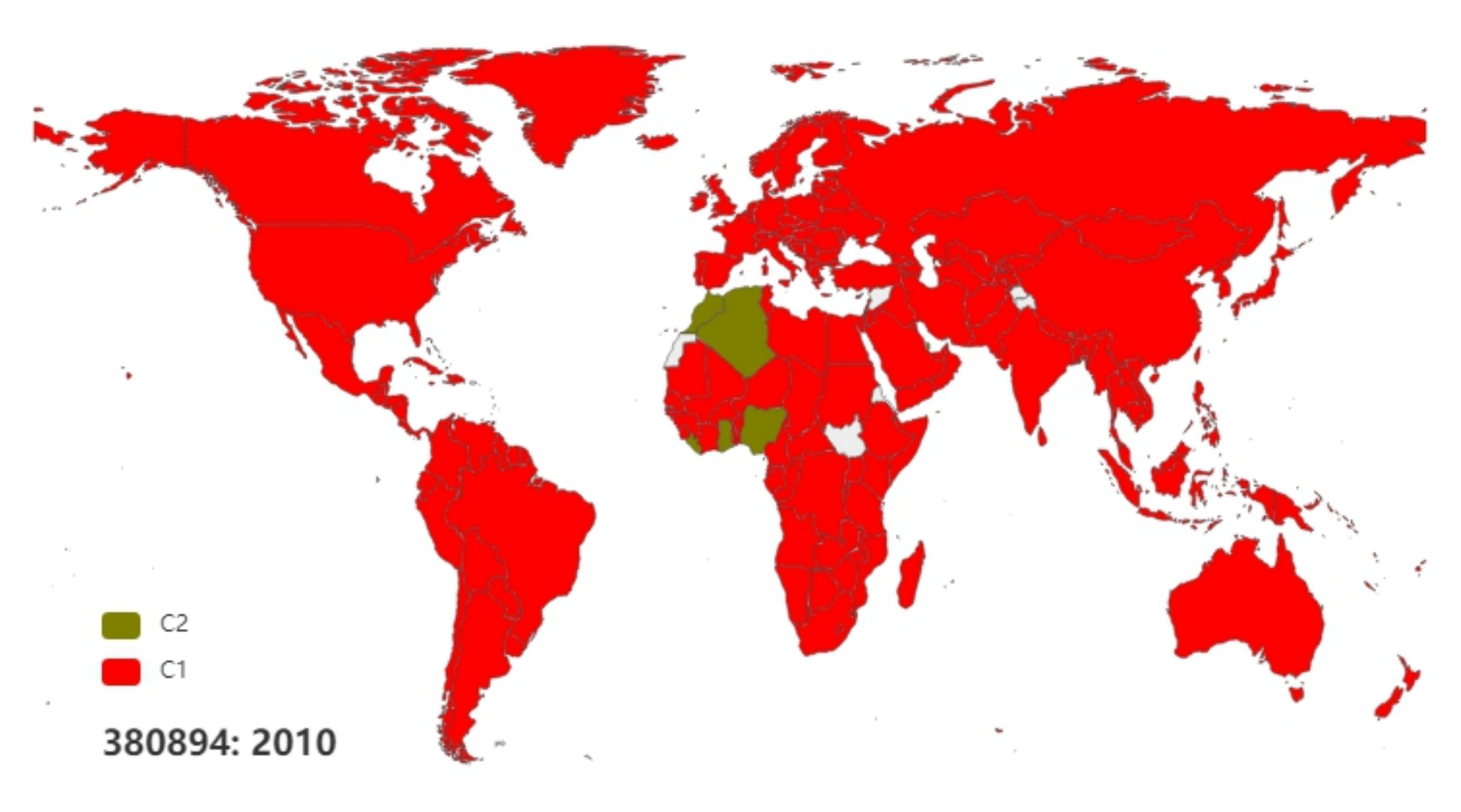}
    \includegraphics[width=0.321\linewidth]{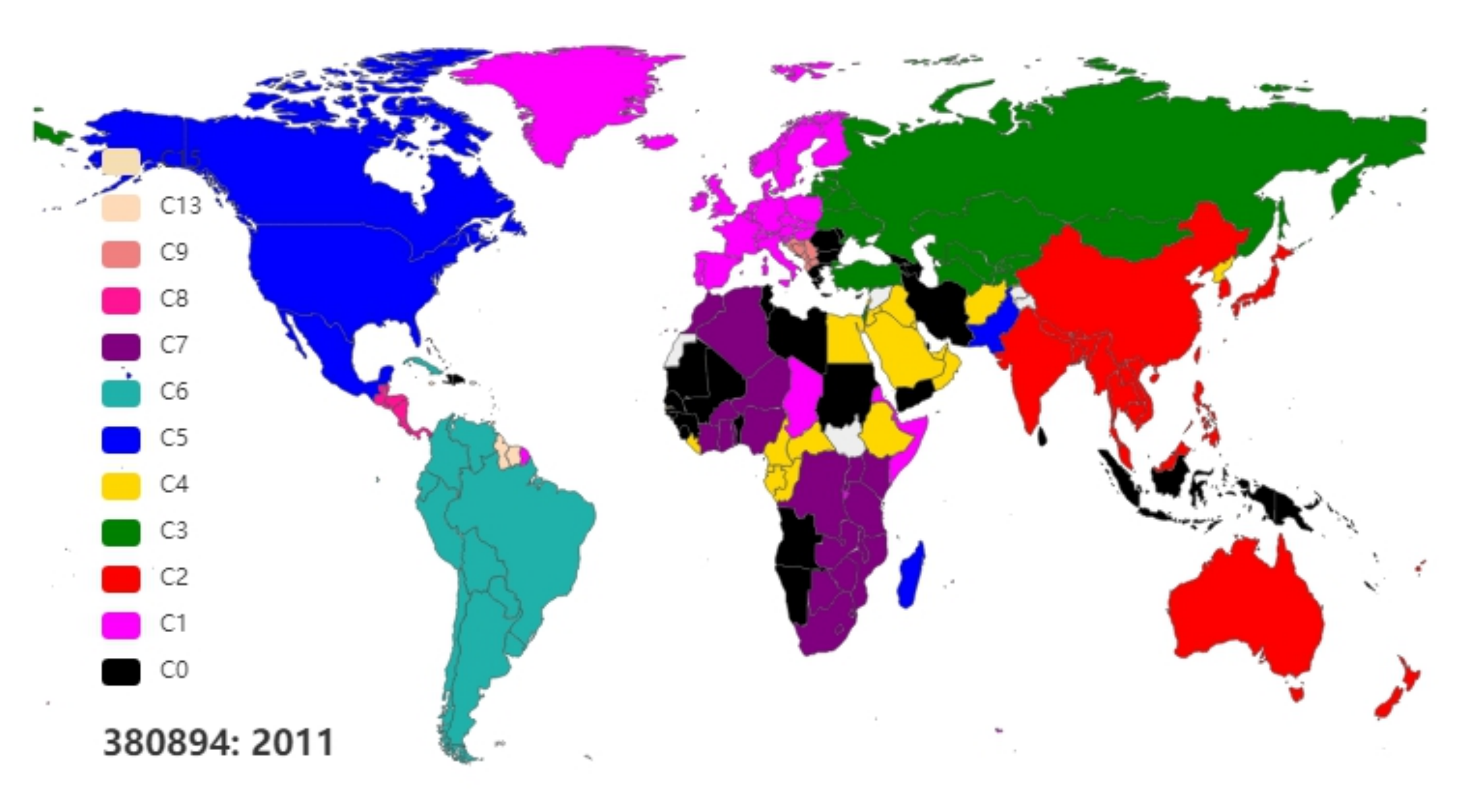}
    \includegraphics[width=0.321\linewidth]{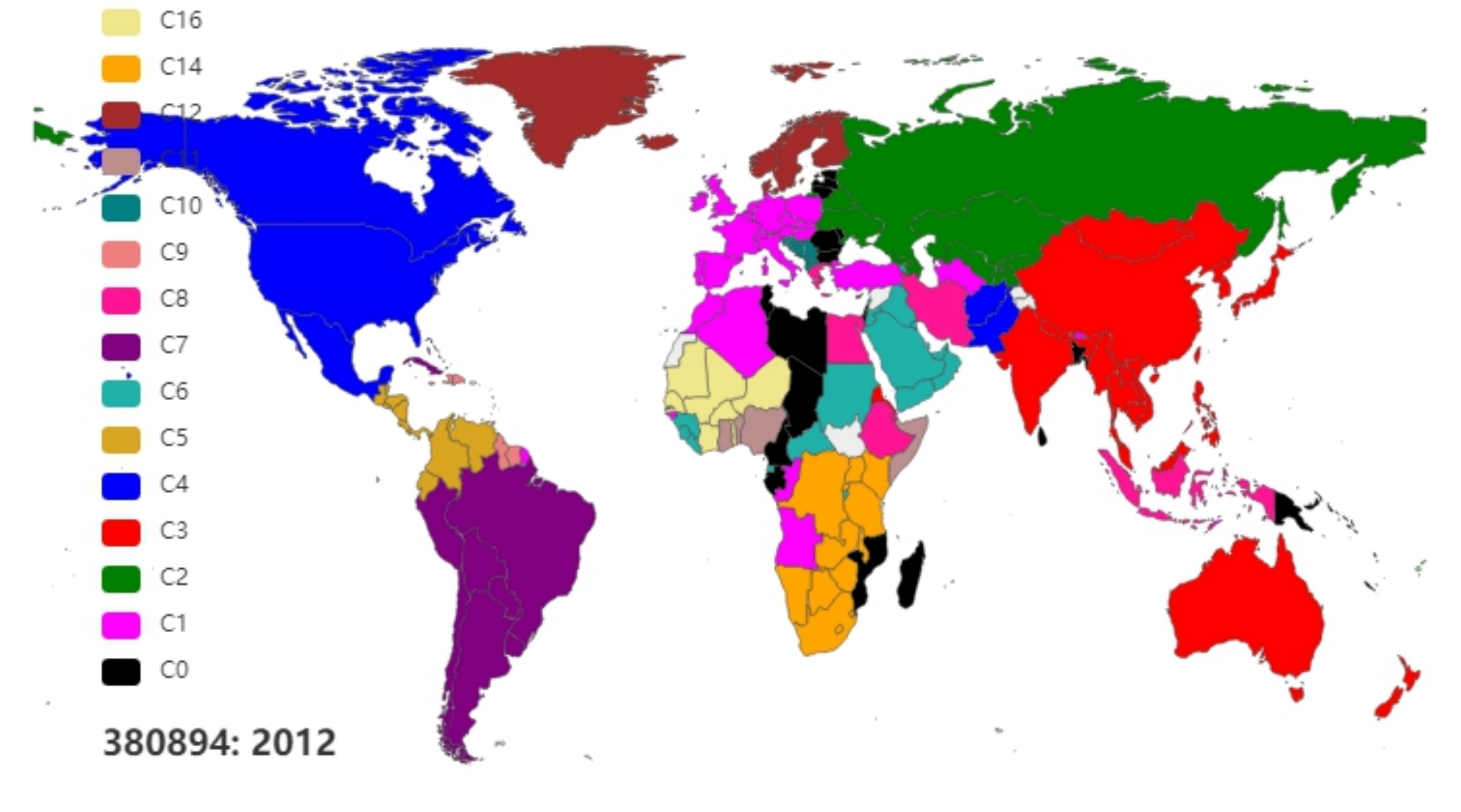}
    \includegraphics[width=0.321\linewidth]{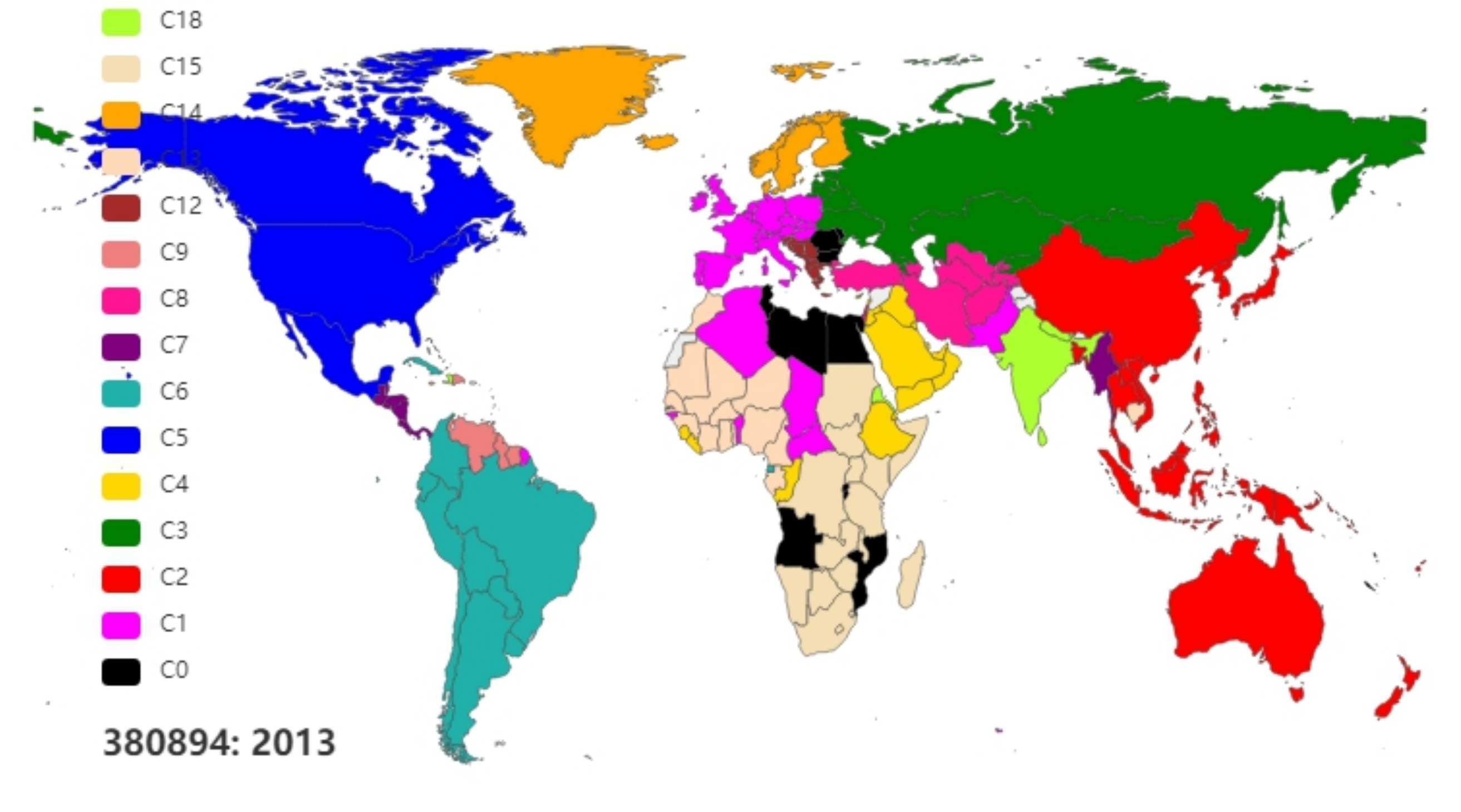}
    \includegraphics[width=0.321\linewidth]{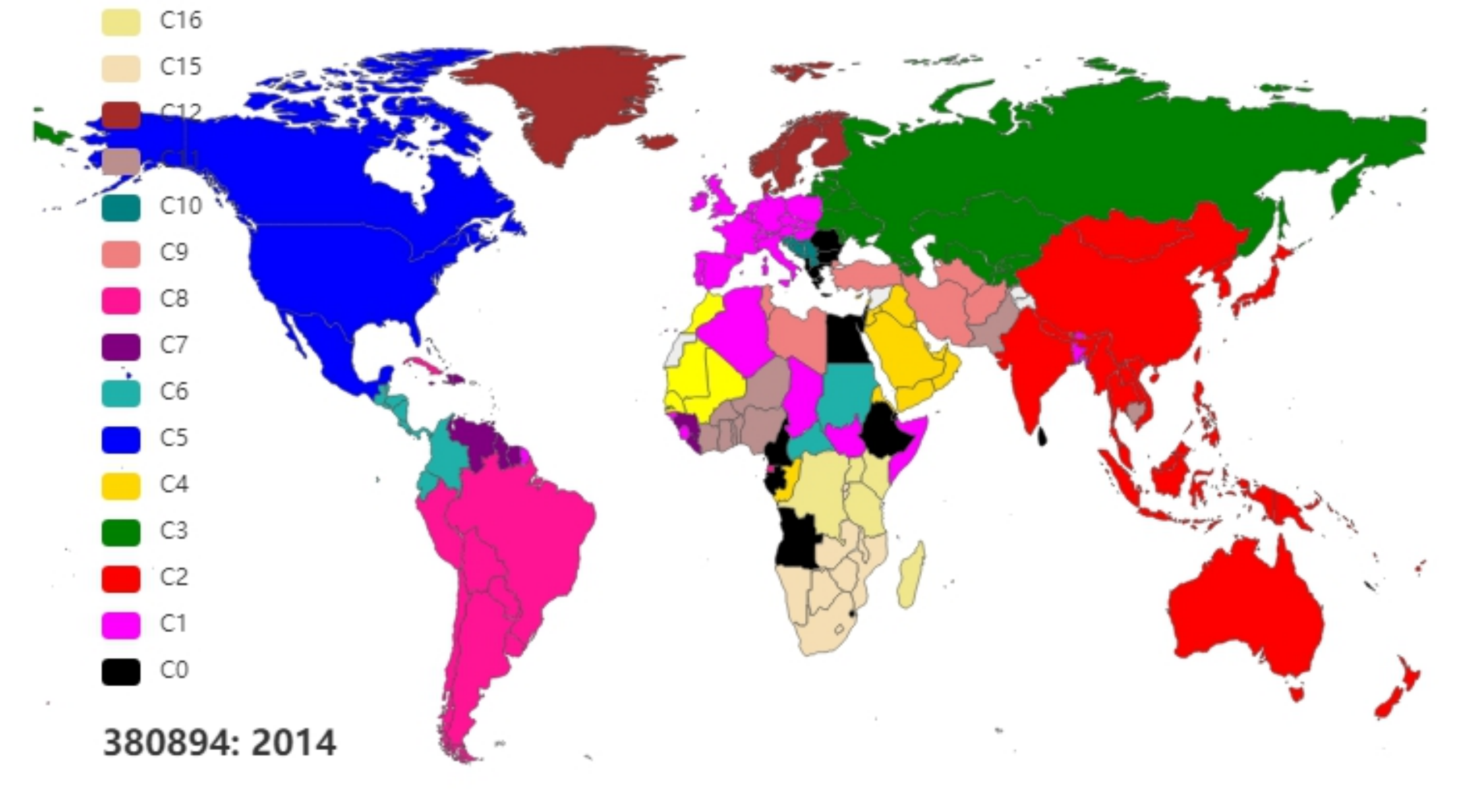}
    \includegraphics[width=0.321\linewidth]{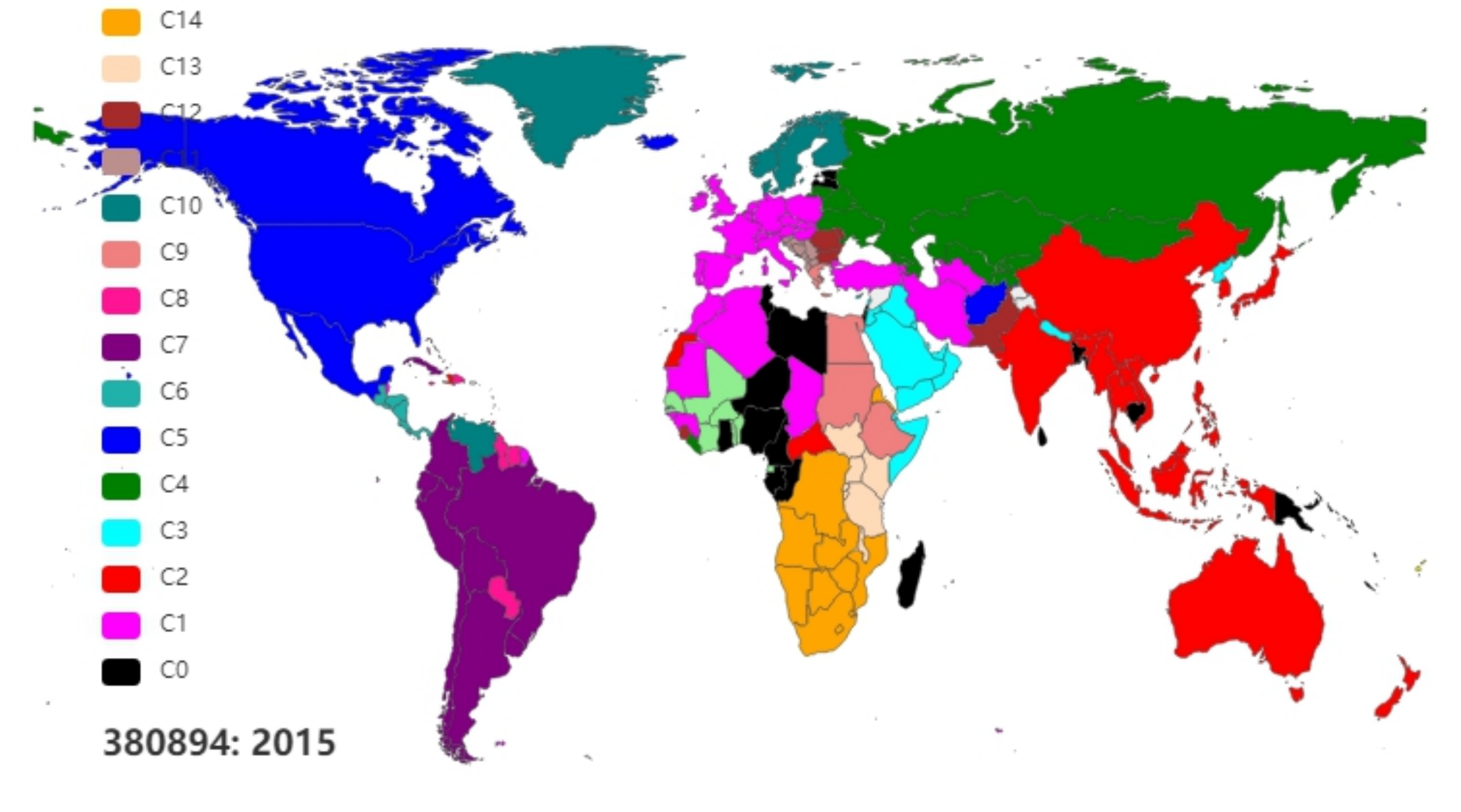}
    \includegraphics[width=0.321\linewidth]{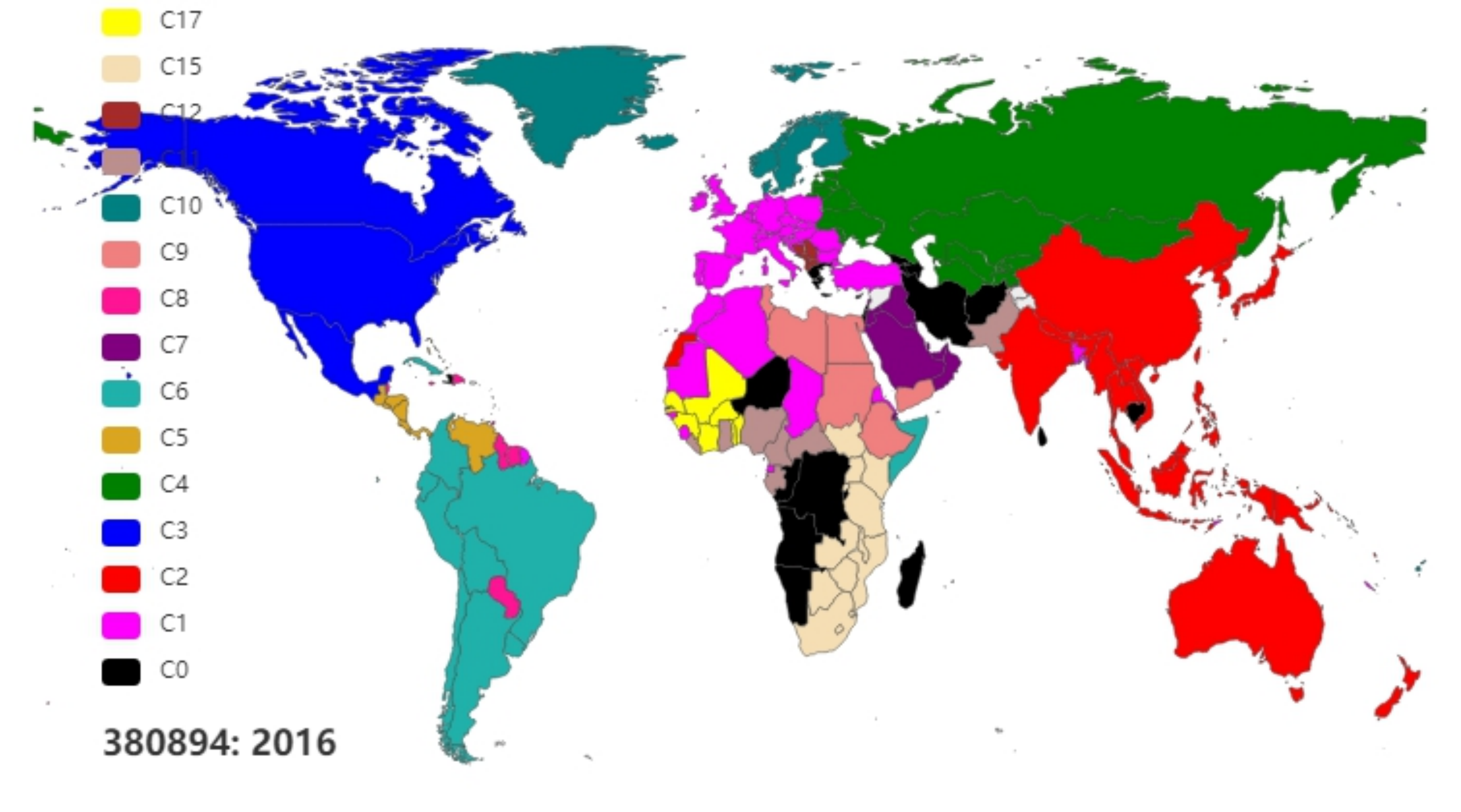}
    \includegraphics[width=0.321\linewidth]{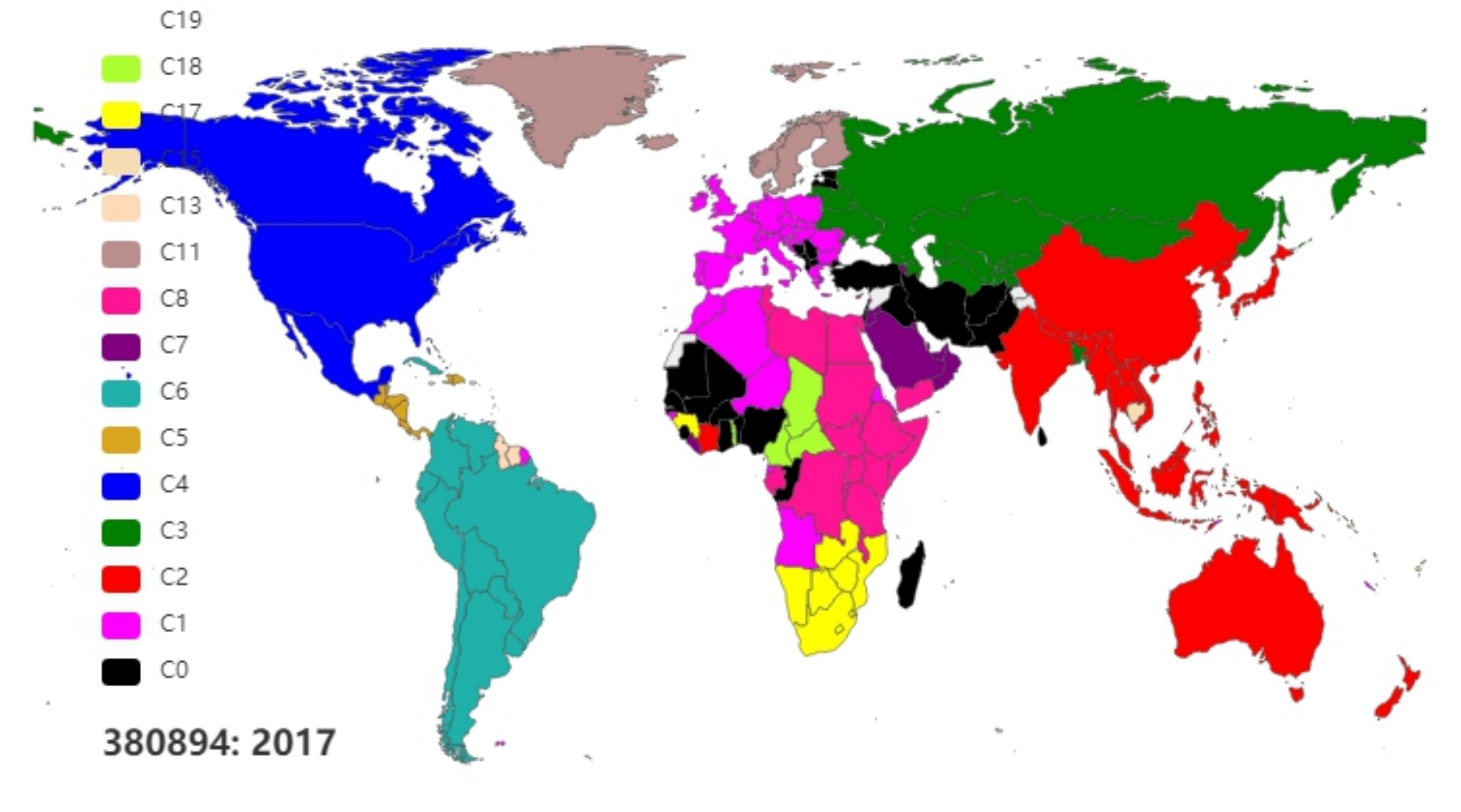}
    \includegraphics[width=0.321\linewidth]{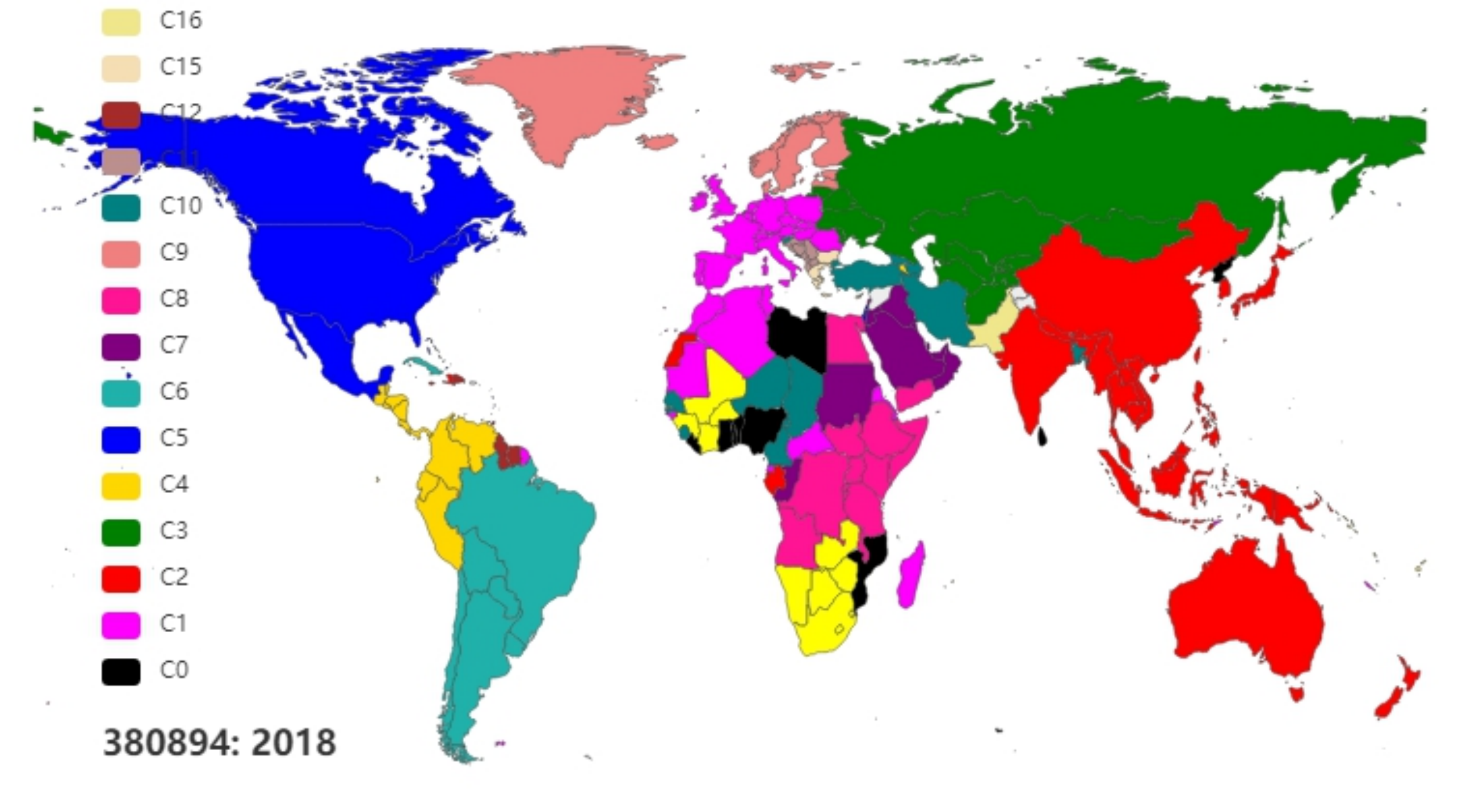}
    \caption{Community evolution of the directed iPTNs of disinfectants (380894) from 2007 to 2018. For better visibility, the non-trivial communities containing less than 5 economies and trivial communities are merged to $C_0$.}
    \label{Fig:iPTN:directed:CommunityMap:380894}
\end{figure}

Figure~\ref{Fig:iPTN:directed:CommunityMap:380894} illustrates the evolution of communities of the directed iPTNs of disinfectants (380894) from 2007 to 2018. For better visibility, the non-trivial communities containing less than 5 economies and trivial communities are merged to $C_0$. All networks but the one in 2010 have more than 13 non-trivial communities. The network in 2010 has two communities with the small one $C_{2}$ containing 5 economies in Africa (Algeria, Nigeria, Ghana, Liberia, and Morocco), 1 economy in the North Caribbean (Antigua and Barb.), and 1 economy in West Asia (Qatar).

The other 11 networks share some common features in the community structure. We describe below several big, stable communities. Most economies in East Asia, South Asia, Southeast Asia and Oceania form a community. Russia and several economies in East Europe and Middle Asia belong to a community, most of which were former members of the Soviet Union. Most Western Europe economies form a community, often combining several economies in Africa. Most Northern Europe economies belong to the Western Europe community from 2007 to 2011 and form a separate community from 2012 to 2018. In the American Continent, we observe two main communities respectively in North America and South America. Except these communities, the economies in the Caribbean, Middle East and Africa often form different communities that are less stable over time.

\begin{figure}[!ht]
    \centering
    \includegraphics[width=0.321\linewidth]{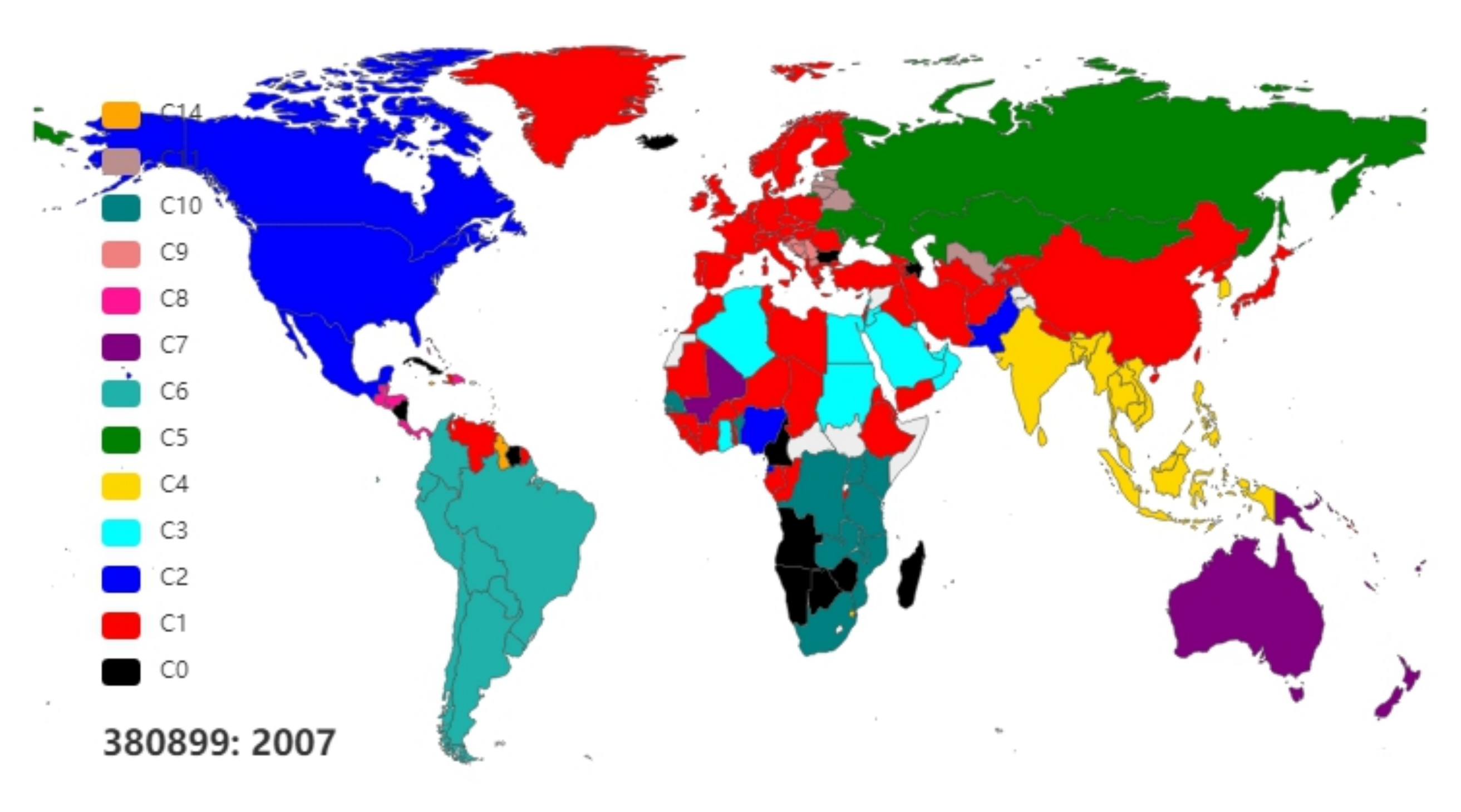}
    \includegraphics[width=0.321\linewidth]{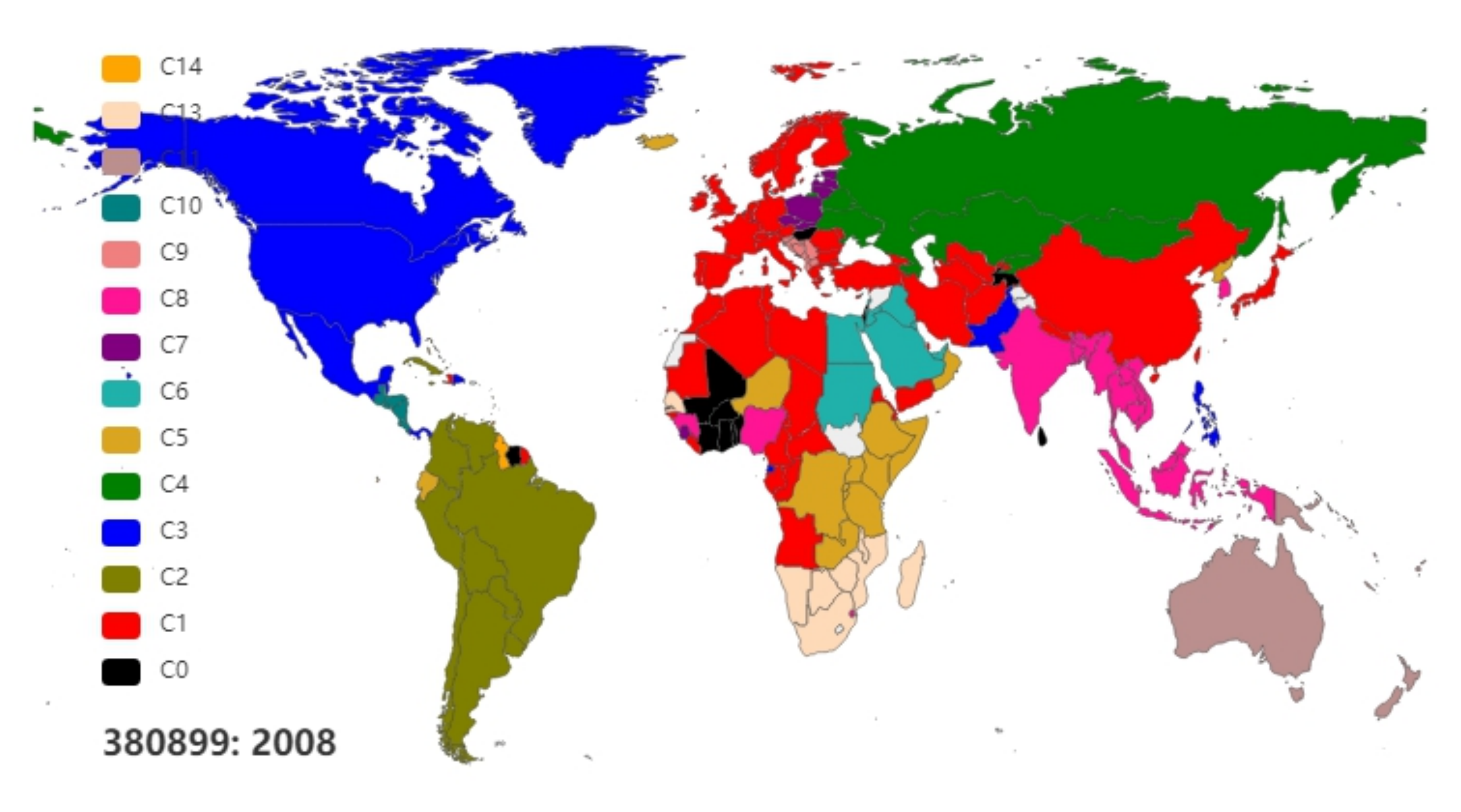}
    \includegraphics[width=0.321\linewidth]{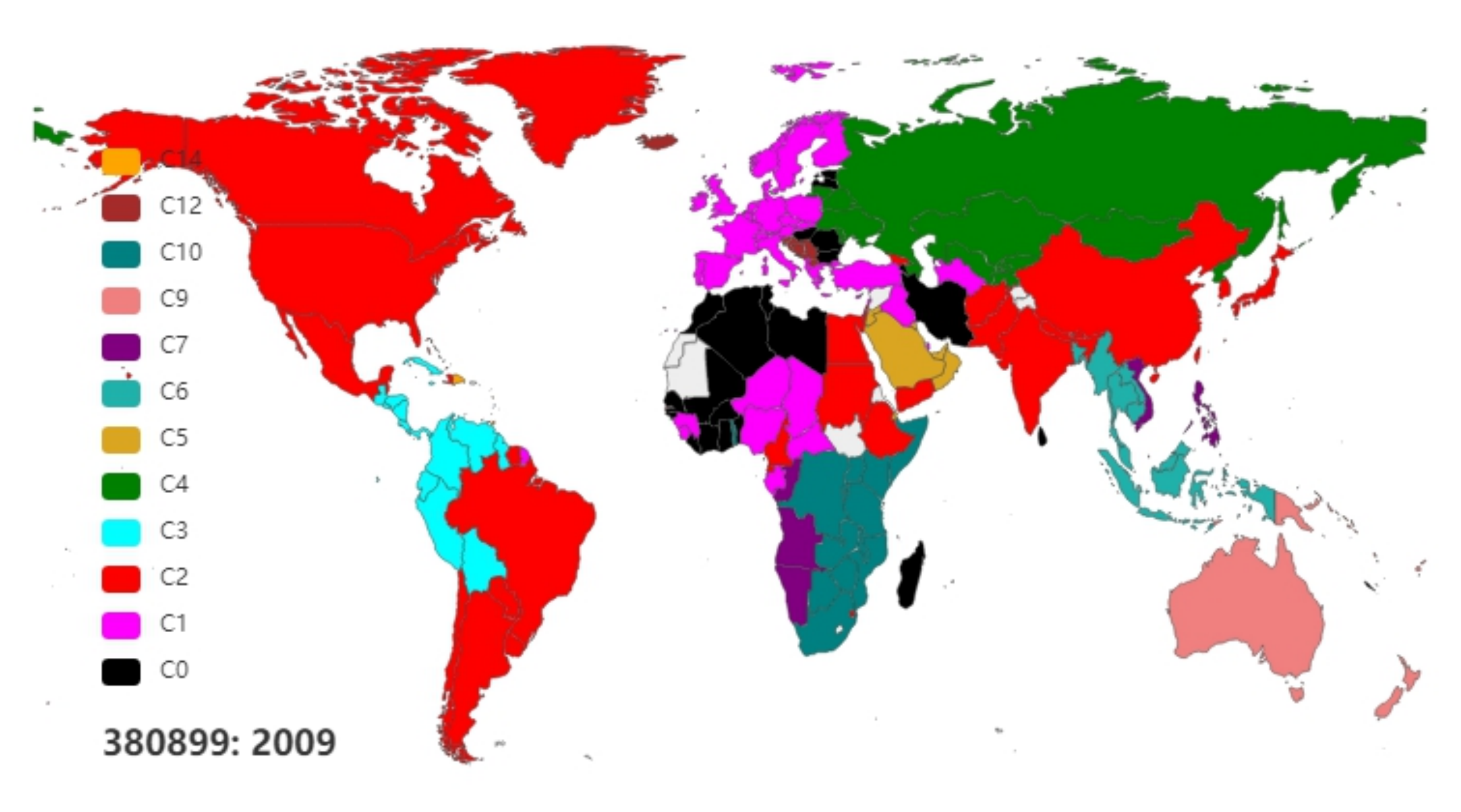}
    \includegraphics[width=0.321\linewidth]{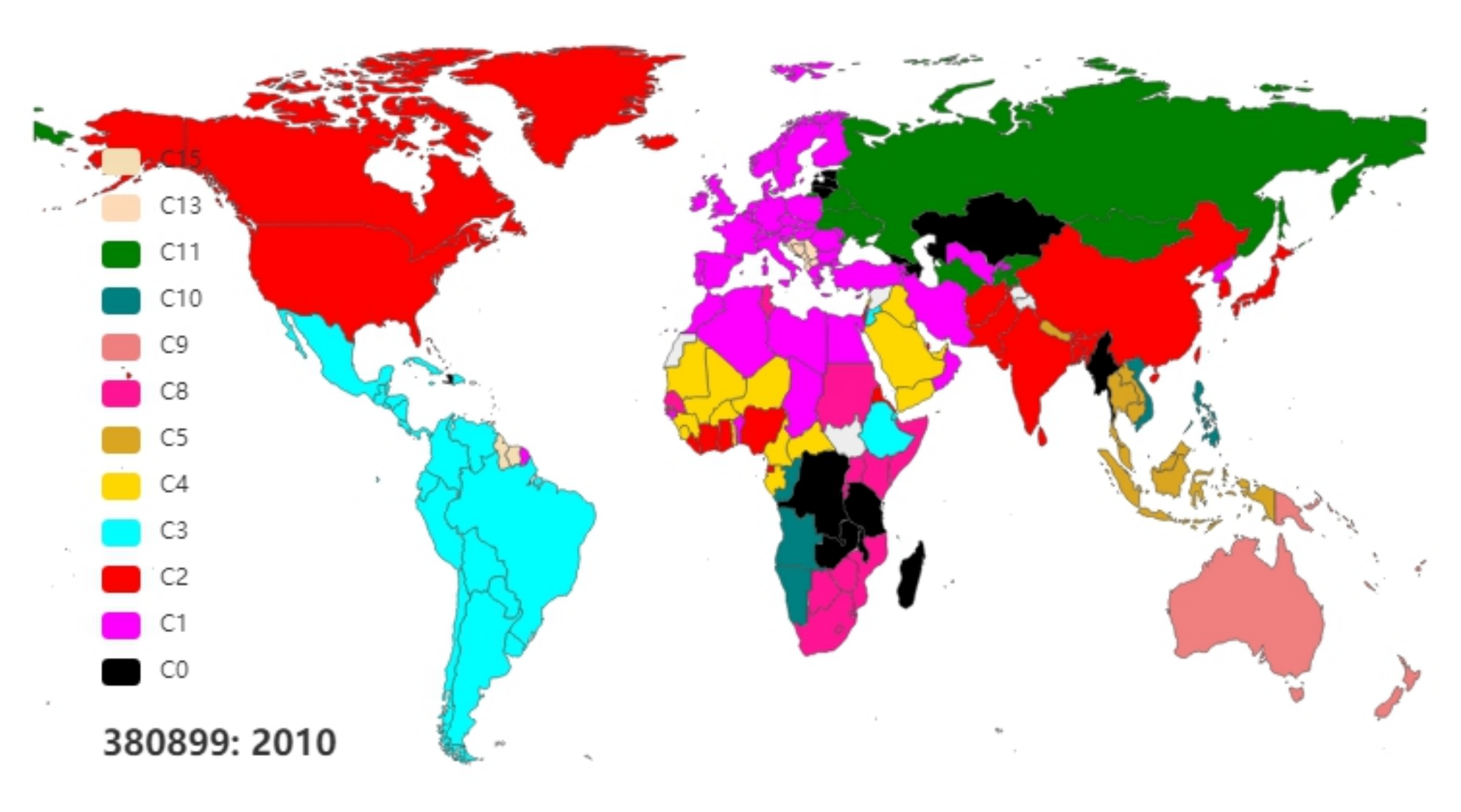}
    \includegraphics[width=0.321\linewidth]{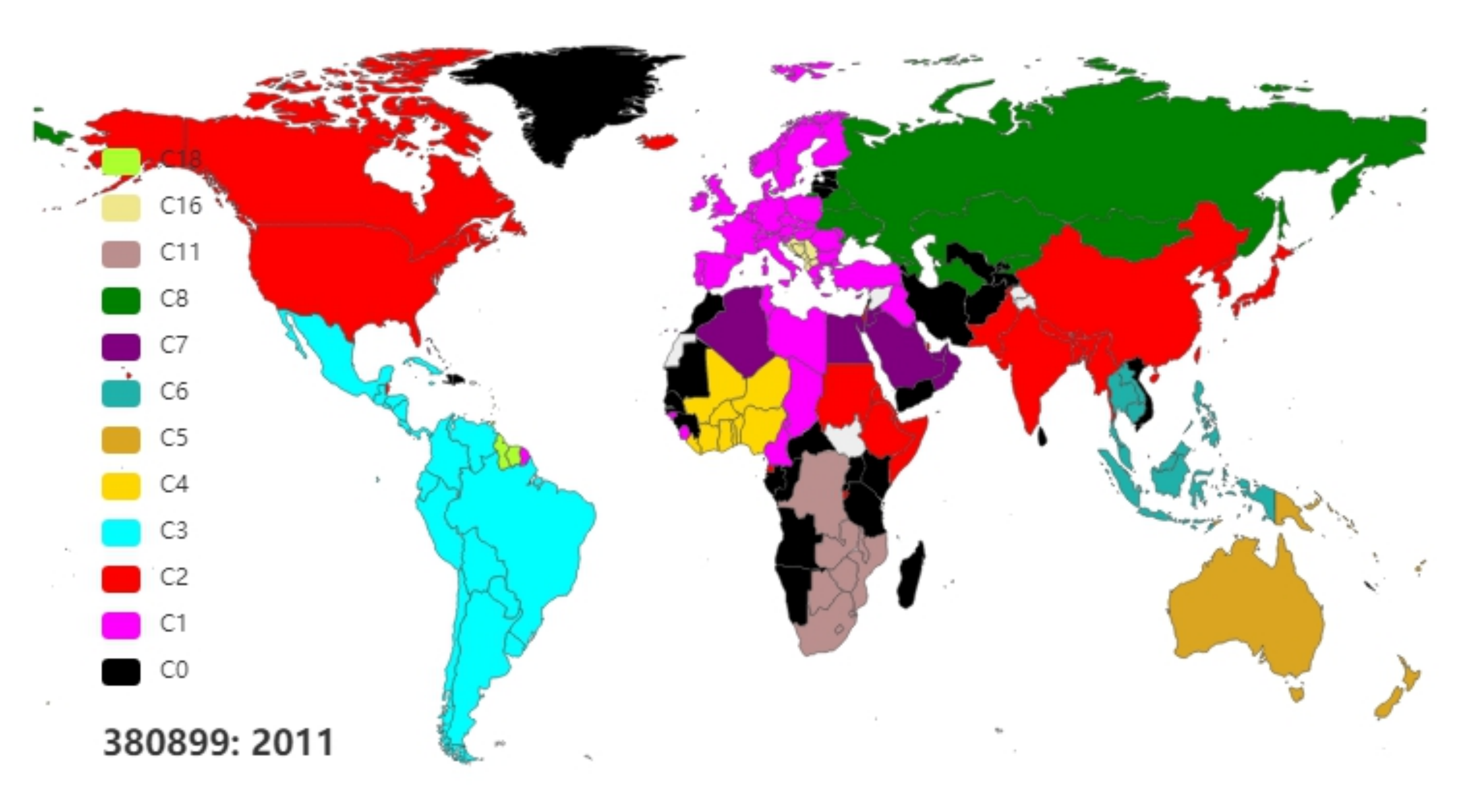}
    \includegraphics[width=0.321\linewidth]{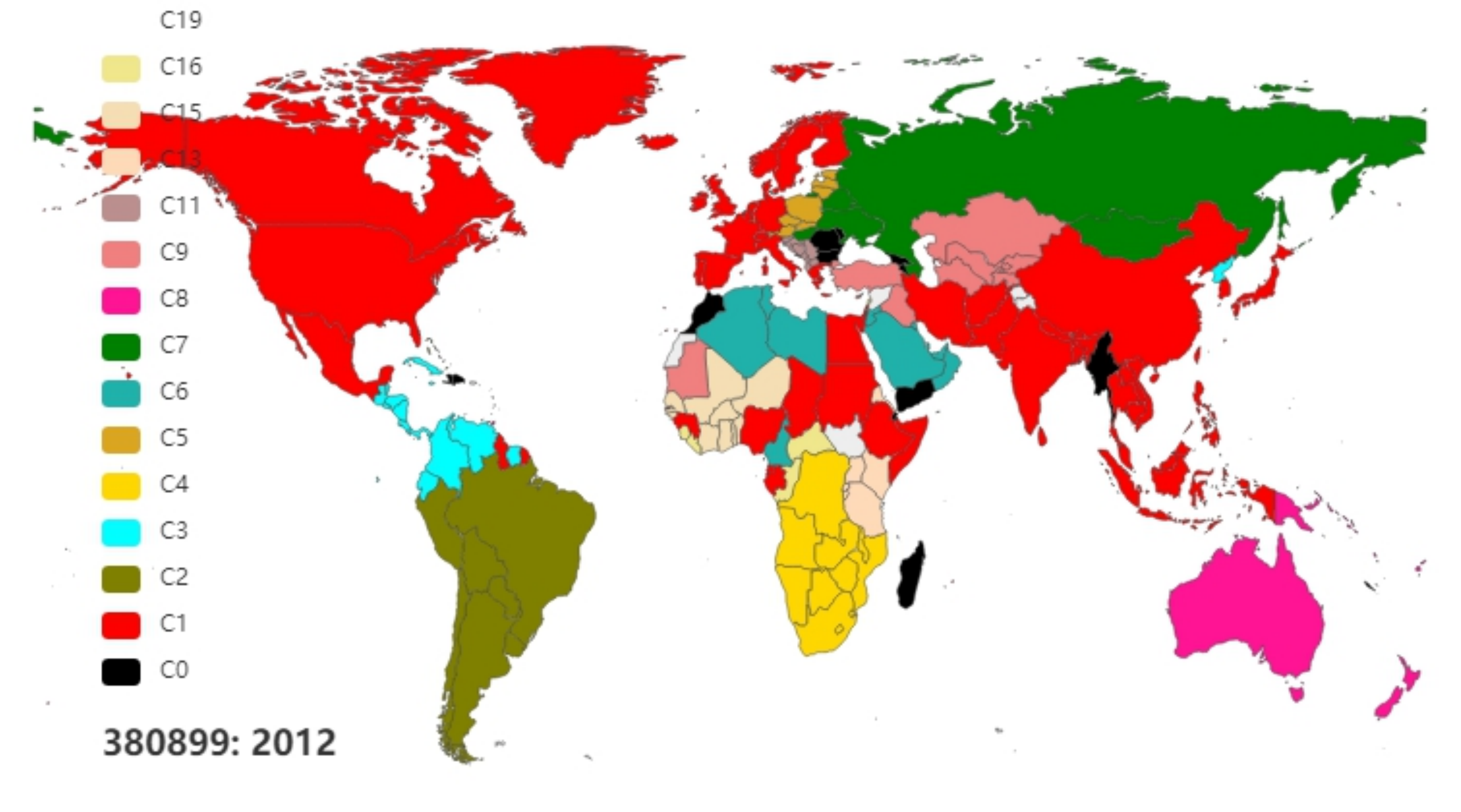}
    \includegraphics[width=0.321\linewidth]{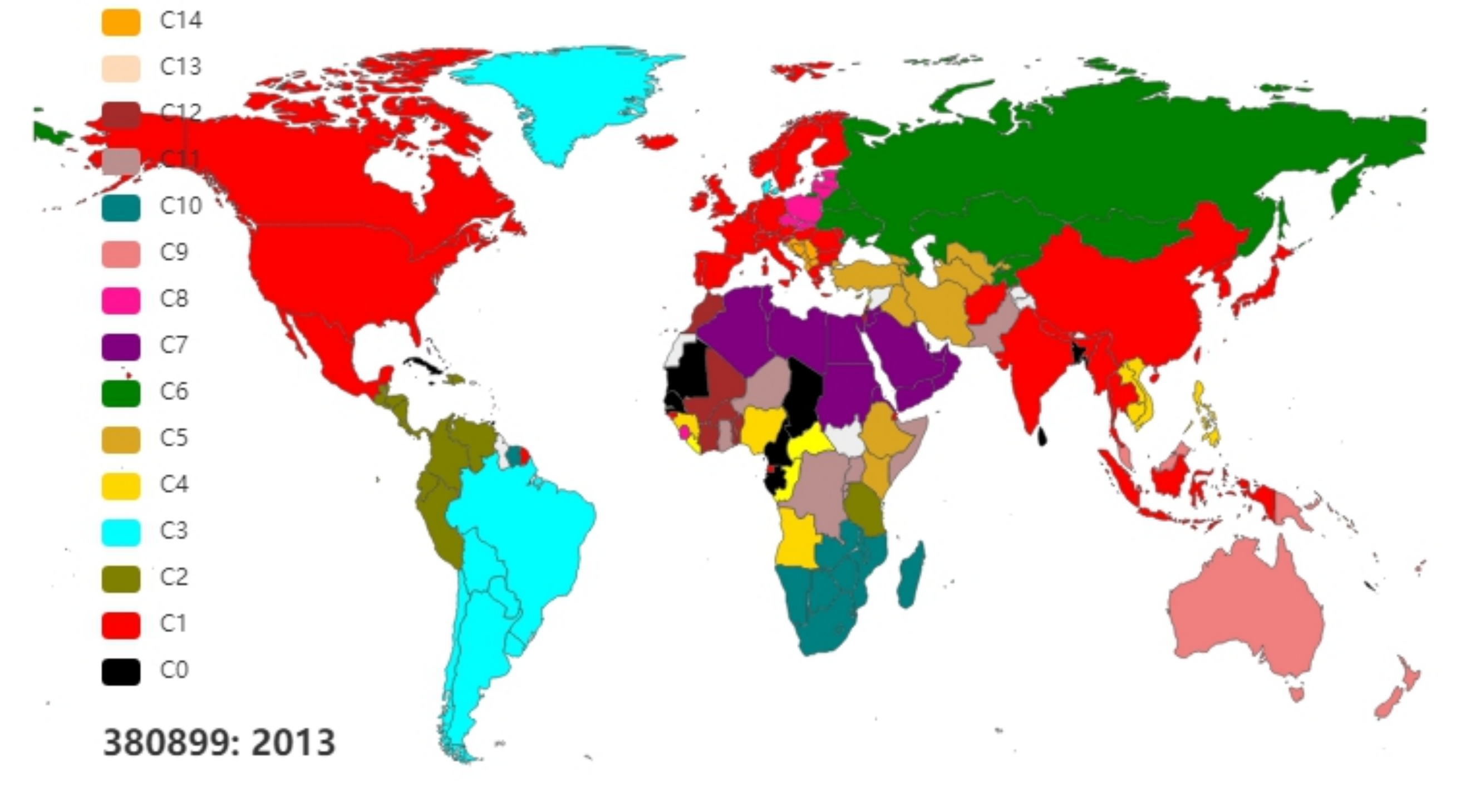}
    \includegraphics[width=0.321\linewidth]{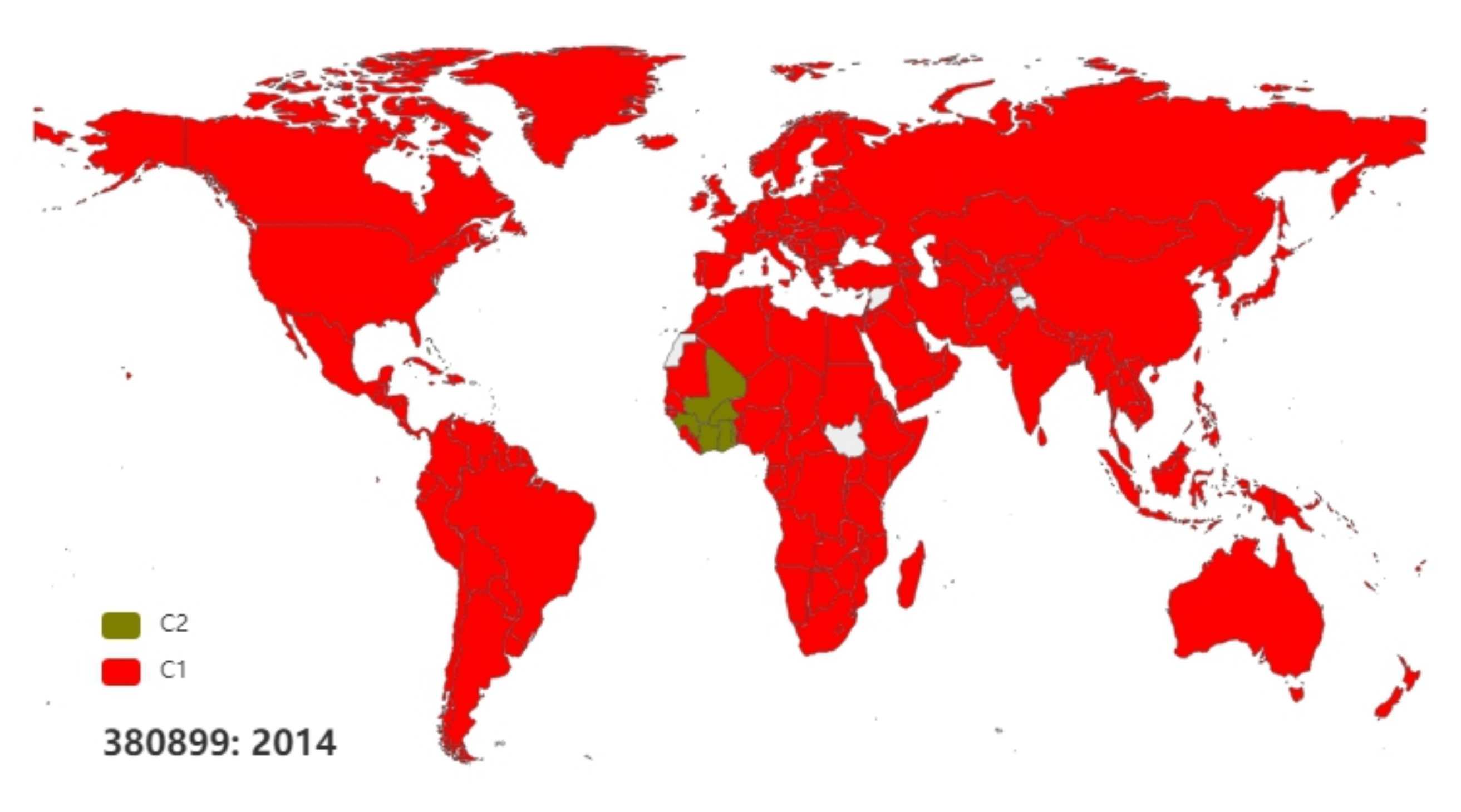}
    \includegraphics[width=0.321\linewidth]{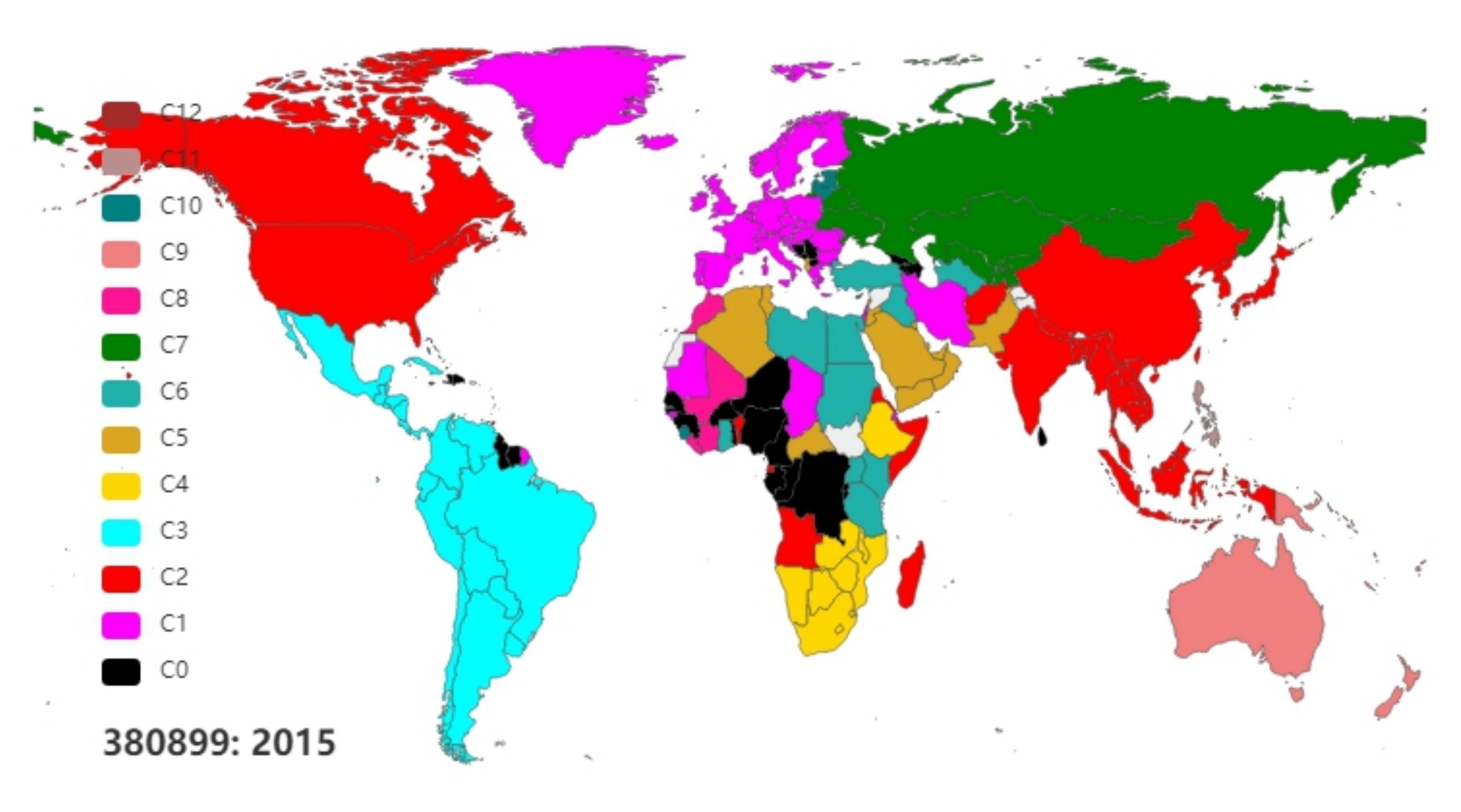}
    \includegraphics[width=0.321\linewidth]{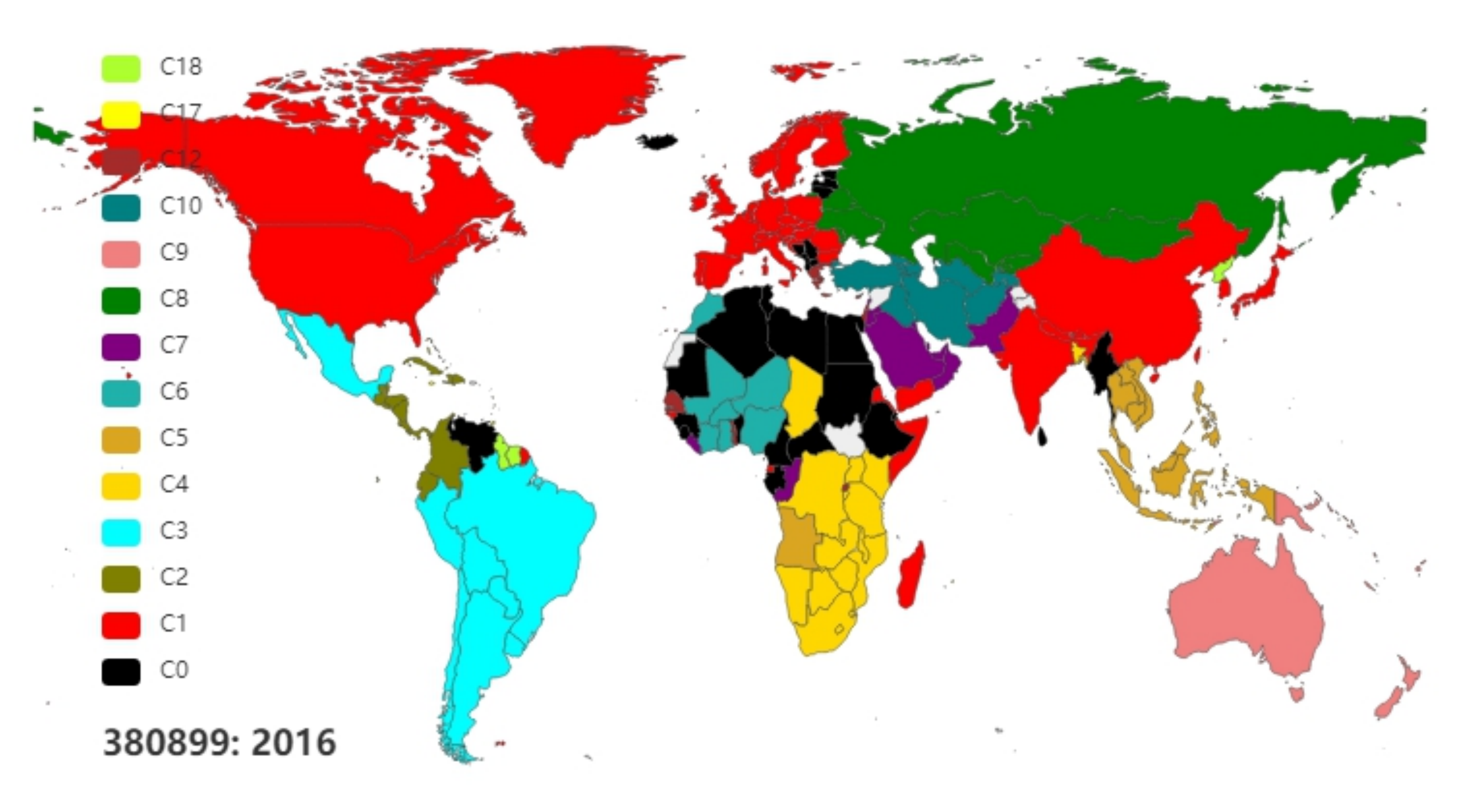}
    \includegraphics[width=0.321\linewidth]{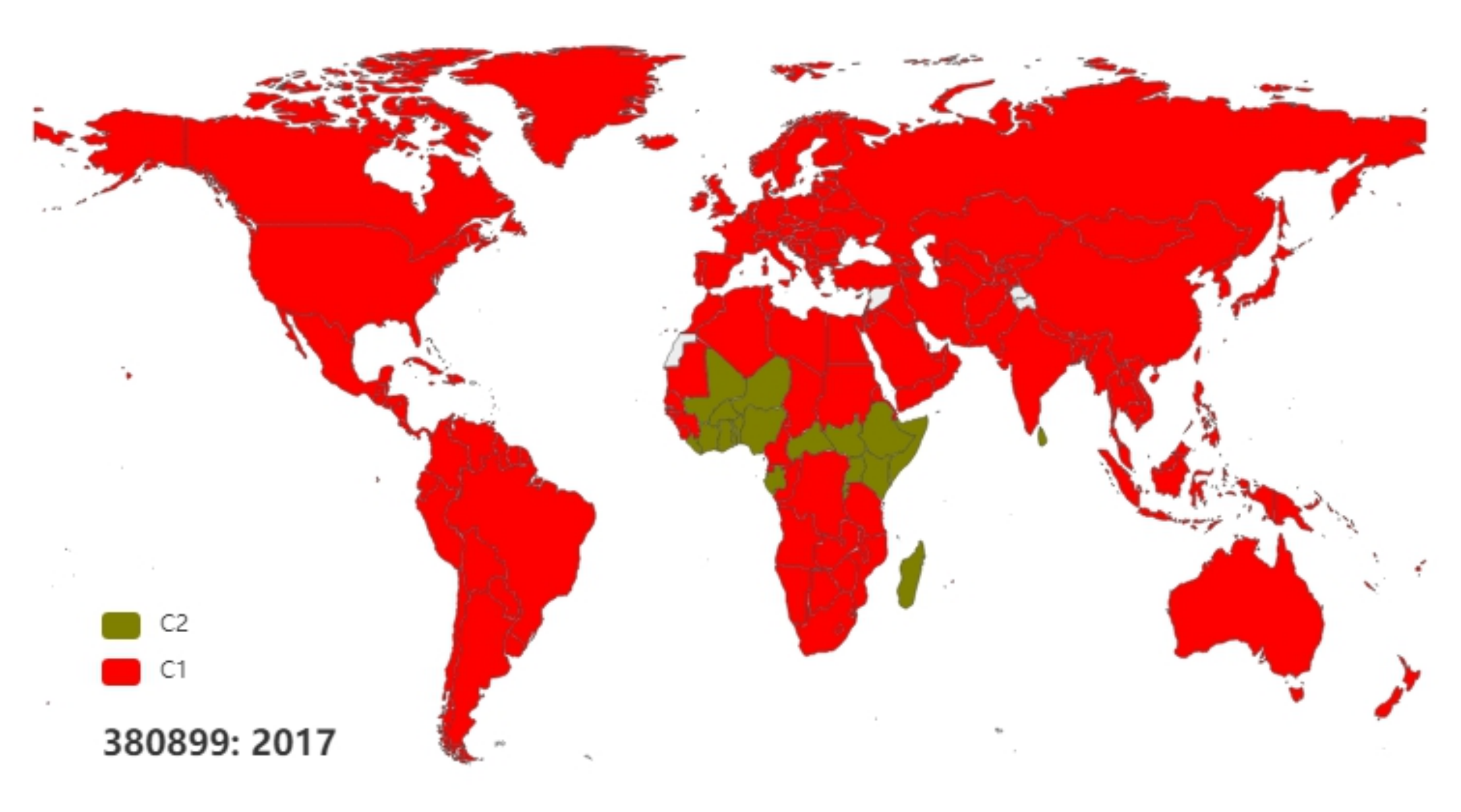}
    \includegraphics[width=0.321\linewidth]{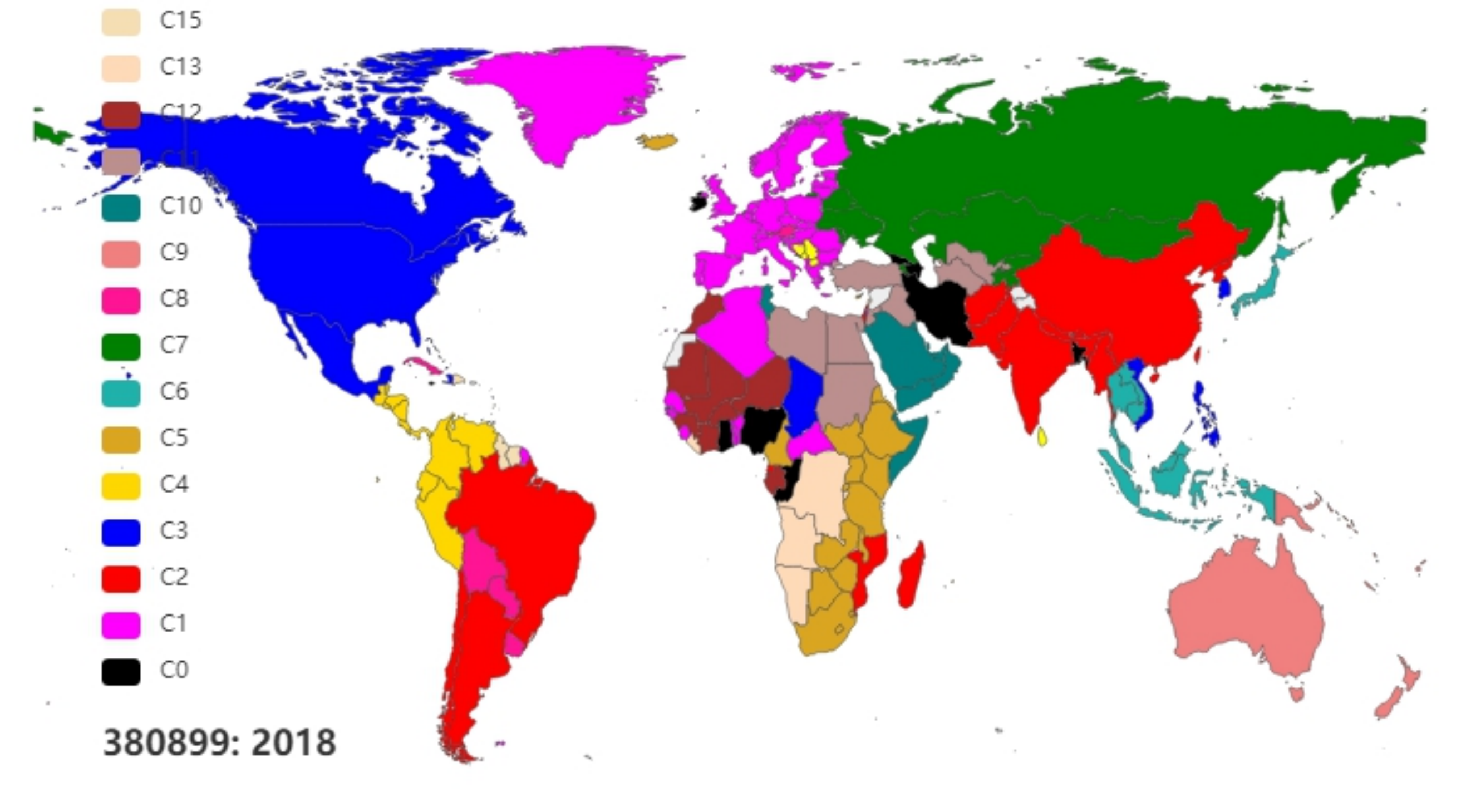}
    \caption{Community evolution of the iPTNs of rodenticides and other similar products (380899) from 2007 to 2018. For better visibility, the non-trivial communities containing less than 5 economies and trivial communities are merged to $C_0$.}
    \label{Fig:iPTN:directed:CommunityMap:380899}
\end{figure}

Figure~\ref{Fig:iPTN:directed:CommunityMap:380899} illustrates the evolution of communities of the directed iPTNs of rodenticides and other similar products (380899) from 2007 to 2018. For better visibility, the non-trivial communities containing less than 5 economies and trivial communities are merged to cluster $C_0$. The networks usually have more than 14 non-trivial communities, except that the two networks in 2014 and 2017 have only two communities. 
The small community $C_{2}$ in 2014 contains 4 African economies (Ghana, Mali,, Guinea, Togo, Burkina Faso, and C{\^o}te d'Ivoire) and Bahamas. 
The small community $C_{2}$ in 2017 contains 19 economies in Africa (Mali, Kenya, Uganda, Burkina Faso, Central African Rep., Benin, Ethiopia, Gabon, Ghana, C{\^o}te d'Ivoire, Madagascar, Nigeria, S. Sudan, Togo, Niger, Rwanda, Liberia, Burundi, and Somalia) and 2 in South Asia (Maldives and Sri Lanka).

In the rest 10 networks, the main part of the Russia community is quite stable, while the Oceania economies form a separate community. China, Japan, Korea and several other Asia economies form the main Asia community, while economies in Southeast Asia may form a separate community. The Western Europe economies form a separate community in 2009, 2010, 2011, 2015, and 2018, merge with the main Asia community in 2007, 2008, 2012, 2013, and 2016. In the American Continent, there are at least two big communities (except in 2009), one in North America and the other in South America or in Latin America and the Caribbean, and main economies in Central America around the Caribbean Sea may form a separate community. The North America community and the Asia community also merge from 2009 to 2017.


\subsection{Temporal similarity of community structure}
\label{S2:iPTN:Similarity}

In the previous subsections, we have found that the community structure of the iPTNs is more or less stable over different years. In order to quantify how stable the community structure is, we calculate the similarity between the partitions in two successive years using the {\it{normalised mutual information}} ($NMI$) measure, which was originally designed for comparing the performance of community structure detection methods when the ``true value'' of the community structure of the network under investigation is known \cite{Danon-DiazGuilera-Duch-Arenas-2005-JStatMech}.
The $NMI$ measure is based on the definition of a confusion matrix $\mathbf{N}$, whose rows correspond to the detected communities of network $\mathscr{G}_{t-1}$ and columns correspond to the identified communities of network $\mathscr{G}_{t}$. The element $N_{ij}$ of the confusion matrix $\mathbf{N}$ is the number of nodes in community $i$ of network $\mathscr{G}_{t-1}$ that appear in community $j$ of network $\mathscr{G}_{t}$. The normalised mutual information ($NMI$) measure between the community partitioning of two successive networks is calculated as follow:
\begin{equation}
    NMI(t-1, t) 
    =\frac{-2\sum^{n^c_{t-1}}_{i=1}\sum^{n^c_t}_{j=1}
    N_{ij}\log\left(\frac{N_{ij}N}{N_{i.}N_{.j}}\right)}
{\sum^{n^c_{t-1}}_{i=1}N_{i.}\log\left(\frac{N_{i.}}{N}\right)
 + \sum^{n^c_t}_{j=1}N_{.j}\log\left(\frac{N_{.j}}{N}\right)},
\end{equation}
where $n^c_{t-1}$ is the number of communities in network $\mathscr{G}_{t-1}$, $n^c_{t}$ is the number of communities in network $\mathscr{G}_{t}$., $N_{i.}=\sum_jN_{ij}$ is the sum over row $i$ of confusion matrix $\mathbf{N}$, and $N_{.j}=\sum_iN_{ij}$ is the sum over column $j$. 
If the two partitions are identical, then $NMI(t-1, t)$ takes its maximum value of 1. If the two partitions are totally independent, we have $NMI(t-1, t)= 0$.

\begin{figure}[!ht]
    \centering
    \includegraphics[width=0.483\linewidth]{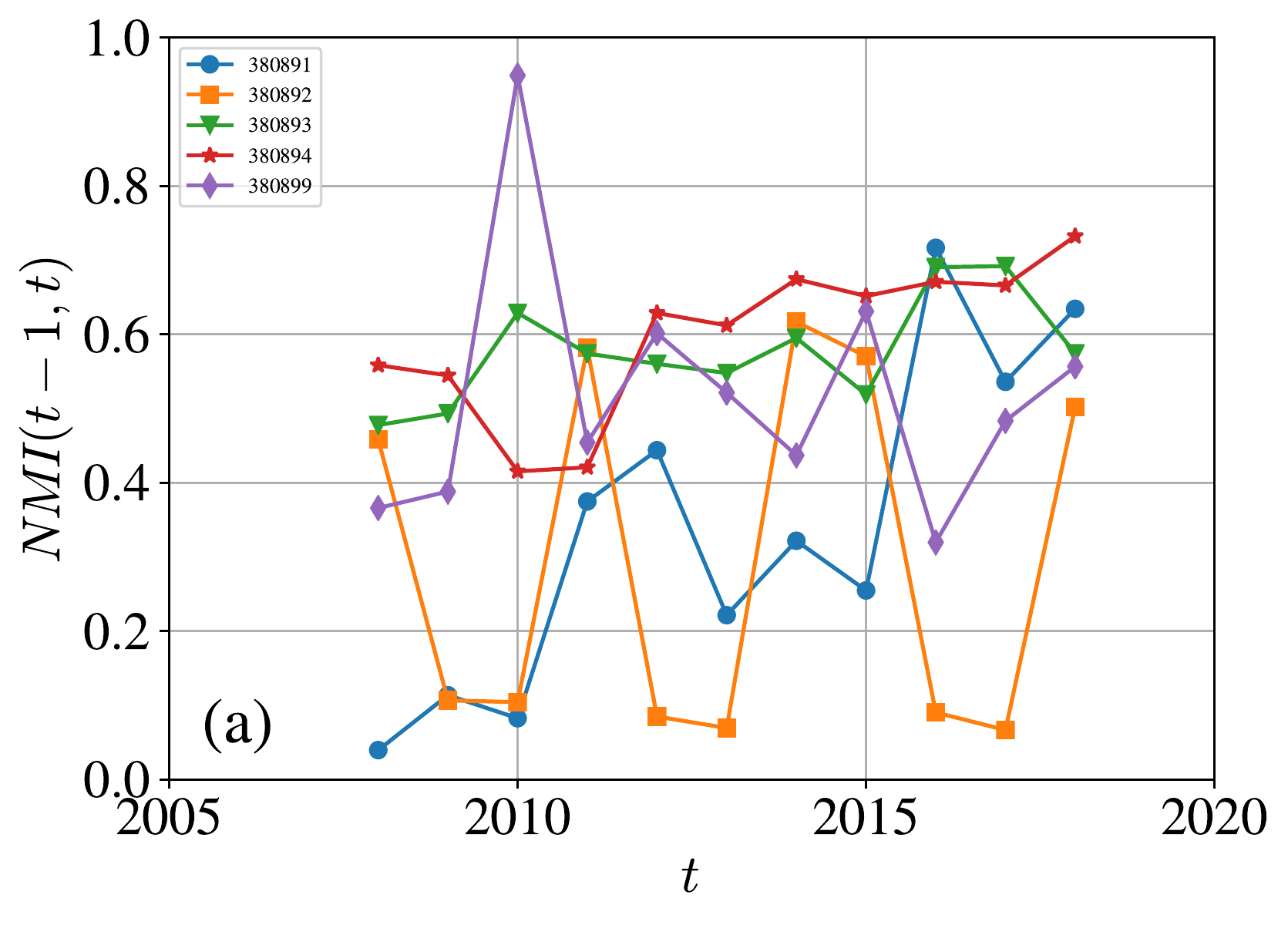}
     \includegraphics[width=0.483\linewidth]{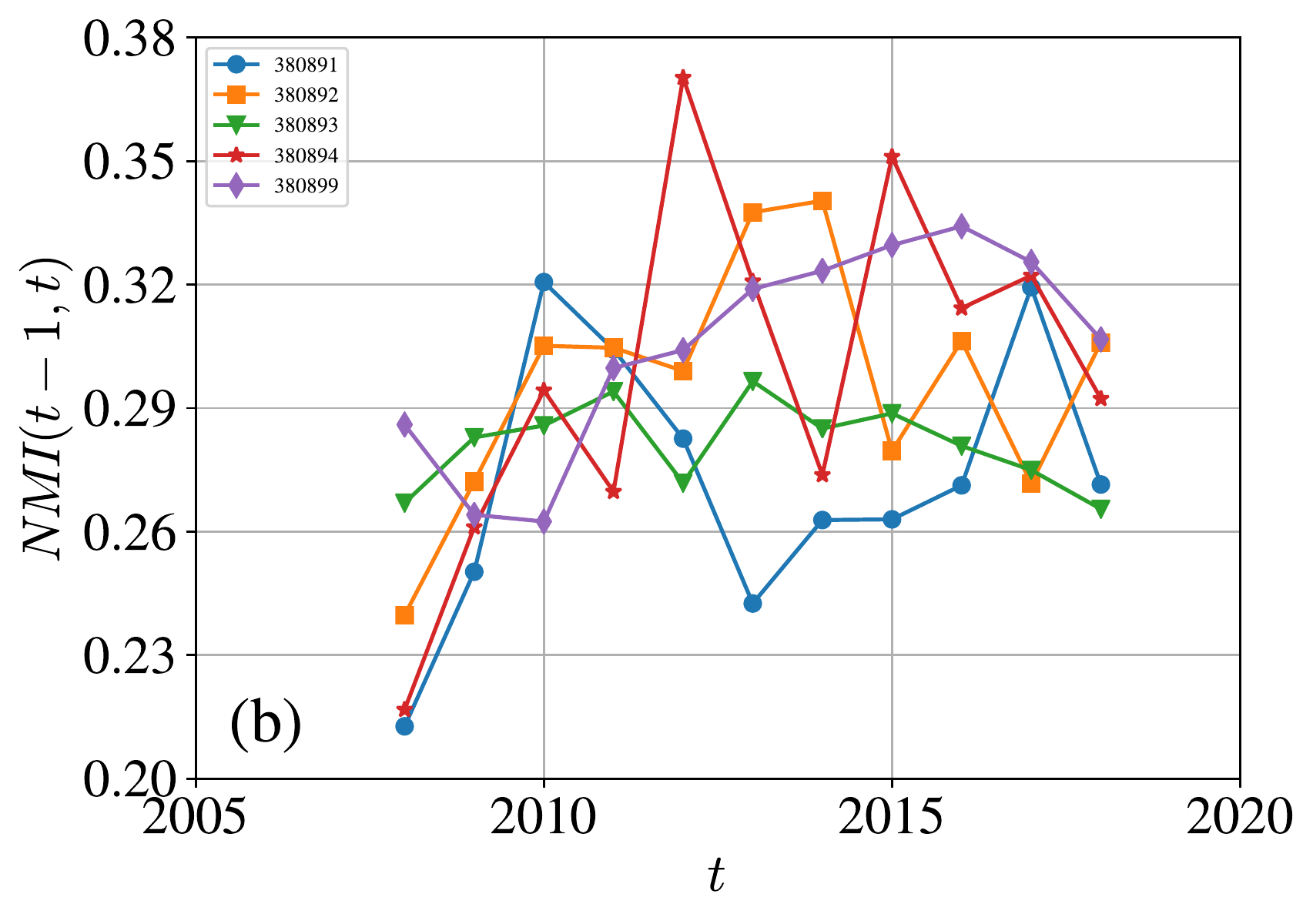}
    \caption{Normalized mutual information $NMI(t-1, t)$ between the partitions of two successive iPTNs. (a) undirected iPTNs, (b) directed iPTNs.}
    \label{Fig:iPTN:undirected:Community:NMI:t}
\end{figure}

Figure~\ref{Fig:iPTN:undirected:Community:NMI:t} shows the normalized mutual information $NMI(t-1, t)$ between the partitions of two successive iPTNs. For the undirected networks in Fig.~\ref{Fig:iPTN:undirected:Community:NMI:t}(a), the $NMI$ curve for fungicides (380892) fluctuates largely, while other four curves exhibit an increasing trend. For the directed networks in Fig.~\ref{Fig:iPTN:undirected:Community:NMI:t}(b), the $NMI$ curves exhibit an increasing trend in early years and become trend-free afterwards. For herbicides (380893),  disinfectants (380894), and rodenticides and other similar products (380899), the $NMI$ value for an undirected network is larger than that for the corresponding directed network:
\begin{equation}
    NMI^{\mathrm{undir}}(t-1,t) > NMI^{\mathrm{dir}}(t-1,t).
\end{equation}
For insecticides (380891) and fungicides (380892), we do not observe any evident relationship between undirected and directed networks.

We also calculate the normalized mutual information $NMI^{\mathrm{undir}}(t_i,t_j)$ and $NMI^{\mathrm{dir}}(t_i,t_j)$ between any two different years $t_i$ and $t_j$ for the undirected and directed iPTNs. For each category of pesticides, we obtain the average and its standard deviation. For the undirected networks, the average normalized mutual information is 
$0.30 \pm 0.19$ for insecticides (380891),
$0.33 \pm 0.20$ for fungicides (380892),
$0.50 \pm 0.09$ for herbicides (380893),
$0.55 \pm 0.09$ for disinfectants (380894), and
$0.32 \pm 0.15$ for rodenticides and other similar products (380899),
For the undirected networks, the average normalized mutual information is 
$0.27 \pm 0.02$ for insecticides (380891),
$0.29 \pm 0.03$ for fungicides (380892),
$0.27 \pm 0.02$ for herbicides (380893),
$0.30 \pm 0.03$ for disinfectants (380894), and
$0.30 \pm 0.03$ for rodenticides and other similar products (380899). 
It is found that the average normalized mutual information are comparable between the undirected and directed iPTNs of insecticides (380891), fungicides (380892), and rodenticides and other similar products (380899), while the undirected iPTNs of herbicides (380893) and disinfectants (380894) have significantly larger average normalized mutual information than the directed iPTNs. The most significant feature is that the directed iPTNs have much smaller standard deviations than the undirected iPTNs, showing that the community structure of the directed iPTNs is much stabler than the undirected iPTNs, which implies that the community structure of the directed iPTNs captures better the underlying economic contents of international pesticide trade.



\subsection{Community blocks in the directed iPTNs}
\label{S2:iPTN:DiGraph:icb}

The results in Sec.~\ref{S2:iPTN:DiGraph:Community} and Sec.~\ref{S2:iPTN:Similarity} have shown that community structure of the directed iPTNs has evolved stably, which suggests that there are intrinsic community blocks that are common to all communities and form the backbone of the identified communities. An intrinsic community block $B_i$ is a maximum set of nodes, in which all the nodes belong to a same community in each iPTN. Hence, starting from an arbitrary node $u$, if node $v$ belongs to the same community of the iPTN in every year, then $u$ and $v$ belong to the same intrinsic community block; Otherwise, $u$ and $v$ belong to two different intrinsic community blocks. We identify the intrinsic community blocks of the temporal directed iPTNs for the five categories of pesticides.

\begin{figure}[!ht]
    \centering
    \includegraphics[width=0.483\linewidth,height=0.32\linewidth]{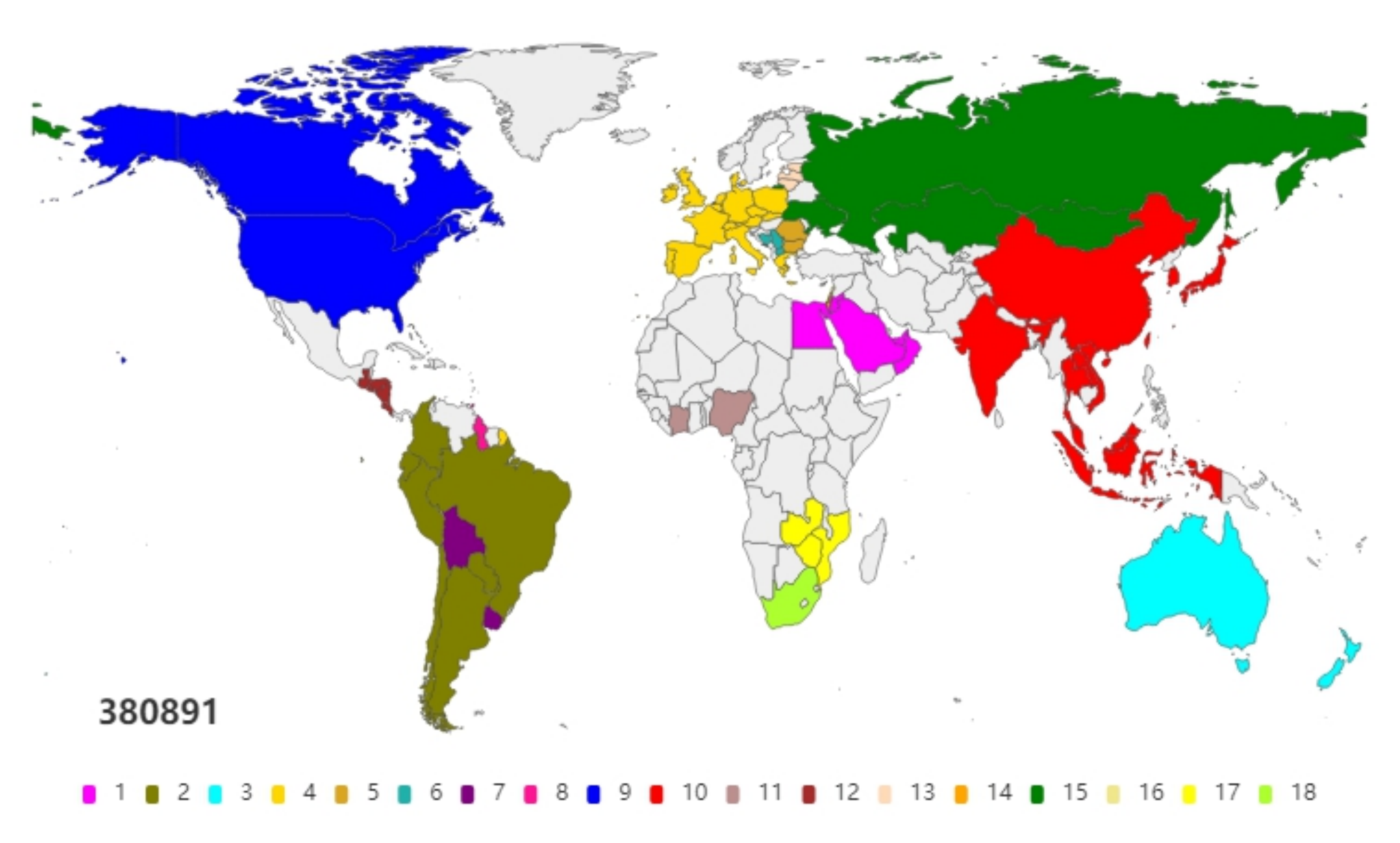}
    \includegraphics[width=0.483\linewidth]{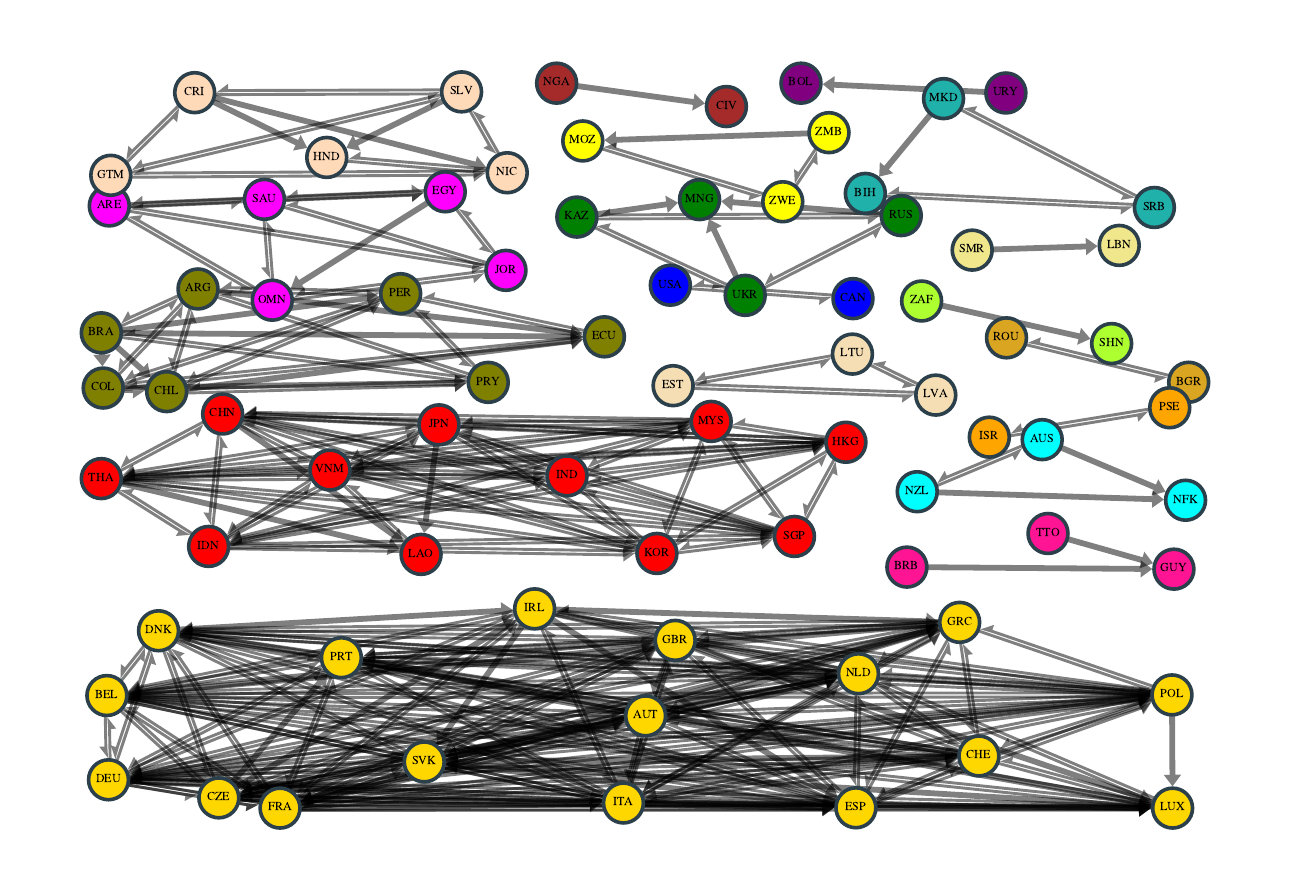}
    \caption{Intrinsic community blocks in the temporal directed iPTNs from 2007 to 2018 of insecticides (380899) in the colored map (left panel) and in colored networks (right panel). The node color of an intrinsic community block is the same in both panels.}
    \label{Fig:iPTN:directed:CommunityBlocks:380891}
\end{figure}

For the temporal iPTN of insecticides (380891), we obtain 18 intrinsic community blocks  containing 78 economies in total, including 
$B_{1}$ (United Arab Emirates, Egypt, Jordan, Oman, Saudi Arabia), 
$B_{2}$ (Argentina, Brazil, Chile, Colombia, Ecuador, Peru, Paraguay), 
$B_{3}$ (Australia, Norfolk Island, New Zealand), 
$B_{4}$ (Austria, Belgium, Switzerland, Czech Rep., Germany, Denmark, Spain, France, United Kingdom, Greece, Ireland, Italy, Luxembourg, Netherlands, Poland, Portugal, Slovakia), 
$B_{5}$ (Bulgaria, Romania), 
$B_{6}$ (Bosnia and Herz., Macedonia, Serbia), 
$B_{7}$ (Bolivia, Uruguay), 
$B_{8}$ (Barbados, Guyana, Trinidad and Tobago), 
$B_{9}$ (Canada, United States), 
$B_{10}$ (China, Hong Kong, Indonesia, India, Japan, Korea, Lao PDR, Malaysia, Singapore, Thailand, Vietnam), 
$B_{11}$ (C{\^o}te d'Ivoire, Nigeria), 
$B_{12}$ (Costa Rica, Guatemala, Honduras, Nicaragua, El Salvador), 
$B_{13}$ (Estonia, Lithuania, Latvia), 
$B_{14}$ (Israel, Palestine), 
$B_{15}$ (Kazakhstan, Mongolia, Russia, Ukraine), 
$B_{16}$ (Lebanon, San Marino), 
$B_{17}$ (Mozambique, Zambia, Zimbabwe), and 
$B_{18}$ (Saint Helena, South Africa).
The map of intrinsic community blocks is shown in the left panel of Fig.~\ref{Fig:iPTN:directed:CommunityBlocks:380891} distinguished by different colors, while the right panel shows the networks of the intrinsic community blocks. The intrinsic community block $B_{4}$ has the largest size with 17 economies in Europe, while the second largest block $B_{10}$ contains 11 economies in East and South Asia. Most of the economies that are not included in any intrinsic community blocks  are in Africa, the Caribbean, and Central and West Asia.

\begin{figure}[!ht]
    \centering
    \includegraphics[width=0.483\linewidth,height=0.32\linewidth]{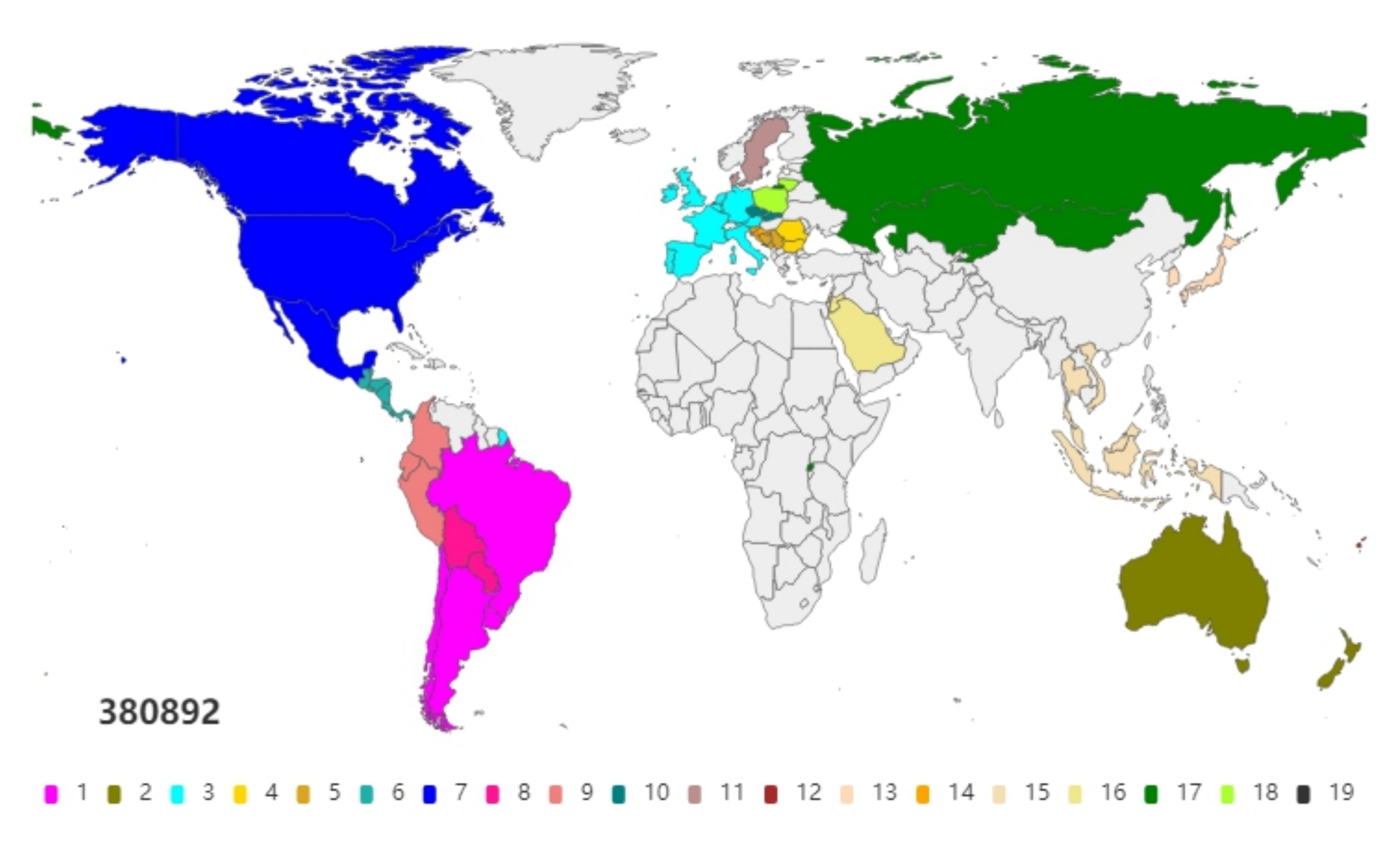}
    \includegraphics[width=0.483\linewidth]{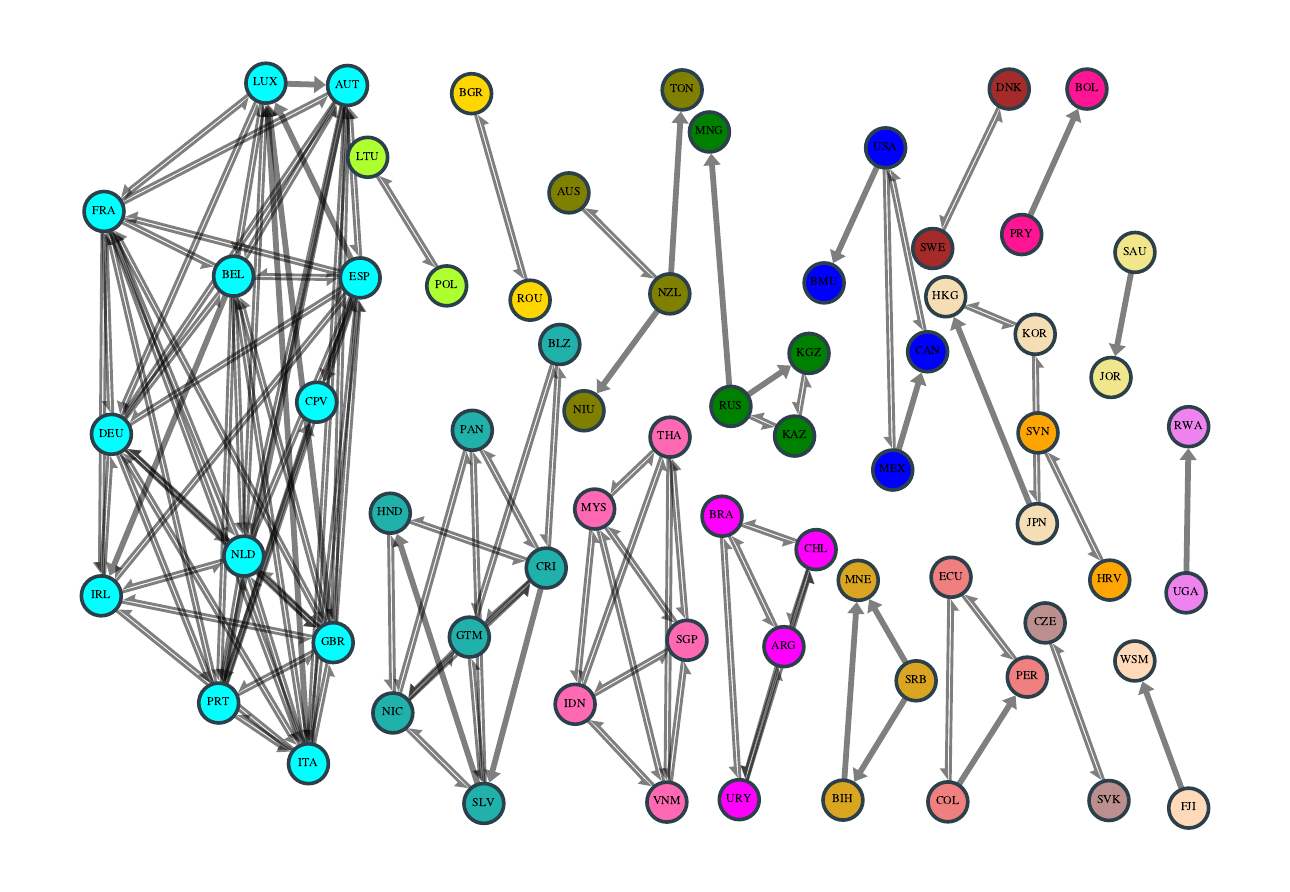}
    \caption{Intrinsic community blocks in the temporal directed iPTNs from 2007 to 2018 of fungicides (380892) in the colored map (left panel) and in colored networks (right panel). The node color of an intrinsic community block is the same in both panels.}
    \label{Fig:iPTN:directed:CommunityBlocks:380892}
\end{figure}

For the temporal iPTN of fungicides (380892), we obtain 19 intrinsic community blocks containing 67 economies in total, including 
$B_{1}$ (Argentina, Brazil, Chile, Uruguay), 
$B_{2}$ (Australia, Niue, New Zealand, Tonga), 
$B_{3}$ (Austria, Belgium, Cabo Verde, Germany, Spain, France, United Kingdom, Ireland, Italy, Luxembourg, Netherlands, Portugal), 
$B_{4}$ (Bulgaria, Romania), 
$B_{5}$ (Bosnia and Herz., Montenegro, Serbia), 
$B_{6}$ (Belize, Costa Rica, Guatemala, Honduras, Nicaragua, Panama, El Salvador), 
$B_{7}$ (Bermuda, Canada, Mexico, United States), 
$B_{8}$ (Bolivia, Paraguay), 
$B_{9}$ (Colombia, Ecuador, Peru), 
$B_{10}$ (Czech Rep., Slovakia), 
$B_{11}$ (Denmark, Sweden), 
$B_{12}$ (Fiji, Samoa), 
$B_{13}$ (Hong Kong, Japan, Korea), 
$B_{14}$ (Croatia, Slovenia), 
$B_{15}$ (Indonesia, Malaysia, Singapore, Thailand, Vietnam), 
$B_{16}$ (Jordan, Saudi Arabia), 
$B_{17}$ (Kazakhstan, Kyrgyzstan, Mongolia, Russia), 
$B_{18}$ (Lithuania, Poland), and 
$B_{19}$ (Rwanda, Uganda).
Figure~\ref{Fig:iPTN:directed:CommunityBlocks:380892} shows the map and the networks of the intrinsic community blocks. The largest intrinsic community block $B_{3}$ contains 12 economies mainly in Europe, while the second largest intrinsic community block $B_6$ contains 7 economies in Western Caribbean. Most economies in Africa and many economies in Asia (including China and India) do not belong to any intrinsic community blocks.

\begin{figure}[!ht]
    \centering
    \includegraphics[width=0.483\linewidth,height=0.32\linewidth]{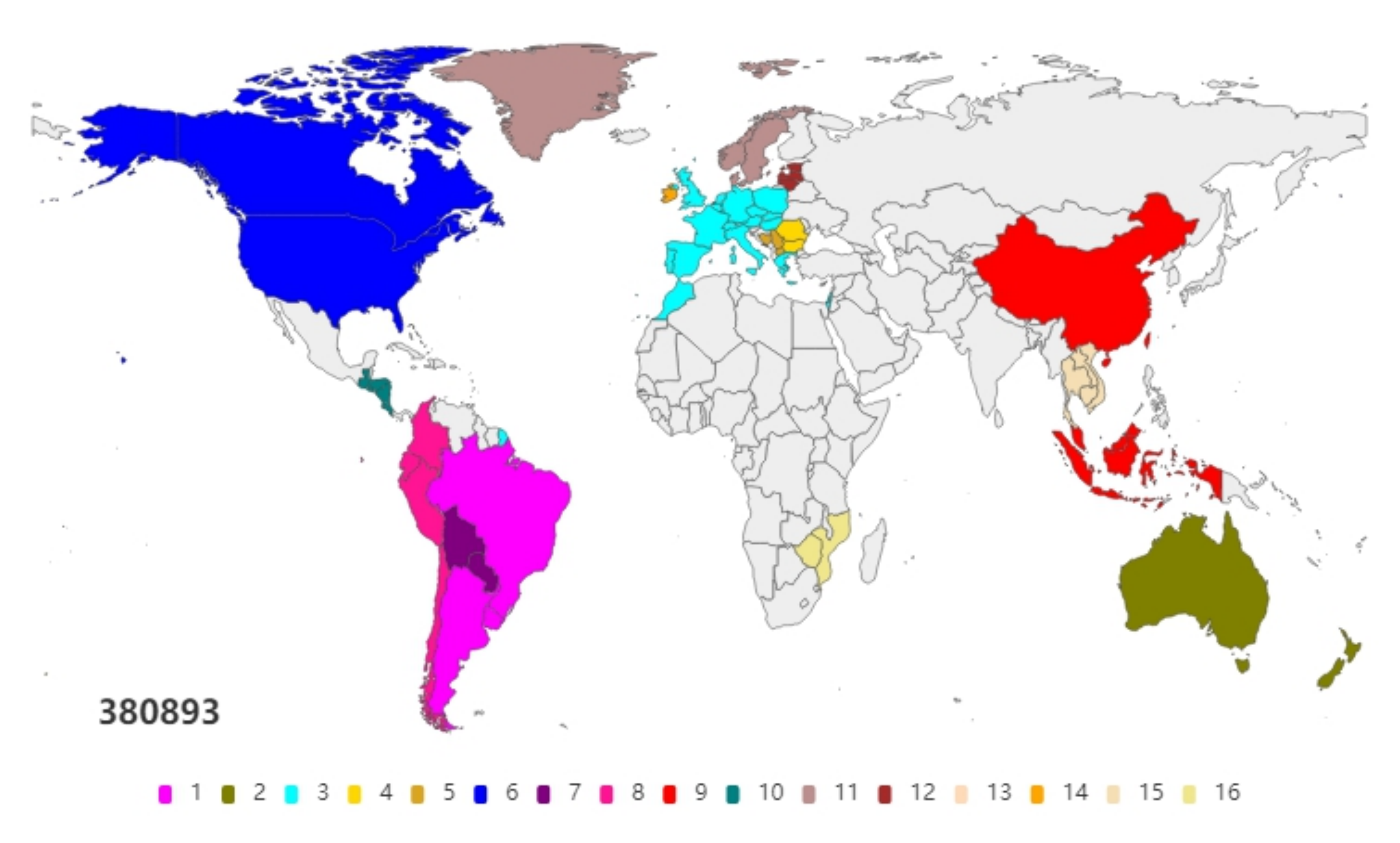}
    \includegraphics[width=0.483\linewidth]{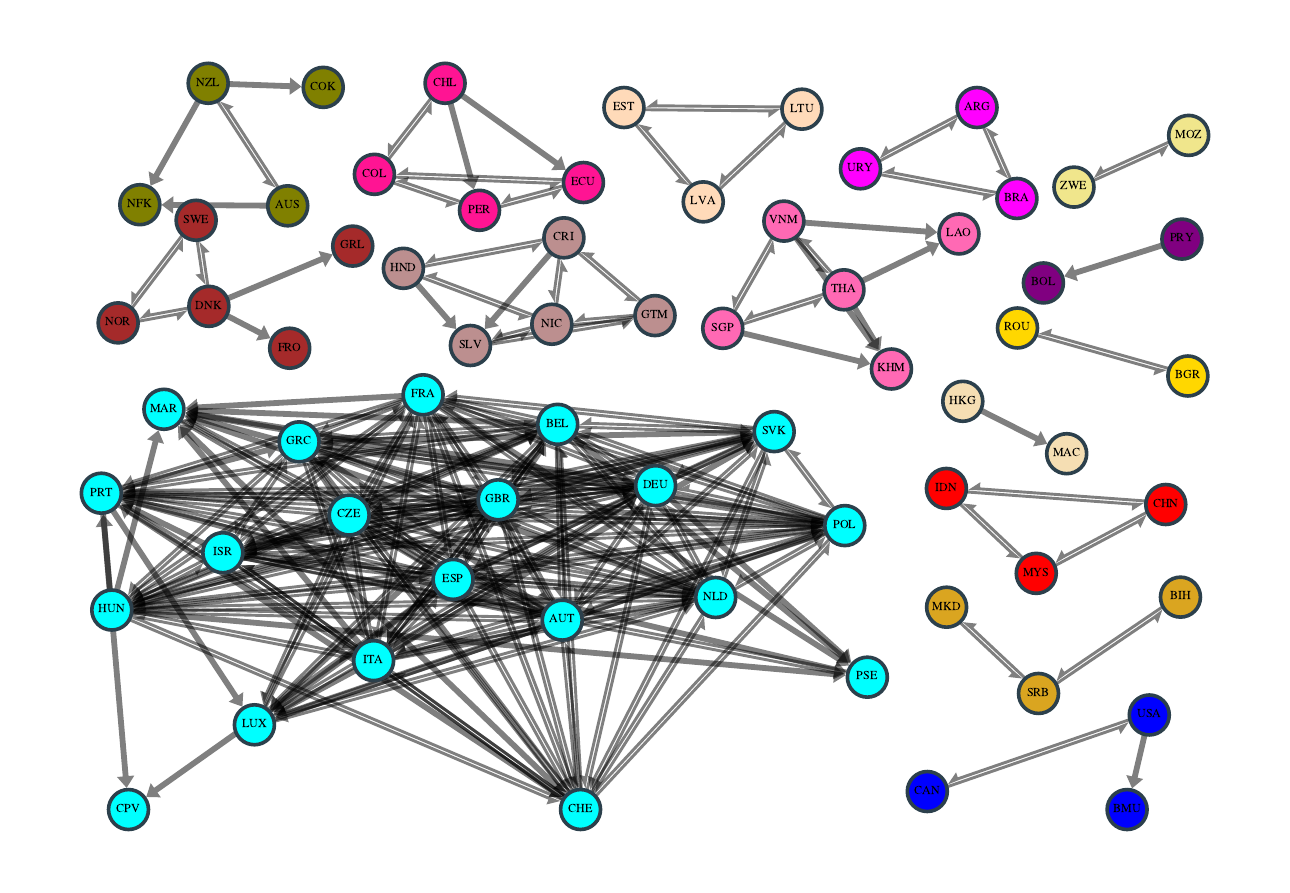}
    \caption{Intrinsic community blocks in the temporal directed iPTNs from 2007 to 2018 of herbicides (380893) in the colored map (left panel) and in colored networks (right panel). The node color of an intrinsic community block is the same in both panels.}
    \label{Fig:iPTN:directed:CommunityBlocks:380893}
\end{figure}

For the temporal iPTN of herbicides (380893), we obtain 16 intrinsic community blocks containing 66 economies in total, including 
$B_{1}$ (Argentina, Brazil, Uruguay), 
$B_{2}$ (Australia, Cook Islands, Norfolk Island, New Zealand), 
$B_{3}$ (Austria, Belgium, Switzerland, Cabo Verde, Czech Rep., Germany, Spain, France, United Kingdom, Greece, Hungary, Israel, Italy, Luxembourg, Morocco, Netherlands, Poland, Portugal, Palestine, Slovakia), 
$B_{4}$ (Bulgaria, Romania), 
$B_{5}$ (Bosnia and Herz., Macedonia, Serbia), 
$B_{6}$ (Bermuda, Canada, United States), 
$B_{7}$ (Bolivia, Paraguay), 
$B_{8}$ (Chile, Colombia, Ecuador, Peru), 
$B_{9}$ (China, Indonesia, Malaysia), 
$B_{10}$ (Costa Rica, Guatemala, Honduras, Nicaragua, El Salvador), 
$B_{11}$ (Denmark, Faroe Is., Greenland, Norway, Sweden), 
$B_{12}$ (Estonia, Lithuania, Latvia), 
$B_{13}$ (Hong Kong, Macao), 
$B_{14}$ (Ireland, Malta), 
$B_{15}$ (Cambodia, Lao PDR, Singapore, Thailand, Vietnam), and 
$B_{16}$ (Mozambique, Zimbabwe).
Figure~\ref{Fig:iPTN:directed:CommunityBlocks:380893} shows the map and the networks of the intrinsic community blocks. The largest intrinsic community block $B_3$ contains 20 economies mainly in Europe. Most economies in Africa, many economies in Asia (including India, Japan and Korea), and Russia do not belong to any intrinsic community blocks. We find that block $B_{14}$ (Ireland, Malta) does not appear in the network plot since Ireland and Malta always belong to the same community in each year but do not have direct international trade of herbicides.

\begin{figure}[!ht]
    \centering
    \includegraphics[width=0.483\linewidth,height=0.32\linewidth]{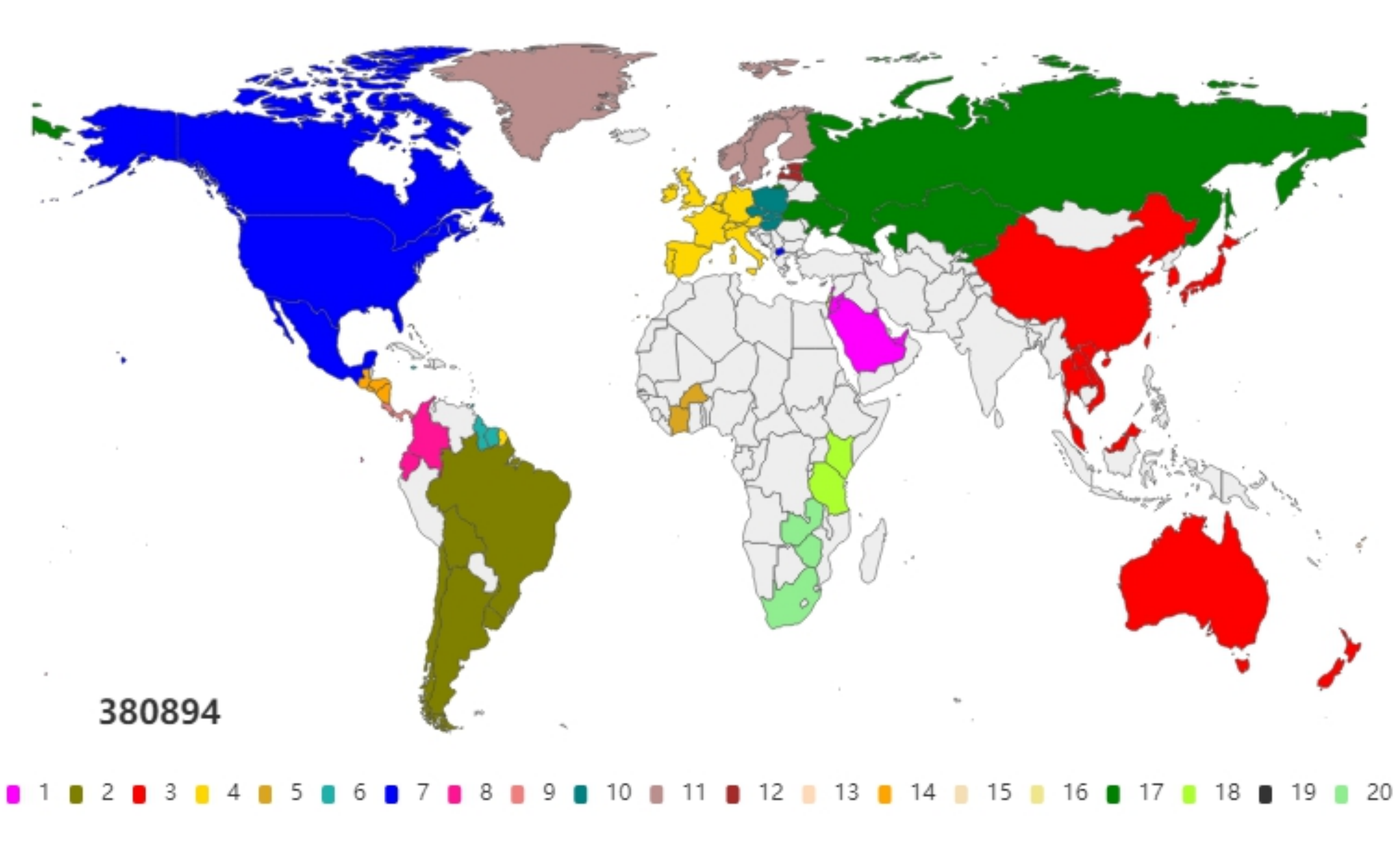}
    \includegraphics[width=0.483\linewidth]{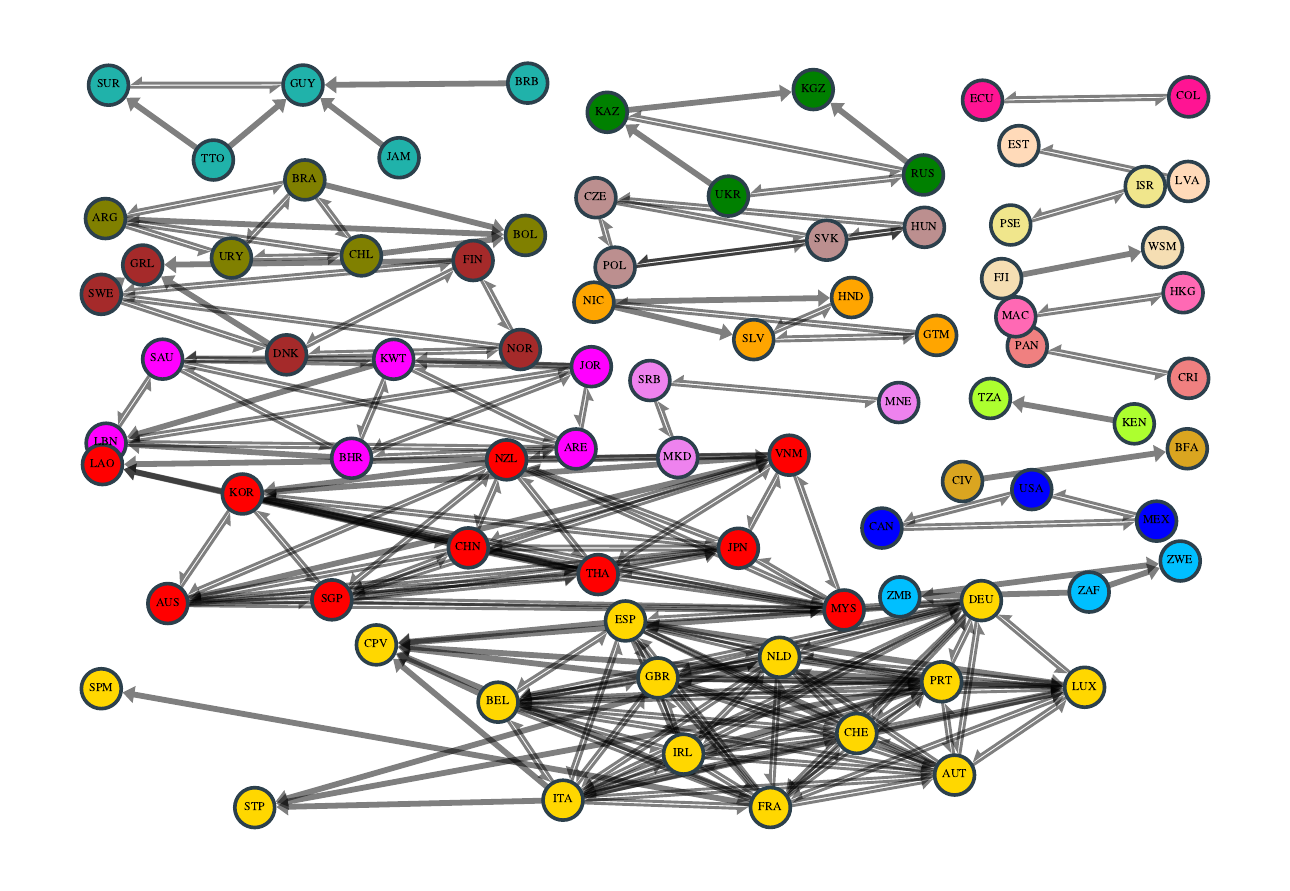}
    \caption{Intrinsic community blocks in the temporal directed iPTNs from 2007 to 2018 of disinfectants (380894) in the colored map (left panel) and in colored networks (right panel). The node color of an intrinsic community block is the same in both panels.}
    \label{Fig:iPTN:directed:CommunityBlocks:380894}
\end{figure}

For the temporal iPTN of disinfectants (380894), we obtain 20 intrinsic community blocks containing 83 economies in total, including 
$B_{1}$ (United Arab Emirates, Bahrain, Jordan, Kuwait, Lebanon, Saudi Arabia), 
$B_{2}$ (Argentina, Bolivia, Brazil, Chile, Uruguay), 
$B_{3}$ (Australia, China, Japan, Korea, Lao PDR, Malaysia, New Zealand, Singapore, Thailand, Vietnam), 
$B_{4}$ (Austria, Belgium, Switzerland, Cabo Verde, Germany, Spain, France, United Kingdom, Ireland, Italy, Luxembourg, Netherlands, Portugal, St. Pierre and Miquelon, S{\~a}o Tom{\'e} and Principe), 
$B_{5}$ (Burkina Faso, C{\^o}te d'Ivoire), 
$B_{6}$ (Barbados, Guyana, Jamaica, Suriname, Trinidad and Tobago), 
$B_{7}$ (Canada, Mexico, United States), 
$B_{8}$ (Colombia, Ecuador), 
$B_{9}$ (Costa Rica, Panama), 
$B_{10}$ (Czech Rep., Hungary, Poland, Slovakia), 
$B_{11}$ (Denmark, Finland, Greenland, Norway, Sweden), 
$B_{12}$ (Estonia, Latvia), 
$B_{13}$ (Fiji, Samoa), 
$B_{14}$ (Guatemala, Honduras, Nicaragua, El Salvador), 
$B_{15}$ (Hong Kong, Macao), 
$B_{16}$ (Israel, Palestine), 
$B_{17}$ (Kazakhstan, Kyrgyzstan, Russia, Ukraine), 
$B_{18}$ (Kenya, Tanzania), 
$B_{19}$ (Macedonia, Montenegro, Serbia), and 
$B_{20}$ (South Africa, Zambia, Zimbabwe)
Figure~\ref{Fig:iPTN:directed:CommunityBlocks:380894} shows the map and the networks of the intrinsic community blocks. The largest intrinsic community block $B_4$ contains 16 economies mainly in Europe, while the second largest intrinsic community block $B_3$ contains 10 economies in Asia and Oceania.

\begin{figure}[!ht]
    \centering
    \includegraphics[width=0.483\linewidth,height=0.32\linewidth]{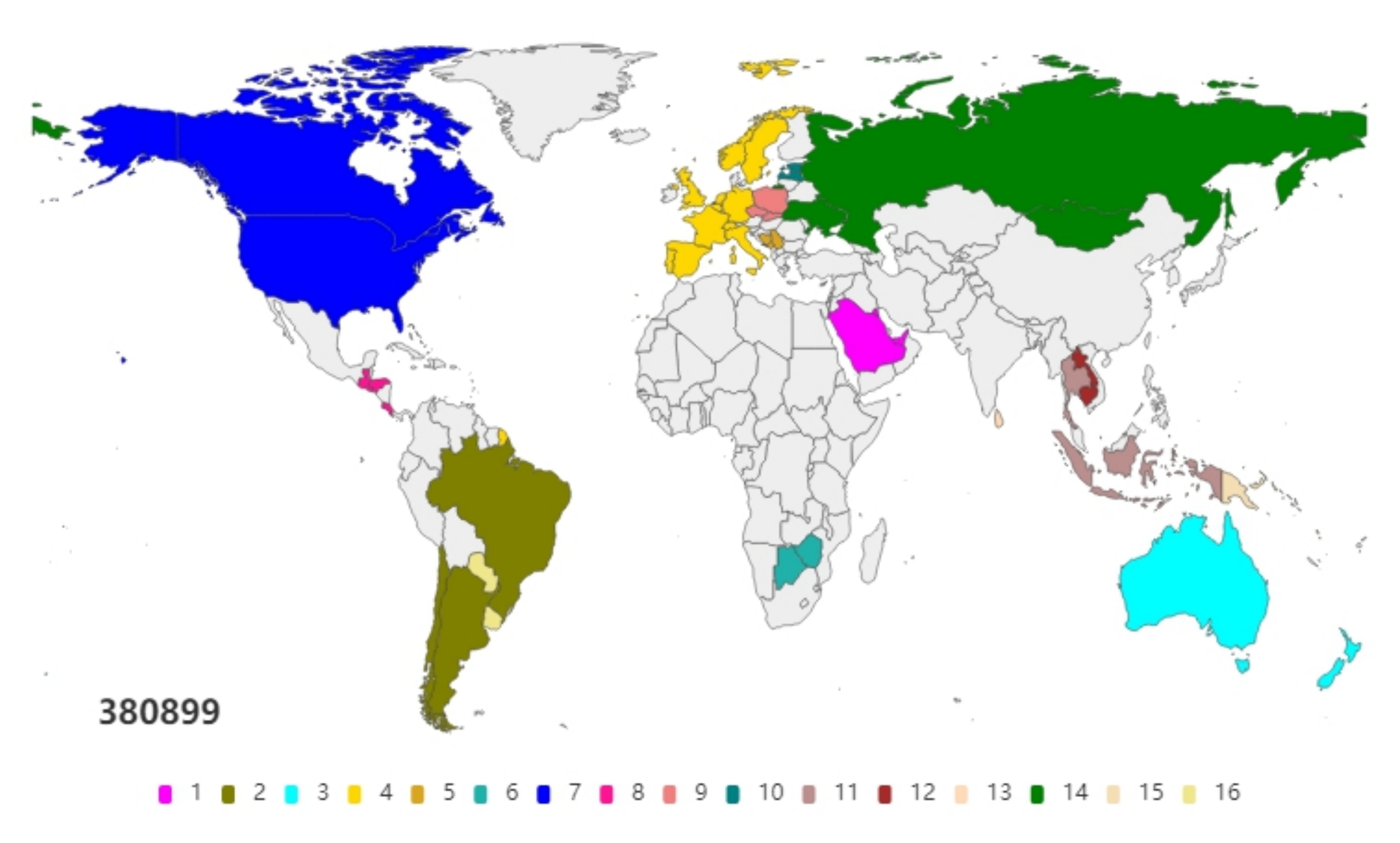}
    \includegraphics[width=0.483\linewidth]{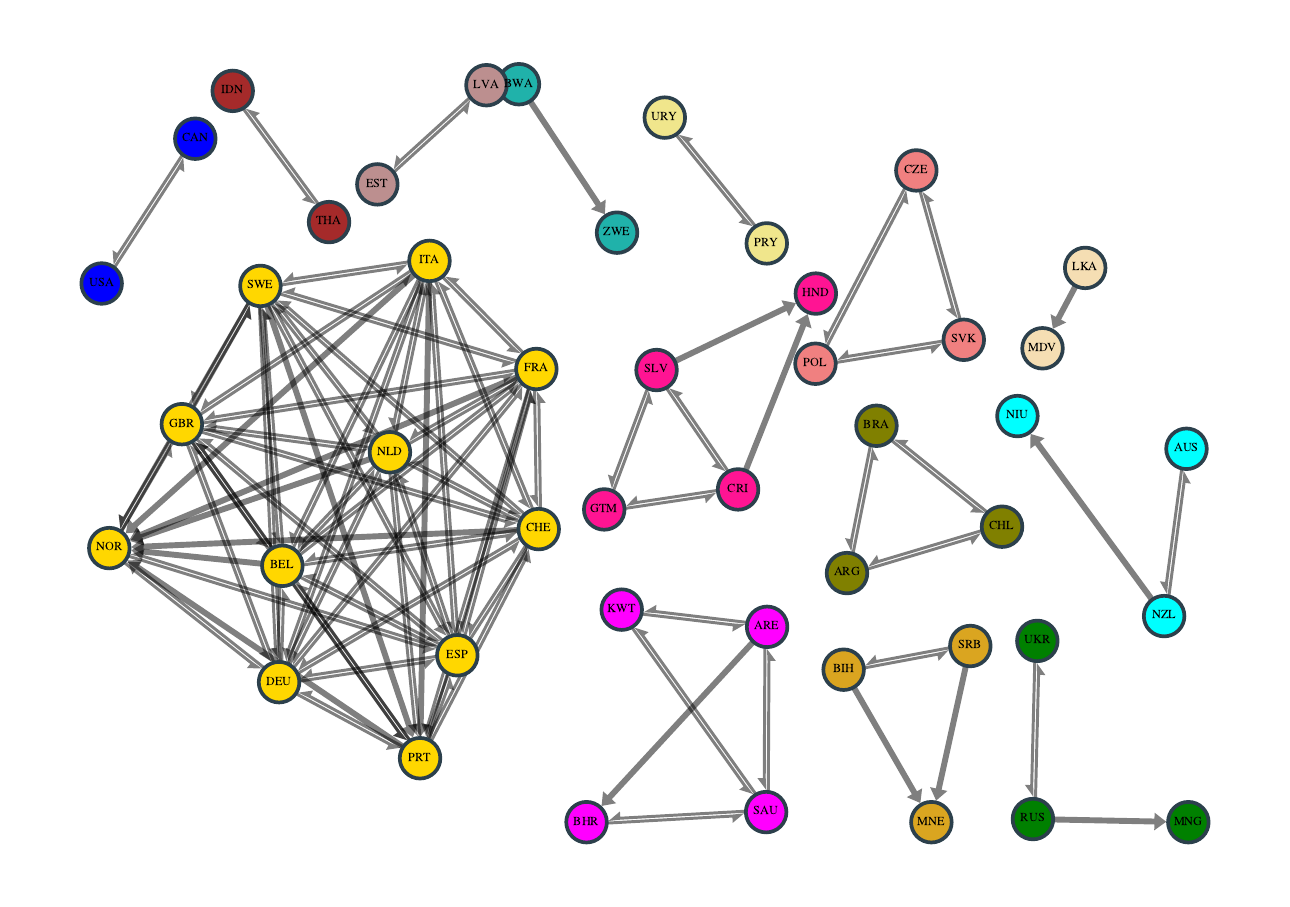}
    \caption{Intrinsic community blocks in the temporal directed iPTNs from 2007 to 2018 of rodenticides and other similar products (380899) in the colored map (left panel) and in colored networks (right panel). The node color of an intrinsic community block is the same in both panels.}
    \label{Fig:iPTN:directed:CommunityBlocks:380899}
\end{figure}

For the temporal iPTN of rodenticides and other similar products (380899), we obtain 16 intrinsic community blocks containing 46 economies in total, including 
$B_{1}$ (United Arab Emirates, Bahrain, Kuwait, Saudi Arabia), 
$B_{2}$ (Argentina, Brazil, Chile), 
$B_{3}$ (Australia, Niue, New Zealand), 
$B_{4}$ (Belgium, Switzerland, Germany, Spain, France, United Kingdom, Italy, Netherlands, Norway, Portugal, Sweden), 
$B_{5}$ (Bosnia and Herz., Montenegro, Serbia), 
$B_{6}$ (Botswana, Zimbabwe), 
$B_{7}$ (Canada, United States), 
$B_{8}$ (Costa Rica, Guatemala, Honduras, El Salvador), 
$B_{9}$ (Czech Rep., Poland, Slovakia), 
$B_{10}$ (Estonia, Latvia), 
$B_{11}$ (Indonesia, Thailand), 
$B_{12}$ (Cambodia, Lao PDR), 
$B_{13}$ (Sri Lanka, Maldives), 
$B_{14}$ (Mongolia, Russia, Ukraine), 
$B_{15}$ (Papua New Guinea, Vanuatu), and 
$B_{16}$ (Paraguay, Uruguay).
Figure~\ref{Fig:iPTN:directed:CommunityBlocks:380899} shows the map and the networks of the intrinsic community blocks. The largest intrinsic community block $B_3$ contains 11 economies in Europe. There are two second largest communities $B_{1}$ and $B_{8}$, each containing 4 economies respectively in Middle East and Western Caribbean. Most economies in Africa and Asia do not belong to any intrinsic community blocks. There are two intrinsic community blocks $B_{12}$ (Cambodia, Lao PDR) and $B_{15}$ (Papua New Guinea, Vanuatu), whose two members do not have direct international trade of rodenticides and other similar products and thus do not appear in the network plot.
 
Comparing the intrinsic community blocks of different pesticides, we find that the largest block is always mainly based in Europe and contains Belgium, Germany, Spain, France, United Kingdom, Italy, Netherlands, and Portugal, the intrinsic community block in North America always contains Canada and the United States, the intrinsic community blocks in South America always contains Argentina and Brazil, and Australia and New Zealand also appear together. In all the iPTNs, the identified intrinsic blocks act as the core of communities.



\section{Conclusion}
\label{S1:Conclusion}

We have investigated the evolving community structure in five categories of iPTNs from 2007 to 2018. We unveil geographic distance plays an important role and the community structures in the undirected and directed iPTNs exhibits regional patterns. These regional patterns are very different for undirected and directed networks and for different categories of pesticide. 
Using the normalized mutual information between successive communities to quantify the stability of community structures in undirected and directed networks, we found that the community structure is stabler in the directed iPTNs. 

We further extracted the intrinsic community blocks for the directed international trade networks of each pesticide category. We found that the largest and stablest intrinsic community block in every pesticide category contains important economies (Belgium, Germany, Spain, France, United Kingdom, Italy, Netherlands, and Portugal) in Europe. Other important and stable intrinsic community blocks are found to be Canada and the United States in North America, Argentina and Brazil in South America, and Australia and New Zealand in Oceania. These findings also indicate the important role the geographic distance plays and imply the importance of pesticide supply and demand complementarity of important adjacent economies in the international trade of pesticides.

The detected community structures and intrinsic community blocks in the iPTNs form the starting point for further research. 
Firstly, it is crucial to uncover influencing factors in the formation of bilateral trades and communities. For instance, the seminal work of Torreggiani et al. found that, besides geographical proximity, trade-agreement
co-membership is another determinant to promote co-presence of economies in communities \cite{Torreggiani-Mangioni-Puma-Fagiolo-2018-EnvironResLett}. 
Secondly, one can use these findings to validate network growth models specific to international pesticide trade, or alternatively to develop network growth models integrating the determinants of community formation. 
Thirdly, the communities and their intrinsic blocks unveil the underlying higher-order interactions among economies and can be used to design attack strategies in the study of network vulnerability. Most previous studies concerned attacks on nodes and/or links and attacks based on higher-order interactions are much less considered. 
Finally, it is interesting to investigate the role of higher-order interactions embedded in the community structure in shock propagation between pesticide and food trade networks, linking pesticide trade patterns and food security.

\section*{Acknowledgements}

This work was partly supported by the National Natural Science Foundation of China (72171083), the Shanghai Outstanding Academic Leaders Plan, and the Fundamental Research Funds for the Central Universities.

\bibliographystyle{ws-acs}
\bibliography{Bib1,Bib2,BibITN,BibRCE}

\end{document}